# Control of Functional Connectivity in Cerebral Cortex by Basal Ganglia Mediated Synchronization

**Daniel Pouzzner**, daniel@pouzzner.name

## Abstract

Since the earliest electroencephalography experiments, large scale oscillations have been observed in the mammalian brain. More recently, episodes of oscillation and bursting have been identified not only in the cerebral cortex and thalamus, but pervasively in the healthy basal ganglia. The *basal ganglia mediated synchronization* model, introduced here, implicates these episodes in the integration of stimulus-response and reinforcement mechanisms in the basal ganglia with cortical association mechanisms. In so doing, the model helps explain how oscillations and synchrony are functionally central, and in particular, how they organize neural activity to exploit the selectivity of coincidence detectors in cortex and beyond. In the core mechanism of the model, salient spatiotemporal activity patterns in cortex are selectively focused by and routed through the basal ganglia to the thalamus. Coherent thalamocortical activity patterns then project back to widely separated areas of cortex, where they establish and facilitate contextually appropriate functional connections, while disconnecting and inhibiting competing ones. Corticostriatal, striatopallidal, and striatonigral conduction delays are crucial to this mechanism. These delays are unusually long, and unusually varied, in arrangements that facilitate learning of useful time alignments and associated resonant frequencies. Other structural arrangements in the basal ganglia show further specialization for this role, with convergence in the inputs from cortex, and divergence in many of the return paths to cortex, that systematically reflect corticocortical anatomical connectivity. The basal ganglia also target the dopaminergic, cholinergic, and serotonergic centers of the brainstem and basal forebrain, and the intralaminar and reticular nuclei of the thalamus, structures broadly implicated in the modulation of network activity and expression of plasticity. By learning to coordinate these various output channels, the basal ganglia are positioned to facilitate and synchronize activity in selected areas of cortex, broadly impart selective receptivity, attenuate and antisynchronize interfering activity, and recurrently process the resulting patterns of activity, channeling cognition and promoting goal fulfillment. This system is the most versatile and flexible example of a repeating architectural motif, in which subcortical and allocortical structures influence functional connectivity in neocortex using spike-timing-dependent gain and neuromodulation. Dysfunctions in the components of these highly distributed systems are associated with syndromes of perception, cognition, and behavior, notably the schizophrenias, some or all of which might fundamentally be disruptions of subcortically mediated neocortical synchronization.

# 1. Introduction and Overview

**In this section:**

### *1.1. The basal ganglia are fundamental to cortical coordination.*

The cerebral cortex has long been styled the seat of higher thought, due to its size and disproportionate growth in mammalian phylogeny (Mountcastle 1998), and its astronomically large dimensionality (Tononi 2004). This size and dimensionality necessitate an exquisitely powerful coordination mechanism.

Here, I present and explore a hypothesis that phase-coherent signal conduction by the basal ganglia (BG) is fundamental to that mechanism. I term this model *basal ganglia mediated synchronization* (BGMS). In the core mechanism of the model, patterns of activity in cortex are selected by the basal ganglia, dimensionally reduced, and routed to the thalamus with delays and targeting that are specific to each learned pattern. These signals are then circulated back through thalamocortical projections to widely separated areas of cortex, synchronizing the targeted areas through superficial facilitation and spike-timing-dependent gain, establishing and sustaining functional connections between them. The BG thus act as essential organizers of cortical activity.

### *1.2. The basal ganglia resolve conflicts and ambiguities with selections informed by goals, context, and expectation.*

A particularly durable account of BG function is that they serve as a selection mechanism, resolving conflicting or ambiguous claims on computational and behavioral resources (Redgrave *et al.* 1999; Mink 1996; Graybiel 1998; Stephenson-Jones *et al.* 2011; Hikosaka *et al.* 2000). Similarly, the BG have been modeled as controllers of gates in cortex, selectively facilitating motor output (Chevalier and Deniau 1990; Hikosaka *et al.* 2000) and establishing contextually appropriate items in working memory (Frank *et al.* 2001; O'Reilly and Frank 2006).

At a more fundamental level, the BG are thought to develop a repertoire of compound stimulus-response relations through reinforcement learning, ultimately forming habits (Graybiel 1998, 2008; Markowitz *et al.* 2018). In this view, the BG transform cortical and subcortical inputs representing bodily, environmental, and cognitive state, elaborately contextualized by other cortical and subcortical inputs representing goals and history, into spatiotemporally complex, precise, widely distributed, often sequential adjustments to brain state, that are expected to promote internal and external (environmental) changes in furtherance of those goals. It has been previously suggested that a fundamental facility of the BG for precisely and flexibly triggered, structured, and directed neurodynamic gestures has far-reaching consequences (Graybiel *et al.* 1994; Graybiel 1997). This facility is at the heart of the proposal advanced here, because—as briefly reviewed below—effective connectivity among the targets of the BG is strongly associated with the precise timing relationships of the activity within them.

### *1.3. Population spike time relations are a pervasive mechanism for selective effective connectivity.*

According to this proposal, the BG and thalamus establish and reinforce effective connections in cortex by distributing precisely timed spike volleys to its feedback-recipient layers, imparting discriminative receptivity by spike-timing-dependent gain control. Cortical microcircuits preferentially respond to coincident spikes (Pouille and Scanziani 2001; Williams and Stuart 2002), providing the basic building blocks for spike-timing-dependent gain control. The proposition that spike synchronies are correlates of functional and effective connectivity, and represent associations, is supported by an array of evidence and integrative theory (von der Malsburg 1981, 1999; Bastos *et al.* 2015b; Bressler 1995; Damasio 1989; Fries 2005, 2015; Friston 2011; Hutchison *et al.* 2013; Lakatos *et al.* 2008, 2019; Kopell 2000; Maris *et al.* 2016; Meyer and Damasio 2009; Salinas and Sejnowski 2001; Siegel *et al.* 2012; Singer 1993, 1999; Singer and Gray 1995; Varela *et al.* 2001; Voloh and Womelsdorf 2016; Wang 2010). In all, temporal codes and mechanisms in the brain are remarkably varied and pervasive (Cariani and Baker 2022; Baker and Cariani 2025; Cariani and Baker 2025; Keitel *et al.* 2025‡).

Synchronous spiking activity normally involves brief oscillatory episodes, and delta (~1-4 Hz), theta (~4-8 Hz), alpha (~8-14 Hz), beta (~14-30 Hz), and gamma (~30-120 Hz) oscillations are particularly prominent (Wang 2010; Keitel *et al.* 2025‡). Measurable signals reflecting these oscillations have been known for nearly a century (Berger 1929; Jasper 1937; Buzsáki *et al.* 2003, 2012; Olejniczak 2006), and local field potentials (LFPs) in particular can serve as proxies for massed neuronal membrane voltage fluctuations and synaptic currents (Haider *et al.* 2016). Networks of loosely coupled neurons, as in cerebral cortex, readily support synchronized oscillation, and these oscillations are thought to be fundamental to the transmission and integration of information in the vertebrate brain (Fields 2025). Loose coupling also characterizes cortical inputs to striatal projection neurons, which exhibit a preference (supralinear response) for widely distributed rhythmic synchrony, while mechanistically disfavoring (sublinear summation) rhythmic stimulation on a single input not synchronized with other inputs (Carter *et al.* 2007; Zheng and Wilson 2002).

Activity-driven plasticity mechanisms are crucially dependent on relationships of precise temporal coincidence among individual spikes and spike bursts (Song *et al.* 2000). These mechanisms can construct axon populations with precisely matched propagation delays (Gerstner *et al.* 1996), and indeed, synchronous spike volleys are thought to propagate coherently through chains of many directly linked neurons (Diesmann *et al.* 1999). In behaving animals, a large variety of stable and behavior-related spatiotemporal patterns of neural activity is observed, with spike jitter of only 1-3 ms (Abeles *et al.* 1993). Precision spike time correlations have been shown to relate particularly to changes in expectation, attention, response latency, and rivalry, in scenarios wherein average firing rates are largely invariant, leading to the suggestion that spike timing dependent mechanisms underlie the selective routing of information, and the alignment of plasticity with attentional orientation (Salinas and Sejnowski 2001). Supporting this proposition, functional control by timing modulations, without prominent rate modulations, has been demonstrated in cortex (Riehle *et al.* 1997; Hatsopoulos *et al.* 1998; Quintana *et al.* 2024‡), thalamus (Eradath *et al.* 2021), the basal ganglia (Wang *et al.* 2021; Fischer *et al.* 2020; Fischer 2021; Holt *et al.* 2019; Gittis *et al.* 2011; Maltese *et al.* 2021), and in dopaminergic modulation of cortex by the basal ganglia (Costa *et al.* 2006).

Neuronal oscillatory periods are intrinsically unstable; thus mutual incoherence is the pervasive and normal condition, except where functional connectivity mechanisms overcome it and establish synchronies (Fries 2005). Even a single modulatory volley can establish synchrony, and it can be sustained by a series of such volleys (Lakatos *et al.* 2008, 2019). Mathematical modeling suggests that synchrony, once established, stabilizes systems against disruption by noise, facilitating information transfer (Tabareau *et al.* 2010). Modeling also suggests that, with population codes that leverage spike synchrony effects to gate information flow, the actual discriminative power of downstream receivers can be enhanced by time correlations in their inputs, despite the marginal loss of channel coding capacity inherent to such correlations (Ibáñez-Berganza *et al.* 2025‡). In principle, coherence coding is intrinsically more robust than rate coding: a neuronal module cannot *a priori* generate oscillations at the precise frequency and phase preferred by a particular receiving area. Rather, a transmitting module must be coordinated through some mechanism with the instantaneous values of dynamic parameters of the receiving module. In contrast, the representation of salience by rate coding *per se* requires no such dynamic system-level information.

Coherence-based access control is particularly apt for generalized cognitive resources, which are subject to "greedy" access strategies by more specialized neuronal modules (van den Heuvel *et al.* 2012). And because spike generation is energetically costly in itself (Moujahid *et al.* 2014), timing-based codes have inherent metabolic advantages over rate codes. With coherence-based access control, a sensible input pattern can be characterized by elaborate and precise spatiotemporal structure that functions like a key, gaining entry to a discriminating target, while senseless patterns are ignored. This arrangement is suggested by the "tidal wave" theory of Braitenberg *et al.* (1997), the "minimal coherence detection" model of Plenz and Aertsen (1994), the "rank order coding" model of Thorpe *et al.* (2001), the spontaneous synchronization avalanche model of Schünemann and Ernst (2023‡), and crucially, by the "polychronous groups" described by Izhikevich (2006), implying an astronomically larger coding capacity than is possible without a coherent neural code. It entails plasticity specific to the relative fine timing of spikes from whole constellations of presynaptic sources, and promotes immunity to noise or dysfunction that could otherwise cause false perceptions, faulty inferences, inapt memory activations, and errant actions.

### 1.4. *Synchronies are crucial in perception, cognition, behavior, and pathology.*

Synchronies are pivotal in perceptual processing. For example, the relationships of oscillatory frequency and phase in interconnected sensory areas, measured by LFPs, are strongly associated with the effective connectivity of those areas (Womelsdorf *et al.* 2007), and the resolution of competitions among sensory inputs can be predicted from the relationship of the LFP frequency and phase within each input to those prevailing within their common target (Fries *et al.* 1997, 2002). In optogenetically manipulated mice, temporal scrambling of spike times in primary visual cortex on the timescale of milliseconds, preserving stimulus-induced spike rates on the timescale of seconds, substantially reduces perceptual performance (Quintana *et al.* 2024 ‡ ). Similarly, oscillatory coherency is a better indicator of sensory surprise than is aggregate firing rate (Sennesh *et al.* 2025 ‡ ). Attentional orientation is accompanied by LFP synchrony between frontal and posterior cortex: strong beta oscillation initiated in frontal cortex is associated with top-down orientation, and strong gamma oscillation initiated in posterior cortex is associated with bottom-up orientation (Buschman and Miller 2007; Bastos *et al.* 2015a), with precise and consistent inter-areal oscillatory phase relations associated directly with task performance (Parto-Dezfouli *et al.* 2023). Top-down beta can enhance bottom-up gamma through cross-frequency interactions, and this phenomenon is suggested to be a fundamental mechanism of attentional focus (Lee *et al.* 2013; Richter *et al.* 2017) and of top-down control in general (Bressler and Richter 2015). Inattention, too, is associated with long range synchronies: alpha and beta synchronization of PFC with somatosensory cortex can be associated with suppression of distracters (Sacchet *et al.* 2015).

Synchronies are also pivotal in the generation of behavior. Long range synchronies can be strongly predictive of behavioral decisions (Verhoef *et al.* 2011; Fiebelkorn and Kastner 2021), and long-range synchronies at opposed phases, with negligible firing rate changes in premotor cortex, are proposed to underlie disconnection and consequent suppression of overt behavior during preparation (Stetson and Andersen 2014). Planning and execution of voluntary movements are associated with characteristic synchronization of activity in shifting ensembles of neurons in primary motor cortex, separate from changes in their firing rates (Riehle *et al.* 1997). Indeed, some of the gesture selectivity of activity in primate motor neurons is apparent only in their synchronies (Hatsopoulos *et al.* 1998). And a recent study in humans (Fischer *et al.* 2020) suggests a similar primacy of spike synchrony in the combined cortico-BG dynamics underlying behavior.

The primacy of timing is also apparent in simpler animals. In the hawkmoth, it has been shown that muscles are coordinated with each other almost entirely by millisecond-scale spike timing relationships, with spike timing pervasively encoding 3 times as much information about behavior as does spike rate (Putney *et al.* 2019; Sponberg and Daniel 2012). Similarly, in *Drosophila melanogaster*, action selections can pivot on spike timing relationships (von Reyn *et al.* 2014).

In humans, the large scale architecture of neural synchronies has clear developmental correlates. Childhood improvements in cognitive performance are accompanied by increases in neural synchrony, while adolescence is accompanied by a temporary reduction in performance and synchrony, followed by oscillatory reorganization and still higher performance and synchrony in adulthood (Uhlhaas *et al.* 2009). Adolescence and young adulthood in humans are marked by uniquely prolonged episodes of myelination, particularly in prefrontal cortex (Miller *et al.* 2012), while cognitive decline associated with senescence in humans is marked by frontoposterior white matter deterioration, and concomitant deficits in the modulatory control of frontoposterior oscillatory synchrony (Hinault *et al.* 2020).

The functional prominence of temporal precision is suggested by a finding that temporal acuity and psychometric *g* (a measure of general cognitive performance) covary, with *g* predicted significantly better by acuity than by reaction time (Rammsayer and Brandler 2007). Similarly, uniformity of cadence in successive gestures within a self-paced rhythm task correlates significantly with performance on a test of general intelligence (Madison *et al.* 2009). In rats, manipulation of striatal processing speed reveals an inverted-U relationship with behavioral proficiency, with natural conditions yielding best performance (Monteiro *et al.* 2023).

Characteristic synchronal abnormalities are associated with diseases such as schizophrenia (Sz), autism, Alzheimer's, and Parkinson's (Uhlhaas and Singer 2006, 2012; Hammond *et al.* 2007), and with anesthetic loss of consciousness (Bardon *et al.* 2025). The reorganization of synchronal architecture in adolescence may be a trigger for the onset of Sz in those at risk (Uhlhaas and Singer 2010; Uhlhaas 2013); Sz is associated with pervasive physiological disruptions of the mechanisms underlying the generation and regulation of, and responses to, spike synchronies and functional connectivity (Friston 1995; Uhlhaas 2013; Pittman-Polletta *et al.* 2015), and multifariously implicates the BG (Robbins 1990; Graybiel 1997; Simpson *et al.* 2010; Wang *et al.* 2015; Grace 2016; Dandash *et al.* 2017; Mamah *et al.* 2007). Abnormal judgment in Sz of time intervals and sensory simultaneity (Martin *et al.* 2013; Schmidt *et al.* 2011; Ciullo *et al.* 2016), and highly significant motor deficits in Sz on tasks as simple as rapidly alternating finger taps (Silver *et al.* 2003), are further evidence of common timing-related mechanisms underlying sensory, motor, and cognitive processing. These abnormalities are likely to implicate the BG directly: for example, evidence suggests that phase-aligned synchronization of the BG with oscillatory activity in cortex is integral to precise judgments of time intervals (Gu *et al.* 2018), and that manipulation of striatal processing speed monotonically affects interval judgment (Monteiro *et al.* 2023).

### *1.5. The thalamus is in an ideal position to control large scale cortical synchronies.*

Much of the large scale oscillatory activity in cortex is not purely intrinsic, and directly implicates subcortical structures, particularly the thalamus. The thalamus is a major target of BG output (Haber and Calzavara 2009), and the proposition that the BG have a prominent role in controlling long range cortical synchronies follows in part from evidence that the thalamus performs this function.

Thalamic control of cortical oscillation and synchronies follows naturally from the developmental relationship of thalamus to cortex. While a ballet of cortically intrinsic developmental processes parcels the cortex into its major cytoarchitectonic areas (Rakic 1988; Wang 2020), thalamocortical axons reach their pallial destinations before neurogenesis and migration of the receiving cortical neurons (López-Bendito and Molnár 2003; Paredes *et al.* 2016), and the architecture of cortical circuitry is thought to develop partly in response to patterns of activity (Katz and Shatz 1996) and modular connectivity (Murakami *et al.* 2022) in these axons. “Developmental exuberance”, entailing the robust proliferation of ephemeral long range corticocortical links, is followed by a postnatal paring process driven in part by early patterns of thalamocortical activity (Innocenti and Price 2005; Price *et al.* 2006). Indeed, the manipulation of thalamocortical input patterns can dramatically alter cortical physiology and function (Rakic 1988). For example, uniquely visual attributes can be induced in cortical areas that normally subserve audition by rerouting retinal inputs to the thalamic auditory nuclei (Sharma *et al.* 2000).

These roles establishing the anatomical connectivity and intrinsic function of cortex position the thalamus uniquely to regulate cortical functional connectivity.

The thalamus is also uniquely positioned anatomically, at the base of the forebrain on the midline. This is an optimal situation for distributing synchronized spike volleys to far-flung loci in cortex, notwithstanding thalamocortical distance disparities due to sulci and gyri. It is striking that postnatally (week 4 in mice), the thalamocortical projection to a given functional area of cortex develops a uniform delay, in many areas less than 1 ms of maximum disparity, despite widely varying axon lengths; even intermodally, thalamocortical delays are often aligned within 2-3 ms (Salami *et al.* 2003; Steriade 1995). The central clustering of thalamic nuclei is noteworthy in itself: absent functional requirements and associated evolutionary pressures to the contrary, many of these nuclei might migrate toward the cortical areas with which they are intimate, realizing physiological efficiencies (Scannell 1999). Moreover, in many mammals the dorsal BG maintain rough radial symmetries centered on the thalamus, suggesting time alignment pressures like those that appear to influence the gross anatomy of the thalamus.

### *1.6. The thalamus can control cortical oscillation and corticocortical synchronies.*

It has been shown clearly that the thalamus can control cortical oscillatory activity (Poulet *et al.* 2012; Lyu *et al.* 2025), and that it can orchestrate lag-free (zero phase shift) long range synchronies in cortex (Ribary *et al.* 1991; Vicente *et al.* 2008; Saalmann *et al.* 2012). Modulatory thalamocortical signals associated with attended stimuli can reset the phase of ongoing oscillations (Lakatos *et al.* 2008, 2007, 2009). “Desynchronization” associated with mental activity in fact consists of focal, high-frequency (20-60 Hz) synchronization of distributed thalamocortical ensembles (Steriade *et al.* 1996). Long distance, multifocal (posterior visual, parietal, and frontal motor), lag-free synchronies in the beta band have been observed in association with visuomotor integration (Roelfsema *et al.* 1997), and similar lag-free beta synchronies, and precise antisynchronies, have been observed among loci in prefrontal and posterior parietal cortex in visual working memory (Dotson *et al.* 2014; Salazar *et al.* 2012).

Indeed, zero-lag synchronies and antisynchronies are pervasive in cortex, particularly between areas with strong structural connections (O’Reilly and Elsabbagh 2021; Mehra *et al.* 2025‡). Simultaneous high-amplitude fluctuations in structurally connected cortical areas are strongly associated with functional connectivity of those areas (Zamani Esfahlani *et al.* 2020; Betzel *et al.* 2022), and the thalamus is positioned to coordinate much of this activity. Phase-locked rhythmic artificial stimulation of widely separated loci in frontal and parietal cortex can significantly improve working memory performance, strongly suggesting that extrinsic synchronous influences on cortical activity—as by the thalamus—meaningfully affect cortical network configuration and associated mental faculties (Violante *et al.* 2017; Alagapan *et al.* 2019).

Projections from single thalamic nuclei to widely separated but directly interconnected cortical areas have been noted (Goldman-Rakic 1988; Saalmann *et al.* 2012), and there is evidence that intralaminar thalamocortical projections systematically reflect corticocortical connectivity, with individual axons branching multi-areally (Kaufman and Rosenquist 1985a; Van der Werf *et al.* 2002). The hypothesis has been advanced that midline and intralaminar thalamic nuclei in particular are the hub of a system to control cortical synchronies and associated effective connectivity (Saalmann 2014; Purpura and Schiff 1997), and the entire population of calbindin-positive neurons in the thalamus (Jones 2001), or indeed the thalamus as a whole (Halassa and Kastner 2017), has been proposed to function in this fashion. The control of functional connectivity in cortex is thought to be mediated particularly by PFC- and BG-connected thalamic nuclei (Phillips *et al.* 2021); at a higher level, the state and contents of consciousness are thought to pivot systematically on activity in distinct thalamic neuron populations (Whyte *et al.* 2024).

Evidence supports these propositions. The mediodorsal nucleus is positioned to gate afferent inputs to PFC through direct connections to cortical interneurons (Delevich *et al.* 2015; Kuroda *et al.* 1998; Cruikshank *et al.* 2012), and has been shown to control sustained and dynamic functional connectivity in PFC (Schmitt *et al.* 2017; Nakajima and Halassa 2017; Mofakham *et al.* 2022) and shifts in PFC representations of rule context (Rikhye *et al.* 2018). Similarly, the pulvinar can synchronize oscillations, establishing functional connectivity associated with attentional engagement, between areas V4 and TEO in visual cortex (Saalmann *et al.* 2012), between V4 and the associative visual cortex in the lateral intraparietal area (LIP) (Saalmann *et al.* 2018‡), and between LIP and the frontal eye field (FEF) (Fiebelkorn *et al.* 2019). This influence of the pulvinar on cortical activity can be confined to modulations of aggregate inter-areal phase coherence, with no significant effects on local firing rates or local aggregate oscillatory power (Eradath *et al.* 2021). Moreover, its influence can be causal (Huang *et al.* 2024).

Evidence suggests that the associative visual thalamus integrates feedforward visual information with representations of behavioral context, particularly from the superior colliculus, so that associative visual thalamocortical pathways bear information that is qualitatively different from that borne by feedforward corticocortical pathways (Blot *et al.* 2021). This is similar to the distinctions found for transthalamic inputs to somatosensory association cortex (Mo *et al.* 2024). Because the superior colliculus is a major target of the BG (Hikosaka and Wurtz 1983), this implies that the BG are involved in contextual contingencies of the associative visual thalamus. These arrangements have parallels in the motor thalamus: corticocortical projections from posterior parietal cortex to premotor cortex carry signals directly associated with movement control, while projections to the striatum from the same parietal area (implicating the motor thalamus) are distinct, reflecting behavioral context (Hwang *et al.* 2019).

### ***1.7. The basal ganglia form loops with cortex that reflect cortical patterns of connectivity and parallelism.***

It has long been appreciated that the cortex, striatum, pallidum/substantia nigra, and thalamus are arranged in loops placing each under the influence of the others (Alexander *et al.* 1986; Parent and Hazrati 1995a; Middleton and Strick 2000). As reviewed in detail later (§7.3), the pyramidal neurons of cortical layer 5 (L5) originate the primary input to the BG “direct path” centrally implicated in these loops, and are among the recipients of the output from the direct path via the thalamus. While subdivision of these loops into parallel circuits and constituent channels has been noted (Alexander *et al.* 1986, 1991), *in toto* the pathways of the BG exhibit remarkably varied patterns of convergence, divergence, and reconfiguration (Joel and Weiner 1994; Zheng and Wilson 2002; Parent and Parent 2006; Hintiryan *et al.* 2016; Korponay *et al.* 2022‡).

Diffuse projection fields from wide areas of cortex exhibit high convergence-divergence, and are thought to supply extensive context throughout the striatum (Calzavara *et al.* 2007; Mailly *et al.* 2013). Projections from interconnected cortical regions, including reciprocally interconnected pairs of individual neurons, systematically converge and interdigitate in the striatum (Yeterian and Van Hoesen 1978; Van Hoesen *et al.* 1981; Selemon and Goldman-Rakic 1985; Parthasarathy *et al.* 1992; Flaherty and Graybiel 1994; Averbeck *et al.* 2014; Lei *et al.* 2004; Morishima and Kawaguchi 2006; Hintiryan *et al.* 2016; Hooks *et al.* 2018), and projections from interconnected areas have been shown to converge on individual striatal fast spiking interneurons (FSIs) (Ramanathan *et al.* 2002). Meanwhile, cortical areas that are not directly connected with each other show little similarity in their projections to the striatum (Yeterian and Van Hoesen 1978). These arrangements show that the BG are particularly concerned with corticocortical connectivity. Even before much of this evidence was uncovered, Mesulam (1990) suggested that arrangements of convergence and interdigitation in the corticostriatal projection position the striatum to integrate, compare, or synchronize neural computations in distant areas of cortex.

By having a sharp view of afferents from directly interconnected areas, simultaneous with a diffuse view of more widespread cortical activity, a striatal neighborhood is supplied with information upon which appropriate cortical activation and corticocortical connectivity decisions can be made as a function of present activity and connectivity, with particular expertise for the functional domains implicated by those focal afferents. And given closed-loop circuitry, a striatal neighborhood convergently innervated by multiple cortical areas can impart oscillation from one of them to the others, with particular significance for directly interconnected areas, and areas linked via a common connectivity hub. Nonetheless, partial segregation of channels through the BG likely facilitates parallel processing of operations that require only partial coordination, with the degrees and directions of segregation tending to reflect the degrees and directions of non-interference and independence.

Parallelism in the BG provides for the simultaneous processing in the striatum of activity at multiple oscillatory frequencies in distinct regions, associated with distinct domains of skill acquisition and performance, with distinct expressions of plasticity in each region, and inter-regional coherence varying task-dependently (Thorn and Graybiel 2014). In cortex, too, evidence suggests that distributed functional networks are largely parallel, and entail interdigitation in circuit nodes, particularly in prefrontal and other associative areas (Goldman-Rakic 1988; Yeo *et al.* 2011; Livingstone and Hubel 1988), even while most areas have direct anatomical connections with each other (Markov *et al.* 2014). fMRI of spontaneous activity in resting humans has demonstrated corresponding integration, regionalization, and parallelism of cortico-BG networks (Di Martino *et al.* 2008).

### ***1.8. The basal ganglia are arranged to regulate cortical activity and functional connectivity in large scale networks.***

Pathways through the basal ganglia exhibit an unusually broad range of conduction delays (Yoshida *et al.* 1993; Kitano *et al.* 1998). This diversity of delays plays a central role in the model introduced here, allowing the BG to meet disparate timing requirements at each stage of the BG-thalamocortical loop, and to tune the preferred frequencies and phases of oscillation at network loci implicated by a selection. Diversity of delays in the corticostriatal projection, the massively convergent-divergent topology noted above, and an enormously diverse population of interneurons, position the striatum to distinguish polychronous groups with exquisite nuance. In the oscillatory regime, as detailed later (§5.9), closed loops through the dorsal striatum and globus pallidus have an average transmission delay corresponding to 40 Hz gamma oscillation, and loops through the dorsal striatum and substantia nigra have an average delay corresponding to 20 Hz beta oscillation, with wide delay ranges among the fiber populations of either path. And as further detailed later (§5.5), paths through the putamen and globus pallidus *pars externa* (GPe) have particularly short delays, positioning them to entrain their targets to phases nearly opposite those imparted by the generally slower direct path, both for gamma targeting the GP *pars interna* (GPi), and beta targeting the substantia nigra *pars reticulata* (SNr).

Prominent oscillatory episodes in the BG are associated with perception, attention, decision making, and working memory (Cannon *et al.* 2014), all of which implicate large scale brain networks. In humans, the coherence of oscillatory activity in the BG, and the relationship of its phase to that in cortex, have been shown in some scenarios to be more predictive of movement than are BG mean firing rates (Fischer *et al.* 2020). Sensory cues are associated with rapid oscillatory phase resets and brief episodes of beta oscillation spanning the BG (Leventhal *et al.* 2012), and mechanisms intrinsic to the BG are posited to underlie the generation of some of these brief oscillatory episodes (Mirzaei *et al.* 2017). Closed-loop networks internal to the basal ganglia are intrinsically capable of generating and sustaining oscillations in the beta range (Bevan 2002; Tachibana *et al.* 2011; Mirzaei *et al.* 2017), and fast spiking interneurons in the striatum exhibit oscillatory tendencies in the theta, beta, and gamma bands, systematically influenced by dopamine (Berke 2009; van der Meer *et al.* 2010; Berke 2011; Chartove *et al.* 2020). These mechanisms might position the BG to generate contextually appropriate oscillatory responses to non-oscillatory inputs, and to tune input frequency preferences state-dependently.

The BG are among the most connected regions of the brain (van den Heuvel and Sporns 2011; McElvain *et al.* 2021), and are densely integrated with cortical hubs (Middleton and Strick 2002; Vatansever *et al.* 2016; Averbeck *et al.* 2014; Schulte *et al.* 2023‡). They participate in a particularly wide variety of large scale synchronized networks, with greater oscillatory specificity than cortical areas (Keitel and Gross 2016), suggesting primary oscillatory selection and generation. Densely BG-recipient association nuclei of the thalamus, such as the mediodorsal and central lateral nuclei, and midbrain BG nuclei (particularly the substantia nigra (SN) and ventral tegmental area (VTA)), have anatomical connectivity that positions them as bridges between major networks such as the default mode network (DMN), ventral attention network (VAN, also known as the salience network), and frontoparietal control network (FPC) (Van der Werf *et al.* 2002; Li *et al.* 2021; Delevich *et al.* 2015; Ray and Price 1993; McElvain *et al.* 2021; Root *et al.* 2015; Aguilar and McNally 2022; Peters *et al.* 2016; Menon 2011; Uddin *et al.* 2019). Parkinson's disease, in which the dopamine cells of the SN are lost, is associated with widespread disruption of these networks. Functional imaging shows a graded association between disease severity and reduced functional coupling between the striatum and salience network, reduced coupling between the salience network and FPC, and elevated coupling between the salience network and DMN (Aracil-Bolaños *et al.* 2019). Structural imaging shows that salience network regions are anatomically intact in Parkinson's disease even while functionally disrupted (Putcha *et al.* 2015), consistent with a model in which BG inputs to the salience network are necessary for normal function.

BG influence on cortical activity is extensive, and can be strong. Cortical oscillatory dynamics, stability, and propensity for synchrony, are profoundly and specifically modulated by central supplies of dopamine, acetylcholine, and serotonin, all of which are integral to BG circuitry (McElvain *et al.* 2021; Yetnikoff *et al.* 2014; Fallon 1988; Ioanas *et al.* 2022; Saunders *et al.* 2018; Yang and Seamans 1996; Towers and Hestrin 2008; Costa *et al.* 2006; Benchenane *et al.* 2010; Mesulam and Mufson 1984; Grove *et al.* 1986; Haber *et al.* 1990; Haber 1987; Sillito and Kemp 1983; Rodriguez *et al.* 2004; Muñoz and Rudy 2014; Howe *et al.* 2017; Baumgarten and Grozdanovic 2000; Neuman and Zebrowska 1992; Gervasoni *et al.* 2000; Carter *et al.* 2005). The substantia nigra *pars reticulata* (SNr) by itself targets not only large portions of the thalamus, but also much of the brainstem reticular formation and the superior and inferior colliculi, with systematic physiological distinctions in the cell populations projecting to different classes of targets (McElvain *et al.* 2021). And the SNr is just one of many BG output structures.

Artificial stimulation of the striatum affects activity spanning the entire cerebral cortex (Lee *et al.* 2016), and rhythmic photostimulation of optogenetically manipulated BG output structures can produce synchronous spike volleys in the motor thalamus and motor cortex (Kim *et al.* 2017). BG input to the thalamus, affecting the temporal structure of activity there rather than its intensity, has been shown to be crucial for pallial burst firing in songbirds (Kojima *et al.* 2013). In monkeys, task-related oscillatory activity in the BG correlates strongly with oscillation in the implicated areas of thalamus (Schwab 2016, chapter 5), and BG oscillations like these appear to induce phase-locked oscillation in frontal cortical areas (Antzoulatos and Miller 2014; Williams *et al.* 2002). There is even evidence that volitional control of cortical oscillation centrally implicates the striatum (Kasahara *et al.* 2022).

The widespread influence of the BG on neocortical activity suggests a general role, integral to the normal operation of cell assemblies. Indeed, normal striatal activity is necessary for the development of normal excitatory-inhibitory balance in cortical microcircuits (Deemyad *et al.* 2024‡). In the normal mature neocortex, this balance is tuned to criticality (Haider *et al.* 2006; Ma *et al.* 2019; Ahmadian and Miller 2021), optimizing sensitivity to inputs, and optimizing dynamic control by those inputs of functional connections and disconnections (van Vreeswijk and Sompolinsky 1996; Vogels and Abbott 2009; Ahmadian and Miller 2021; Finlinson *et al.* 2020; Li *et al.* 2019). Evidence and simulations suggest that even small task-related fluctuations in functional connectivity can have decisive implications for large scale network organization, task-evoked activations, and associated behavior (Cole *et al.* 2021). These arrangements have particular implications for the influence of the BG on the neocortex (Djurfeldt *et al.* 2001; Shine 2021), not least if (as suggested by biophysical simulations) the BG-recipient thalamus is crucial for cortical criticality (Müller *et al.* 2023).

Striatal activity continuously tiles task spaces to follow context (Weglage *et al.* 2021; Arcizet and Krauzlis 2018; Markowitz *et al.* 2018), much as PFC activity does (Schmitt *et al.* 2017).

In the model introduced here, the BG tune the delays of circuits within the cortico-BG-thalamocortical loop (the direct path), so that spike volleys associated with selected signals return to cortex, delayed by one or more cycles, precisely coincident in time and space with subsequent corticocortical spike volleys associated with selected temporally structured signals. Circuits through the indirect path entail shorter delays, such that targeted areas (notably, the thalamic reticular nucleus, and through it, most of the thalamus) are selectively modulated to roughly opposed phase, functionally disconnecting cortex from distracting thalamocortical inputs. Cortical pyramidal cells and cell assemblies are arranged to respond selectively when their thalamocortical and corticocortical inputs are time-coincident (Larkum *et al.* 1999, 2004; Llinás *et al.* 2002; Pouille and Scanziani 2001; Williams and Stuart 2002; Volgushev *et al.* 1998), which arranges for sharp selectivity as a function of alignments in time. And indeed, habit learning is associated with the gradual emergence of widespread task-related spike synchronies and sharpened responses in the striatum (Barnes *et al.* 2005; Howe *et al.* 2011; Desrochers *et al.* 2015).

Sensitivity to widespread synchronies is also intrinsic to striatal physiology (Zheng and Wilson 2002; Carter *et al.* 2007), and subcortical projections from the BG-recipient thalamus to the striatum (Sidibé *et al.* 2002; Smith *et al.* 2004; McFarland and Haber 2000; Mandelbaum *et al.* 2019; Lemke *et al.* 2021) suggest that dynamics and plasticity in the BG are driven in part by the synchronies present at their output. If the coherent relay of oscillatory signals with precise selectivity for time relations is a key function of the BG, as proposed here, then these arrangements are central to that facility: a contextually contingent, sharp, coordinated, widely distributed striatal spike volley could then evoke or reinforce widely distributed synchronous rhythmic spiking, and consequent functional connectivity, in downstream structures.

Excitatory input from thalamus to the L1 (distal apical) dendrites of L5 pyramidal neurons, synchronous with somatic layer inputs to those neurons, promotes burst firing in those cortical cells (Larkum *et al.* 1999, 2004; Larkum 2013). Cortical burst firing, in turn, establishes long range synchronies (Womelsdorf *et al.* 2014).

These sorts of burst-induced transient long range synchronies are thought to provide for flexible information routing (Palmigiano *et al.* 2017; Besserve *et al.* 2015). Dysfunctions of this mechanism, resulting in spurious routing of information, plausibly underlie the delusions and hallucinations associated with psychedelic drugs and schizophrenia (Carter *et al.* 2005; González-Maeso *et al.* 2007; Preller *et al.* 2018; Geyer and Vollenweider 2008; Ji *et al.* 2019; Giraldo-Chica *et al.* 2018), and plausibly implicate the BG (Walpola *et al.* 2020; ffytche *et al.* 2017; Schmack *et al.* 2021). Autism has been associated with cortical dysconnectivity (Belmonte *et al.* 2004), and particularly with functional hyperconnectivity in children, with a graded relationship with symptom severity (Supekar *et al.* 2013). The matrix compartment of the striatum is expanded in autism, dramatically so in profoundly autistic individuals (Waugh *et al.* 2025). This co-occurence suggests an intimacy of the striatal matrix with functional connectivity in cortex, with abnormal elevation of the latter prompting development of abnormal proportions in the former.

van Schouwenburg *et al.* (2010b) showed with fMRI data that, in tasks with shifting stimulus-response contingencies, the human BG establish appropriate functional connections between prefrontal and posterior visual cortex. Similarly, Nestor *et al.* (2024‡) used fMRI to show that basal ganglia influence precedes large scale cortical network reconfiguration, shifting from modularity to integration. Using MEG data, Portoles *et al.* (2022) showed that control of functional connectivity in cortex is consistent with impulsive reorganization at successive task stage boundaries. They focused on long range synchronies in the theta band, with directionality from the phase-leading to the phase-trailing area. In their view, which they support with simulations, the impulses that prompt the reorganization originate in the BG and propagate via the thalamus.

### *1.9. The physiology of the basal ganglia, thalamus, and cortex, suggest that the basal ganglia can mediate synchronization in cortex.*

Drawing on these findings, I propose the *basal ganglia mediated synchronization* (BGMS) model, and detail its mechanistic components and their relations below. In the BGMS model, the BG learn to recognize salient patterns of distributed, phase-correlated cortical activity, responding with synchronized spike volleys with functionally optimal delays, relayed via the thalamus, to the feedback-recipient layers of other areas of cortex, and back to those of the cortical areas of origin. These spike volleys reinforce activity in the areas of origin, promote contextually appropriate activity in allied areas, and establish and sustain selective long range oscillatory synchronies and consequent effective connectivity with other areas, both by elevating receptivity in the other areas by superficial facilitation and spike-timing-dependent gain modulation, and—with stronger and more coherent activity—by promoting burst firing.

In the BGMS model, the core function of the striatum is to discern and direct the moment-to-moment large scale network configuration of the brain—particularly, though not exclusively, that of the cerebral cortex. The corticostriatal projection can be viewed as a transformation from the microscopically specific feature maps of the cortex (Huth *et al.* 2012, 2016; Simmons and Barsalou 2003; Rajalingham and DiCarlo 2019; Rao *et al.* 1999; Lettieri *et al.* 2019; Zhang *et al.* 2020) to mesoscopic connectivity maps in the striatum. In these connectivity maps, each striatal projection neuron (SPN) is associated with *recognition* of a handful of network states and constituent representational contents by their idiosyncratic spike signatures, and with *activation* and *inhibition* of various related networks, via downstream structures and pathways. These striatal maps have a jumbled and fractured relationship with cortical maps (Flaherty and Graybiel 1994; Hintiryan *et al.* 2016), and this model implies that they are organized such that spatial proximity is proportional to topological similarity of the associated functional networks.

If so, feedforward inhibition by SPNs on other SPNs in their local neighborhood (Twedell *et al.* 2024‡; Tepper *et al.* 2008; Wilson 2013) functions as lateral inhibition of competing and conflicting *networks*, promoting unconflicted functional network activations.

The intralaminar thalamus is particularly intimate with the BG (Parent and Parent 2005), and much of it has a regular topographic relationship with the striatum (Mandelbaum *et al.* 2019; Sadikot *et al.* 1992a; Sidibé *et al.* 2002), suggesting it too is organized as connectivity maps, not as feature maps, involving similar network-centric principles.

Striatal output directed to matrix thalamus via basal ganglia output nuclei is transformed from connectivity maps back to feature maps, again with mesoscopic spatial specificity. This mesoscopic influence of the BG over cerebral cortex constitutes a context signal in the sense meant by the “spatial computing” proposal (Lundqvist *et al.* 2023; Chen *et al.* 2025‡). This proposal entails a separation of content from control, so that top-down control mechanisms can operate decoupled from the microscopic particulars of the affected content, providing for generalization and flexibility (Miller *et al.* 2024). The BG play a key role in this scheme, flexibly transforming microscopically detailed cortical state into appropriate mesoscopically detailed control signals.

The striosomes of the striatum—the principal origin of direct and indirect striatal projections to midbrain dopamine output nuclei (Jiménez-Castellanos and Graybiel 1989; Crittenden *et al.* 2016; Lazaridis *et al.* 2024)—likely function in a roughly similar fashion, but with macroscopic spatiotemporal specificity.

As detailed later (§8.8), the interaction of physiologically distinct but spatiotemporally coincident inputs to cortex from intralaminar and non-intralaminar BG-recipient thalamus is a key mechanism within the model proposed here. In short, GABAergic output fibers from the BG, bearing phasically synchronized silent episodes and rhythms, appose the distal dendrites of projection neurons in the intralaminar thalamus. These phasic inputs act as frequency- and phase-selective filters, favoring corticothalamic inputs with the preferred frequency and phase, while disfavoring others. The associated intralaminar thalamocortical projections then carry signals that broadcast those frequency and phase preferences to wide areas of cortex, with high temporal specificity. The multi-areal matrix cells (Jones 2001; Clascá *et al.* 2012) in non-intralaminar nuclei receive powerful, enveloping somatic inputs from collaterals of the same population of BG output fibers; thalamocortical projections from this population mainly target superficial layers, reinforcing selected activity and disfavoring unselected activity, in spatially delimited predominantly frontal cortical areas, with considerably less temporal specificity than the intralaminar paths. These distinct inputs to cortex interact, locally and inter-areally, with each other and with intrinsic cortical activity, arranging for dynamic recruitment of specific, contextually appropriate large scale cortical networks, and for their contextually appropriate dissolution.

The long and diverse conduction delays of corticostriatal projection fibers provide for the temporal focusing of widely distributed, phase-locked but phase-dispersed cortical activity, rendering it coincident as it converges on sparse subsets of striatal spiny projection neurons (SPNs). Emergent task-related synchronies and response sharpening in the striatum, noted above (Barnes *et al.* 2005; Howe *et al.* 2011; Desrochers *et al.* 2015), suggest this dynamic. Over the course of learning, sparse subsets of SPNs develop focused expertise for a given task, and unrelated SPNs, showing in-task activity at the start of learning, fall silent (Barnes *et al.* 2005).

Long and diverse delays of striatopallidal/striatonigral pathways provide for the temporal focusing of phase-skewed SPN outputs on sparse subsets of pallidal/nigral output neurons, and provide for additional phase shifting of BG outputs to the thalamus, to meet coincidence criteria in cortex associated with inter-areal phase relationships. The resulting combined delays through the BG direct path define preferred oscillatory periods and inter-areal phase relations.

Structures associated with the BG “indirect” and “hyperdirect” paths inhibit, desynchronize, or antisynchronize conflicting, aborted, irrelevant, and completed activity, and in general, delimit network activity, consistent with functions already proposed and demonstrated for these structures (Smith *et al.* 1998; Parent and Hazrati 1995b; Schmidt *et al.* 2013; Lee *et al.* 2016; Nakajima *et al.* 2019).

Open loops through the indirect path can entrain targeted areas to oppose the phase of a dominant direct path output, functionally disconnecting distracting activity, as in sensory nuclei of the thalamus via the thalamic reticular nucleus. Closed loops through the direct path, with tuned delays, inherently promote oscillations at preferred frequencies, while the indirect path must disconnect and terminate these oscillations when the connections they effect are no longer useful. Additionally, indirect path structures are considered (through the STN) to consolidate selections, amplifying localized BG activity (in particular, oscillations) to influence large surrounding areas in the BG.

A diversity of delays through the BG, focusing activity from distributed networks on particular striatal and pallidal projection neurons, is analogous to the "tidal wave" timing mechanism proposed by Braitenberg *et al.* (1997). This mechanism centers on the dynamics of granule cell input to the Purkinje cells of the cerebellar cortex via the system of parallel fibers, and has been demonstrated *in vitro*, *in vivo*, and in simulation (Braitenberg *et al.* 1997; Braitenberg 1961; Heck 1993, 1995; Heck *et al.* 2001; Heck and Sultan 2002; Sultan and Heck 2003). Indeed, a BGMS-like mechanism centered on the cerebellum has been demonstrated directly (Popa *et al.* 2013; Liu *et al.* 2022; McAfee *et al.* 2019).

BG influences on dopaminergic, cholinergic, and serotonergic centers, and the thalamic reticular nucleus, are proposed to be coordinated with (and partly by) these direct and indirect path outputs, promoting activity that contributes to the selected effective connections, attenuating or functionally disconnecting activity that conflicts with them, and modulating the dynamics within effective connections, to promote gainful computation and motor output. These mechanisms also greatly influence the expression of plasticity in the implicated structures, orienting neurophysiological investments to favor salient stimuli and behaviors, aligning selections with expectations and goals, and improving the immediacy, precision, and thoroughness of those selections.

### *1.10. Basal ganglia mediated synchronization suggests mechanistic explanations for several physiological mysteries.*

As detailed throughout this paper, the BGMS model helps explain several historically mysterious aspects of BG and related physiology, among them:

- Unusually long and diverse delays and unusually high convergence and divergence of paths through the BG;
- Unusual discoid dendritic plexuses and temporally inverted spike-timing-dependent plasticity at the striatopallidal interface;
- Special sensitivity in the striatum to large scale synchronies, large scale oscillatory synchronies spanning BG components, and striatal oscillations that induce synchronous cortical oscillations;
- Rapid statistically independent tonic discharge by projection neurons in BG output structures;
- Large scale lateral inhibition in the basal ganglia, producing selections;
- Widespread, diffuse projections from the intralaminar nuclei of the thalamus to cortex, dense topographic projections from these nuclei to the striatum, and loss of consciousness from inactivation of these nuclei;
- Paradoxical results from lesions of BG output structures, and permanent loss of normal consciousness by their bilateral destruction;
- The function of corticothalamic projections;
- Stereotyped rhythmicity of spiking in effective corticomotor signaling;
- Dense integration into BG circuitry of highly associative and abstractly cognitive areas of cortex; and
- The etiology and ontology of schizophrenia.

These phenomena, among others, are in principle explained by the proposition that the intact BG recognize and select useful patterns of synchronized cortical activity, and route their constituent spike volleys back to cortex chiefly via the thalamus, activating and synchronizing widely separated areas, and broadly promoting precisely discriminative receptivity to selected activity.

### *1.11. BGMS is one in a family of models in which the BG control functional connectivity.*

The BGMS model is hardly the first to ascribe to the basal ganglia the control of functional connectivity in cerebral cortex.

O'Reilly and Frank (2006), mentioned above, propose that the BG adjust large scale functional connectivity to fit context, controlling the formation, activation, and extinction of working memories. Stocco *et al.* (2010) propose that the BG act to control the routing of information within cortex, dynamically establishing bridges between "source" and "destination" regions to facilitate goal-directed cognition. In a similar vein, Hayworth and Marblestone (2018‡) propose such a role for the BG within a biologically inspired machine learning model, formulated to emphasize gating actions by the BG through the inhibition-disinhibition mechanism described by Chevalier and Deniau (1990), and inter-areal routing of information via the thalamic relay mechanism described by Guillery and Sherman (2002).

While the above models do not expressly consider alignment of spike volleys or modulation of oscillatory synchronies in the control of information routing, others do.

Shine (2021) proposes that the BG, acting through thalamic matrix, flatten the attractor landscape of cortex to elevate neural variability and facilitate formation of broader networks when context calls for integrative, deliberative, and exploratory processing. He relates this dynamic to the mechanism described by Larkum (2013), with BG-activated thalamic matrix supplying the apical excitatory input to promote burst firing, activating additional circuits.

Fountas and Shanahan (2017) propose and simulate a model in which coalitions of oscillating cortical inputs to the BG change the course of information flow in the latter in a frequency-specific fashion, and are crucial to selection dynamics. Population-level synchronies, band pass filtering, and dynamically selective reinforcement of oscillatory frequencies have been shown to be a plausible mechanism for flexible, selective signal routing and functional connectivity in and beyond neocortex, even absent the direct involvement of subcortical structures such as the BG and cerebellum (Akam and Kullmann 2010; Sherfey *et al.* 2020). In particular, the model articulated by Sherfey *et al.* (2018, 2020), in which frequency-specific reinforcement of oscillations in PFC govern large scale functional connectivity underlying working memory, is compatible with the BGMS model. Indeed, as explored later (§5.3), this mechanism likely interlocks intimately with BGMS as part of the larger architecture of functional connectivity control.

The fullest antecedent of the BGMS model is the "minimal coherence detection" model proposed presciently by Plenz and Aertsen (1994). Here, the striatum is viewed as akin to a "retina" observing the totality of cortical activity, responding rapidly and selectively to synchronized patterns embedded in a wider din of activity; these responses, acting through the thalamus, rapidly adjust the network topology of cortical effective connections. They summarize: "the formation of cell assemblies in the cortex is accompanied by spatiotemporal changes of input activity to the neo-striatum. The correlations of this input activity are evaluated under dopaminergic control in a way similar to local "movement" detection in the visual system. The very moment a certain minimal amount of coherence is detected — compare the "pop out" of "figure" from "ground" — the basal ganglia output results in a general rise of activity in cortical (pre)-motor areas, leading to a motor action. The specificity of this general rise of activity mainly emerges from the dynamic linking of neurons to currently active cell assemblies." Their account finds direct support in the results reported by Oberto *et al.* (2022). Plenz and Aertsen (1994) propose that, as in BGMS, "dynamic linking" depends on the nonlinear summatory interaction of coincident activity in pyramidal cells, as likewise the synchrony of idiosyncratically convergent corticostriatal inputs is crucial for striatal activations.

Plenz and Aertsen (1994) also associate the BG with dimensionality reduction and decision making, and ascribe paramount importance to the timing of BG output. They explain: "The overall anatomy of the neostriatal/pallidal complex strongly suggests that spatial integration over local correlation detectors is used to pass a highly complex judgement on the change of the cortical activity distribution ('minimal coherence'). Nevertheless, the information carried by the coherence signal itself (e.g. in the firing rate) is low. Its real information content resides in the time at which this signal raises the general population activity in — spatially not necessarily restricted — cortical premotor and motor areas."

### *1.12. Introducing BGMS*

My approach in introducing BGMS here is the obvious—to review the functional physiology of the BG and thalamocortical systems, much of it established in studies conducted in previous decades, recontextualized to the BGMS model. Following the foregoing introduction, I continue with a description of the essential mechanisms of BGMS at an intermediate level of detail ("in a nutshell"), followed by a discussion of the general relationship of the BG to cortex and thalamus from a signal processing perspective. Following this is a short review of the roles of the BG in gating motor output, emphasizing the relevance of precision spike timing. Next is a review of BG path delays, proposed delay plasticity mechanisms, and patterns of convergence and divergence in these paths, all fundamental to the BGMS model. Following this is a detailed review of the areas of thalamus receiving BG direct path output, the areas of cortex receiving output from the BG via the thalamus, the principal pathways through the BG (direct, indirect, and striosomal), the modulatory functions of dopamine, acetylcholine, and serotonin in the thalamocortical system, and their integration into BG circuitry. Finally, I consider the roles of the BG in sensory perception, general cognitive coordination, and disorders of consciousness and thought, and discuss some relationships and contrasts with the cerebellar, hippocampal, and other analogously positioned systems, and extend the foregoing explorations to their logical conclusion, that the BG are mechanistically integral to and necessary for consciousness.

Within the theoretical framework of synchrony-mediated effective connectivity, von der Malsburg (1999) mused that "If there were mechanisms in the brain by which connections could directly excite or inhibit each other, fast retrieval of associatively stored connectivity patterns could be realized." The BGMS model is a proposal that the BG, with the thalamus, implement such a mechanism, enabling patterns of effective corticocortical connectivity to excite and inhibit other connections with nearly arbitrary flexibility. And as discussed later (§14), this system may be just one instance of a general architectural motif in the mammalian brain, with homologues in other vertebrates (and perhaps beyond), in which subcortical and allocortical structures influence functional connectivity in cortical structures by spike-timing-dependent gain.

# 2. The BGMS Mechanism in a Nutshell, Step By Step

**In this section:**

### *2.1. Integrating the BG with the thalamocortical system—The challenge*

At the heart of the BGMS model is an apparently simple proposition: the BG-thalamus system learns to process cortical input volleys to produce well-timed modulatory output spike volleys, which are relayed back to cortex with precisely the timing and targeting to favor selected inputs to cortex, boosting them relative to competing signals, and helping the selected signals entrain the associated cell assemblies, overcome the threshold for pyramidal cell activation, and initiate or sustain effective transmission to receiver structures further on. In short, the BG-thalamus system instigates the effects described by Larkum *et al.* (1999, 2004) Llinás *et al.* (2002), Lakatos *et al.* (2009), and Womelsdorf *et al.* (2014), within the broader dynamic described by Singer (1993) and Fries (2005, 2015), and anticipated by von der Malsburg (1981).

Larkum (2013) showed that pyramidal neurons in deep layers of cortex intrinsically function as an association mechanism, with burst firing contingent on simultaneous apical and proximal inputs. Womelsdorf *et al.* (2014) showed that cortical burst firing establishes long range corticocortical synchronies associated with functional connectivity. Lakatos *et al.* (2008, 2007), 2009 showed that well-timed apical inputs amplify middle layer inputs, inducing prolonged oscillatory synchronization of the local ensemble with the attended input.

For the BG to integrate with these thalamocortical and corticocortical mechanisms, they must control the thalamus on the timescale of cortical coincidence windows, which range from 2 ms (Pouille and Scanziani 2001; Volgushev *et al.* 1998) to 10 ms (Williams and Stuart 2002) depending on the particular targeted cell and compartment, and 5 ms for the intercompartmental mechanism described by Larkum *et al.* (1999).

The proposition that the basal ganglia can process spike patterns with the necessary temporal and oscillatory specificity and control is not new (Plenz and Aertsen 1994; Barnes *et al.* 2005; Howe *et al.* 2011; Thorn and Graybiel 2014; Desrochers *et al.* 2015; Leventhal *et al.* 2012; Schmidt *et al.* 2013; Pouzzner 2017 ‡ (this work's initial version); Oberto *et al.* 2022; Banaie Boroujeni and Womelsdorf 2023; Fischer 2021; Fischer *et al.* 2020; Portoles *et al.* 2022; Holt *et al.* 2019; Wang *et al.* 2021; Kasahara *et al.* 2022). Nonetheless, the arrangements whereby the BG and thalamus meet these criteria implicate historically mysterious and underappreciated aspects of their physiology.

### *2.2. A nutshell in a nutshell*

Specific, stable, yet diverse delays in BG pathways are a crucial ingredient in the model proposed here. Indeed, an unusually broad range of conduction delays characterizes axon populations in the corticostriatal and striatopallidal projections (Yoshida *et al.* 1993; Kitano *et al.* 1998). (See Figure 1.) Furthermore, axodendritic plexuses in the pallidonigral system (Goldberg and Bergman 2011; Difiglia *et al.* 1982), entailing arrays of apparently redundant synapses distinguished only by their respective conduction delays and efficacies, appear positioned to provide for fine tuning of the delay of a given axodendritic trajectory.

Patterns of cortical activity are distributed across large scale functional networks (Bressler 1995; Varela *et al.* 2001), with significant inter-areal delays (Sorrentino *et al.* 2022; van Blooijs *et al.* 2023; Nowak and Bullier 1997; Schmolesky *et al.* 1998). In the model proposed here, the combinatorial topology of the corticostriatal projection (Zheng and Wilson 2002; Hintiryan *et al.* 2016), and its diversity of delays, focus distinct patterns of widely distributed cortical activity upon sparse and idiosyncratic combinations of striatal cells. The focusing of these patterns on striatal cells has previously been described as “dynamic convergence” (Plenz and Aertsen 1994), with “pop out” salience for patterns of synchronized activity that the striatum has learned to recognize. In this arrangement, not only does dispersed coherent cortical activity converge on single striatal cells, but crucially, cortical areas with earlier activity travel to the responding striatal cells through slower fibers, while those with later activity travel to them through faster ones, so that the convergence is spatiotemporal. These delicate arrangements can form because plasticity can be as specific to the conduction delays of pathways as to their connectivity, provided the conduction delays of the implicated fibers are stable, and have a substantial impact on outcomes and the associated expression of plasticity, as suggested below. Evidence suggests that the corticostriatal projection is indeed structurally stable from the postnatal epoch onward (Mesías *et al.* 2023).

Once a task is well-learned, spatiotemporally converged spike volleys associated with task-relevant input patterns can overcome the high activation threshold of these cells, inducing firing (Zheng and Wilson 2002). Striatal projection axons have yet longer and more varied conduction delays than corticostriatal ones (Yoshida *et al.* 1993; Kitano *et al.* 1998), and by way of these additional delays, striatal output can be focused spatiotemporally upon sparse subsets of pallidal and nigral cells, interrupting their tonic inhibition of task-specific populations in thalamus (Chevalier and Deniau 1990). These phasic disinhibitions facilitate thalamocortical bursts (Bodor *et al.* 2008) with contextually appropriate delays, targeting cortex in widely divergent patchy distributions with mesoscopic spatial specificity (Rubio-Garrido *et al.* 2009). Notably, the precise timing of the output from striatal SPNs is generally determined by striatal FSIs (explored in detail later (§6.6)); thus it is temporally decoupled from the precise arrival time of corticostriatal patterns at SPNs. The SPNs can thus specialize in integrating widely dispersed activity with immense specificity (further suggested by their lack of intrinsic frequency preferences (Beatty *et al.* 2015) and their low pass characteristic (Stern *et al.* 1997)), while the FSIs specialize in appropriately gating and timing the outputs associated with those integrations.

That the basal ganglia systematically form closed loops with the cerebral cortex is canonical (Alexander *et al.* 1986, 1991); the BGMS proposal extends consideration of these loops to the effects of spike timing and oscillatory phase relationships downstream from the basal ganglia. Particular cumulative conduction delays through the basal ganglia and thalamus, along axonal trajectories that target particular cortical ensembles, can impart selective facilitation of particular inputs to those cortical ensembles. That is, modulatory thalamocortical spike volleys mediated by the basal ganglia can be temporally aligned to amplify the effectiveness of driving corticocortical and thalamocortical spike volleys constituting favored inputs. For a disfavored primary input, in contrast, the implicated BG trajectories arrange for the primary input to arrive non-coincident with modulatory thalamocortical spike volleys, or coincident with modulatory inhibition mediated by cortical interneurons, or at disfavored phases of ensemble oscillations, or indeed all of these at once.

The plastic formation of these time-aligned pathways is suggested by the gradual emergence during learning of widespread task-related spike synchronies and sharpened responses in the striatum (Barnes *et al.* 2005; Howe *et al.* 2011; Desrochers *et al.* 2015). These pathways activate in association with synchronized cortical activity, and when they do, the implicated cortical and striatal cells are fused tightly into ephemeral assemblies with well-defined temporal structure (Oberto *et al.* 2022; Banaie Boroujeni and Womelsdorf 2023). Evidence suggests that the precise delay of these pathways is functionally paramount, with behavioral proficiency exhibiting an inverted-U relationship with experimentally manipulated striatal latency (Monteiro *et al.* 2023).

BG-induced thalamocortical spike volleys are modulatory, systematically avoiding middle layer (L4) neurons, and mostly targeting apical dendrites of pyramidal neurons (Kuramoto *et al.* 2009; Jinnai *et al.* 1987; Herkenham 1979; Glenn *et al.* 1982; Rubio-Garrido *et al.* 2009; Jones 2001; Vitek *et al.* 1996; Buford *et al.* 1996) and inhibitory interneurons (Delevich *et al.* 2015; Kuroda *et al.* 1998; Rikhye *et al.* 2018). But this does not imply that this influence on cortical activity is weak. A single modulatory thalamocortical volley, coincident with a forward input, can admit that forward input and reset ongoing local oscillation to favor it (Lakatos *et al.* 2008, 2007, 2009; Tiesinga and Sejnowski 2010; Reyner-Parra and Huguet 2022). After phase reset, ongoing oscillation in the targeted area can align with (and follow) a favored forward input, relaying it preferentially without additional resets (Lakatos *et al.* 2007, 2009).

Trans-BG pathways entail significant delays, typically 10s of milliseconds. Given cortical coincidence windows of 2-10 ms, the BG can mediate reinforcement of a favored primary input signal only when an input-related corticostriatal volley is followed by at least one additional input-related volley targeting cortex, with a consistent, characteristic inter-volley delay — a criterion met by, but not strictly limited to, oscillatory inputs. Only then can the BG learn the optimal delay to impart to the triggering spike volley as it is routed from the source, through the BG, to the thalamus, and on to a joint cortical target. As described above, after the BG have learned an optimal delay, a BG-mediated modulatory thalamocortical volley associated with a favored input is delayed to arrive at a cortical target at just the right moment to amplify a second (or otherwise subsequent) volley traveling from a favored source through thalamocortical or corticocortical paths to a joint target, and to blunt the influence of competing inputs from disfavored sources.

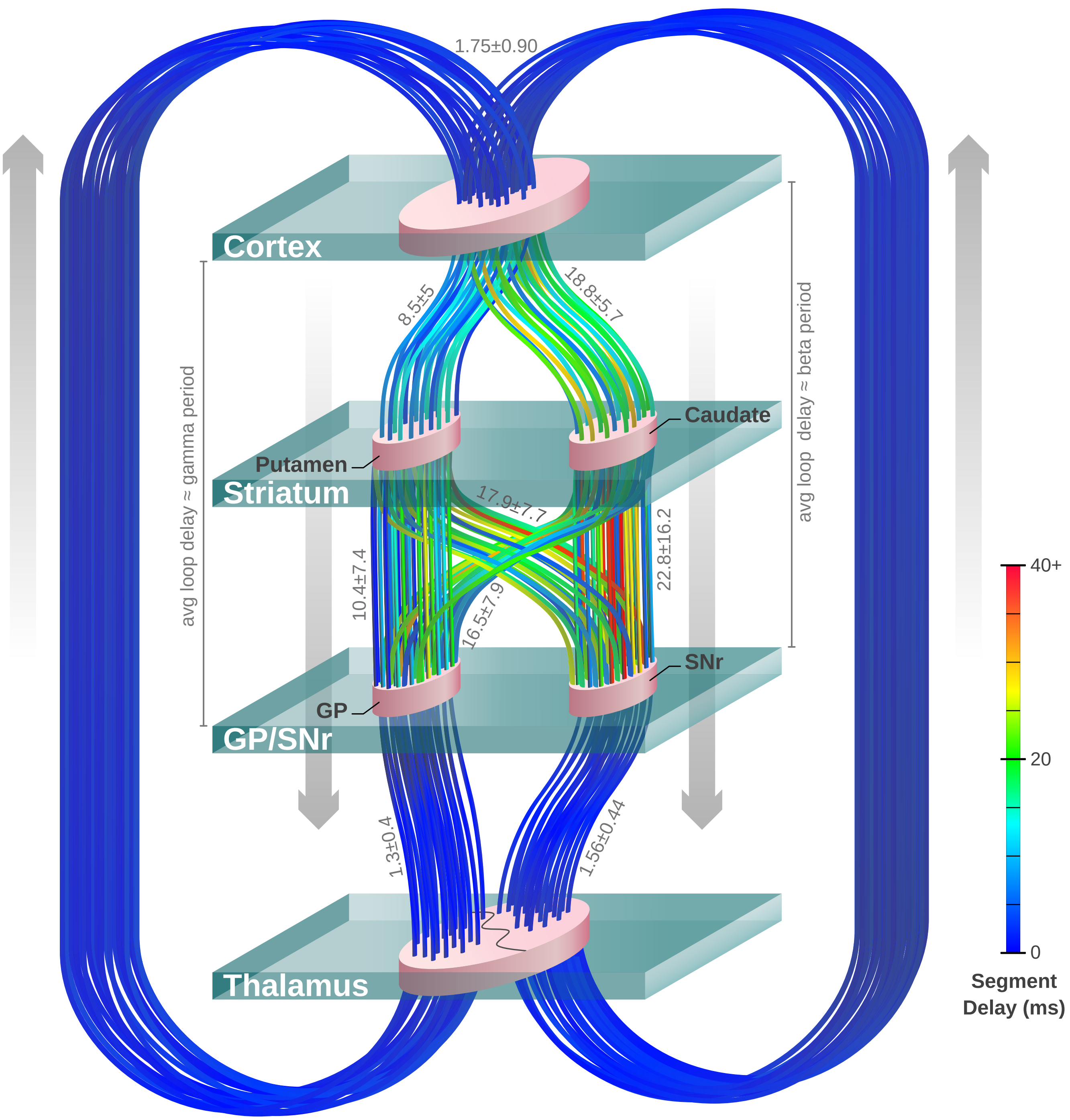

***Figure 1: Conduction delays through the main segments of the BG direct path, showing slow and varied propagation to and from the striatum, and relatively instantaneous and uniform propagation to and from the thalamus. Color-coding is for illustration purposes only and is simulated using empirical conduction statistics from Harnois and Filion (1982), Yoshida et al. (1993), Kitano et al. (1998), Turner and DeLong (2000), and Kurata (2005))***

This arrangement brings a remarkable implication: thoroughly integrated responses can be selected almost instantaneously through corticostriatal convergence upon sparsely responding cells, and can then be widely distributed, integrating that selection into corticothalamic activity patterns at the very first opportunity, as proposed by Plenz and Aertsen (1994). Because the corticothalamic projection is itself convergent and implicitly integrative (Whyte *et al.* 2024), the overall architecture appears to prioritize rapid integration. It has been proposed that neural firing patterns carry much information in their temporal fine structure, and that most of the information associated with a stimulus is available in the spikes from the first 1% (~15 ms) of responding neurons, whose characteristic firing pattern can be learned and subsequently recognized in just one oscillation cycle (Rullen and Thorpe 2001; Thorpe *et al.* 2001; Masquelier *et al.* 2009; Luczak *et al.* 2015; Sotomayor-Gómez *et al.* 2025; Zhu *et al.* 2025a ‡ ). According to the model proposed here, the striatum indeed learns to recognize these spiking patterns the instant they occur, immediately distributing contextually appropriate responses. Effectively, the striatum is a library of "reader" neurons, in the sense meant by Pompili *et al.* (2022‡). The robust stability of task-specific neural dynamics (Oby *et al.* 2025), and characteristic region-specific spike patterns (Tolossa *et al.* 2024‡), presumptively arrange for sensical and stable relationships between large scale spatiotemporal patterns of spiking activity and appropriate responses, making these relationships learnable. The implications for the representational power of the combined system are awesome (Izhikevich 2006).

These contextually appropriate responses can be either facilitatory or suppressive, and indeed are usually a combination of the two (Cui *et al.* 2013; Oldenburg and Sabatini 2015). For favored inputs, the BG direct path is more implicated, targeting thalamic matrix and intralaminar nuclei and the brainstem tectum (McElvain *et al.* 2021). The resulting modulatory volleys align with those of favored inputs, facilitating post-synaptic output that reflects the selected input. But for inputs deemed distracting or disruptive, the faster BG indirect and hyperdirect paths are more implicated (Schmidt *et al.* 2013; Nakajima *et al.* 2019). In the BGMS model, indirect path signals to the thalamus, via projections to the thalamic reticular nucleus, reset or entrain distracter oscillations at the thalamic level to a phase disfavored by targeted cortical areas, suppressing responses. These indirect path regulatory pathways notably encompass the primary sensory relay nuclei (Nakajima *et al.* 2019). Internal to the BG, the indirect and hyperdirect paths can regulate direct path output structures, and through them, reset or entrain their cortical targets to a disfavored phase, effecting functional disconnection of the targets (Fischer 2021; Schmidt *et al.* 2013). The subthalamic nucleus also targets the intralaminar thalamus directly (Castle *et al.* 2005), and this pathway may function like the GPe pathway to the reticular nucleus, albeit with chemical and circuit distinctions.

### *2.3. Temporal specificity of BG actions in the dynamic control of cortical gates*

Projection cells in the BG output nuclei activate with various stereotypical, dynamically context-dependent patterns (Tremblay and Filion 1989; Yoshida *et al.* 1993; Kitano *et al.* 1998; Yoshida *et al.* 2025). (See Figure 2.) The responses of pallidal and nigral cells to artificial cortical stimulation often feature early and pronounced acceleration of spiking, followed by the precipitous onset of a silent episode lasting 10-20 ms, followed by a sharp resumption of activity at or above that accelerated rate (Yoshida *et al.* 1993; Kitano *et al.* 1998; Yoshida *et al.* 2025). Similar patterns are seen in natural and simulated BG activations (Wongmassang *et al.* 2021; Mirzaei *et al.* 2017). Corticothalamic activity is also seen to accelerate immediately before the onset of BG activation (Schwab *et al.* 2020).

In the BGMS model, accelerations in BG output spike rates before and after silent episodes, and the assiduous independence of tonic output spikes (Stanford 2002; Wilson 2013), are functionally crucial. According to the BGMS proposal, BG output nuclei must relay activity with utmost agility, minimizing the jitter added to signals arriving at unpredictable and often rapidly successive moments. Absent this agility, the BG would be unable to meet cortical coincidence criteria, and would consequently fail to open cortical gates for the intended signals, and moreover would open them for contextually inappropriate ones. Indeed, the activity of projection neurons in healthy BG output nuclei has unusually high diversity and dimensionality, exceeding that of functionally related areas of frontal cortex, striatum, and cerebellum (Zur *et al.* 2024).

In mammals, several BG output cells are tightly coupled to each thalamic cell (Ilinsky *et al.* 1997; Kultas-Ilinsky and Ilinsky 1990; Bodor *et al.* 2008; Nejad *et al.* 2021). This makes it likely that striatal interruption of ongoing pacemaking output, regardless of its exact timing, will be immediately preceded by GABA spikes to all of the targeted thalamic cells. This assures that the timing of silent episodes tracks the timing of striatal spike volleys with minimal jitter.

As noted above, the BG-recipient thalamus targets cortical fast-spiking interneurons (Delevich *et al.* 2015; Kuroda *et al.* 1998; Rikhye *et al.* 2018), which tightly govern pyramidal cell activity (Hasenstaub *et al.* 2005). Striatal spike volleys closely follow the timing of striatal FSIs (Howe *et al.* 2011; Gage *et al.* 2010; Berke 2011), which themselves closely follow the timing of corticostriatal projection (pyramidal) neurons (Sharott *et al.* 2009, 2012; Howe *et al.* 2011). And BG output neurons follow striatal inputs over a wide range of frequencies, with negligible jitter (Connelly *et al.* 2010). Thus the entire BG-thalamocortical loop can preserve precise timing. This arrangement presumptively underlies findings of precise task-related spike time relations between PFC, ACC, and striatum (Oberto *et al.* 2022; Banaie Boroujeni and Womelsdorf 2023).

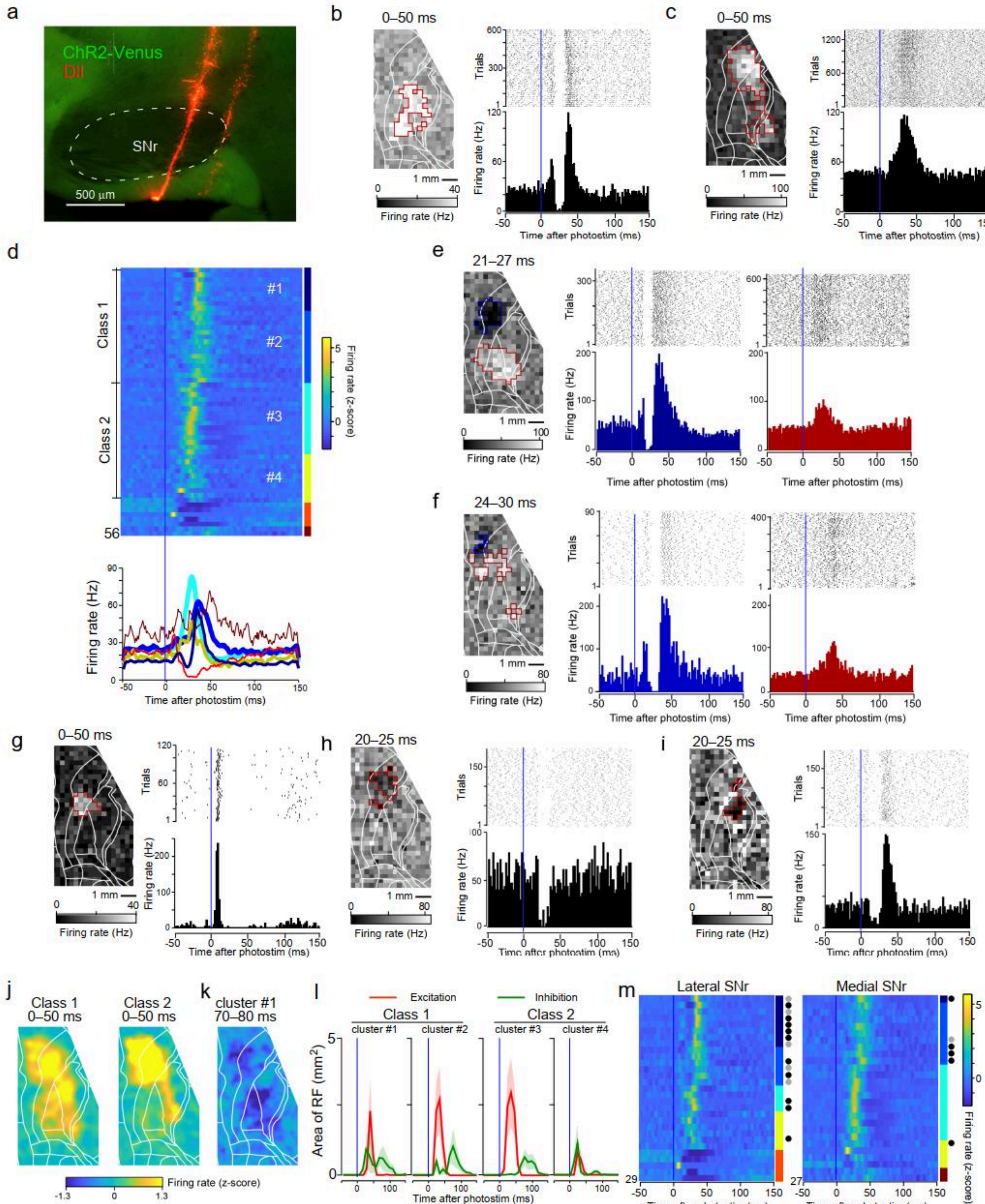

***Figure 2: Illustrations of stereotypical activity patterns in a BG output nucleus, from figure 4 of Yoshida et al. (2024‡) (CC-BY 4.0)****, now published as Yoshida et al. (2025)*

Figure legend (*op. cit.*): "Analysis of single neuron response patterns of substantia nigra pars reticulata (SNr) to cortical stimulation. a. A sagittal section showing electrode traces with DiI. b. A map, raster plot, and PSTH of a typical SNr neuron with class 1 response (a triphasic response). c. A typical neuron with class 2 response (monophasic excitatory response). d. Z-scored PSTH of all responsive SNr neurons clustered by K-means clustering (top) and the mean PSTH of each cluster (bottom). Colors correspond to cluster numbers. The thickness indicates the numbers of neurons in the cluster. e,f. Examples of SNr neurons that were classified as class 1 but also received class-2-like

monophasic excitatory input. g,h,i. Neurons showing a very fast excitatory monophasic response (g), monophasic inhibitory response (h), and inhibition-excitation pattern (i). j. Average input maps of neurons with class 1 (left) and class 2 (right) responses in SNr. k. An inhibitory response map of cluster #1 neurons in SNr. l. Time courses of the responsive areas. m. The same analysis as d, but separated according to recording location. Black and gray circles indicate triphasic and inhibition-excitation patterns, respectively."

### *2.4. The functional implications of convergence-divergence, diversity, and redundancy in BG paths*

The BG must rapidly integrate related information regardless of its representational topography in cortex, and distribute the resulting responses to the many specific cortical areas implicated in cognitive and behavioral production associated with a particular response. Extensive convergence, divergence, and reconfiguration in BG pathways (Flaherty and Graybiel 1994; Joel and Weiner 1994; Hintiryan *et al.* 2016; Korponay *et al.* 2022‡; Parent *et al.* 2001; Parent and Parent 2006; Zheng and Wilson 2002) is thought to underlie this topological flexibility. Absent it, the BG would not be positioned to take full advantage of information encoded in its inputs, restricting the latitude of behavioral and cognitive adaptation.

Indeed, as explored in greater detail later (§12.9), the BG are consistently found to underpin cognitive flexibility and agility (Leber *et al.* 2008; van Schouwenburg *et al.* 2010b, 2012, 2014; Vatansever *et al.* 2016; Weerasekera *et al.* 2023; Wan *et al.* 2012, 2011; Nestor *et al.* 2024‡), and BG dysfunction in Parkinson's disease is associated with bradykinesia, cognitive rigidity, perseveration, and contracted repertoire (Hammond *et al.* 2007; Sorrentino *et al.* 2021; O'Callaghan *et al.* 2017; Stoffers *et al.* 2008; Olde Dubbelink *et al.* 2014).

In birds, each thalamopallial projection neuron receives only a single input fiber from the pallidum (Luo and Perkel 1999; Person and Perkel 2005), suggesting that the avian pallidothalamic projection is highly specific but relatively inflexible. In mammals, the pallidothalamic projection has the more flexible arrangement described above, in which several pallidal neurons project to each thalamocortical projection neuron (Ilinsky *et al.* 1997). A thalamic neuron is fully activated only if all of its pallidal afferents are simultaneously silenced (Bodor *et al.* 2008; Kim *et al.* 2017) simultaneous with excitatory input. Mammals are thus endowed with an additional dimension of combinatorial richness, with several apparent advantages: a higher density of tonic inhibitory input with a corresponding improvement in signal fidelity, an improved capacity for bootstrapping during learning, an intrinsic capacity for partial and weak responses, and greater tolerance for localized physiological insults to the BG.

### *2.5. Control of thalamocortical oscillatory dynamics by the basal ganglia*

Implicit in the dynamics described above is that striatal projection cells (SPNs) activate in at least two general patterns, impulsive and oscillatory. Ultimately this is a dual view of a single underlying dynamical continuum of activity. Coherent oscillations across the constituent nuclei of the BG are apparent in LFP recordings of behaving animals (Leventhal *et al.* 2012 (see particularly figure 2)), and learning entails the emergence of oscillatory ensembles synchronized across wide areas in the BG (Howe *et al.* 2011). But also, learning entails the emergence of temporally sharpened responses marking moments of contextual transition in tasks (Barnes *et al.* 2005; Desrochers *et al.* 2015). In BGMS, the interpretation of these dynamics is that momentary bursts at contextual transitions induce shifts of configuration in downstream structures—particularly, of functional topology. Oscillatory activations of the BG, meanwhile, stabilize and sustain configurations underlying contextually aligned action, attention, and working memory. An intriguing possibility is that widespread synchronized oscillatory BG activity, by evidencing the consolidation of a task-related stimulus-response relation (Desrochers *et al.* 2015), may inherently signal that the relation has been adequately represented and operationalized, inhibiting further investment in its representation.

The BG, by tuning signal delays to meet cortical coincidence criteria, implicitly reinforce selected fundamental frequencies and their harmonics. This is a corollary of the inherent requirement, noted above, that facilitatory BG output via the thalamus aligns with the second (or later) in a series of spike volleys. In short, each inter-volley period of such a series has a corresponding frequency which the BG reinforce when they select that signal. In closed loops, which are common in the BG (Alexander *et al.* 1986; Parent and Hazrati 1995b; Smith *et al.* 2004), it is particularly clear that learned delays through the BG correspond to reinforcement of specific frequencies in the implicated thalamocortical channels. Mean-field simulations of the parkinsonian brain exhibit this dynamic clearly, showing strong resonant frequency modulation from conduction delay shifts of 1-9 ms (Asadi *et al.* 2024); in BGMS, sparse assemblies (which are not modeled by course-grained simulations) ephemerally resonate with context-appropriate connectivity and frequency. Given convergence in the corticostriatal projection, this has the interesting implication that even segregated closed loops can have an integrative and articulately selective influence on cortical targets, due to temporal/phase specificity. That is, these closed loops do not simply reinforce whatever oscillation is ongoing in the implicated thalamocortical module, but rather favor oscillation *at a particular frequency* there, as a function of which particular closed loop is active.

As noted above, the SPNs of the striatum exhibit no intrinsic frequency preferences (Beatty *et al.* 2015), further suggesting a capacity to integrate arbitrary signals converging from the cortex at large. However, when an SPN activates, it exhibits an ephemeral preference for afferent inputs matching the firing rate of the activation, and tracks the phase of those matching inputs (Beatty *et al.* 2015). Similarly, striatal FSIs preferentially respond to inputs that align with their ongoing local oscillation (Mohapatra *et al.* 2025). These ephemeral preferences lock a striatal cell assembly to activity in the network that activated it (with functional similarities to an electrical device known as a "lock-in amplifier"), positioning it to reinforce that network, to ignore extraneous inputs, and to modify that network through synchronization, antisynchronization, and neuromodulation of other areas, through the downstream effects described above.

The activation of an SPN reflects the influences of various striatal interneurons. Striatal interneurons *do* exhibit characteristic frequency preferences (Beatty *et al.* 2015; Tepper *et al.* 2018)—different input frequencies preferentially activate different classes of interneuron, each with its own efferent pattern. These frequency preferences are mechanisms whereby the striatum's response can be contingent on the frequencies of its inputs. Computational modeling suggests that, even in simplified models of the striatum restricted to SPNs and FSIs, input frequencies strongly affect BG input-output relations (Fountas and Shanahan 2017).

As detailed later (§5.9), closed cortico-BG-thalamocortical loops through the putamen and globus pallidus (GP) have an average cumulative transmission delay corresponding to 40 Hz gamma oscillation, and loops through the caudate nucleus and substantia nigra have an average delay corresponding to 20 Hz beta oscillation, with wide delay ranges along either path. The BG can also state-dependently generate beta intrinsically (Pittman-Polletta *et al.* 2018). These arrangements implicate the BG in beta-mediated attentional spotlighting by PFC, which has been shown to selectively enhance gamma activity encoding stimuli (Lee *et al.* 2013; Richter *et al.* 2017; Bressler and Richter 2015). Moreover, paths through the putamen and globus pallidus *pars externa* (GPe) have particularly short delays, positioning them to entrain their targets to phases nearly opposite those imparted by the generally slower direct path, both for gamma targeting the GP *pars interna* (GPi), and beta targeting the substantia nigra *pars reticulata* (SNr). Such phase opposition is associated with orchestrated functional disconnection (Stetson and Andersen 2014; Dotson *et al.* 2014; Maris *et al.* 2016; Schmidt *et al.* 2013; Helfrich *et al.* 2014; Zhu *et al.* 2025b‡).

Key neuromodulatory influences (§10) that promote intrinsic oscillation and modulate the resonant frequencies and other dynamical parameters of cortical cell assemblies — particularly dopamine, acetylcholine, and serotonin — are influenced by the striatum, providing additional channels for flexible orchestration of circuit dynamics (Bargmann 2012). Bidirectional control of DA and ACh by the BG has been demonstrated, structured around direct and indirect path structures and populations (Lazaridis *et al.* 2024; Chen *et al.* 2024‡; Fallah *et al.* 2024‡), suggesting articulate influences like those in the thalamic paths. The striatum is thus positioned both to adjust the dynamical regime of cortical areas for episodes of oscillation at a contextually appropriate frequency, and to initiate or reset that oscillation with well-timed excitatory impulses via the thalamus. Beyond neuromodulators, changes to the intensity of apical (L1) excitatory bombardment of cortex — largely determined by the BG-recipient thalamus — can shift the resonant frequency of cortical assemblies from moment to moment to select appropriate inputs while rejecting distracters (Sherfey *et al.* 2020).

As oscillatory volleys enter the BG, the BG can impart various precise delays to selected spike volleys over diverging pathways, widely distributing them to cortex (chiefly via the thalamus) to arrange for context-appropriate inter-areal oscillatory synchronies and phase relationships at both fundamental and higher frequencies. In this way, the BG can arrange the various phase relationships in an oscillatory hierarchy to activate and stabilize functional networks spanning the brain, with the cerebellum functioning synergistically to fine-tune them (McAfee *et al.* 2022). Indeed, some learning depends on system-wide oscillatory synchrony of cerebellum, BG, and cerebral cortex (Yoshida *et al.* 2025).

This hierarchical orchestration obviously relates to the proposition that lower frequencies in the mammalian brain carry top-down control signals (Bressler and Richter 2015; Miller *et al.* 2018). In contrast, invertebrates — behaviorally dominated by reflexive input-output reactions — are devoid of autonomous synchronized low frequency neural oscillations (Bullock and Başar 1988), with the notable exception of the octopus, wherein vertebrate-like behavioral and oscillatory patterns are apparent (Bullock 1984).

The centrality of these cognitive rhythms in mammals is evident in theta oscillations generated in lateral PFC underlying preparation for conflicted perceptual decision making (Martínez-Molina *et al.* 2024), in an increase in phase-amplitude coupling that accompanies acquisition of cognitive expertise (Yagura *et al.* 2024‡), in coordination of the PFC and hippocampal system by coherent delta oscillations (Fujisawa and Buzsáki 2011), and in the breakdown of nested gamma-delta dynamics, and the cortical hierarchy, in psychosis (Missonnier *et al.* 2020; He *et al.* 2024).

### *2.6. Activation of contextually appropriate functional networks by the basal ganglia*

As reviewed above, when forward and modulatory inputs are coincident, the criteria are met for burst generation (Larkum *et al.* 2004), and synchronization with areas further downstream (Womelsdorf *et al.* 2014). The BG and thalamus are thus positioned to control, from moment to moment, which inputs to a cortical area are conveyed as outputs from that area. According to this narrative, a single coherent volley from an array of BG-recipient neurons in the thalamus, widely distributed with target-specific delays, can activate topologically complex large scale functional networks appropriate for momentary context. These new patterns of activation are then reflected in corticostriatal input. Thus, the BG are positioned to iterate learned sequences rapidly and precisely, with each stage in the sequence associated with a particular pattern of cortical activation and directed functional connectivity, and each transition between stages associated with a particular impulsive modulation by the BG-thalamus system (Nestor *et al.* 2024 ‡ ; Graybiel 1998; Portoles *et al.* 2022).

Patterns of cortical activation are nuanced — for example, the neural subspaces (more generally, manifolds) of ensembles are dynamic and systematically structured, shifting with context to effect appropriate integration and segregation (Miller *et al.* 2024). Orchestration of these dynamic subspace shifts is presumptively a core function of the BG. Importantly, full characterization of neural subspaces pivots on the temporal structure of the implicated spikes (Guidolin *et al.* 2022) — indeed, this is effectively a restatement of the *Communication through Coherence* proposal (Fries 2015), and has long been recognized in hippocampal phase codes (Climer *et al.* 2013; Siegle and Wilson 2014) and similar timing-based mechanisms noted earlier (§1.3). As noted above, single thalamocortical volleys can reset the phase of oscillation in targeted ensembles, impulsively shifting their patterns of receptivity (Lakatos *et al.* 2008, 2007, 2009; Tiesinga and Sejnowski 2010; Reyner-Parra and Huguet 2022). In the BGMS model, this is a core mechanism whereby the BG rapidly and flexibly shift neural subspaces in cortex.

Impulsive activation in the thalamus can initiate episodes of large scale theta oscillation in cortex (Lyu *et al.* 2025) that can root oscillatory hierarchies. This clearly relates to evidence for theta initiation and phase modulation, and associated synchronization, by the human intralaminar and medial thalamus at the earliest moment of visual awareness (Fang *et al.* 2024‡), and also relates to evidence for distinct network activations as a function of theta frequency modulations in entorhinal cortex (Salvan *et al.* 2021). Areas of cortex implicated in "multiple demand" cognitive control also exhibit low frequency (delta and theta) oscillations synchronized across large scale networks, with distinct roles for different frequency bands, and mid-frontal theta power tracking demand (Lu 2025); the cortical areas most implicated are among the densest targets of BG output. As detailed later (§5.3), distinct functional networks have characteristic large scale spatiospectral signatures (Keitel and Gross 2016; Becker and Hervais-Adelman 2020; Vezoli *et al.* 2021; Lyu *et al.* 2025), implicating the frequency-selective mechanisms of the BG described above.

Modeling suggests that large scale cortical activity is self-organizing, with reciprocal long range connections in cortex stabilizing self-consistent network configurations, and destabilizing inconsistent configurations (Javadzadeh *et al.* 2024 ‡ ). This implies that the crucial roles of the basal ganglia in network orchestration are initial activation of contextually appropriate networks, and promotion of one among several contextually implicated and spontaneously competing candidate networks, each self-consistent but conflicting with the others.

The outcomes that follow activation of large scale networks contribute to the potentiation of axodendritic paths through the BG (Kreitzer and Malenka 2008; Reynolds *et al.* 2022; Barnes *et al.* 2005; Shan *et al.* 2015). In the predictive routing model (Bastos *et al.* 2020; Sennesh *et al.* 2025 ‡ ), context-appropriate (dynamic) directed network topology is central to successful prediction, which in predictive coding theory is central to mental activity writ large (Clark 2013). Following this proposal, and considering the present model of BG function, the natural inference is that routing control by the BG is driven by the predictive success of the resulting networks. Put more directly, the BG learn to activate, for any given context, the directed network which is most effective at generating descending signals that accurately predict ascending signals. And since many of the implicated modulatory signals (dopamine, acetylcholine, serotonin) are themselves under BG influence or control §10, the BG have a key role in cortical plasticity too. Predictive coding theory thus implies that local predictive success in cortex drives neuromodulation via the BG, and the BG drive local plasticity and dynamics in cortex, through innumerable loops, evolving the system as a whole toward ever-improving capacity for accurate prediction—or, in pathology (particularly schizophrenia), toward ever-deteriorating capacity. Notably, in predictive coding theory, it is not just percepts that are predicted, but (through a sort of reframing) cognition and action too (Clark 2013). This dynamic also relates to the "spatial computing" proposal (Lundqvist *et al.* 2023; Chen *et al.* 2025 ‡ ), according to which top-down influences (particularly, rule and goal context) propagate as oscillations in the alpha and beta bands, imparting "inhibitory stencils" that dynamically organize cortical activity.

### 2.7. Plasticity in the basal ganglia

To function as described here, the system of the cerebral cortex, basal ganglia, and thalamus, must overcome a daunting challenge. Before learning, and aside from inborn circuits (e.g. Cromwell and Berridge 1996), there is likely to be little or no preferential activation, and little if any a priori knowledge, of the axodendritic trajectories with the requisite delays and topologies to effect contextually appropriate stimulus-response relations. They are hidden among a vastly larger population of irrelevant trajectories that, ideally, remain untouched by the plasticity mechanisms at work.

There are, then, three closely related mysteries: (1) How are trajectories first activated in connection with the contexts wherein they are useful? (2) Once usefully activated (effectively, found), how are they strengthened to facilitate subsequent reactivation as the context repeats, without plastic perturbation of other trajectories? (3) Once the stimulus-response relation is consolidated by plastic changes along suitable trajectories, how is the search for useful trajectories slowed and ultimately stopped, preventing wasteful and disruptive over-representation of the relation?

Results from recordings of single neurons in rats (Barnes *et al.* 2005) offer clues to the dynamics of the underlying mechanisms: “Early in training, the spike activity of the task-responsive population was spread throughout task time, as though all task events were salient (neural exploration). Even neurons without detectable phasic task-responsive activity fired at low rates during the task. Then, with continued training, this widespread spiking of the task-responsive population diminished, and their spike activity became focused (neural exploitation). At the same time, the non-task-responsive population fell silent, further reducing the task-irrelevant firing of the total projection neuron population.”

Noise generation endemic to the basal ganglia, discussed above, and with greater depth later in §12.6 and §16.9, is likely central to the pathway discovery process. Indeed the intralaminar thalamus is ideally positioned to recirculate noisy inputs from the BG back to the striatum and cortex, bidirectionally “fuzzing” plasticity and receptivity there (Huerta-Ocampo *et al.* 2014; Smith *et al.* 2004; Steriade *et al.* 1993; Pakhotin and Bracci 2007). This “fuzzing” plausibly allows weak, or weakly recognized, signals to occasionally activate SPNs in quiescent striatal matrix, some of which will have connectivity and delays useful for the task being learned, resulting in a performance bump, phasically elevated dopamine, cholinergic pause, associated plastic reinforcement (explained below), and consequently, a higher likelihood of subsequent activation in the same context. Noise injection in the learning process also provides for more robust representations.

Striosomes are thought to bootstrap striatal learning, with spatiotemporally crude modulations representing task contingencies early, while matrix plasticity refines task representation later, within and consistent with the broad outlines established by early striosomal learning (Graybiel and Matsushima 2023). In striatal matrix, plasticity is fully expressed only with a threefold co-occurrence of (1) recent SPN activation, (2) phasic dopamine modulation, either from midbrain DA centers or endogenously in association with synchronized cholinergic interneuron activity, and (3) a cholinergic pause, likely under intralaminar thalamic control (Reynolds *et al.* 2022; Morris *et al.* 2004; Bradfield *et al.* 2013; Threlfell *et al.* 2012; Cover *et al.* 2019; Pakhotin and Bracci 2007; Ding *et al.* 2010). When this tripartite gate on plasticity is disrupted, dystonia results, due to pathological expression of plasticity (Gemperli *et al.* 2025).

In striosomes, cholinergic innervation is greatly attenuated or absent (Dautan *et al.* 2014; Graybiel and Ragsdale 1978) — effectively, acetylcholine is always paused there, so the additional contingency does not exist. This positions the striosomes to learn early. The striosomes, through their control of matrix dopamine release, can then dynamically drive spatiotemporal patterns of activation — acting as a scaffolding — often decoupled from structural connectivity patterns (Korponay *et al.* 2022‡).

Phasic dopamine has opposite effects on striosome and matrix direct path SPNs, so that as an appropriate phasic dopaminergic response is learned, activity progressively shifts from striosomes to matrix (Prager *et al.* 2020), implicitly shifting plasticity from striosomes to matrix too. Importantly, midbrain DA is controlled not only by striosomes, but also by orbitofrontal and anterior cingulate cortex, which project to midbrain DA centers directly and reciprocally, in addition to direct projections to the striosomes, positioning cortical motivation centers to direct striatal learning.

Much of the intralaminar thalamus has been shown to project to the striatum with systematic topography (Mandelbaum *et al.* 2019; Sadikot *et al.* 1992a; Sidibé *et al.* 2002), selectively targeting cholinergic interneurons (Bradfield *et al.* 2013). This suggests modularity: locally synchronous activity in the intralaminar thalamus generates a locally synchronized cholinergic pause signal that is specific to the striatal locus linked with the locally synchronized thalamic population, providing a mechanistic substrate for learning new time-aligned trans-BG pathways while leaving unrelated pathways untouched. Indeed, targeting of the striatal cholinergic population by intralaminar thalamus has already been experimentally and causally implicated in striatal learning that preserves and integrates with existing representations (Bradfield *et al.* 2013), and in its disruption (Gemperli *et al.* 2025). The brainstem pedunculopontine nucleus (PPN) is densely targeted by basal ganglia output, and contains a cholinergic population that exogenously targets the thalamus and basal ganglia, presumptively with similar functional correlates contingent on synchrony.

Spike timing dependent plasticity (STDP) in the corticostriatal projection is quirky. In inputs to SPNs, synapses are strengthened that activate in the ~20 ms *after* other inputs, and weakened when their activity *precedes* other inputs (Fino *et al.* 2005, 2008), opposite the patterns normally seen in STDP (Markram *et al.* 1997; Bi and Poo 1998, 2001; Song *et al.* 2000). These unusual dynamics are reviewed in more detail later (§5.4). The arrangement seems tuned to maximize the variety, and the consequent breadth of associativity, of SPN afferents, consistent with the integrative role ascribed to SPNs above. It also stabilizes the availability of potentiated inputs at delays ultimately (polysynaptically) aligned with subsequent cycles of oscillatory activity at lower frequencies, as required for BGMS.

### *2.8. The Big Picture*

There is a long history in science and medicine, even a tradition, wherein a very different picture of basal ganglia function has prevailed. According to this tradition, the basal ganglia are functionally specialized for motor performance, act principally by opening permissive gates in subcortical relay structures, are highly restricted in their capacity for integration, are extraneous to consciousness, and are implicated in oscillatory dynamics only in pathology. It is an interesting question how this perspective gained traction in the first place, and how it spread as far and endured as long as it did. Evidence has mounted over time for a starkly different and multifariously conflicting narrative, one in which the basal ganglia are functionally expansive, nuanced and coherent in their influence on the thalamocortical system, specialized for integration, indispensable to consciousness, and integral to oscillatory coordination in a state of health. With these revisions, the basal ganglia can be seen as a crucial piece in the puzzle of cognition and indeed consciousness itself. Some of the rationale for this conclusion is above, and the rest of it is below.

# 3. The General Nature of Basal Ganglia Direct Path Inputs, Transforms, and Outputs

**In this section:**

### *3.1. Basal ganglia outputs constitute decisions, and only incidentally relay information.*

In this chapter, I review recent and current thinking on the transfer function of the basal ganglia, focusing on the direct path, adding detail to the condensed treatment above, and laying the groundwork for more detailed treatments that follow.

According to the BGMS model, the information represented by a pattern of activation in a particular cortical area passes to other receptive cortical areas chiefly via direct and indirect corticocortical connections between them. Information passing through the BG to the thalamus arrives there in drastically reduced and fragmentary form, fundamentally transformed by the process. While the striatum is continually supplied with inputs that span the entire cortex (Parent and Hazrati 1995a; Hintiryan *et al.* 2016; Peters *et al.* 2021; Grandjean *et al.* 2017), only a small fraction of the information borne by these inputs can emerge from the BG direct path, due to the >1000:1 reduction in neuron count from the corticostriatal population to the output neuron populations in the GPi and SNr (Yelnik 2002; Kincaid *et al.* 1998; Zheng and Wilson 2002; Goldberg and Bergman 2011).

By a similar rationale, noting a 100:1 ratio of visual cortex neurons to pulvinar neurons in macaque, Van Essen (2005) suggested that the associative thalamus itself generally operates in a modulatory role, managing information transfers that are fundamentally corticocortical. This proposition is further supported by results, noted earlier (§1.6), indicating that the thalamic mediodorsal nucleus regulates functional connectivity in PFC rather than acting as an information relay (Schmitt *et al.* 2017).

All of these accounts support the view that the dimensionality at which cognitive control mechanisms operate is vastly smaller than that of representation in neocortex (MacDowell *et al.* 2022; Miller *et al.* 2024). That the basal ganglia might be at the center of such a mechanism is the essence of the reinforcement-driven dimensionality reduction (RDDR) model of Bar-Gad *et al.* (2003), and is an implication of the BGMS model.

### *3.2. Basal ganglia output, and cortical activity patterns, are highly stochastic, implicating populations of neurons.*

Due to the general irregularity and independence of firing patterns in individual BG projection cells, the entropy of the BG paths is substantial (Wilson 2013), suggesting that decisions represented by BG output are highly flexible and can be quite nuanced.

The neurons projecting from the BG to the thalamus are noted for their continual and independent high frequency discharge patterns, averaging ~70 Hz in humans, fluctuating continuously under the influence of intrinsic noise and background synaptic barrage (Brown *et al.* 2001; Stanford 2002; Wilson 2013; Zur *et al.* 2024). This activity must be functionally crucial, given its inherent metabolic burden, simultaneous with remarkable evolutionary stability, spanning hundreds of millions of years and all known vertebrate taxa (Stephenson-Jones *et al.* 2012).

It has been suggested that these signals are particularly suited to act as carriers for motor commands (Brown *et al.* 2001); in the BGMS model these signals act as carriers for control signals spanning all domains. Rapid independent pacemaking activity by BG output cells positions them for agile high fidelity transmission, despite the vastly larger projecting cell populations in upstream structures. Moreover, the transthalamic BG influence on cortical activity is continual, because thalamocortical neurons are themselves tonically active during waking and paradoxical (REM) sleep, with some (e.g. in the rostral intralaminar nuclei) capable of following high frequency (100-300 Hz) spike volleys (Steriade and Llinás 1988; Glenn and Steriade 1982). The intralaminar thalamus also projects directly and comprehensively to the striatum (Sadikot *et al.* 1992b; Kaufman and Rosenquist 1985a; Lacey *et al.* 2007; Sidibé and Smith 1999), so that noise in BG output is recirculated back to BG input.

Because each BG output neuron tonically oscillates at an independent frequency, aggregate tonic BG output statistically resembles Gaussian noise, suggesting that oscillatory modulation (by inputs from the striatum, in particular) can produce output signals with high oscillatory and pulsatile fidelity. This is akin to audio signal dithering techniques that use additive noise with a triangular probability distribution to reduce waveform distortion, in systems that represent intrinsically continuous signals using quantized digital schemes (Lipshitz 1992). Convergence in mammals of several BG output neurons to single thalamocortical neurons (Ilinsky *et al.* 1997), and the multitude of thalamocortical neurons innervating each neighborhood in cortex (Rubio-Garrido *et al.* 2009), comport with such an arrangement. In the BGMS model, the resulting high temporal resolution lets the BG produce aggregate thalamocortical activity that is precisely coincident with converging corticocortical activity. Indeed, computational simulations suggest that convergence and mutual independence in BG output neurons are indispensable for precisely timed BG-induced spike generation in the thalamus, and for the avoidance of spurious spiking (Nejad *et al.* 2021).

### 3.3. *Waveform fidelity is functionally significant.*

These arrangements are closely related to the "stochastic resonance" mechanism suggested by simulations, whereby neural network sensitivity and waveform fidelity may be enhanced by the pervasive injection of noise tuned to effect network criticality (McDonnell and Ward 2011; Vázquez-Rodríguez *et al.* 2017; Krauss *et al.* 2019). There is evidence that oscillatory waveforms in brains are often non-sinusoidal, conforming to various source-specific stereotypes (Cole and Voytek 2017); functional significance has been ascribed to the fine time structure of spike "packets" exhibiting source-specific stereotypes over time spans of 50-200 ms (Luczak *et al.* 2015), and in general, to the information-carrying capacity of dynamic variations in inter-spike intervals (Li and Tsien 2017). To the degree that waveform harmonics and the fine time structure of spiking are functionally significant, waveform fidelity is likewise significant.

Stereotyped non-sinusoidality in cortical oscillatory waveforms, such as the sawtooth waveforms of motor cortical beta oscillations (Cole and Voytek 2017), may facilitate the learning and production of sharply time-coincident spike volleys in striatum (discussed in detail later (§5.6)). But beyond the facilitation of tightly synchronized spike volleys, waveform structure on short timescales might be exploited by the striatum to selectively filter inputs, because the diverse delays of the corticostriatal projection (Yoshida *et al.* 1993; Kitano *et al.* 1998), to which synchronized and converging cortical inputs are subject, in concert with plastic variations in corticostriatal synaptic efficacy, might realize finite impulse response (FIR) filters, engendering preferences that favor some waveforms while disfavoring others. Because of similar arrangements in the cerebellum (Heck and Sultan 2002), it too might realize FIR filters with associated selectivities.

### 3.4. *Oscillations evident in LFP are functionally significant, but often hardly evident in individual neurons.*

In neocortex, individual neurons in a state of wakefulness exhibit almost completely random discharge patterns (Softky and Koch 1993; Stiefel *et al.* 2013). Computational modeling suggests that top-down synchronizing influences on a population of cortical neurons (of the sort exerted by thalamocortical projections, reviewed in detail later (§7)) profoundly impact their aggregate oscillation, evident in the LFP, with highly selective attentional effects, even while individual cells within the population continue to exhibit nearly Poissonian random firing patterns (Ardid *et al.* 2010). Indeed, simulations and evidence suggest that the stochasticity and brevity of synchronies characteristic of biological neural networks result in particularly effective modulation of information flow among the synchronized areas; even brief episodes of synchrony, lasting only a few cycles, may suffice for efficient information transfer, with directionality from phase-leading to phase-lagging areas (Palmigiano *et al.* 2017; Besserve *et al.* 2015). Recent evidence suggests that brief episodes of coherent event-related bursting, amidst a background of random bursting, are associated with working memory operations (Lundqvist *et al.* 2022). More generally, as explored in some detail later (§12.6), noisiness in the brain may crucially aid problem solving.

Behaviorally consequential aggregate oscillatory synchronies, in the absence of significant correlations in the spiking activity of the individual contributing neurons, are apparent in the relationship of the BG to the thalamus. In recent experiments with monkeys, it was found that movement-related LFP oscillations in GPi and its target area in thalamus (ventral lateral, anterior part, VLa) were strongly and likely causally correlated, for the duration of each trial, with a time lag from GPi to thalamus shorter than 10 ms, even while individual neuronal firing patterns in GPi showed little correlation to GPi LFP, and virtually no correlation to LFP in thalamus (Schwab 2016, chapter 5). These results suggest that the neurons discharging synchronously are sparsely embedded within a much larger population of neurons whose discharges are not correlated, or that the LFP synchrony is due to a coherent but weak influence on large numbers of those neurons, or some combination. While this is expected from the known physiology of the GPi, discussed above, and at greater length later (§6.15), it is doubtless methodologically frustrating.

In any case, because the BG form closed loops with cortex and with themselves, they are well-positioned to select and reinforce, or indeed generate and sustain, large scale aggregate oscillations. It is suggestive that even without tunable delays, artificial recurrent neural networks can learn to oscillate at various specific frequencies as a precise function of non-oscillatory input patterns (Sussillo and Barak 2013). And as explored later (§5), the immense diversity of BG path delays suggests that closed-loop circuits through the BG can be readily tuned to prefer particular frequencies. This in itself might be an important selection mechanism (Akam and Kullmann 2010).

### *3.5. The basal ganglia preserve the temporal structure of afferent cortical activity.*

A variety of evidence suggests that the BG process and preserve oscillatory time structure, positioning them to manipulate cortical synchronies. For some time it has been appreciated that cortico-BG circuits in a state of health show synchronized oscillations across the full spectrum of power bands, from the “ultra-slow” (0.05 Hz) to the “ultra-fast” (300 Hz), with robust oscillatory activity in the striatum and STN of alert behaving animals (primate and rodent) that is modulated by behavioral tasks (Boraud *et al.* 2005). In PD patients treated with levodopa, STN oscillation in the high gamma band, starting immediately before and accompanying movement, appears to entrain cortex, with the BG leading cortex by 20 ms (Williams *et al.* 2002; Litvak *et al.* 2012). Coherent response in the STN to auditory stimuli reaches at least to 333 Hz (Hnazaee *et al.* 2024‡), underscoring the remarkable fidelity of signal transmission through the STN. In normal monkeys, task-related beta band oscillations in PFC follow and, according to Granger analysis, are caused by, activity in the striatum; this striatal activity, and that of its targets, sustain a spatially focused phase lock, with no inter-areal delay at beta (Antzoulatos and Miller 2014). In task phases preceding a choice, spiking activity in striatum closely and coherently follows that in PFC, while at the moment of choice, PFC neurons with activity lagging striatal activity show distinct and stronger firing responses than neurons leading the striatum (Banaie Boroujeni and Womelsdorf 2023).

BG output responds quickly to sensory stimuli, accompanies and is sustained during delays, and precedes behavioral responses (Nambu *et al.* 1990). The striatum synchronizes with cortical theta (Berke *et al.* 2004) and gamma (Jenkinson *et al.* 2013; Berke 2009) oscillation, and populations of neurons within each of the successive and parallel nuclei of the BG can synchronize with cortical beta oscillation, each nucleus exhibiting a task-related characteristic phase relationship with cortical oscillation that becomes consistent and precise with task mastery, and is most pronounced at the moment of task-crucial decision (Leventhal *et al.* 2012). Moreover, BG beta synchrony with cortical oscillation associated with a task-relevant sensory cue is established with an entraining phase reset that is sharp and immediate, within tens of milliseconds following presentation of an auditory stimulus (Leventhal *et al.* 2012). At the output stage, *in vitro* evidence suggests that the SNr follows striatal inputs closely at frequencies of 10, 50, and 100 Hz, with a measured spike jitter of 0.21 ± 0.02 ms, and a strong preference for repetitive bursts due to pronounced paired-pulse facilitation (Connelly *et al.* 2010).

### *3.6. GABAergic neurons can precisely control activity in their targets.*

In cortex, GABAergic fast spiking inhibitory interneurons (FSIs) play a dominant role in the induction and control of oscillatory activity in the beta and gamma bands, exerting fine control over phase (Hasenstaub *et al.* 2005). Projections from the BG-recipient thalamus to these cortical FSIs (Delevich *et al.* 2015; Kuroda *et al.* 1998; Rikhye *et al.* 2018; Peyrache *et al.* 2011) provide a path implicating the BG directly in these dynamics. Similarly in thalamus, GABAergic projections from the reticular nucleus (TRN) are believed to be crucial for the induction of the intense, globally synchronized spike bursts constituting sleep spindles (Contreras *et al.* 1997), and extensive BG inputs spanning the TRN (Hazrati and Parent 1991; Shammah-Lagnado *et al.* 1996; Antal *et al.* 2014; Nakajima *et al.* 2019) implicate the BG directly in TRN regulatory mechanisms.

GABA, classically viewed as an inhibitory neurotransmitter, has a biphasic excitatory effect in certain circumstances, as a function both of the intensity of GABAergic release, and of the timing relationship between that release and the post-synaptic activity with which it interacts; GABA activity can either inhibit or enhance NMDA-dependent synaptic plasticity as a function of that timing relationship (Staley *et al.* 1995; Lambert and Grover 1995). Consistent with these *in vitro* and *in vivo* results, computer simulation suggests that synchronized rhythmic activity in cortical FSIs can substantially raise the sensitivity or gain of their pyramidal targets, even to constant (non-rhythmic) current injections (Tiesinga *et al.* 2004).

There is evidence of some of these effects, particularly biphasic activation and entrainment, in the GABAergic innervation of the thalamus by the BG (Goldberg *et al.* 2013; Bodor *et al.* 2008; Kim *et al.* 2017). These effects are particularly accessible to experimental probing in songbirds, where BG-recipient neurons in thalamus exhibit physiological similarity to mammalian thalamocortical cells, but unlike mammals, each receives only a single pallidal/nigral fiber, terminating in a calyx enveloping the soma (Luo and Perkel 1999). Studies in songbird thalamus have found coherent oscillatory entrainment at pallidothalamic terminals, and synchronous post-synaptic oscillation driven by pallidal input in the absence of excitatory presynaptic input (Person and Perkel 2005; Doupe *et al.* 2005; Leblois *et al.* 2009).

### *3.7. BG input to the thalamus is not purely inhibitory.*

Simultaneous phasic intensification of ostensibly inhibitory pallidal and nigral output and activity in their thalamic targets has also been noted (Goldberg *et al.* 2013; Lee *et al.* 2016; Guo *et al.* 2017 “Extended Data Figure 10”). This has several possible explanations, among which are the effects described above, concurrent corticothalamic acceleration, and the actions of dopaminergic, cholinergic, and serotonergic nuclei, which facilitate responsive oscillation, and are integral to BG circuitry (these paths and effects are reviewed later (§10)). It may also be partly explained by coexpression of excitatory neurotransmitters in the pallidothalamic projection, or indeed within the terminal processes of individual axons therein, which could be particularly effective at entraining a target. Indeed, several studies have found a glutamatergic component within the pallidothalamic and nigrothalamic projections (Kha *et al.* 2000, 2001; Conte-Perales *et al.* 2011; Yamaguchi *et al.* 2013; Antal *et al.* 2014).

The tonic level of activity in BG-recipient thalamus is similar to that in cerebellum-recipient thalamus, even though the latter is subject to tonic excitatory input, and the two compartments show no apparent distinctions in cholinergic or TRN innervation (Nakamura *et al.* 2014). This apparent paradox may be explained not only by the effects described above, but by systematic cytological preferences, in which the BG and cerebellum target cytologically distinct thalamic populations, with distinct physiology and connectivity (Kuramoto *et al.* 2009; Jones 2001). However, it seems clear that much of the explanation is in the nature of the BG input itself, given findings explored later (§4.1) that no excess of movement follows from PD treatments in which BG inputs to thalamus are removed (Brown and Eusebio 2008; Marsden and Obeso 1994; Kim *et al.* 2017).

Notably, just as thalamic activity increases simultaneous with increases in GPi activity, GPi metabolism and spiking activity increase simultaneous with activation of the direct path spiny projection neurons (SPNs) in the striatum that target it (Lee *et al.* 2016; Phillips *et al.* 2020), despite similar ostensibly inhibitory chemistry in the striatopallidal projection. Moreover, physiologically realistic modeling suggests that striatal FSI activation, ostensibly inhibiting connected SPNs, increases the firing rates of those SPNs (Humphries *et al.* 2009). These are the relationships needed for oscillatory relay through the successive stages of the BG, and are incompatible with models in which BG actions are limited to inhibition and release. Remarkably, even in the striosomal path through the striatum to the dopaminergic (DA) centers of the ventral midbrain, there is evidence that GABA acts by a non-inhibitory mechanism, with striosomes preferentially encoding reward-predictive cues (as does DA, discussed later (§10.2)) (Bloem *et al.* 2017), while DA cells in the midbrain show coherent theta oscillatory activity associated with task phase (Oberto *et al.* 2023).

### *3.8. Pallidothalamic LFP and unit spiking are a paradox.*

The dynamics of the BG output structures, and their interfaces with the striatum and thalamus, remain among the most mysterious in the vertebrate brain (Goldberg *et al.* 2013; Nambu 2008; Schwab *et al.* 2020). While measurements of LFP in the GP and SN clearly demonstrate coordinated activity associated with ongoing cognition and behavior (Leventhal *et al.* 2012; Mirzaei *et al.* 2017; Schwab 2016, chapter 5), individual spikes generated by experimentally sampled subsets of cells there are seen to be mostly or entirely uncorrelated, both tonically and phasically (Brown *et al.* 2001; Stanford 2002; Wilson 2013; Deister *et al.* 2013; Hammond *et al.* 2007; Nevet *et al.* 2007; Schwab *et al.* 2020; Wongmassang *et al.* 2021).

A plausible explanation for this paradox is that BG activation in a state of health implicates sparse subsets of neurons not only in the corticostriatal population (Turner and DeLong 2000) and striatum (Kincaid *et al.* 1998; Zheng and Wilson 2002), but also in output structures and their targets in thalamus, through sparsely embedded axodendritic trajectories. This is strongly implied by evidence of profound independence and parallelism in spiking there (Schwab *et al.* 2020; Wongmassang *et al.* 2021). Indeed, independence there seems to be fundamental to high fidelity information transfer (Nejad *et al.* 2021), and a breakdown of that independence is thought to be fundamental in Parkinson's disease (Hammond *et al.* 2007; Wilson 2013). Nonetheless, recent electrophysiological evidence demonstrates cofluctuations in SN that are precisely coincident, and highly specific to task parameters, based on recordings and optogenetic manipulations of hundreds of individual neurons in healthy behaving mice (Wang *et al.* 2021). Consistent with an arrangement of phasically active pathways sparsely embedded in a vastly larger population of inactive pathways, Wang *et al.* (2021) found that disruption of SNr activity “only weakly modulated thalamic activity. However, this weak modulation strongly reduced [cortical] selectivity, indicating that nonlinear amplification is involved in the circuit. The basal ganglia possibly function through the thalamus as an external input to modulate [cortical] activity to form discrete attractors.”

# 4. Precision Timing in Motor-Related Basal Ganglia Output

**In this section:**

### *4.1. The basal ganglia are integral to movement, but selective disinhibition is an inadequate model for their involvement.*

It has long been recognized that the BG are integral to movement performance (DeLong and Georgopoulos 2011; Chevalier and Deniau 1990). Selective disinhibition of tonically inhibited motor centers, concurrent with enhanced inhibition of unselected motor centers, is a prominent model for this involvement (Chevalier and Deniau 1990; Hikosaka *et al.* 2000). However, firing rate models do not fully describe the implicated mechanisms (Goldberg *et al.* 2012; Kojima *et al.* 2013).

Various lines of evidence underscore the complexity of these mechanisms. Removal of ostensibly inhibitory pallidal input to thalamus for treatment of Parkinson's disease (PD) does not result in an excess of movement (Brown and Eusebio 2008; Marsden and Obeso 1994; Kim *et al.* 2017), and manipulation of the oscillatory phase of STN stimulation optimizes alleviation of parkinsonian symptoms, without affecting STN unit firing rates (Holt *et al.* 2019). Direct and indirect path activation have effects on activity levels in BG direct path output structures opposite those predicted by the inhibition-release model (Lee *et al.* 2016), and BG direct path output structures and receiving thalamic structures often show simultaneous movement-related rate increases (Schwab *et al.* 2020).

### *4.2. Production of motor behavior implicates the basal ganglia at fine time scales.*

All cortical output to the brainstem and spinal cord arises from pyramidal neurons in L5, whose apical dendrites ascend to L1 (Deschênes *et al.* 1994), apposed directly by the terminals of BG-recipient thalamic projection neurons (Kuramoto *et al.* 2009; Jinnai *et al.* 1987). These appositions are excitatory. The apical and proximal dendritic processes of these pyramidal neurons are thought to interact as a coincidence detector (or “vertical associator”) mechanism, with a window width of 20-30 ms, particularly gating burst generation (Larkum *et al.* 2004; Larkum 2013), and the apical dendrites themselves function intrinsically as coincidence detectors, with a 10 ms half-width window (Williams and Stuart 2002) and the possibility of coincidence detection on time scales an order of magnitude shorter (Softky 1994). This implies that BG facilitation of thalamocortical spiking has temporal specificity as fine as these time scales. Somatic coincidence detection in pyramidal neurons is subject to an even tighter window, ~4 ms (Pouille and Scanziani 2001; Volgushev *et al.* 1998), and evidence is reviewed later (§8.5) suggesting BG alignments at this much finer time scale, particularly implicating the intralaminar nuclei.

### *4.3. Production of motor behavior entails long range oscillatory synchronies.*

Motor performance entails patterns of synchronized activity in motor neurons (Riehle *et al.* 1997; Hatsopoulos *et al.* 1998), though interestingly, these synchronies are not consistently driven by motor neurons. For example, Granger causality analysis of LFPs in sensorimotor cortex suggests that sensory and inferior posterior parietal cortex (PPC) drive sustained beta oscillation in motor cortex during a sustained gesture (maintenance of a hand press) (Brovelli *et al.* 2004). Beta oscillatory synchrony in premotor cortex during delay periods appears to be extrinsically driven; this activity is selective for specific features of the forthcoming gesture, and is displaced by simultaneous bursting immediately before the onset of movement (Lebedev and Wise 2000). The BG are understood to be integral to these phenomena, and recent findings suggest the nature of this integration: The PPC has distinct projections to premotor cortex and striatum, with premotor cortex receiving signals that control movements, while signals to striatum reflect task-historical context, informing decision making (Hwang *et al.* 2019). This arrangement suggests that BG output to premotor cortex in such tasks is likewise driven by input from PPC. According to BGMS, for facilitatory decisions, this input will be phase-locked, and after learning, phase-aligned, to corticocortical inputs from PPC to premotor cortex.

Preparatory and sustained activity in premotor cortex depends crucially on excitatory inputs from the BG-recipient motor nuclei of the thalamus, and activity in those nuclei is likewise largely dependent on activity in premotor cortex (Guo *et al.* 2017). That the BG are directly implicated in these dynamics is suggested by the finding that oscillatory activity in the subthalamic nucleus (STN) couples coherently to gamma oscillations in motor cortex, in apparent preparation for action, with performance improving with increased anticipatory phase coupling (Fischer *et al.* 2020).

Control of gates in these corticocortical connections by the BG appears to depend on consistent rhythmicity in the implicated cortical activity. The delays of paths through the BG (e.g. 50 ± 15 ms via the substantia nigra to the frontal eye field, reviewed in detail later (§5)) are significantly longer than the corresponding corticocortical delays (e.g. 8-13 ms between area V4 in visual cortex and the frontal eye field (Gregoriou *et al.* 2009)), so that trans-BG spike volleys triggered by a given cortical spike volley return to cortex outside the coincidence window of that same cortical spike volley traveling corticocortically. Moreover, activation of the BG is dependent on synchronized cortical activity, due to physiology in the striatum (reviewed later (§6.8)). Thus, the BG can open a corticocortical gate, at the earliest, for the second in a series of synchronized spike volleys.

Volleys from a particular efferent area, in a particular scenario, must have a consistent characteristic time structure, even if only for a spike volley doublet, in order for consistently coincident arrival to be possible (and, as proposed later (§5.6), learnable) for volleys traveling both corticocortically and through the BG to the same target area. Of particular relevance to this mechanism, behavior-correlated spiking in premotor and primary motor cortex has been found to always endure for at least 1 cycle of oscillation, and to often endure for only 1 (Churchland *et al.* 2012), representing the parsimonious spiking pattern for integration with the BG. In sensory systems, there is evidence that responses are organized into structurally stereotyped episodes lasting ~50-200 ms, with finer differences in spike timing and density that consistently represent stimulus dimensions (Luczak *et al.* 2015). This similarly provides for ready integration with the BG.

Sustained activity has also been proposed to be necessary for conscious cognition, entailing “dynamic mobilization” of long range functional networks (Dehaene and Naccache 2001; Dehaene and Changeux 2011). Irreducible delays in BG responses to preconscious cortical and thalamic activity might figure prominently in this dependency.

#### ***4.4. Production of motor behavior can be prevented by a single spike volley directed to a BG output structure.***

That the BG generally facilitate effective connectivity using multi-areally synchronized spike volleys is suggested indirectly by the finding that a solitary, precisely timed spike volley from the subthalamic nucleus (STN) to the substantia nigra reticular part (SNr) can be effective in stopping (preventing) behavior (Schmidt *et al.* 2013). This might be evidence that the disruption of the timing of BG output is enough to abolish its facilitatory effect, so that its timing is implicitly crucial. Along these lines, it was recently proposed that the BG modulate the timing of thalamic output so that it will arrive in cortex at the precise phase instant needed to modulate targeted motor cycles to better match context (Magnusson and Leventhal 2021).

STN axons appose neurons in the SNr throughout their input processes, both somatically and dendritically (Bevan *et al.* 1994; Tiroshi and Goldberg 2019), suggesting that paths through the STN are arranged to strongly modulate other inputs. The precise timing of STN spikes has recently been shown to be more broadly significant. As noted above, STN activity couples to cortical gamma activity: the STN and motor cortex exhibit phase relationships specific to behavioral scenarios, shifting by 180 degrees for ipsilateral versus contralateral gripping, and measures of these phase relationships better predict behavioral performance than do STN mean firing rates, which were found to not change significantly over the course of behavior (Fischer *et al.* 2020).

As suggested in earlier accounts (e.g. Parent and Hazrati 1995b), the STN appears arranged to act as a crossroads, through which activity can spread from localized sectors of the BG to wider areas, implying competition and facilitating the completion of selections. Through its strong and highly divergent appositions on projection neurons in BG output nuclei, the STN might broadly entrain BG output to a winning rhythm, maximizing or indeed minimizing the efficacy of cortical activity converging with BG output. Once a winning rhythm is no longer useful (not contextually appropriate), the STN is positioned to disrupt it throughout the BG.

The intralaminar nuclei of the thalamus also target the STN, and through it, the SNr and its telencephalic homologue, the globus pallidus internal part (GPi) (Sadikot *et al.* 1992a; Parent and Hazrati 1995b). As explored in depth later (§8), the intralaminar nuclei are themselves positioned to distribute oscillatory activity to cortex and BG very broadly and with high temporal fidelity, reflecting the combined effects of interacting inputs arising in cortex and the BG.

# 5. Delay Mechanisms in Basal Ganglia-Thalamocortical Circuits

**In this section:**

### *5.1. Timing structure and frequencies in mammals are highly conserved, while fiber conduction velocities vary widely within and between species.*

Among mammalian species, conduction velocities (CVs) for a given homologous projection vary widely, while alpha, beta, and gamma oscillatory frequencies are roughly constant, despite a 17,000-fold variability in brain volume (Buzsáki *et al.* 2013). Geometrically proportional scale-up of axonal propagation velocities appears to arrange for similar long range delays regardless of size, maintaining the compatibility of circuit synchrony mechanisms with the conserved and intrinsic dynamics of neurons and their microcircuits, with few exceptions (Buzsáki *et al.* 2013; but see Caminiti *et al.* 2009).

Simulations and evidence suggest that small inter-areal phase delays of ~4 ms can be decisive in determining the direction of information transfer in reciprocal long range links, from phase-leading to phase-lagging areas (Palmigiano *et al.* 2017; Besserve *et al.* 2015). Broad and systematic diversity in the delays attending cortical responses to sensory stimuli, apparent in the visual system of the monkey (Schmolesky *et al.* 1998), plausibly allow the BG to bias the salience of a selected dimension of the stimulus, by aligning the phase of BGMS signals to selectively reinforce activity associated with that dimension, and decouple activity associated with other dimensions.

More generally, as noted in the introduction (§1.3), the brain is fundamentally composed of myriad interlocking time-domain processes and mechanisms (Cariani and Baker 2022; Baker and Cariani 2025).

### *5.2. The basal ganglia must accommodate widely varying long range timing requirements.*

The BG are as beholden to the intrinsic dynamics of neurons, assemblies, and circuits, as is the rest of the brain, but according to the BGMS model, they must additionally align their responses to meet the timing requirements in each *learned* combination of scenario, efferent areas, and recipient areas, necessitating enormous spatiotemporal flexibility and precision. Precision appears to be paramount in this function: behavioral proficiency exhibits an inverted-U relation to experimentally manipulated striatal latency, with best skeletomotor performance under natural conditions matching those of the learning epoch, while striatal acceleration and retardation are associated with similar (and significant) performance deterioration (Monteiro *et al.* 2023).

In cortex, collateral targeting of inhibitory fast spiking interneurons arranges for feed-forward perisomatic inhibition of pyramidal cells, and concomitantly narrow (4 ms) coincidence windows (Pouille and Scanziani 2001), while apical dendrites are intrinsic coincidence detectors with a half-width of 10 ms (Williams and Stuart 2002), and the possibility of coincidence detection on timescales an order of magnitude shorter (Softky 1994). Thus, thalamocortical inputs that proximally appose L5 pyramidal cells and related FSIs must be coordinated on the 4 ms time scale, likely implicating the heavily BG-recipient intralaminar nuclei (Parent and Parent 2005), while apical inputs, implicating the entire BG-recipient thalamus, are most effective when aligned at the 10 ms time scale, and may be able to differentially favor a competing input using spike timing effects at the 1 ms time scale.

In the other direction, from neocortex to the BG, structurally intrinsic inter-areal delays (Sorrentino *et al.* 2022; van Blooijs *et al.* 2023; Gregoriou *et al.* 2009; Nowak and Bullier 1997; Schmolesky *et al.* 1998) must be compensated to render coincident the spikes associated with synchronized activity in these distributed areas, as their projections converge on individual striatal cells. Interestingly, patterns of myelination in corticocortical projections arrange to keep delays within a significantly narrower range than is implied by fiber length alone (Sorrentino *et al.* 2022). This keeps inter-areal delays in a range that can be compensated by corticostriatal fibers, as discussed below (§5.4).

Context-specific distributed episodes of neocortical activity conform to stereotyped timing relations, with jitter of only 1-3 ms in many of the repeating firing patterns (Abeles *et al.* 1993). In the representations of stimuli by large (>200 neuron) ensembles, it is fine spike timing that is most stable and informative, not fluctuations of spike rates (Sotomayor-Gómez *et al.* 2025; Zhu *et al.* 2025a ‡). Because of their consistency, these temporal patterns can drive the expression of plasticity, particularly as they pass through the corticostriatal projection, by which they are subject to extensive spatiotemporal rearrangement.

In the BG, the obvious substrate for meeting time alignment requirements is the enormous variety of paths, delays, and time constants of striatal neurons, inputs, and outputs. A multiplicity of paths, exhibiting a multiplicity of delays, may assure that for any two cortical loci, there exist polysynaptic paths to the implicated thalamocortical neurons, exhibiting nearly optimal delays, that need only be strengthened to effect learning of appropriately selective, timed, and directed responses. In this arrangement, the population of projection fibers and synapses from cortex through the BG to thalamus are a vital reservoir, with some portion held in latent reserve for accommodation of future path and delay requirements. Accordingly, evidence suggests that the projection from PFC to striatum is guided by early and precise endogenous growth and guidance factors, and is then held through adulthood, not subject to experientially driven pruning (Mesías *et al.* 2023). This pattern is opposite that seen in the internal circuitry of the PFC, which is subject to extensive axonal and dendritic pruning in adolescence (Riccomagno and Kolodkin 2015). Moreover, the conduction velocity of long distance corticocortical fasciculi roughly doubles from childhood to adulthood (van Blooijs *et al.* 2023), underscoring the need to retain a reserve to accommodate new timing requirements.

The general mechanisms whereby the BG accommodate these diverse timing requirements likely endow them with particularly rich representational power: When similar arrangements in cortex were simulated, an unanticipated result was that the number of distinct ephemeral neuronal assemblies greatly exceeded the number of neurons, and might even exceed the total number of synapses in the network (Izhikevich 2006).

Plasticity mechanisms in the central nervous system are exquisitely sensitive to timing relationships, at time scales of several or even fractional milliseconds within a ±20 ms window, so that in many neurons, faster paths of communication are consolidated, and slower paths are culled (Markram *et al.* 1997; Bi and Poo 1998, 2001; Song *et al.* 2000). Spike-timing-dependent plasticity (STDP) in conjunction with coherent oscillatory activity may build temporally coherent circuits by grouping axons with precisely matching delays (Gerstner *et al.* 1996). However, the relationship of these mechanisms to the delay of polysynaptic BG paths is complicated, given evidence that STDP in striatal SPNs is reversed (Fino *et al.* 2005). Nonetheless, striatal FSIs exhibit typical STDP (Fino *et al.* 2008). The implications of this are explored below.

### *5.3. The basal ganglia must accommodate widely and dynamically varying oscillatory frequencies.*

Particular large scale functional networks exhibit characteristic profiles of prevailing oscillatory frequencies, with the frequencies of oscillatory episodes in a particular area dynamically dependent on the network with which that area is functionally connected during that episode (Keitel and Gross 2016; Becker and Hervais-Adelman 2020; Vezoli *et al.* 2021; Lyu *et al.* 2025), so that reinforcement of particular frequencies in cortex in itself can establish and transiently stabilize corresponding large scale networks characterized by those frequency patterns.

A closely related proposal, mentioned earlier (§1.11) and supported with simulations, is that microcircuits in PFC respond preferentially to afferent activity near their resonant frequencies, which can be modulated by nonspecific, subthreshold excitation of pyramidal cells, with the resulting preferences promoting particular patterns of functional connectivity (Sherfey *et al.* 2018, 2020).

As detailed later (§7.5), the BG, through the associative, motor, limbic, and intralaminar thalamus, are among the subcortical systems positioned to modulate the intensity of apical dendritic bombardment, modulating the resonances of mesoscopic regions of cortex. Coordinated shifts of neuromodulator activity by the BG, also reviewed later (§10), also contribute. Importantly, neuromodulatory projections from the brainstem and basal forebrain project not only to cortex and thalamus, but extensively to the BG. Thus oscillatory acceleration in cortex and thalamus is likely accompanied by acceleration in the BG. This might arrange to preserve the applicability of timing and frequency relationships learned by the BG at widely varying levels of arousal.

In BGMS, the oscillatory frequency of thalamocortical inputs intrinsically matches that of the selected corticocortical inputs, so that tuning of cortical microcircuit resonance for selective corticocortical receptivity implicitly arranges for diffuse proximal thalamocortical inputs (particularly from the intralaminar nuclei) to capture subthreshold membrane oscillation (Richardson *et al.* 2003; Hutcheon *et al.* 1996), defining a preferred phase.

There is evidence that synchronous oscillatory stimulation of direct path output structures such as GPi systematically modulates the amplitude of oscillation in related, indirectly connected structures such as STN, as a function of the phase of oscillations in the output structures relative to that in the related structures; favored phases produce large increases in the amplitude of oscillations in the related structures, while outputs at antiphase to the favored phases attenuate the related oscillations (Escobar Sanabria *et al.* 2020). While these effects may be due to experimental artifacts (antidromic activations) (Escobar Sanabria *et al.* 2020), they suggest that coherent relay of oscillations through the BG can have bidirectionally selective effects, pivoting on the precise timing of BG output. Simulations similarly suggest that the precise timing relationship of stimulus-related to movement-related BG output can be decisive, with thalamocortical rebound facilitation by BG sensory responses in the 40 ms preceding movement-related disinhibition, versus rebound suppression when BG sensory responses follow movement-related disinhibition by 10-40 ms (Nejad *et al.* 2021).

### *5.4. Corticostriatal and striatopallidal delays are long and diverse, so that particular patterns of cortical activation can be focused on particular striatal, pallidal, nigral, thalamic, and cortical targets.*

Corticostriatal fibers exhibit fairly slow average CV, measured to be about 3 m/s in macaque (delay range 2.6 - 14.4 ms), in marked contrast to corticopeduncular fibers, measured to average greater than 20 m/s (delay range 0.75-3.6 ms) (Turner and DeLong 2000). Striatopallidal fibers are markedly slower still, measuring under 1 m/s in macaque (Tremblay and Filion 1989). The typical striatopallidal CV is so slow that at peak spike rate (~80 Hz (Kimura *et al.* 1990)), apparently more than one action potential can be propagating simultaneously on the same axon. Not only are corticostriatal and striatopallidal/striatonigral CVs notably slow, the implicated delays are also highly varied. In a study of macaques investigating paths through the GP (Yoshida *et al.* 1993), delays from the caudate and putamen portions of the striatum to GP averaged 16.5 ± 7.9 ms and 10.4 ± 7.4 ms respectively, and overall delays of the corticostriatopallidal path from motor cortex to GP (identified as the studied cortical area with shortest delay) averaged 15.5 ± 4.2 ms. A subsequent companion study investigating the paths through the SNr (Kitano *et al.* 1998) found even greater delays and variance; delays from the caudate and putamen to SNr averaged 22.8 ± 16.2 ms and 17.9 ± 7.7 ms respectively, and delays from frontal cortex to caudate nucleus averaged 18.8 ± 5.7 ms, ranging from 7-31 ms. Combined delay of the corticostriatonigral path was 39.8 ± 14.8 ms, ranging from 12-90 ms, i.e. nearly a full order of magnitude. Evidence from rat suggests conduction in the ventral pathway, through the nucleus accumbens and SNr to the thalamus, that is even slower than in the dorsal pathways (Deniau *et al.* 1994).

It is significant that CVs are slow and diverse in both the corticostriatal and striatopallidal/striatonigral projections. Locus-specific phase disparities, associated with converging corticostriatal inputs from widely separated but functionally connected loci, can be compensated by distinct conduction delays in their respective corticostriatal projection fibers. Activation of a multi-areal cortical ensemble can then produce spatiotemporally coincident activity at particular striatal FSIs and SPNs, despite phase skews in the ensemble at the cortical level. In essence, the spatiotemporal pattern of activation in cortex is convolved with the function embodied by the corticostriatal projection, so that particular cortical activation patterns are focused on particular cells in the striatum, which can learn to respond to them. Separately and subsequently, the output from SPNs is subjected to slow and diverse CVs in the striatopallidal projection, by which additional phase corrections can be applied to align outputs from converging but phase-skewed striatal neighborhoods, and by which additional delays can be inserted to optimally phase-align BG output as transmitted to the thalamus. By these delays, according to the BGMS model, BG output spikes are temporally aligned to promote thalamocortical activity associated with selected connections, and inhibit competing activity. The BG then promote oscillation at the period characteristic of the activated path delay (obliquely supported by simulations in the parkinsonian brain (Asadi *et al.* 2024)), and recognize and promote that oscillation with the various and specific nonzero inter-areal phase relationships characteristic of a given large scale functional network, by shifting back input from phase-leading areas to align in striatum with that from phase-trailing areas, then in the striatonigral/striatopallidal stage, adding more delay to paths looping back to phase-trailing areas. As noted above, simulations and evidence suggest that such control of inter-areal phases in itself controls the direction of information flow in functional networks (Palmigiano *et al.* 2017; Besserve *et al.* 2015).

As suggested above, diverse striatopallidal delays allow for coactivated SPNs to align their inputs to jointly targeted pallidal and thalamic cells. The primacy of synchrony rather than aggregate rate in these relations likely has profound consequences: fragmentary and spurious activation of BG projection neurons is overwhelmingly likely to produce robustly incoherent and thus ineffectual inputs to the thalamus and other BG targets, assuring that only synchronized and unconflicted outputs can promote thalamocortical activations. These relations, which imply a moderate tolerance for spurious activations, may maximize the versatility of BG projection neurons by dynamic functional pluripotentiality mechanisms of the sort studied in cortex (Izhikevich 2006; Rigotti *et al.* 2013).

As mentioned above, the STDP of SPNs is apparently reversed: synapses are strengthened that activate in the ~20 ms *after* activity in other synapses has induced postsynaptic discharge, and synapses are weakened that bear activity in a similar time window *preceding* discharge (Fino *et al.* 2005). This arrangement seems to systematically maximize the delay of paths through striatum, and it may also tend to maximize the variety, and the consequent breadth of associativity, of SPN afferents. Striatal FSIs show normal STDP relations, tending to minimize delays by strengthening the synapses that bear the earliest activity correlated with discharge (Fino *et al.* 2008). Striatal physiology thus appears to promote dispersion, while minimizing the delay of FSIs, which—as reviewed later (§6.6)—consistently activate before SPNs in their vicinity, and precisely control the timing of SPN discharge through powerful appositions. Selective reinforcement of slow cortico-SPN paths and fast cortico-FSI paths might arrange so that SPNs are rebounding from a GABAergic cortico-FSI-SPN spike volley precisely when that same spike volley arrives at the soma (with greater dispersion) via glutamatergic cortico-SPN paths.

### 5.5. *The indirect path can impart phase advancement relative to the direct path, which allows for systematic entrainment to antiphase.*

The external globus pallidus (GPe), through its projections to the thalamic reticular nucleus (TRN) (Hazrati and Parent 1991; Asanuma 1994; Kayahara and Nakano 1998), has been shown in mice to mediate PFC-directed inattention to distracters (Nakajima *et al.* 2019). In the BGMS model, the BG can mediate inattention by imparting shorter conduction delays via the GPe, entraining distracting or otherwise conflicting thalamic activity such that it is silent during the time windows when target cortical areas, under direct path influence, are receptive to input. The direct and indirect paths relay the same oscillatory activity, often coactivating (Cui *et al.* 2013; Klaus *et al.* 2017), but in the BGMS account these paths are not only spatially divergent, but also have differing delays that place their ultimate targets at or near phase opposition. Striatal ensembles projecting to both paths indeed coactivate synchronously (Klaus *et al.* 2017), as required for the fixing of phase relationships between the two paths. Coordinated coherency in the spiking patterns of SPNs is also an effect of striatal FSIs governing neighborhoods of SPNs (Parthasarathy and Graybiel 1997; Koós and Tepper 1999).

Consistently faster signal conduction through the indirect than the direct path is a fundamental implication of the geometry of striatal projections. Striatal axons reach the GPe at a relatively short distance from their origins, before continuing to the more distant GPi and SNr (Parent *et al.* 1995). As detailed above, striatopallidal projection axons are particularly slow-conducting; in contrast, pallidothalamic projections are fast-conducting, as detailed below (§5.8). Moreover, somatic appositions by striatopallidal fibers are prevalent in the GPe, arranging for rapid signaling, whereas only a third of such appositions in the GPi are of striatal origin, and indeed half of somatically apposed inputs to GPi arise from the GPe (Shink and Smith 1995). In the projection from the GPe to the SNr, somatic appositions are predominant, kinetics are much faster, and average conduction velocity is much (>3 times) faster than those of striatonigral fibers (Connelly *et al.* 2010). Due to the combined effect of shorter length, relatively proximal appositions, faster kinetics, and faster conduction velocity, striatal paths through the GPe presumptively entail significantly shorter total delays than paths through the GPi/SNr, shifting the phase of oscillation in GPe-recipient areas relative to that in direct-path-recipient areas. The length disparity is even greater for the SNr, which might arrange to facilitate phase opposition for beta activity transiting the SNr and GPe, just as the GPi and GPe might arrange phase opposition for low gamma activity. This comports neatly with evidence presented below (§5.9) that the average full loop cumulative delay of the putamen-GP path is roughly one gamma period, while the cumulative delay of the caudate-SNr path is roughly one beta period.

Pathological distractibility in Huntington's disease (Lasker *et al.* 1988) and schizophrenia (Grillon 1990) may be explained by dysfunction in different components of this same combined mechanism. In Huntington's disease, cell loss is most pronounced in the GPe-projecting cells of the striatum (Reiner *et al.* 1988; Deng *et al.* 2004). This deprives the GPe of the inputs that normally drive inattention via the TRN, so that distracter areas of the thalamus are not inhibited during preferred-phase moments in cortical targets. In schizophrenia, cortical coincidence windows are pathologically loosened (Lewis *et al.* 2005; Gonzalez-Burgos *et al.* 2015), so that pyramidal neurons are pathologically receptive to off-phase inputs. Either etiology leads to similar distractibility. Indeed, Huntington's disease, even years before diagnosis, is associated with many of the same deficits as schizophrenia (Duff *et al.* 2010), and like schizophrenia, sometimes involves psychosis (Rocha *et al.* 2018).

The model proposed here naturally relates to other pathways associated with the indirect path. Projections from the GPe to GPi, SNr, and back to the striatum (Parent and Hazrati 1995b; Sato *et al.* 2000) might modulate the phase of activity in output structures to effect inattention and disconnection, suggested by evidence of a race relationship between facilitatory signals in the direct path, on the one hand, and signals effecting cancellation via the indirect and hyperdirect paths, on the other (Schmidt *et al.* 2013). The GPe densely targets the STN, which in turn targets the striatum (Parent and Hazrati 1995b). This projection of the STN selectively targets FSIs (Kondabolu *et al.* 2023), through which it can modulate the phase of oscillatory activity generated in striatal SPNs, as detailed later (§6.6).

STN targeting of the GPi has been proposed to be crucial to the inhibition of competing motor programs, realizing a timing-intensive center-surround arrangement in which gamma oscillations in the surround are entrained by the STN, while oscillation in the selected center is phase shifted under the influence of direct path striatal SPNs whose axons focally converge and interact with the STN inputs (Fischer 2021).

The GPe is significantly larger than the GPi or SNr (Yelnik 2002; Kreczmanski *et al.* 2007; Hardman *et al.* 2002; Oorschot 1996), which comports with a central role in the coordination of inattention, by continually directing signals to all areas *not* relevant to the current context. Task-related reductions in functional connectivity (Ito *et al.* 2020; Cole *et al.* 2021) may relate to task-related coactivation of the direct and indirect paths during action initiation (Cui *et al.* 2013; Donahue *et al.* 2018‡; Oldenburg and Sabatini 2015; Klaus *et al.* 2017), with the indirect path effecting functional disconnections.

### 5.6. Learning and extinction establish and dissolve context-specific synchronous spike responses in the striatum.

In rats trained on a maze task until habit formation, then given extinction training, and finally retrained on the original task, ensembles of SPNs in the dorsal striatum formed, narrowed, and changed their responses to fire synchronously at the beginning and end of the task, then reverted, and finally reestablished their synchronous responses, respectively, with a high correlation of response synchrony to behavioral performance (Barnes *et al.* 2005). Similarly in monkeys, over the course of self-initiated, reward-motivated learning, large numbers of neurons in the dorsal striatum developed phasic responses aligned with the beginning and end of saccade sequences (Desrochers *et al.* 2015). In rat ventral striatum, a shift in the patterns of phasic activity from local islands of high gamma synchrony, to beta synchrony spanning wide areas and both SPNs and FSIs, accompanies skill acquisition and habit formation (Howe *et al.* 2011). These studies provide strong evidence that striatal plasticity entails the formation of widely distributed constellations of FSI-SPN assemblies that learn to discharge in synchrony as a function of context.

### 5.7. Intralaminar thalamus and cholinergic striatal interneurons may be key components of the mechanism whereby the BG learn to generate context-appropriate synchronous output.

Presumed cholinergic interneurons in the striatum, recognized electrophysiologically by their tonic firing patterns, may be key components of a time alignment learning mechanism in the BG. Over the course of skill acquisition, progressively larger proportions of these sparsely distributed interneurons, over very wide areas of striatum, have been seen to pause in brief, precise synchrony in response to salient sensory stimuli, with this response dependent on dopamine supply (Graybiel *et al.* 1994; Ding *et al.* 2010). In general, corticostriatal long-term potentiation (LTP) depends on the temporal coincidence of cholinergic interneuron pause, phasic dopamine activation, and SPN depolarization (Reynolds *et al.* 2022). These interneurons have been implicated in the learning of changes in instrumental contingencies, and that learning is dependent on activity in thalamostriatal projections originating in the intralaminar nuclei (Bradfield *et al.* 2013). Moreover, precisely synchronized stimulation of these interneurons directly induces dopamine release through cholinergic receptors on dopaminergic axons, independent of somatic activation of midbrain DA neurons (Threlfell *et al.* 2012), suggesting that synchronous BG output *per se*, as measured by activity in intralaminar afferents, is intrinsically reinforced in the striatum.

As discussed in greater detail later (§7.12), striatal matrix is extensively and preferentially targeted by inputs from intralaminar thalamus, apposing both SPNs and FSIs. Thus, mechanisms of striatal plasticity are positioned to monitor and respond to the synchronies that the BG generate in thalamus, and so presumptively in cortex. Significantly, evidence suggests that the large scale synchronized spike volleys of sleep spindles, which also arise from the intralaminar thalamus and neighboring nuclei (Contreras *et al.* 1997), may be central in the expression of corticostriatal plasticity (Lemke *et al.* 2021). Dopamine-dependent and dopamine-inducing activity in striatal cholinergic interneurons, innervated by these thalamostriatal projections, might act to strengthen striatal synapses that contribute to the production of synchronous thalamocortical activity associated with reward. Related mechanisms may similarly drive plasticity in other BG structures targeted by the intralaminar nuclei, notably the GP and STN (Sadikot *et al.* 1992a).

### 5.8. Pallidothalamic and thalamocortical delays are short and uniform.

Consistent with the proposition that pallidothalamic and nigrothalamic axons collateralize to orchestrate tightly coherent long range synchronies via the thalamus, the delay of these segments is comparatively short and uniform: in macaques, antidromic response from thalamus to SNr was found to average 1.56 ± 0.44 ms (Kitano *et al.* 1998), and an earlier study (Harnois and Filion 1982), on squirrel monkeys, found similar antidromic delays from thalamus to GPi, tightly clustered about an average of 1.3 ms from ventral anterior (VA) and ventral lateral (VL) sites, and 1.6 ms from centromedian (CM) sites, arising from a CV of 6 m/s. Similarly, as noted earlier (§1.5), the thalamocortical projection appears to be tuned for rapidity and exquisitely precise (sub-millisecond) alignment of the projection to any given area of cortex; selective myelination of the portion of thalamocortical axons within cerebral white matter, the length of which varies two-fold within a target area, appears to account for this (Salami *et al.* 2003).

In macaque, the delay from motor thalamus to the supplementary motor area (SMA) averages 1.75 ± 0.90 ms, in a range of 0.8 ms to 5.0 ms (Kurata 2005). In cats, the delays for antidromic stimulation of thalamocortical projections from VA-VL and ventromedial (VM) thalamic nuclei to Brodmann areas 4, 6, 8, and 5 were found to average from 2.3 ms (VA/VL to area 4, primary motor) to 4.2 ms (VM to area 8, motor association cortex), with almost all measured delays falling below 6 ms, and significant and systematic, but small, shifts in delay as a function of thalamic nuclear origin (Steriade 1995). In the intralaminar nuclei, which in the BGMS model are a crucial broadcast hub for timing information (reviewed in detail later (§8)), antidromically measured thalamocortical delay is less than 500 µs, indicating conduction velocities (CVs) of 40-50 m/s (Glenn and Steriade 1982; Steriade *et al.* 1993).

Curiously, evidence from rat suggests that nigrothalamic conduction in the ventral pathway, through the nucleus accumbens, is relatively slow, with an antidromic delay of 6.1 ± 0.6 ms to the mediodorsal (MD) nucleus and 4.3 ± 0.6 ms to the ventromedial (VM) nucleus (Deniau *et al.* 1994). Nonetheless, the dispersion of these delays is similar to those seen in the nigrothalamic pathway of primates, suggesting BGMS dynamics similar to those in primate, and in the dorsal BG.

### *5.9. Total delay from a corticostriatal neuron, through the basal ganglia direct path to thalamus, back to cortex, is roughly one gamma cycle through GPi, and one beta cycle through SNr.*

In songbirds, an average delay of 5.1 ± 0.48 ms, range 3.5-7.9 ms, has been measured between arrival of a pallidothalamic terminal spike and the next discharge by the targeted thalamic projection neuron (Goldberg *et al.* 2012). Assuming this figure is roughly representative of the figure in mammals, and given the results summarized above from Yoshida *et al.* (1993), Kitano *et al.* (1998), and a thalamocortical delay of 1.75 ± 0.90 ms (Kurata 2005), the total transmission delay for a closed loop through the BG can be estimated. This delay averages about 25 ms (40 Hz) for paths through putamen, GP, and VA/VL to primary motor, and 50 ms (20 Hz) for paths through SNr to a frontal eye field in area 8, with a large range of possible delays, roughly 19-31 ms (32-53 Hz) and 35-65 ms (15-29 Hz) respectively for average ±1 standard deviation.

These relationships suggest that cortical activity routed through the BG and thalamus is typically delayed by a single cycle upon its return to cortex.

Moreover, as discussed above (§5.5), paths through the GPe appear arranged systematically to introduce delays that are shorter than those of the direct path, so that their thalamic targets (particularly via the TRN) are entrained to approximate phase opposition relative to direct path targets. The shortening of the conduction delay is greatest for GPe relative to SNr, with a ~20 ms average delay from striatum to SNr reduced to almost no delay (perhaps <5 ms) from striatum to adjacent GPe, leading to a relative shift of roughly 15 ms, sufficient to shift downstream activity well outside cortical coincidence windows.

### *5.10. Multiple mechanisms might underlie delay plasticity in paths from cortex to thalamus via the basal ganglia.*

Because position along the distal-proximal dimension of dendritic processes introduces a graded delay, thereby altering the phase shift imparted by the inputs upon the neuron's output (Goldberg *et al.* 2007), fine tuning of BG path delays may be possible within the spatially extensive terminal and dendritic processes of corticostriatal (Mailly *et al.* 2013) and striatopallidal (Levesque and Parent 2005) neurons. Striatopallidal axons penetrate perpendicular to the dendritic disks of pallidal output neurons, emitting thin (diameter 100-200 nm), unmyelinated (hence particularly low CV) collaterals parallel to the disks, repeatedly synapsing with the same target neuron (Goldberg and Bergman 2011; Difiglia *et al.* 1982). The diameters and termination patterns of these thin ramifications match those of the similarly positioned parallel fibers of the cerebellar cortex, particularly in the upper molecular layer, where nearly all fibers are 100 to 250 nm in diameter (Sultan 2000).

In the substantia nigra, striatal inputs constitute a large majority of inputs, and appose dendrites at various distances from the soma, with only a small fraction apposing the soma directly, and most apposing small (distal, slow-conducting) dendrites (Bevan *et al.* 1994). Conduction through nigral dendrites entails delays of up to ~12 ms relative to proximally apposed inputs (which predominantly arise from GPe and STN) (Tiroshi and Goldberg 2019), which is a plausible range for phase tuning of direct path outputs, and closely matches the range of phase tuning by parallel fibers of the cerebellar cortex (Heck and Sultan 2002). Moreover, the frequency preferences of networks may respond in a nonlinear fashion to much smaller adjustments of path delays, with synchrony depending strongly on precise matching of delays (Ivanov *et al.* 2019). In short, localized selective strengthening of appositions distributed along the length of dendrites might adjust path delays and associated oscillatory frequency preferences, in natural response to reinforcement.

Axonal CV plasticity, which has only recently been appreciated (Fields 2015), and which is only just being illuminated in its mechanistic particulars (Pajevic *et al.* 2023), might also operate in the BG, likely with peculiar distinctions from its actions in cortex.

Whether optimization of conduction delays is by competition between distinct fiber paths, or between distinct synapses along the same fiber path, reinforcement-driven persistent modulation of synaptic efficacy could optimize not only the output rates (the efficacy with which a particular input evokes an output), but the fine time structure of the outputs, to the degree that reinforcement is a function of fine time structure. Axonal CV plasticity might operate in conjunction with these mechanisms, responding to the same (or to coordinated) reinforcement signals. Moreover, the traversed neurons themselves may exhibit a diversity of intrinsic time constants, similar to an arrangement that has been described in PFC (Bernacchia *et al.* 2011). Indeed, the activation of both PFC and striatal neurons shows a finely graded diversity of delays, though path variety may be the underlying mechanism (Jin *et al.* 2009).

Behaviors must be precisely paced to meet contextual requirements, and the BG are clearly integral to the performance of these behaviors. Moreover, as discussed here, BG circuitry entails diverse time constants and diverse, often lengthy conduction delays. However, the complex delay mechanisms of the BG direct path appear to not be central to the mechanisms underlying precise and variable pacing of overt behavior: patterns of proportional temporal scaling in neural activity, putatively associated with pacing, are likely generated in cortex, and are much less apparent in thalamus (Wang *et al.* 2018).

# 6. The Basal Ganglia as a Flexible Oscillation Distribution Network

**In this section:**

### *6.1. BGMS entails the coherent transmission of cortical spike volleys through the basal ganglia.*

BGMS crucially entails the coherent transmission of corticostriatal spike volleys, through BG and thalamic relays, back to cortex, with routing and delays providing for spatiotemporal coincidence with corticocortical spike volleys associated with the selected effective connections. Above is a discussion of the various mechanisms that might underlie these temporal alignments. Below, I discuss the myriad patterns of convergence and divergence in the BG that underlie the capacity of the BG to distribute spike volleys coherently, widely, flexibly, and specifically.

### *6.2. Activity in a single corticostriatal neuron can influence activity in large expanses of cortex in a single loop through the basal ganglia.*

Divergence in the paths from corticostriatal neurons through striatal fast spiking interneurons and spiny projection neurons, pallidal and nigral projection neurons, and thalamocortical projection neurons, suggest geometric expansion of activity from a single cortical column to a scope encompassing large areas of cortex. Well over $10^9$ cortical neurons might be influenced by the output of a single striatally projecting neuron in cortex. The axonal processes of each corticostriatal neuron distribute sparsely through large regions of striatum, spanning on average 4%, and up to 14%, of total volume, forming on average ~800 synaptic boutons, likely apposing nearly as many distinct striatal neurons (Zheng and Wilson 2002; Parent and Parent 2006).

While more than 90% of striatal neurons are spiny projection neurons, 3-5% are fast spiking interneurons (Koós and Tepper 1999). If corticostriatal neurons innervate SPNs and FSIs with similar preference, this suggests that each innervates on average ~24 FSIs (though there are indications of specialization in corticostriatal targeting of FSIs (Ramanathan *et al.* 2002)). Each FSI projects to ~300 SPNs (Koós and Tepper 1999), each SPN projects to ~100 pallidal neurons (Yelnik *et al.* 1996; Goldberg and Bergman 2011), each pallidal neuron projects to ~250 thalamic neurons (Parent *et al.* 2001), and each thalamic neuron projects to more than 100 cortical neurons (Parent and Parent 2005) (likely far more (Rubio-Garrido *et al.* 2009)). With a cortical neuron population in lower primates of approximately $10^9$ (Herculano-Houzel *et al.* 2007; Azevedo *et al.* 2009), these divergence ratios suggest that a single corticostriatally projecting neuron can influence all of the cortical neurons within the relevant bounds of segregation. This influence is further fortified by intrinsic mechanisms in superficial cortical layers, described later (§7.8), that horizontally spread oscillations.

### *6.3. Afferents from interconnected cortical areas converge in the striatum.*

Striatal activity closely reflects cortical activity (Peters *et al.* 2021; Groot *et al.* 2023). As emphasized earlier (§1.7), interconnected cortical regions systematically converge and interdigitate in striatum, even while the projection of each cortical region diverges in a spotty, widely distributed pattern (Van Hoesen *et al.* 1981; Selemon and Goldman-Rakic 1985; Parthasarathy *et al.* 1992; Flaherty and Graybiel 1994; Hintiryan *et al.* 2016; Hooks *et al.* 2018). Direct path SPNs are preferentially innervated by neurons in cortex that are reciprocally interconnected over long ranges at the single unit level (Lei *et al.* 2004; Morishima and Kawaguchi 2006), and projections from these “intratelencephalic” cortical neurons, with widely separated but interconnected origins, show a particular tendency to converge in striatum (Hooks *et al.* 2018).

Convergence of densely interconnected cortical areas to single striatal FSIs is common; a study in rats found that nearly half of FSIs innervated by primary somatosensory or primary motor cortex receive projections from both (Ramanathan *et al.* 2002). Moreover, FSIs show a significant preference for direct path SPNs, with functional connectivity demonstrated for roughly half of identified direct path FSI-SPN pairs, but roughly a third of indirect path pairs (Gittis *et al.* 2010).

### *6.4. Convergence at a single FSI of afferents from interconnected cortical areas positions the striatum to respond appropriately to the dynamic functional connectivity of those areas.*

As reviewed earlier (§1.3), synchronization of activity in two areas signifies that those areas are functionally connected, and asynchrony or antisynchrony signifies functional disconnection. This prompts the expectation that functionally connected areas projecting convergently to an FSI robustly entrain that FSI, which imparts their shared cortical rhythm to the SPNs it innervates. By this mechanism, effective connections might act through the BG to directly excite further connections, or to inhibit connections, as envisioned by von der Malsburg (1999). On the other hand, when the afferents to an FSI are active but unsynchronized, the FSI is likely arrhythmically activated, imparting an incoherent inhibitory spike pattern to those SPNs, thereby preventing rhythmic discharge. Indeed, as discussed in greater detail later (§7.11), individual cortical cells subject to conflicting synchronies are themselves likely to exhibit arrhythmic spiking patterns (Gómez-Laberge *et al.* 2016). Antisynchronized afferent activity might have similar results, activating the FSI at twice the fundamental frequency, likely imparting a spike pattern to the SPNs that is particularly efficient at inhibiting discharge. In a third mode of operation, afferents to the FSI from one area bear strong oscillatory activity, while other afferents bear significantly weaker activity that may or may not be rhythmic. In this case, the FSI might impart the strong oscillatory activity to the SPNs, while the weak afferent activity has relatively little effect on FSI spiking, so that strong localized cortical oscillation is selected for effective connection to other areas.

### *6.5. Striatal FSIs are tightly coupled to projection neurons in cortex and striatum, and FSI activity is idiosyncratic and independent.*

The physiology of striatal FSIs in normal behaving animals, and their relationships with cortical and striatal projection neurons, are complex, specialized, and nuanced (Berke 2011). The temporal structure of FSI spiking closely conforms to that of afferent activity, aligning precisely with the trough of extracellular afferent LFP, regardless of band (Sharott *et al.* 2009, 2012; Howe *et al.* 2011), though there is also evidence that FSIs can respond preferentially to inputs that align with their ongoing local oscillation (Mohapatra *et al.* 2025). The phasic activation of each FSI is strongly but idiosyncratically related to ongoing behavior, and in particular, is independent of activity in other FSIs (Berke 2008). Nearly half of corticostriatal synaptic inputs to FSIs are robust, apposing somata or proximal dendrites (Lapper *et al.* 1992), and corticostriatal axons commonly form several synaptic boutons targeting a single FSI, indicating selective innervation and stronger coupling (Ramanathan *et al.* 2002). Consistent with the observed idiosyncrasy and independence of FSI responses to cortical activity, synaptic inputs to FSI somata are few, and FSI dendrites are almost entirely devoid of spines (Kita *et al.* 1990)

### *6.6. Striatal FSIs regulate the spike timing of SPN activity, with profound and apparently causal impact on behavior.*

FSI projections to SPNs are robust (Koós and Tepper 1999), but FSI activation has been found to modulate SPN activity, rather than simply inhibiting or releasing it (Gage *et al.* 2010). Natural FSI activity promotes SPN activation, which is reduced if FSI activity is artificially decreased below, or increased above, normal levels (Lee *et al.* 2017). In behaving rats, FSIs and nearby SPNs are simultaneously active in various stages of task learning and performance, at precisely opposite phases, at both beta and gamma frequencies (Howe *et al.* 2011). Simulation of normal *in vivo* conditions in the striatum shows formation of small assemblies of synchronized SPNs, with FSI activation increasing the firing rates of connected SPNs (Humphries *et al.* 2009).

Pharmacological blockade of FSIs in sensorimotor striatum does not substantially change the average firing rates of nearby SPNs, but induces severe dystonia (Gittis *et al.* 2011), demonstrating that FSI regulation of the temporal structure of SPN activity is crucial to normal behavior.

Tourette syndrome is associated with abnormally low density of presumed FSIs in the striatum (Kalanithi *et al.* 2005). Indeed, the cancellation of inapt behaviors has been associated with GABAergic feedback projections from the GPe that selectively target FSIs (Mallet *et al.* 2016; Deffains *et al.* 2016). Pathological sparseness in FSI afferents to an SPN might result in entrainment of that SPN to cortical activity that would normally be inhibited by another FSI. The resulting spurious SPN discharges, phase-locked to localized cortical activity, then might induce spurious reinforcement and connections in cortex, manifesting as tics and other compulsions.

Whereas SPNs exhibit highly heterogeneous responses to dopamine, FSIs exhibit a largely uniform, dose-dependent response to drugs that manipulate DA, reducing firing rate in response to DA antagonism, and increasing it in response to DA agonism; likewise, FSI activity is positively correlated with locomotor activity, while SPN activity shows highly variable relations (Wiltschko *et al.* 2010).

FSIs exhibit significantly lower firing thresholds than do SPNs relative to the intensity of cortical activity; consequently, activation of SPNs is preceded by, and spatially embedded within, an encompassing area of activated FSIs governing their output (Parthasarathy and Graybiel 1997). In a study inducing focused, synchronized activity in primary motor cortex, nearly all (88%) of the FSIs in the center of the zone of striatal activation were activated, and nearly as many (78%) of the FSIs in a penumbra were activated; FSIs showed a robust and disproportionate response, comprising 22% of the responding striatal neuron population, while representing <5% of striatal neurons (Berretta *et al.* 1997).

Each SPN receives inputs from several (estimated 4-27) FSIs (Koós and Tepper 1999), suggesting that FSI recruitment in a striatal neighborhood reliably imparts strong modulatory input to all of the SPNs in that neighborhood. While there is some evidence that FSI prevalence in the striatal population follows a gradient, with highest concentration in the dorsal and lateral striatum and lowest in the medial and ventral striatum (Kita *et al.* 1990; Bennett and Bolam 1994; Berke *et al.* 2004), more recent evidence demonstrates FSI effects and connectivity in VS similar to those in dorsal striatum (Taverna *et al.* 2007; Howe *et al.* 2011), and the appearance of a striatal FSI density gradient may be an artifact of spatially correlated cytological heterogeneity in the FSI population (Tepper *et al.* 2008).

It has been shown in awake behaving rats that the activity of FSIs in the sensorimotor striatum rises shortly before, and peaks during, initiation of behavior reflecting a decision, and that FSI activity precedes that of coactivating neurons in primary motor cortex (Gage *et al.* 2010). FSI activity is erratic and bursty in the resting animal, but transitions to rhythmically regular activity at the moment a cue is presented, and remains regular throughout the delay period until movement execution, at which point the erratic bursting resumes (Berke 2011; Lau *et al.* 2010).

There is some evidence that FSI activity is more prominent early in learning, and subsides as learning progresses (Lee *et al.* 2017). This is congruent with the gamma resonance of FSIs (Beatty *et al.* 2015), and the transition over the course of task acquisition from locally synchronized gamma to widely synchronized beta (Howe *et al.* 2011). These dynamics suggest FSIs might function as a learning scaffold for SPNs.

FSIs are electrically woven together into a loose, sparse continuum by gap junctions (Kita *et al.* 1990; Koós and Tepper 1999), that in simulation modestly encourage synchronization of neighboring FSIs, while modestly damping their activity unless afferent activity is well-synchronized (Hjorth *et al.* 2009). Gap junctions are also thought to be crucial for the regularization of FSI activity during cued delay periods, noted above, in response to transitions of corticostriatal input from random to patterned (Berke 2011; Lau *et al.* 2010). These phenomena further suggest an arrangement in which FSIs operate as a matrix, comprehensively regulating the temporal structure of SPN spike activity, with particular sensitivity to synchrony in the corticostriatal projection.

### *6.7. Striatal FSI physiology facilitates high fidelity relay.*

Striatal FSIs contain parvalbumin, can sustain firing rates of 200 Hz with little or no adaptation, have narrow action potentials (shorter than 500 µs), do not feed back to the inputs of other FSIs, and do not receive inputs from SPNs (Koós and Tepper 1999; Mallet *et al.* 2005; Taverna *et al.* 2007). SPNs and FSIs produce similar inhibitory post-synaptic currents (Koós *et al.* 2004), but FSI inputs to SPNs are directed to somata and proximal dendrites, where they can exert a more decisive and precise effect on the target, whereas corticostriatal inputs to SPNs, and SPN inputs to other SPNs, are directed to distal dendrites (Bennett and Bolam 1994).

SPNs may exhibit a low pass characteristic (Stern *et al.* 1997), so that even while the SPNs are highly sensitive to synchrony in their excitatory afferents (discussed below), the fine timing of the spikes they produce could be determined almost entirely by the FSIs. The influence of FSIs on the SPNs they target entails not only retardation of SPN phase, but phase advancement, through a rebound effect that reduces the firing threshold of the targeted SPN; the effect is most pronounced 50-60 ms after the FSI spike; SPN depolarization is advanced by ~4 ms when FSI spikes reach the SPN 30-70 ms before excitatory afferent spiking reaches the SPN (Bracci and Panzeri 2005).

### *6.8. The projection from cortex to striatal spiny projection neurons is massively convergent, and SPNs fire only when their inputs are substantial and synchronous, reflecting robustly synchronized cortical activity.*

Convergence is physiologically inescapable in paths through cortex, striatum, and pallidum. There are roughly ten times as many pyramidal cells in cortex projecting to the striatum, as there are medium spiny projection neurons, with each SPN afferented by roughly 10,000 distinct cortical neurons (Kincaid *et al.* 1998; Zheng and Wilson 2002). The massive convergence to single SPNs, and their high firing threshold, arrange so that SPNs fire only when their afferent activity is substantial, synchronous, and distributed broadly across dendrites (Carter *et al.* 2007; Zheng and Wilson 2002; Mailly *et al.* 2013). As noted earlier (§1.8), the sensitivity of the striatum to synchrony in its inputs is particularly consequential if striatal output induces synchronies (as in the BGMS model), because the striatum is then positioned to iteratively process information encoded as patterns of synchrony.

Though relatively sparse, direct corticostriatal projections from GABAergic interneurons to SPNs (Rock *et al.* 2016; Melzer *et al.* 2017; Bertero *et al.* 2020), particularly from cortical FSIs, are an additional mechanism positioned to regulate SPN activity. As reviewed in detail later (§8.5), cortical FSIs have been shown to be part of a coincidence detection mechanism with a very narrow window (Pouille and Scanziani 2001). Depending on the pattern of apposition of these cortical FSIs, they might also function like striatal FSIs, precisely controlling the timing of SPN output, as described above.

### *6.9. Striatal projection neuron activation during resting wakefulness is sparse.*

Corticostriatal neurons seldom fire periodically, but rather, their activity is aperiodic but phase-locked to oscillation in their cell membranes (Stern *et al.* 1997); thus their converged input to SPNs can exhibit substantial periodicity, but only when cortical activity is robust and synchronized. The aperiodicity, low spontaneous rates, and narrowly discriminative activity of individual corticostriatal neurons, suggest that few SPNs will be active at a given moment, and many will be silent (Turner and DeLong 2000). In the awake, resting animal, a large majority of SPNs are silent (Sandstrom and Rebec 2003), and some SPNs remain silent even in the awake, behaving animal, with no apparent physiological distinctions to explain the silence (Mahon *et al.* 2006).

### *6.10. SPNs exhibit no frequency preference, but the fine timing of an SPN's activity can be determined by that of a narrow and dynamic subset of its excitatory afferents.*

SPNs exhibit no persistent or membrane-intrinsic frequency preference; rather, the aggregate intensity of afferent activity (simulated *in vitro* by injection of constant current) establishes a firing rate, and the phase of that firing preferentially follows that of afferent components at frequencies near the established rate, with particularly sharp frequency selectivity at beta frequencies (Beatty *et al.* 2015). This signifies that, as the aggregate afferent activity to an SPN increases to and beyond the firing threshold, the phase of that firing will be preferentially determined by progressively higher-frequency synchronized components of that afferent activity. It also suggests a lock-in dynamic in which an activated SPN preferentially follows activity in the network that activated it, as described earlier (§2.5).

### *6.11. SPN receptivity depends on Up and Down states.*

During slow wave sleep and drowsiness, SPNs fluctuate between “Up” and “Down” states, characterized by subthreshold depolarization and hyperpolarization respectively, and they only discharge when in the Up state (Wilson and Kawaguchi 1996; Wilson 1993; Mahon *et al.* 2006; Kitano *et al.* 2002). Impulsive dendritic bombardment, largely that arising from the corticostriatal population, pushes SPNs to the Up state, and quiescence in those inputs returns SPNs to the Down state (Kasanetz *et al.* 2006). Because SPNs can only discharge when in the Up state, patterns of SPN activation reflect the initial pattern of input bombardment, with the associated population of Up SPNs responding to their cortical inputs as long as they are Up, while Down SPNs are silent (Kasanetz *et al.* 2006). With this arrangement, trans-striatal routes can be activated by a single strong impulse, then be held open by weaker activity originating in the same corticostriatal population, passing rhythmic energy from those inputs, sculpted by striatal FSIs, to BG output structures.

However, the functional significance of this arrangement is unclear, because Up/Down bimodality seems to be characteristic of drowsiness, slow wave sleep, and anesthesia, and not of the awake state (Mahon *et al.* 2006). Indeed, evidence that striatal activity in the awake state continually tracks context (Arcizet and Krauzlis 2018; Weglage *et al.* 2021) is consistent with an arrangement in which SPNs are tonically Up in the awake state. In the NREM sleep state, evidence suggests that sleep spindles are crucial for the expression of corticostriatal plasticity (Lemke *et al.* 2021); the Up/Down dichotomy implicitly relates to this dynamic.

As noted above, evidence suggests that FSIs, which have a powerful influence on SPN activity, are themselves preferentially responsive to inputs that align with their ongoing local oscillation, constituting another gating mechanism in the striatum (Mohapatra *et al.* 2025), also relating to the lock-in dynamic described earlier (§2.5).

Curiously, the Purkinje cells of the cerebellum also exhibit Up and Down states, with transitions triggered by impulsive input currents, and discharge only in the Up state (Loewenstein *et al.* 2005). And like striatal SPNs, Purkinje cells receive enormously convergent inputs, respond only when that input is synchronized (Sultan and Heck 2003), and are thought to be key to a BGMS-like mechanism (Popa *et al.* 2013; McAfee *et al.* 2022; Liu *et al.* 2022; McAfee *et al.* 2019).

### *6.12. SPNs are almost entirely independent of each other.*

SPNs exhibit uncorrelated activity even when they are immediate neighbors, suggested to be due to an arrangement in which each SPN receives axons from a unique, sparse subset of corticostriatal neurons (Kincaid *et al.* 1998; Zheng and Wilson 2002; Wilson 2013). This also follows from the firing patterns of direct path corticostriatal neurons, which are highly idiosyncratic (Turner and DeLong 2000). Consistent with these arrangements, simulations suggest that the responses of SPNs have a high signal to noise ratio when correlated inputs converge on single SPNs, but this degrades when multiple SPNs share correlated inputs (Yim *et al.* 2011).

SPN axon collaterals synapse upon the distal dendrites of other SPNs, with each SPN inhibited by up to 500 other SPNs, but these inputs are sparse, weak, unreciprocated, and asynchronous, and do not induce correlated activity among neighborhoods of interconnected SPNs (Tepper *et al.* 2008; Wilson 2013). Instead, their location suggests they interact with excitatory afferents, enhancing the combinatorial power of the corticostriatal and thalamostriatal projection systems.

### *6.13. The striatopallidal projection is massively convergent.*

Convergence in the projections from striatum to the pallidal segments and SNr is inevitable given the further reduction in volume and cell count. In human, volume ratios from the striatum are 12:1 to the GPe, 21:1 to the GPi, 24:1 to the SNr, and 6:1 from the striatum to GPe, GPi, and SNr combined (Yelnik 2002), with cell count ratio estimates of 97:1, 400:1, and 210:1, respectively, for a combined ratio of 57:1 overall, and 138:1 in the direct path (Kreczmanski *et al.* 2007; Hardman *et al.* 2002). In rats, the volume ratios are 7:1, 112:1, and 19:1, for GPe, GPi (EP), and SNr, respectively, for a combined ratio of 5:1, and the cell count ratio estimates are 61:1, 880:1, and 106:1, respectively, for a combined ratio of 37:1 (Oorschot 1996).

Each pallidal projection neuron forms a large 1.5 $mm^2$ dendritic disk perpendicular to incident striatopallidal axons, innervated by 3,000-10,000 SPNs (Yelnik *et al.* 1984; Goldberg and Bergman 2011). The dendritic processes of projection neurons in the SNr have variable forms, with an extent similar to that of pallidal dendrites, resulting in similar convergent innervation by SPNs (François *et al.* 1987). These arrangements imply a notional convergence ratio from corticostriatal neurons, to SPNs, to pallidal projection neurons, as high as $10^8$. There is only slight convergence in the pallidothalamic projection, but extensive convergence in the thalamocortical projection (Rubio-Garrido *et al.* 2009) suggests a notional convergence ratio substantially greater than $10^9$ for the full loop back to cortex.

The striatopallidal projection, like the corticostriatal projection, also entails divergence. Each SPN axon forms 200-300 synapses, sparsely distributed through a large volume of pallidum, with 1-10 synapses formed with a given pallidal dendrite (Yelnik *et al.* 1996; Goldberg and Bergman 2011), implying that an SPN projects to 20-300 pallidal neurons. Moreover, with a tracer injection in the striatum encompassing a small cell population, a hundredfold increase is seen in the volume of pallidum labeled by the tracer, while larger injections increase the density, but not the volume, of the labeled area (Yelnik *et al.* 1996), confirming an arrangement of simultaneous, extensive divergence and convergence like that of the corticostriatal projection.

Experiments in primates show that the projection from striatum to the GPi entails reconvergence, such that divergence in the projection from a cortical locus to multiple loci in the striatum is followed by convergence from those striatal loci to a single pallidal locus (Flaherty and Graybiel 1994). Graybiel (1998) suggested that the striatum thus acts as a dynamically configurable hidden layer. The BGMS model further proposes that this arrangement subserves selection and activation of effective connections in cortex. The path from a cortical locus, by diverging to many distinct FSI neighborhoods, then reconverging to a single pallidal locus, can be subjected to any of a variety of spike timings, representing a variety of candidate effective connections, while suppressing action through that pallidal locus when multiple SPNs impart conflicting activity upon it, as suggested above.

### *6.14. Striatal input to BG output neurons controls their timing, and coherence in afferent activity to BG output neurons may be crucial to their effective activation.*

Experiments *in vitro* demonstrate that striatal afferents to GP can control the precise timing of firing by the targeted cells, and that these cells can follow striatal oscillatory inputs up to the gamma range (Rav-Acha *et al.* 2005; Stanford 2002).

The relationship of FSIs to SPNs may help elucidate the role proposed for the BG in competitive selection (Redgrave *et al.* 1999). An output neuron in GPi or SNr bombarded by mutually incoherent SPNs would be incapable of imparting a coherent temporal pattern to its thalamic targets. If coherence in this path is crucial, as suggested by the BGMS model, then only those GPi/SNr neurons with predominantly coherent activity in their afferents can participate in activation of a motor or cognitive connection, while activation of clashing connections tends to be suppressed.

GPi activity increases when direct path SPNs are activated, and decreases when indirect path SPNs are activated, even while direct and indirect path activation are associated with widespread cortical activity increases and decreases, respectively (Lee *et al.* 2016). This suggests that oscillatory modulation and facilitation by spike-timing-dependent gain is among the core functions of the GABAergic output neurons of the BG direct path.

### *6.15. Direct path output neurons are tonically and phasically independent in the normal brain.*

As discussed earlier (§3.2), spike timing within, and activation of, pallidal output neurons is almost completely independent in the healthy brain (Nini *et al.* 1995; Nevet *et al.* 2007; Stanford 2002), so any coordination or synchronization must be sparsely distributed. Preliminary results from a study of functionally connected cortical, pallidal, and thalamic areas, demonstrates as much, with strong phasic correlation of LFPs but almost no correlation of individual neuron spiking with those LFPs (Schwab 2016, chapter 5). According to the BGMS model, broadly and densely synchronized BG output would produce spurious and pathologically persistent effective connectivity in cortex, arresting task progress. Indeed, the independence of BG output neurons breaks down in parkinsonism, in which task progress is retarded or arrested (Stanford 2002; Wilson 2013; Deister *et al.* 2013; Hammond *et al.* 2007; Nevet *et al.* 2007).

PD has been shown in magnetoencephalography (MEG) studies to be associated with progressively greater functional connectivity in cortex, determined by measures of synchronization likelihood, both intra- and inter-areal, in both the alpha and, in moderate and advanced disease, the beta bands (Stoffers *et al.* 2008). In fMRI, the effects of PD are particularly evident in measures of *directed* functional connectivity (Mijalkov *et al.* 2022). PD is associated with global disruption and progressive inefficiency of functional connectivity (Olde Dubbelink *et al.* 2014), and with the gradual emergence of psychosis as the disease progresses (ffytche *et al.* 2017). MEG evidence further suggests that a key consequence of pathological synchrony in parkinsonism is a contracted repertoire of functional constellations, progressively associated with deficits of cognitive and behavioral flexibility (Sorrentino *et al.* 2021). Pathological synchrony in parkinsonism may be rooted in the striatum: *in vitro* experimentation with, and physiologically realistic simulation of, the dopamine-depleted striatum has demonstrated spontaneous pervasive formation of clusters of synchronized SPNs (Humphries *et al.* 2009).

### *6.16. The pallidothalamic projection is powerful, and its constituent axons are especially divergent.*

The main BG output projections from GPi and SNr appose neurons in the motor and association thalamus in giant inhibitory terminals, with multiple synapses, exerting powerful and precise inhibitory control of individually targeted cells, with GPi and SNr projections well-compartmented from each other (Bodor *et al.* 2008). Many of these projection neurons contain parvalbumin, with especially high density in the dorsal GP (Cote *et al.* 1991).

Pallidal axons branch extensively within thalamus, into 10-15 collaterals with highly confined terminal varicosities (Parent *et al.* 2000), so that each pallidal neuron projects to 200-300 neurons in thalamus; these are the most widely arborized neurons in the BG (Parent *et al.* 2001). The somata and primary dendrites of GPi- and SNr-recipient thalamocortical neurons outside the intralaminar nuclei are contacted almost exclusively by these afferents, with very dense terminal processes, suggesting that the GPi and SNr exercise predominant control over activity in these cells (Kultas-Ilinsky and Ilinsky 1990; Ilinsky *et al.* 1997; Bodor *et al.* 2008). As discussed in detail later (§7.4), BG projections to the intralaminar thalamus do not follow this pattern, but instead predominantly appose small and medium dendrites (Sidibé *et al.* 2002).

In the pallidothalamic projection, a degree of convergence on single thalamocortical cells has been noted (Ilinsky *et al.* 1997). As discussed length earlier (§3.2), convergence of independently oscillating pallidal projection neurons produces aggregate input that bears the characteristics of Gaussian noise, so that pallidal output can reflect inputs to pallidum with high fidelity. Perhaps this is a crucial advantage that evolutionarily stabilizes this arrangement in mammals, which have a surficial cerebral cortex and closely aligned thalamocortical conduction delays (Salami *et al.* 2003; Steriade 1995), and exceedingly narrow cortical coincidence windows due to feed-forward inhibition (Pouille and Scanziani 2001), while birds lack a surficial pallium, their thalamic projection cells receive only a single calyceal BG input, and their thalamic response to BG disinhibition entails a 40-60 ms latency (Luo and Perkel 1999).

Given the high typical tonic discharge rate of pallidal and nigral neurons, thorough disinhibition or entrainment of targeted thalamocortical neurons requires coordination of multiple pallidal projection neurons. As with the convergence of several corticostriatal neurons on a single FSI, several FSIs on a single SPN, or several SPNs on a single pallidal projection neuron, this suggests several activation scenarios. If all BG output neurons targeting a thalamocortical projection neuron are phasically silenced coincidentally, their common target would likely be activated directly by rebound, and thereafter would tend to be activated synchronous with corticothalamic input. If instead those several inputs remain active, but are phasically synchronized, this would likely entrain their common target, even in the absence of corticothalamic input, and with great vigor in the presence of excitatory input exhibiting the favored frequency and phase. In a third scenario, some of the inputs are phasically silenced, coincident with others that are phasically synchronized, with the likely result that their common target is entrained by the active BG inputs. Until one of these coordinated arrangements is learned, modulation of one or several GABAergic afferents likely has lesser but significant post-synaptic effects, which can bootstrap learning in paths associated with the other afferents.

The dynamic in the intralaminar thalamus is somewhat different: because pallido- and nigrothalamic terminals there predominantly appose dendrites, their likely effect is to modulate receptivity to particular corticothalamic inputs with which they share a dendrite, so that only those inputs with the selected timing (frequency and phase) can affect the soma and, consequently, affect the targeted L3 and L5 pyramidal neurons (some of which are apposed somatically by these projections). As explored later (§8.6), this arrangement may maximize combinatorial power in the relationship of the BG to the intralaminar thalamus.

### *6.17. Basal ganglia influence on cortex can be viewed as biasing the probabilities that functional connections will be established or continued.*

The various BG activation scenarios and pathways can be viewed as biasing the probabilities that various selected functional activations and connections will be established or continued, as suggested by Plenz and Aertsen (1994) ("the task of tuning the transitional probabilities between (sets of) cell assemblies"). When BG input to a thalamic area is thoroughly synchronized, that biasing is strong, whereas partial synchrony produces weaker biasing. If an SPN activation is a vote for the establishment or continuation of some family of functional connections and activations, then the GPi/SNr, BG-recipient thalamus, and BG-recipient neuromodulatory centers of the midbrain and basal forebrain, are positioned to tally those votes, due to convergence. Thus, cortical connectivity decisions that are broadly and consistently supported within the striatum (particularly, habits) are strongly biased, while narrowly supported decisions weakly bias connectivity, and conflicts are akin to mutual vote cancellation. Breadth and consistency of support is representative of decision confidence. And because of divergence in the BG, activation of an SPN can simultaneously contribute to strong biasing of some connections, but weak biasing of others.

### *6.18. The BG-recipient thalamic projection to cerebral cortex is massively divergent.*

The path from BG-recipient thalamus back to cortex exhibits striking divergence. As reviewed later (§7.5), neurons of the rat VL, VA, and VM nuclei project profusely and with massive overlap to L1, with individual neurons collateralizing to widely separated areas (Rubio-Garrido *et al.* 2009; Kuramoto *et al.* 2009), and the intralaminar nuclei in all common laboratory mammals innervate nearly the entire cerebral cortex (Van der Werf *et al.* 2002; Scannell 1999). Centromedian and parafascicular (PF) axons that reach cortex arborize diffusely and widely, with a single axon from CM forming on average over 800 synaptic boutons in cortex, a count that may miss many poorly stained axonal processes in L1 (Parent and Parent 2005). Even considering the possibility of extensive multiple terminations on the same target neuron, which is relatively unlikely in a diffuse projection, the number of cortical neurons innervated by a single BG-recipient thalamocortical neuron might be conservatively estimated at >100. As estimated above, the cumulative notional divergence ratio from a single corticostriatal neuron, through striatal FSIs, SPNs, pallidal projection neurons, and thalamocortical neurons, may exceed $10^9$. Neurons are targeted throughout the cortical column, with a bias for superficial layers (Middleton and Strick 2002; Clascá *et al.* 2012; Markov and Kennedy 2013), and include both pyramidal neurons and GABAergic interneurons (Delevich *et al.* 2015; Kuroda *et al.* 1998; Cruikshank *et al.* 2012). These appositions may themselves be differentially implicated in learning: in auditory learning, for example, it is the apical dendrites of L2/L3 pyramidal neurons that show enhanced responses after learning, while the basal dendrites do not (Godenzini *et al.* 2022).

# 7. The Cortical Origins and Targets of the Direct Path, and Associated Nuclei and Territories of the Thalamus

**In this section:**

### *7.1. Direct path output is directed to most of the cerebral cortex, emphasizing frontal areas but including posterior areas.*

In the BGMS model, it is through the direct path that signals arising in cortex are focused, selected, and coherently distributed, establishing long range connections implicating the areas originating the selected activity. The circuitry and scope of the direct path are thus of paramount importance in the model.

Large areas of the motor, limbic, association, and intralaminar thalamus are BG-recipient (Haber and Calzavara 2009; Groenewegen and Berendse 1994; Sidibé *et al.* 2002; Smith *et al.* 2004; Mandelbaum *et al.* 2019; Phillips *et al.* 2021), projecting to feedback-recipient layers in cortex, particularly L1, L3, L5, and L6 (Clascá *et al.* 2012; Markov and Kennedy 2013). The primate pulvinar extensively influences visual cortex (Saalmann *et al.* 2012, 2018 ‡ ; Fiebelkorn *et al.* 2019), and is directly influenced by the superior colliculus (Stepniewska *et al.* 2000; Wurtz *et al.* 2005), itself systematically BG-recipient (Hikosaka and Wurtz 1983; Parent and Hazrati 1995a).

Direct path targets include the near entirety of frontal cortex, with some variation by phylogeny, via thalamic mediodorsal (MD), ventral anterior, ventral lateral, ventromedial, and ventral posterolateral *pars oralis* (VPLo) nuclei (Middleton and Strick 2002; Sidibé *et al.* 1997; Haber and Calzavara 2009; Sakai *et al.* 1996; Herkenham 1979; Phillips *et al.* 2021). In primates, agranular insular and anterior cingulate cortex are targeted via MD (Ray and Price 1993), and additional targets include high-order visuocognitive areas in inferotemporal cortex via the magnocellular part of VA (VAmc) (Middleton and Strick 1996; Saint-Cyr *et al.* 1990), and anterior intraparietal (Clower *et al.* 2005) cortex. These latter two areas are at the end of the ventral and dorsal visual streams, respectively, proposed to be associated with perceptual cognition relating to physical objects (Baizer *et al.* 1991), and have been found to show choice-predictive beta band synchronization in a 3D-shape discrimination task (Verhoef *et al.* 2011), plausibly implicating the BG directly (Leventhal *et al.* 2012). In chimpanzee, limited experiments have demonstrated projections from the MD, VA, and VL nuclei to posterior cortical areas 19 and 39 (Tigges *et al.* 1983). In humans, the dorsal and ventral visual streams in these areas are densely interconnected (Takemura *et al.* 2016), suggesting a role for BG-mediated inter-stream effective connectivity control in the middle stages of visual processing.

Extensive but diffuse projections from BG-recipient intralaminar nuclei (centromedian, parafascicular, paracentral (PC), and central lateral (CL)), discussed in detail later (§8), have been found to reach nearly the entire cerebral cortex, and furthermore project intrathalamically and to the basal forebrain (Kaufman and Rosenquist 1985b; Scannell 1999; Van der Werf *et al.* 2002).

Consistent with these widespread projections of BG-recipient thalamus, artificial stimulation of direct path SPNs in the striatum significantly and consistently increases activity throughout the entire cortex (Lee *et al.* 2016).

### *7.2. The thalamus is sharply divisible into exclusive zones of influence, each associated with a particular coherent topographically projecting control structure.*

The thalamus exhibits a natural parcellation into zones of predominant influence, each associated with a single subcortical structure. In the BGMS model, this is understood to arrange for parcellation into coherent timing domains, even while temporally non-disruptive inputs converge and overlap to facilitate integration.

Percheron *et al.* (1996) show how the primate motor thalamus can be divided into fully separable afferent territories associated with the cerebellar nuclei, the GPi, and the substantia nigra, noting particularly that the boundaries they describe are not primarily cytoarchitectonic, but rather, are associated with topographic territories and origins. Nonetheless, there is substantial correspondence between parcellations by connectivity and traditional cytoarchitectonic parcellations.

The VA is in the zone of the SNr, the VL *pars oralis* (VLo) is in the zone of the GPi, and the VL *pars compacta* (VLc) is in the zone of the cerebellum (Kuramoto *et al.* 2009). The mediodorsal nucleus is also divisible into territories — in the rat, the GPi (entopeduncular nucleus) is primarily associated with the MD *pars lateralis* (MDl), the SNr and VTA primarily with the MD paralamellar part (MDpl), and the ventral pallidum (VP) and basal forebrain primarily with the MD *pars medialis* (MDm) (Groenewegen *et al.* 1991; Groenewegen 1988; Mitchell and Chakraborty 2013).

Even in the intralaminar nuclei, long considered non-specific (Herkenham 1986), there is evidence that the GPi and SNr have complementary territories (Sidibé *et al.* 2002; Van der Werf *et al.* 2002). Indeed, while their cortical projections are diffuse, the intralaminars have topographically specific projections back to the entire striatum, which provide for segregation into associative, limbic, and somatosensory domains (Mandelbaum *et al.* 2019; Sadikot *et al.* 1992a; Sidibé *et al.* 2002), and corticothalamic inputs to the intralaminar thalamus are topographically organized, with cortical and connected intralaminar thalamic areas co-targeting striatum such that domain segregation is maintained (Mandelbaum *et al.* 2019).

The segregation of cerebellar from pallidal and nigral territories is distinct (Kuramoto *et al.* 2009), though more recently, neurons in a territory at the border between the VL and ventromedial (VM) nuclei of the mouse have been shown to be co-targeted by the SNr and the cerebellum (Roth *et al.* 2024‡). If the cerebellum is involved in the control of forebrain functional connectivity (McAfee *et al.* 2022), but the precise timing and periodicity of glutamatergic output from the deep cerebellar nuclei is not controlled by those of its inputs (McAfee *et al.* 2022), then the convergence of that output with the GABAergic output of the SNr can synergistically generate vigorous thalamocortical output with spike volleys timed for optimal effect on downstream structures.

The thalamic nuclei associated with the hippocampal system are particularly well-segregated, chiefly implicating the midline nuclei (Van der Werf *et al.* 2002) and the anterior nuclear group and lateral dorsal nucleus (Saunders *et al.* 2005).

The compartmentation motif extends even further, encompassing the entire thalamus. The superior colliculus has a territory consisting of the pulvinar and lateral posterior nucleus (Stepniewska *et al.* 2000; Schmidt *et al.* 2001). The sensory inputs have exclusive territories -- primary visual the lateral geniculate nucleus, primary auditory the medial geniculate nucleus, and primary somesthetic and gustation the ventral posterolateral and ventral posteromedial nuclei.

According to the “binding by synchrony” (BBS) theory (von der Malsburg 1999; Singer and Gray 1995; Womelsdorf *et al.* 2007; Jia *et al.* 2013; Barth and MacDonald 1996; Siegel *et al.* 2008), fine timing information is representationally crucial. The strict segregation in the thalamus of the BG direct path from the primary sensory pathways is a joint prediction of the BBS and BGMS theories (see later (§11.2) discussion). In primary sensory cortical areas, modeling suggests that attention modulates the dynamics of cortical assemblies to promote spontaneous synchronous activity in response to input stimuli, with synchrony promoting subsequent routing through target areas (Schünemann and Ernst 2023 ‡ ). Millisecond-scale derangement of spike timing in primary sensory cortex, preserving average firing rate over a longer one second time window, substantially reduces perceptual performance (Quintana *et al.* 2024‡).

Vestibular sensation is an interesting exception. Ascending inputs from the vestibular periphery target many thalamic nuclei, overlapping other inputs, providing for the integration of vestibular with other sensory information at the thalamic level (Wijesinghe *et al.* 2015). Overstimulation of the vestibular sense results in general and even overwhelming disorientation (Bronstein *et al.* 2013), consistent with the exceptionally broad targeting of these ascending signals.

Sensory inattention—the deliberate disruption of sensory input—involves paths through the striatum, GPe, and thalamic reticular nucleus, to the lateral and medial geniculate nuclei (Nakajima *et al.* 2019). The implications of this arrangement were explored in depth earlier (§5.5).

### *7.3. The laminar hierarchical architecture of the cerebral cortex implies a role for the BG in orienting attention, biasing competition, and other roles associated with corticocortical feedback projections.*

The laminar functional architecture of cortex, and the layer-specific targeting of corticocortical projections, can help explain the functional relationship of the BG to cortex. An arrangement of cortical areas in primates has been described in which hierarchies are defined by long range feedforward and feedback relations, with characteristic laminar origins and destinations that grow more distinct as hierarchical distance grows (Barone *et al.* 2000). In primates, feedforward projections tend to arise from L5 and deep L3, adjacent to L4 (the middle granular layer), and project to L4, while feedback projections arise from L6 and upper L3, and project to L1/L2, upper L3, and L6 (Barone *et al.* 2000; Markov and Kennedy 2013).

The activity borne by hierarchical projections tends to conform to stereotyped oscillatory frequency bands, with gamma and theta in the feedforward direction and beta in the feedback direction; feedback activity can enhance feedforward activity (Bastos *et al.* 2015a; Bressler and Richter 2015; Richter *et al.* 2017; Lee *et al.* 2013). The functions ascribed to feedback projections include attentional orientation by biasing competition, and disambiguation and hypothesis-driven interpretation of high resolution feedforward inputs ("biasing inference"), while feedforward projections introduce environmental state information into hypothetical representations (models), promoting rectification of their inaccuracies and inadequacies (Markov and Kennedy 2013).

Recent evidence and analysis indicates that activity in the alpha and beta bands is more prominent in deep cortical layers (L5/L6) than is gamma band activity, and mediates top-down control and inhibitory mechanisms, while the gamma band is more prominent in superficial layers (L2/L3), and underlies bottom-up inputs and the local maintenance of working memory representations (Bastos *et al.* 2018; Lundqvist *et al.* 2018a; Miller *et al.* 2018; Mendoza-Halliday *et al.* 2024; Chen *et al.* 2025 ‡ ; Sennesh *et al.* 2025 ‡ ). Consistent with these laminar specializations, evidence suggests that neurons in L5 have broad receptive fields, while narrower receptive fields predominate in L2/L3, providing for extensive combinatorial coverage of stimulus dimensions (Xie *et al.* 2016; Li *et al.* 2016). As discussed in greater detail later (§8.5), the prominence of lower band activity in deep layers may be significant for inputs to those layers from intralaminar thalamic nuclei, which appear arranged for temporally precise but spatially diffuse distribution of spike volleys. Such inputs are likely to be particularly selective when interacting with lower frequency activity. There is, however, also evidence from human stereo-electroencephalography that long range phase coupling at low frequencies (<30 Hz) is preferentially localized to superficial layers, while very high gamma activity (> 100 Hz) correlations are localized to deep layers (Arnulfo *et al.* 2020), suggesting that temporal precision in thalamocortical projections to deep layers may be crucial for these paths to match phases into the high gamma frequency range.

Cortical inputs to striatal matrix, including the direct and indirect paths, arise from L3 and L5 (Gerfen 1989; Kincaid and Wilson 1996; Reiner *et al.* 2003), like corticocortical feedforward paths. This suggests that the BG receive detailed, highly specific, environmentally and contextually informative inputs. However, BG-recipient thalamocortical axons terminate chiefly in L1, L3, deep L5, and L6, and completely avoid L4 (Kuramoto *et al.* 2009; Jinnai *et al.* 1987; Parent and Parent 2005; Kaufman and Rosenquist 1985a; Berendse and Groenewegen 1991). Thus, the termination pattern of BG-recipient thalamus in cortex is like that of corticocortical feedback paths, consistent with the putative role of the BG as a modulator and selection mechanism, and suggests a key role in hypothetical modeling of the environment. Corticostriatal inputs, while originating in the same cortical layer as cortical projections to the pyramidal tract, are physiologically and functionally distinct, consistent with specialization for highly nuanced selection; in particular, corticostriatal projections more prominently represent context, and activate far more sparsely (Hwang *et al.* 2019; Turner and DeLong 2000; Bauswein *et al.* 1989; Lei *et al.* 2004; Morishima and Kawaguchi 2006).

The cortical input to BG-recipient thalamus arises both from L6, and from collaterals of motor output to the brainstem and spinal cord arising from L5 (Deschênes *et al.* 1994), pooling afferents whose origins resemble those of corticocortical feedback and feedforward projections. This might have consequences for interactions with BG output, discussed later (§7.11). Also discussed later (§7.12), projections from the thalamus back to striatum relay this pooled input, which intermingles convergently with the more plentiful cortical inputs.

Curiously, the deepest pyramidal neurons of the neocortex, those of layer 6b, are mostly driven by long range intracortical projections originating in layers 5 and 6a, and are almost devoid of thalamocortical (and therefore of BG) inputs, even while some of them strongly target the thalamus, through which they may form large scale thalamo-cortico-cortical circuits (Zolnik *et al.* 2020). It nonetheless seems likely that these neurons are targeted by diffuse intralaminar thalamic projections (Parent and Parent 2005), subjecting them to BG influence by spike-timing-dependent gain.

Some projections of the PFC to posterior areas have been found to resemble those of the BG-recipient thalamus, terminating most densely in L1 and avoiding L4, in the pattern of feedback projections (Selemon and Goldman-Rakic 1988). Transthalamic paths from cortex through "core" neurons in the thalamus, on the other hand, have been found to originate chiefly in L5, and to terminate in L4 and deep L3 (Jones 2001; Rouiller and Welker 2000), like corticocortical feedforward paths. Similarly, input to the cerebellum from the neocortex arises from deep L5 (Glickstein *et al.* 1985; Schmahmann and Pandya 1997), and its output targets the core population in thalamus and, through them, middle layers in cortex (Kuramoto *et al.* 2009; García-Cabezas and Barbas 2014), so that many paths through the cerebellum also resemble corticocortical feedforward paths.

### *7.4. BG output in the normal brain is modulatory.*

Movement is difficult or impossible to evoke by electrical stimulation of BG-recipient loci in the motor thalamus, even while such movement can be readily evoked from nearby loci receiving cerebellar output (Vitek *et al.* 1996; Buford *et al.* 1996; Nambu 2008; but see Kim *et al.* 2017). As noted above, cerebellum-recipient cells project mainly to middle cortical layers, while BG-recipient cells project mainly to deep and superficial layers. In primate, BG-recipient neurons in the motor thalamus may be consistently within the calbindin-positive population, associated with widely distributed and divergent cortical modulation, while cerebellum-recipient neurons are consistently within the parvalbumin-positive population, associated with specific and narrowly circumscribed topographic projections (Jones 2001; Bodor *et al.* 2008; Kuramoto *et al.* 2009).

While movement can be evoked by microstimulation of the intralaminar nuclei (Schlag *et al.* 1974), BG input there, as in the nigrotectal projection, is mostly directed to the dendrites of projection neurons (Sidibé *et al.* 2002; Behan *et al.* 1987), where it can only modulate other, excitatory inputs. This contrasts with calyx-like BG terminals in the motor thalamus, that exercise predominant control over their targets, and can directly induce rebound firing and bursting (Bodor *et al.* 2008; Kim *et al.* 2017).

### 7.5. *The BG target apical dendrites in cortex, through thalamic projections with mesoscopically and multi-areally branching axons.*

Outside the intralaminar nuclei, projections of the BG-recipient thalamus terminate most densely in L1 (the molecular layer, consisting mostly of the apical dendrites of pyramidal neurons), while projections of cerebellum-recipient thalamus terminate chiefly within L2-L5; the two terminal fields overlap and intermingle in cortex so that pyramidal neurons are simultaneously under the influence of the BG and cerebellum (Kuramoto *et al.* 2009; Jinnai *et al.* 1987). The rat and cat ventromedial nucleus, a major recipient of SNr output, has been noted for directing its output, covering large portions of the cerebral cortex, almost exclusively to L1 (Herkenham 1979; Glenn *et al.* 1982). Recent experiments in the rat have demonstrated that thalamic projections to superficial cortex are comprehensive, extensively overlap, broadly arborize tangentially, and are variously intra- and inter-areally divergent and convergent (Rubio-Garrido *et al.* 2009). Each square mm of superficial cortex was found to be innervated by an average of ~4500 thalamocortical neurons, and the most profuse nuclei of origin were the VL, VA, and VM, each of which was noted for terminal fields targeting widely separated areas in cortex. Primary sensory nuclei were found to be completely absent from the projection to superficial cortex. Also largely absent in L1 are the local inputs that dominate in deeper layers: instead, roughly 90% of synaptic inputs there have distant origins associated with modulatory feedback paths (Larkum 2013).

Broadly branching axons, preference for superficial cortex, and the absence of contributions by primary sensory thalamus, suggest a mesoscopic modulatory function for the BG-recipient thalamocortical projections. Electrophysiological experiments and simulations support this. *In vivo* experiments on optogenetically manipulated mice show that the VM nucleus is directly implicated in large scale cortical activation (Honjoh *et al.* 2018). *In vitro* experiments in rat suggest that apical inputs to L5 neurons have a negligible direct effect on the soma, due to severe attenuation, and shunting by back-propagating action potentials (Larkum *et al.* 2004). Mathematical modeling of these projections suggests that they can serve a purely modulatory function, biasing inference by selectively amplifying activity in the pyramidal neurons they target, without disrupting information flowing through their more basal inputs, and without contributing information to their outputs beyond that implicit to selective amplification (Kay *et al.* 2019 ‡ ). Beta (and higher) oscillation in the BG, reaching superficial cortex, apparently does not in itself manifest as oscillation at the somata of receiving cortical projection neurons, but rather appears to be low-pass filtered with a cutoff frequency of 5 Hz (Rivlin-Etzion *et al.* 2008).

### 7.6. *Thalamocortical projections to apical dendrites control spike-timing-dependent gain.*

When thalamocortical projections provide subthreshold input to the L1 apical dendrites of L5 pyramidal neurons, that input appears to selectively increase the effective gain of the pyramidal neurons, such that temporally coincident input (within 20-30 ms) to their somata induces bursting output (Larkum *et al.* 2004; Kay *et al.* 2019‡). According to the BGMS model, BG-influenced activity in these thalamocortical projections, exhibiting multi-areal synchrony, reinforces activity in selected cortical areas, both those that triggered the BG response, and associated areas that are contextually relevant. If corticocortical bursting itself effects long range synchronization (Womelsdorf *et al.* 2014), then the BG may instantiate long range synchronies by controlling and coordinating the location and timing of cortical bursting and burst receptivity. Paths through the intralaminar nuclei to deep cortical layers are likely crucial to the temporal coordination and selectivity of these responses, as discussed in detail later (§8).

### 7.7. *Thalamic inputs to cortical FSIs may be a crucial path for precisely discriminative selection among conflicting inputs.*

Recent evidence demonstrates that projections of the mediodorsal nucleus to PFC directly drive activity in cortical fast spiking interneurons, suggesting a mechanism whereby MD can mediate inhibition of conflicting or contextually extraneous activity (Delevich *et al.* 2015; Kuroda *et al.* 1998; Rikhye *et al.* 2018). As discussed earlier (§3.6), cortical FSIs exert a robust and precise influence on oscillatory activity in their targets, and may elevate receptivity to activity with the preferred frequency and phase. Thus, paths from the BG to cortical FSIs via the thalamus are a likely substrate for the precise selections that have long (Redgrave *et al.* 1999) been attributed to the BG.

### *7.8. Intrinsic mechanisms in cortex facilitate semantically valid mesoscopic modulation and selection.*

The density and overlap of the thalamocortical projection to L1 (Rubio-Garrido *et al.* 2009) suggest that BG output associated with consolidated skills recruits comprehensive modulation of targeted cortical areas. This bears directly on the "spatial computing" proposal, according to which modulatory ("contextual") signals organize information processing through spatial patterns of alpha/beta oscillatory activity across the surface of the cerebral cortex (Lundqvist *et al.* 2023; Chen *et al.* 2025‡). Moreover, there are intrinsic mechanisms in cortex that may heal intra-areal gaps in modulation. It has been proposed that activity entering cortex through L1 spreads horizontally through L2 and L3, yielding mesoscopic facilitation of firing in L5 pyramidal cells, thereby controlling long range effective connectivity (Roland 2002). Lag-free lateral spread of oscillation up to high gamma has been demonstrated *in vitro*, working through interactions in L2 and L3, entailing an ensemble dynamic of gap junctions and GABAergic fast spiking interneurons (Tamás *et al.* 2000). And there are additional mesoscopic mechanisms underlying coherent modulation of cortex, beyond the neural propagation of action potentials and associated activation of synapses. Glial cells (astrocytes) continuously parcellate cortex, coupling as many as 2 million synapses per astrocyte into non-overlapping mesoscopic domains (Bushong *et al.* 2002; Fields 2025). Ephaptic coupling may similarly pervade cortex, whereby extracellular electrical field oscillations strongly and coherently modulate the responses of nearby neurons (Pinotsis *et al.* 2023).

The neurons of L2 (the external granular layer) have larger receptive fields and a higher incidence of combined feature selectivity than L3 neurons, and their projections are exclusively corticocortical (Gur and Snodderly 2008; Markov and Kennedy 2013). The apical dendrites of the small pyramidal neurons of L2 intermingle in L1 with the apical dendrites of L5 pyramidal neurons, spreading 100-200 µm laterally (Meller *et al.* 1968; Noback and Purpura 1961). Because L1 and L2 are adjacent, the implicated thalamocortical appositions are more proximal, so likely subject to markedly less of the attenuation and filtration characterizing apical inputs to L5 neurons. Putative high frequency BG modulation of thalamocortical projections to L1 might therefore induce correlated discharges in L2 pyramidal neurons, spreading this high frequency activity laterally within the superficial layers. Moreover, *in vitro* experiments with pyramidal neurons from these layers have demonstrated nonlinear coincidence detection dynamics, with windows only 4-7 ms long (Volgushev *et al.* 1998), suggesting temporal specificity in L2/L3 mechanisms sufficient to bias competition among conflicting gamma oscillations.

Discontinuities have been found in the local horizontal linkages of cortex, particularly in L2 and L3, that are posited to tessellate sensory areas along boundaries of similar function and features (Rockland and Lund 1983; Ojima *et al.* 1991; DeFelipe *et al.* 1986; Juliano *et al.* 1990). Similar tessellation, into "stripes" 2-3 mm long and 200-400 µm wide, has been described in PFC (Levitt *et al.* 1993; Pucak *et al.* 1996), and is posited to define the limiting spatial resolution with which the BG can modulate cortical activity (Frank *et al.* 2001). Spreading, but spatially restricted, synchronized activation of superficial modulatory layers has semantically valid effects with a cortical layout in which mental categories and analogs are represented by precise, spatially graded semantic continuities—feature maps—not only in sensory receptive fields, but throughout the cerebral cortex, as evidence suggests (Rao *et al.* 1999; Huth *et al.* 2012; Simmons and Barsalou 2003; Rajalingham and DiCarlo 2019; Lettieri *et al.* 2019; Zhang *et al.* 2020).

The precise significance of these areal activations is complicated by multidimensionality and mixed selectivity in feature maps. A canonical example of a multidimensional map is the primary visual cortex, wherein the representations of ocular dominance, orientation, and spatial frequency, are folded into small local maps embedded within an overall map arranged in registration with the two major dimensions that represent visual field position (Yu *et al.* 2005). In associative areas of cortex, which are more densely BG-recipient, multidimensionality and mixed selectivity are yet more prominent (Rigotti *et al.* 2013; Huth *et al.* 2016).

### *7.9. Dopamine under BG control modulates the vertical and horizontal dynamics to which effective connections in cortex are subject.*

Striosomal BG paths (reviewed later (§9.3)) are major modulators of the supply of dopamine (DA) to the forebrain. While DA promotes oscillatory responses to activity in proximally apposed afferents to pyramidal neurons, it has been found to attenuate receptivity to inputs on apical dendrites, which may "focus" or "sharpen" the effects of inputs to those cells (Yang and Seamans 1996). Moreover, the GABAergic interactions among L2/L3 basket FSIs that are crucial for the elimination of phase lags in laterally spreading oscillation (Tamás *et al.* 2000) are depressed by DA (Towers and Hestrin 2008). This suggests that phasic DA induces phase lags in L2/L3 that increase with distance from the locus of excitatory input, perhaps producing a temporal center-surround effect that effectively focuses cortical responses. Because expressions of plasticity are pervasively spike-timing-dependent (Song *et al.* 2000), this phase lag control mechanism may also have important consequences for the formation and refinement of cortical feature maps.

These arrangements suggest a corollary to the central proposition of the BGMS model: not only do the BG control effective connectivity in cortex, they also separately control the dynamic characteristics of effective connections. BG influences on cholinergic and serotonergic centers, reviewed later (§10), extend this control.

### *7.10. Projections to thalamus from cortical layer 6 exhibit topography, and bear activity, suited to mesoscopic modulation by BG direct path output.*

Most of the corticothalamic population arises from L6, with small terminals apposing distal dendrites in thalamus, and reciprocation is particularly prominent in this projection (Rouiller and Welker 2000). The corticothalamic projection topographically and comprehensively reciprocates the thalamocortical projection, with consistent rules such that each thalamic locus that originates a given type of projection to cortex has a corresponding corticothalamic type reciprocated (Deschênes *et al.* 1998). The population of corticothalamic fibers is considerably more numerous than the thalamocortical population, by roughly a factor of 10 (Deschênes *et al.* 1998), and some of these fibers exhibit terminal fields that spread laterally, to encompass neighboring reciprocal receptive fields (Rouiller and Welker 2000).

In an *in vitro* study in rat, the delay of the corticothalamic projection from L6, and its variability, were found to be significantly greater than those of the thalamocortical projection, 5.2 ± 1.0 ms and 2.1 ± 0.55 ms respectively (Beierlein and Connors 2002). While thalamocortical delays are essentially fixed and very tightly aligned (Salami *et al.* 2003), L6 corticothalamic delays evidence supernormality. This entails reduction of delay and threshold below baseline after the relative refractory period (Swadlow *et al.* 1980). Supernormality was found to persist for roughly 100 ms following a discharge, and with a 40 Hz stimulus it reduced corticothalamic delay by up to 12% (Beierlein and Connors 2002). While the functional significance of supernormality in the corticothalamic projection is elusive, it might arrange for advancement of spike timing in cortex as activity intensifies, either matching supernormality in implicated corticocortical projections, or producing other useful timing-related effects, such as phase-of-firing intensity encoding (Masquelier *et al.* 2009).

The numerosity, structure, and variability of the L6 corticothalamic projection suggest it may be subject to some of the same pressures producing convergence, divergence, and variability in the corticostriatal and striatopallidal projections—particularly, the need for a supply of inputs with appropriate characteristics to meet complex and widely varying topological and spike alignment requirements in paths through thalamus terminating in feedback-recipient layers of cortex. These arrangements were discussed at much greater length earlier (§5).

Dissociation of the functions of corticothalamic and corresponding thalamocortical projections has recently been demonstrated in rats. Using narrowly targeted pharmacological manipulations of the mediodorsal thalamic nucleus and the reciprocally linked area of frontal cortex, Alcaraz *et al.* (2018) show that inhibition of the corticothalamic cell population disrupts behavioral responsiveness to changes in reward magnitudes, while inhibition of the thalamocortical population disrupts behavioral responsiveness to changes in cause-effect associations. These patterns appear to be consistent with the proposition that corticothalamic projections to motor, associative, and limbic thalamus carry a supply of convergent inputs that is modulated by extrinsic inputs, particularly from the BG, producing selective and sensical thalamocortical responses.

### *7.11. Intrinsic cortical network dynamics assure a supply of activity to thalamus that can be effectively entrained by converging BG inputs.*

A full account of the function of the corticothalamic projection has proved elusive (Goldberg *et al.* 2013). According to the BGMS model, the corticothalamic projections from L5 and L6 to BG-recipient thalamus arrange for BG output to be able to select, in each channel, which frequency and phase of cortical activity is to be reinforced and which are to be inhibited. Evidence and modeling suggest that conflicting rhythms in the afferent activity to a cortical neuron shift its discharge pattern away from rhythmic regularity and toward randomness (Gómez-Laberge *et al.* 2016). Such a response might work to assure that corticothalamic afferents from conflicted cortical loci, converging with BG afferents, can produce postsynaptic activity entrained by those BG afferents, for any particular frequency and phase of BG-selected activity. Moreover, as noted above, corticothalamic fibers are far more numerous than thalamocortical ones, and pool afferents from L5 and L6. This suggests a relatively high degree of convergence in the corticothalamic projection, further working to assure a supply of suitable excitatory afferent activity, and including activity associated with both feedforward and feedback projections.

### *7.12. Motor and intralaminar thalamus relay BG-modulated cortical activity to BG inputs, with functional distinctions suggesting contextualization and dynamical regulation.*

While the main input to the matrix compartment of striatum arises from cortical L3 and L5, as reviewed above, there are profuse projections from BG-recipient thalamus to striatum, implicitly relaying input from L6. Inputs from cortex and thalamus converge on individual SPNs, with similar axodendritic patterns, but with cortical inputs more numerous (Huerta-Ocampo *et al.* 2014). Projections from thalamic VA/VL to striatum converge with functionally corresponding projections from cortex (McFarland and Haber 2000). Evidence from primates shows that striatal FSIs are likely broadly targeted by projections from intralaminar thalamus (Sidibé and Smith 1999), and that projections of the intralaminar nuclei target the entire striatum, with specific topography (Sidibé *et al.* 2002;

Mandelbaum *et al.* 2019). Intralaminar projections have been noted for supplying the striatum with information relating to salient sensory events (Matsumoto *et al.* 2001), and are implicated in the learning of changes in instrumental contingencies, through projections to cholinergic interneurons (Bradfield *et al.* 2013). *In vitro* experiments demonstrate projections from the thalamus to striatal SPNs, with distinctive synaptic properties, such that postsynaptic response generation is likelier than for corticostriatal synapses, but repetitive stimulation depresses postsynaptic depolarization (Ding *et al.* 2008). In awake monkeys, activation of projections from intralaminar thalamus to striatum has complex effects, with SPN discharge induced only by rapid bursts from thalamus, and long latencies peaking 100-200 ms after intralaminar stimulation (Nanda *et al.* 2009).

Significantly, intralaminar thalamostriatal projections strongly prefer the matrix compartment (Sadikot *et al.* 1992b) to which the direct and indirect paths are confined. Thalamostriatal and thalamocortical projection neurons in the intralaminar nuclei are intermingled, and many axons branch to innervate both striatum and cortex (Deschênes *et al.* 1996; Parent and Parent 2005; Kaufman and Rosenquist 1985a), so that synchronies in the striatal projections of these nuclei are presumptively representative of synchronies in their cortical projections. Thus, thalamostriatal inputs implicitly reflect the current effect of BG output upon the cortex. These subcortical feedback loops may facilitate regulation of BG output to bring modulatory results into conformity with intentions, both dynamically and, as discussed earlier (§5.7), by driving the expression of plasticity.

They may also provide for sequential elaboration of BG output, adjusting thalamocortical modulations with greater speed and precision than is possible within cortico-BG loops. Intralaminar thalamic projections to the globus pallidus, substantia nigra, and subthalamic nucleus (Sadikot *et al.* 1992a) may serve similar roles, exploiting loop delays that are much shorter than the propagation delays of the corticostriatal and striatopallidal projections (Kitano *et al.* 1998; Harnois and Filion 1982). Indeed, for some tasks, cortex is crucial for initial acquisition but not necessary for subsequent performance (Kawai *et al.* 2015; Wolff *et al.* 2022; Dhawale *et al.* 2021). Moreover, tight and bidirectional integration of the BG indirect path with the cerebellum through subcortical pathways has been noted (Bostan and Strick 2010; Bostan *et al.* 2013; Bostan and Strick 2018; Milardi *et al.* 2016), and seems likely to be prominent in mechanisms underlying performance of rapid, precise, sequential cognition and behavior, including the production of rhythmic behavior with cadences that are precisely consistent, but continuously variable.

# 8. The Role of the Intralaminar Nuclei in the Direct Path

**In this section:**

### *8.1. The intralaminar nuclei are a uniquely important link between the basal ganglia and cerebral cortex.*

Compared to the motor and association nuclei, the intralaminar nuclei of the thalamus are small, but they have exceptional characteristics and functions placing them at the very center of cognitive coordination and awareness. Saalmann (2014) implicates them centrally in the regulation of cortical oscillations and associated synchronies. Moreover, uniquely among thalamic nuclei, the intralaminars project topographically to the entire striatum (Mandelbaum *et al.* 2019; Sadikot *et al.* 1992a; Sidibé *et al.* 2002), and indeed have been described as an integral part of the BG system (Parent and Parent 2005). In the BGMS model, the intralaminar nuclei, through their broad projections and dynamical characteristics, work as a high fidelity broadcast mechanism whereby long range effective connectivity, and therefore cognition, are oriented by spike-timing-dependent gain.

### *8.2. Intralaminar thalamus in primates projects to pyramidal somatic layers.*

Some studies in rat and cat report intralaminar thalamocortical projections principally targeting L1 (Royce and Mourey 1985; Royce *et al.* 1989), like the projections of the BG-recipient populations in the MD, VA, VL, VM, and VPLo nuclei, whereas other studies in primate, rat, and cat report intralaminar projections principally to L5 and L6, where individual axons branch widely and arborize massively to appose the somata and proximal dendrites of great numbers of pyramidal neurons, and may terminate in L1 only more sparingly (Parent and Parent 2005; Deschênes *et al.* 1996; Kaufman and Rosenquist 1985a; Berendse and Groenewegen 1991; Llinás *et al.* 2002). The disparities among these studies have been suggested to relate to actual physiological distinctions among the species at issue, made all the more likely by the particularly active recent evolutionary history of the intralaminar nuclei and cerebral cortex (Royce and Mourey 1985), but may simply be methodological artifacts.

### *8.3. The thalamocortical projections of the BG-recipient intralaminar nuclei reach nearly the entire cortex.*

While intralaminar projection fibers to frontal cortex are greatly outnumbered by those from non-intralaminar BG-recipient thalamic nuclei (Barbas *et al.* 1991; Schell and Strick 1984), intralaminar projections are strikingly widespread, encompassing nearly the entire neocortex. Experiments in rats and cats demonstrate that the CM/PF nuclei, comprising the caudal group, project to motor, frontal eye fields (FEF), orbitofrontal, anterior limbic, cingulate, parietal, and visual cortex, and to many structures of the medial temporal lobe, though not to the hippocampus proper (Royce and Mourey 1985; Berendse and Groenewegen 1991). In the same two species, the rostral CL and PC nuclei project widely and without consistent topography to the FEF, anterior cingulate, insular, parietal areas 5 and 7, visual, and auditory cortex (Kaufman and Rosenquist 1985a; Royce *et al.* 1989; Berendse and Groenewegen 1991). Studies in cat (Cunningham and Levay 1986) and macaque (Doty 1983) have identified sparse but distinct projections from the rostral intralaminar nuclei to L1, L5, and L6 of primary visual cortex (area 17). Collating many of these results, a metastudy pooling thalamocortical and corticothalamic projections in cat concluded that the intralaminar nuclei connect very widely with most of visual, auditory, motor, and prefrontal cortex; though nearly all of these connections were characterized as weak or sparse, of 53 cortical areas studied, only 7 (the contiguous primary, posterior, ventroposterior, and temporal auditory fields, the posterior suprasylvian area of visual cortex, and the hippocampus/subiculum) were not reported to be connected with any of the BG-recipient intralaminar nuclei (Scannell 1999).

Single axons from CL and PC have been noted to branch multi-areally to innervate visual and parietal association cortex, suggesting a general function for the intralaminar nuclei, rather than specific functions in the spatial processing of visual information (Kaufman and Rosenquist 1985a).

The broad cortical projection field of the intralaminar nuclei, their extreme divergence, and their intimacy with oscillatory dynamics, were demonstrated by the "recruiting response" reported in early experiments in cats. Oscillatory activity spanning nearly the entire cerebral cortex, most strongly in frontal areas, was evoked with electrical stimulation centered anywhere within the intralaminar region (Morison and Dempsey 1941; Dempsey and Morison 1941). The ventral anterior, mediodorsal, and ventromedial nuclei, prominent in the system of superficially projecting BG-recipient thalamus detailed earlier (§7), exhibit similar indications of large scale connectivity. The VA nucleus in particular has also been implicated in the generation of the recruiting response (Skinner and Lindsley 1967).

CNS insults that bilaterally destroy not only the rostrocaudal extent of the intralaminar nuclei, but also the adjacent MD nucleus, are consistently associated with the permanent vegetative state (Schiff 2010). Activity in these nuclei has been shown in humans to have an especially close association with conscious perception (Fang *et al.* 2024‡). Pharmacological manipulation of the intralaminar nuclei can rapidly abolish or restore wakefulness (Alkire *et al.* 2008), and a special indispensability to consciousness has been proposed for these nuclei (Bogen 1995; Baars 1995). Indeed, primates rendered unconscious with propofol anesthesia can be promptly roused to wakefulness solely by high frequency electrical stimulation of the intralaminar thalamus (Redinbaugh *et al.* 2020; Bastos *et al.* 2021).

Sleep spindles, which entail tightly synchronized responses spanning large areas of cortex, also demonstrate the broad scope of intralaminar projections. In spindling, activity in the corticothalamic projection and thalamic reticular nucleus are thought to drive thalamocortical cells to simultaneous discharge in nuclei spanning much of the thalamus, particularly through highly divergent projections from the rostral reticular nucleus through the BG-recipient intralaminar and association nuclei (Contreras *et al.* 1997).

### *8.4. Unlike other BG-recipient thalamic areas, the intralaminar nuclei of the thalamus are not purely modulatory.*

The CL and PC nuclei in cats contain neurons whose activity is uniquely related to all kinds of eye movements, fast or slow, self-initiated or evoked, to stimuli and movements characterized visuotopically, allocentrically, by direction of gaze, and various combinations thereof, to eye position, and to polysensory context and vigilance (Schlag *et al.* 1974, 1980). Activity in these neurons precedes saccade onset by 50-400 ms, and continues during the saccade, whether the saccade is self-initiated or visually evoked, with each neuron showing a consistent but idiosyncratic pattern (Schlag *et al.* 1974; Schlag-Rey and Schlag 1977; Schlag *et al.* 1980). While each completed saccade is accompanied by a consistent pattern of activation in some of these neurons, the reverse is not always the case—the same pattern of activation in an intralaminar neuron is sometimes seen in the absence of an executed saccade (Schlag *et al.* 1974). Nonetheless, microstimulation in the CL and PC nuclei consistently evokes conjugate saccades with a delay of 35 ms for large deviations, suggesting primary involvement in saccade generation (Maldonado *et al.* 1980).

Similar to CL and PC, evidence suggests that activation of the parafascicular nucleus can by itself generate turning and orienting movements, though this apparently occurs not via its projections to cortex, but rather, via its projections to the STN, implicating STN and other BG components in movement initiation (Watson *et al.* 2018‡).

### *8.5. The characteristics of the BG-recipient intralaminar nuclei suggest high fidelity relay of precisely timed activity.*

Unlike other BG-recipient populations in thalamus, the CM and PF nuclei are densely parvalbumin-positive (Jones and Hendry 1989). In other thalamic nuclei, as noted earlier (§7.4), parvalbumin is associated with putative "core" or "driving" neurons, which are not BG-recipient. In other brain organs, notably the cerebral cortex, striatum, GP, and SNr, parvalbumin is associated with fast-firing, fatigue-resistant neurons. Via the caudal intralaminar nuclei, the BG complete loops within which spike timing is largely determined by parvalbumin-containing, fast-firing, non-fatiguing neurons (Mallet *et al.* 2005; Bennett and Bolam 1994; Cote *et al.* 1991), targeting somata and proximal dendrites of pyramidal neurons in deep cortex as described above.

The rostral intralaminar nuclei are densely calbindin-positive (Jones and Hendry 1989), like non-intralaminar BG-recipient thalamus. The laterodorsal part of the CL nucleus has been shown in adult cats to contain a population of neurons projecting to parietal association cortex that, during wakefulness and REM sleep, regularly emit bursts of 3-4 spikes with interspike intervals (ISIs) shorter than 1.3 ms, at a burst rate of 20-40 Hz, with no apparent signs of fatigue, and an antidromic thalamocortical delay ⩽ 500 µs, indicating conduction velocities (CVs) of 40-50 m/s (Glenn and Steriade 1982; Steriade *et al.* 1993). During the spindling characteristic of stage 2 sleep, bursts in these cells were found to be even more intense, 8-9 spikes with ISIs as low as 1 ms. The CVs of their axons, uniquely fast among thalamocortical cell classes (Steriade *et al.* 1993), are roughly 50 times those of striatopallidal axons, which in awake cynomolgus monkeys exhibit CVs under 1 m/s (Tremblay and Filion 1989).

Because BG inputs to intralaminar nuclei are collaterals of inputs to other thalamic nuclei (Parent *et al.* 2001), the information received from the BG by the intralaminar nuclei presumably duplicates that received by non-intralaminar cells. But high fidelity relay by neurons of the intralaminar thalamus, combined with pyramidal somatic layer targeting, appears to arrange for particularly narrow selectivity through spike synchrony effects. Indeed, feed-forward inhibition in cortex, implicating fast-spiking interneurons, arranges for an extremely narrow coincidence detection window for proximally apposed afferents to pyramidal neurons, -1.5 to +2.4 ms for effective spike summation, even while the coincidence requirement in distal inputs was found to be much looser, -8.6 to +12.3 ms (Pouille and Scanziani 2001). Even absent the influence of FSIs, pyramidal neurons stimulated somatically *in vitro* have been shown to act as nonlinear coincidence detectors with windows only 4-7 ms wide, that become narrower with rising oscillatory frequency, with the timing of discharges tightly correlated to the timing of somatic membrane potential oscillation (Volgushev *et al.* 1998).

Evidence discussed earlier (§7.3) implicating L5/L6 in top-down control, with prominent activity in the alpha and beta bands (Bastos *et al.* 2018), underscores the significance of temporally precise spike relay and distribution by the intralaminar nuclei to these layers. Oscillatory periods ⩾ ~50 ms predominate in L5/L6, much longer than the ~4 ms coincidence window for proximally apposed inputs, so that intralaminar thalamic inputs with high temporal fidelity appear arranged to enable precise selections.

### 8.6. *Most BG output to the intralaminar nuclei is non-somatic, increasing combinatorial power and decoupling BG output from cortical somatic inputs.*

As noted above, BG inputs to primate caudal intralaminar thalamus overwhelmingly appose dendrites, not somata (Sidibé *et al.* 2002). These appositions are not homogeneous, in that over 80% of SNr inputs to PF in monkey were found to appose small or medium, mostly distal, dendrites, with none apposing somata, while over 75% of GPi inputs to CM were found to appose medium or large, mostly proximal, dendrites, and 5% to appose somata. These patterns of apposition clearly result in looser coupling between the BG and intralaminar thalamus than does the tight coupling through perisomatic, quasi-calyceal appositions seen in non-intralaminar BG-recipient thalamus (Bodor *et al.* 2008; Kim *et al.* 2017). Perhaps more important, because corticothalamic inputs to these nuclei are also predominantly through small distally apposed terminals (Rouiller and Welker 2000), BG inputs are positioned as frequency- and phase-selective filters, imposing alternating permissive and prohibitive time windows on afferent corticothalamic activity, with the potential for extensive presomatic nonlinear computation (Murphy-Baum and Awatramani 2022; Mehaffey *et al.* 2005; Burger *et al.* 2023 ‡ ), enhancing computational power and combinatorial flexibility. Thus a single intralaminar neuron might participate in a vast variety of scenarios characterized by distinct corticothalamic and nigrothalamic input patterns, each producing somatic discharges, but by different combinations of dendritic inputs.

### 8.7. *Cortical projections to CM/PF predominantly arise in layer 5, as do many corticocortical projections.*

While inputs from L6 predominate in BG-recipient motor/association thalamus, in BG-recipient caudal intralaminar thalamus it is L5 inputs that predominate (Van der Werf *et al.* 2002; Balercia *et al.* 1996; Cornwall and Phillipson 1988; Royce 1983a, 1983b). This mirrors targeting of L5 in thalamocortical projections from this area, discussed above, and moreover shares its laminar origin with many corticocortical projections (Reiner *et al.* 2003). This is significant, because it suggests that corticocortical projections are systematically accompanied by trans-intralaminar paths, sharing exactly the same origins and targets, and subject to temporally precise gating by the BG direct path, in which L5 is similarly predominant in inputs to striatum, as reviewed earlier (§7.3).

### 8.8. *Intralaminar and non-intralaminar projections from BG-recipient thalamus have complementary functions.*

Widespread intralaminar projections appear arranged to broadcast a temporally precise but spatially diffuse signal to most of cortex, while non-intralaminar projections to superficial layers have mesoscopic spatial specificity, more restricted (principally frontal) areal targets, and relatively crude temporal specificity (though their appositions on cortical FSIs (Delevich *et al.* 2015; Kuroda *et al.* 1998; Rikhye *et al.* 2018) likely provide for temporal precision). At the heart of the BGMS model is the proposition that the BG coherently modulate these two influences, so that their convergence and inter-areal linkage in cortex provide for spatiotemporal specificity and consequent precision in the control of effective connectivity. By interacting with intrinsic cortical activity, these inputs rapidly and dynamically recruit specific large scale networks. A corollary of this view is that the nuclei of the thalamus act as attentional spotlights, with selectivity rooted in both temporal and spatial specificity, while the BG are prominent in the orientation of those spotlights.

The BGMS proposal can be summarized as follows: When an input pattern triggers a selection in the striatum, the timing of striatal output tracks the prevailing timing of the input pattern, and the GPi, VP, and SNr impart that timing to the thalamus, with striatopallidal and striatonigral delays tuned for optimum effect (optimality being a function of cortical rhythms and corticocortical conduction delays, discussed in detail earlier (§5)). The intralaminar nuclei, through widespread diffuse projections throughout the cortical column (excepting only L4), impart discriminative receptivity to any activity that is precisely synchronous with that prevailing in the input pattern that stimulated the BG response, and narrowly reinforce its generation in its loci of origin. The non-intralaminar nuclei, through dense, mesoscopically specific, largely closed-loop projections, chiefly to L1, fortify activity in selected areas, particularly those contributing to the input pattern. When this fortification is strong, and coincident with substantial activity in the corresponding proximally apposed afferents, bursting is promoted (Larkum *et al.* 1999, 2004), further promoting establishment of effective connections (Womelsdorf *et al.* 2014).

Closed-loop paths through non-intralaminar nuclei largely implicate areas in frontal cortex, which are the densest targets of non-intralaminar BG-recipient thalamocortical projections, but other association areas in primates, notably in parietal and temporal cortex, are also implicated. All of these areas are thought to originate feedback signals with top-down control over their targets. By this narrative, the BG direct path establishes and fortifies top-down control connections from both ends, with the MD, VA, and VL nuclei fortifying the top end of the connection, and the CM, PF, PC, and CL nuclei tuning both ends to complete the connections. An additional function of BG-modulated thalamocortical afferents to L1 is that they open gates for feedforward signals, suggested by evidence (Bastos *et al.* 2018; Lundqvist *et al.* 2018a) that activity in L2/L3 is associated with bottom-up inputs.

Open loop direct paths through non-intralaminar nuclei may serve to complete activation of a distributed cortical ensemble that is only partly activated when it first triggers a striatal response, particularly when the triggering pattern largely originates in sensory cortex. Closed loop paths through intralaminar nuclei may tighten synchrony throughout the selected ensemble, and provide reinforcement that is highly selective, due to the narrow coincidence windows associated with proximal inputs to pyramidal neurons.

Notably, trans-thalamic inputs may actively inhibit and disconnect activity that is not synchronous (particularly, that is antisynchronous) with the thalamocortical signal, by feedforward inhibition via cortical FSIs. Evidence noted earlier (§7.7) strongly suggests that the path from MD to PFC entails such a mechanism (Delevich *et al.* 2015; Kuroda *et al.* 1998; Rikhye *et al.* 2018). And it is clear from the response to sleep spindles (Peyrache *et al.* 2011) that both pyramidal neurons and FSIs are targeted by thalamocortical projections. Feedforward inhibition associated with this arrangement enforces extremely short windows of summational receptivity (Pouille and Scanziani 2001).

It may be important that intralaminar projections, which target most of the cortex, are subject to extremely narrow coincidence windows. With wider windows, the intralaminar broadcast mechanism seems prone to establishment of spurious connections. Indeed, schizophrenia involves abnormal enlargement of these coincidence windows (Lewis *et al.* 2005; Gonzalez-Burgos *et al.* 2015), while lesioning and deactivation of intralaminar nuclei has been found to relieve hallucinations and delusions associated with Sz and other psychoses (Hassler 1982).

Sz is also characterized by enlargement of the time window within which visual stimuli are judged to be simultaneous (Schmidt *et al.* 2011), and by abnormalities in the simultaneity criteria for implicit audiovisual fusion (Martin *et al.* 2013). Beyond Sz, loosening of simultaneity criteria, and deficient perception of short time intervals, may be characteristic of psychosis generally (Schmidt *et al.* 2011; Ciullo *et al.* 2016).

### *8.9. The BG-recipient intralaminar nuclei are most developed in humans.*

As evident from their function in vision and saccades, the BG-recipient intralaminar nuclei are a jumble of perceptual and motoric function, with activity in individual neurons highly correlated with both. Roles for these nuclei in executive control, working memory, and general cognitive flexibility—capacities that are most developed in humans—have also been shown (Van der Werf *et al.* 2002). Over the course of mammalian evolution, the intralaminar nuclei, particularly the posterior group, have undergone relative expansion and elaboration, reaching their greatest extent in primates, and in humans particularly (Macchi and Bentivoglio 1986; Royce and Mourey 1985; Herkenham 1986).

### *8.10. The BG-recipient intralaminar nuclei may be crucial to the expression of pathology in Tourette syndrome, OCD, and schizophrenia.*

Psychosurgical results in humans give further evidence that these nuclei can originate driving inputs to motoric, perceptual, cognitive, and motivational centers. Treatment of Gilles de la Tourette syndrome (GTS) by stereotactic ablation or rhythmic electrical stimulation of the rostral (Rickards *et al.* 2008) or caudal (Houeto *et al.* 2005; Servello *et al.* 2008) intralaminar nuclei has produced substantial and sustained abatement, in some cases almost complete remission, of compulsive behavior (tics) in many patients. Similarly, severe or extreme symptoms of obsessive compulsive disorder (OCD) have been substantially, consistently, and sustainably alleviated by unilateral lesioning of the right intralaminar nuclei (Hassler 1982), or by rhythmic electrical stimulation localized to the inferior thalamic peduncle, inactivating connectivity between intralaminar nuclei and orbitofrontal cortex (Jiménez-Ponce *et al.* 2009). GTS and OCD involve extensive BG abnormalities (Graybiel and Rauch 2000; Albin and Mink 2006; Kalanithi *et al.* 2005), so alleviation of symptoms by IL inactivation suggests functional prominence of the intralaminar nuclei in BG dynamics, and may be evidence of key involvement in the transmission of BG output to cortex.

Functional deficits in Sz are intimately related to the functional roles of the intralaminar nuclei. Eye tracking and saccade control are dysfunctional, suggesting particular deficits in anticipatory control and the suppression of distracters (Levy *et al.* 1994; Fukushima *et al.* 1988; Hutton *et al.* 2002), and aberrant connectivity between the intralaminar nuclei and PFC has also been described (Lambe *et al.* 2006). Sz has been found to be associated with significant relative reduction in volume and metabolic hypofunction in the centromedian nucleus, in addition to the MD nucleus and pulvinar, in a study that found no significant effects by these measures in other thalamic nuclei (Kemether *et al.* 2003; Hazlett *et al.* 2004).

The BG-recipient intralaminar thalamus expresses $D_2$ dopamine receptors at particularly high density (Rieck *et al.* 2004), and these receptors are targeted by antipsychotic drugs, usually with ameliorative effect for positive symptoms (Nordström *et al.* 1993; Kay *et al.* 1987). There is evidence from experimental clinical practice that lesioning of the mediodorsal and rostral intralaminar nuclei can permanently eliminate delusions and somatosensory, auditory, and visual hallucinations associated with Sz, while rhythmic (20 and 50 Hz) electrical stimulation of these areas can abolish symptoms promptly (Hassler 1982).

That some hallucinations and visuocognitive deficits in Sz may involve BG interaction with the intralaminar nuclei is further suggested by the common occurrence in PD of visual and other hallucinations and delusions (Barnes and David 2001; ffytche *et al.* 2017) and impaired shifting and maintenance of visual attention (Wright *et al.* 1990). PD is marked by abnormally strong coupling within BG loops (Hammond *et al.* 2007), and hallucinations incidental to PD appear to be associated with pathological coupling of visual areas with the “default mode network” (Yao *et al.* 2014; Shine *et al.* 2015; Walpola *et al.* 2020). In short, hallucinations and delusions incidental to PD might be due in large part to pathologically synchronized BG output, inducing pathological persistence and widespread synchronization of neocortical activity, which could functionally connect spurious activity in visual cortex to hub areas. Consistent with this account, extensive thalamic cell loss in PD specific to the caudal intralaminar nuclei (Henderson *et al.* 2000) suggests that, among thalamic areas, these nuclei bear the brunt of the abnormal dynamics characteristic of the disease. Remarkably, unusually high BG dopamine levels are also associated with hallucination: experimental elevation of dopamine level in mouse visual striatum causes hallucination-like perception, which is alleviated by systemic administration of the $D_2$ antagonist haloperidol (Schmack *et al.* 2021). PD and Huntington's disease are also both associated with voluntary saccade deficiencies, including abnormal distractibility in Huntington's (Bronstein and Kennard 1985; Lasker *et al.* 1987, 1988), resembling some of the oculomotor abnormalities associated with Sz.

Auditory hallucinations are commonly associated with Sz (de Leede-Smith and Barkus 2013; McCarthy-Jones *et al.* 2014), and also sometimes occur in advanced PD (ffytche *et al.* 2017). Many of the brain areas implicated in these hallucinations are within or intimate with the BG (Shergill *et al.* 2000). In cat, connections of the parafascicular nucleus with secondary auditory cortex and the anterior auditory field have been demonstrated (Scannell 1999), but as noted above (§8.3), no direct connections have been found between the intralaminar nuclei and the primary and several adjoining auditory fields (Scannell 1999). This lacuna is intriguing, in that it suggests that intralaminar input may be detrimental to signal integrity there, outweighing the benefits that evolutionarily stabilize intralaminar innervation elsewhere.

Instead, it is plausible that cell assemblies in primary sensory areas are largely self-synchronizing in response to stimuli, under attentional modulation (Schünemann and Ernst 2023‡). Indeed, evidence from EEG studies indicates that activity in cortical auditory areas synchronizes precisely with regular features of rapidly changing auditory stimuli, independent of attention and indeed even during propofol-induced unconsciousness; beyond auditory cortex, in widely distributed areas such as frontal and parietal cortex, attention is required for sustained increases in activity in response to such auditory patterns (Herrmann and Johnsrude 2018; Tauber *et al.* 2024). Consistent with local, automatic processing, activity in secondary auditory cortex shows a delay relative to primary cortex consistent with direct, feedforward signal propagation (Barth and MacDonald 1996). Bringing these disparate facts together: perhaps the prominence of auditory hallucinations in Sz is a result of top-down regulatory influences that are particularly weak (both normally and in Sz), intrinsically dysregulated activity in non-BG-recipient auditory areas, and intrinsic and extrinsic dysregulation in BG-recipient auditory areas, pervasively implicating GABA signaling (discussed at greater length later (§13.5)).

### *8.11. Disruption in schizophrenia of sleep spindling and prefrontal FSI activity likely grossly disrupt BGMS.*

As noted above, sleep spindling, generating broadly synchronized responses in cortex, particularly implicates the BG-recipient intralaminar and association nuclei of the thalamus (Contreras *et al.* 1997). Spindling is thought to be crucial for consolidation during sleep of new associations (Tamminen *et al.* 2010; Genzel *et al.* 2014). Moreover, a direct association has been demonstrated between the prevalence of fast parietal spindles during stage 2 and slow wave sleep, and fluid intelligence (Fang *et al.* 2017).

A consistent pattern of deficient spindle activity in stage 2 sleep has been demonstrated in Sz, with severity of symptoms correlated to degree of deficiency (Ferrarelli *et al.* 2007, 2010a; Wamsley *et al.* 2012). Sz is correspondingly characterized by a profound deficiency of sleep-dependent motor skill consolidation (Manoach *et al.* 2004). Because synaptic homeostasis mechanisms largely operate at the level of individual microcircuits, neurons, and synapses (Turrigiano 2011), spindling deficits may cause progressive deterioration of the long range circuits that are the physiological basis of effective connectivity in wakefulness. Such deterioration is, in any case, characteristic of Sz (Lim *et al.* 1999; Mori *et al.* 2007; Collin *et al.* 2014; de Leeuw *et al.* 2015). As reviewed later (§13.5), it is the hub areas of cortex that are most implicated in the circuit deterioration characteristic of Sz. These are the areas most clearly implicated in fluid intelligence, as explored later (§16.4).

Sleep spindles have been found to preferentially recruit FSIs in PFC, more than pyramidal projection cells there (Peyrache *et al.* 2011). This is likely a consequence of feedforward inhibition in response to the lengthy ultra-high frequency bursts associated with spindling, importantly demonstrating that thalamocortical projections appose both FSIs and pyramidal cells in cortex. Impairment of GABA synthesis in intrinsic FSIs of DLPFC, and consequent deficiencies in cortical projection neuron synchronization and loosening of spike coincidence criteria, have been implicated in Sz (Lewis *et al.* 2005; Gonzalez-Burgos *et al.* 2015). PFC FSI response patterns are also modified by dopamine inputs (Tierney *et al.* 2008), which are abnormal in Sz (Grace 2016). The consequences of severe deficiencies in sleep spindling, simultaneous with disruption of feedforward inhibition by cortical FSIs, may disrupt BGMS with particular potency. Indeed, as noted earlier (§5.7), evidence suggests that sleep spindles are central to the expression of corticostriatal plasticity (Lemke *et al.* 2021). Whether spindle and PFC FSI deficiencies are part of the etiology of Sz, or are sequelae, remains to be determined and may vary. It is probably significant that both can result directly from GABA dysfunction.

### *8.12. Reports on the functional correlations of the intralaminar nuclei, and their physiological relationships with the basal ganglia and cortex, likely supply some of the best available evidence supporting the BGMS model.*

Evidence that the intralaminar nuclei are profusely innervated by the BG and integral to BG circuitry, that they are innervated by and proximally appose L5 pyramidal neurons, that these appositions are subject to stringent (<4 ms) coincidence requirements, and that spike bursts from highly energetic intralaminar neurons in a state of wakefulness last only 4-5 ms and recur at a rate of 20-40 Hz, suggest that BG output associated with well-practiced behavior and cognition is precisely aligned on this timescale. While the timing of spikes in projections to superficial cortex is surely significant, it is in the projections to somatic layers that timing appears most critical, and that the potential for timing-based selectivity is most apparent.

## 9. The Direct, Indirect, and Striosomal Paths in the Regulation of Cortical Dynamics

**In this section:**

### *9.1. The regulation of cortical dynamics implicates all BG circuitry, and the striatum is the linchpin.*

While the direct path is particularly prominent in BGMS, due to its crucial role in precision activation of effective connections, BG circuitry beyond the direct path is also functionally crucial, and indeed is even more extensive and broadly connected than the direct path. The striatum is the common component in all these circuits. The striatum is a particularly complex brain organ, structured simultaneously along multiple schemes overlaid upon, and interacting with, each other in intricate patterns (Graybiel 1990; Kreitzer 2009; Tepper *et al.* 2010; Bolam *et al.* 2000; Märtin *et al.* 2019). Its striosome-matrix compartmentation, and its direct-indirect dichotomy, both bear upon the present hypothesis.

### *9.2. SPNs in the direct path are preferentially innervated by cortical neurons with reciprocal corticocortical connectivity.*

Among corticostriatal projection neurons, there is evidence that most direct path cells, but not most indirect path cells, are reciprocally connected over long ranges at the single unit level, and are a specialized population dedicated to intracortical connectivity and striatal innervation ("intratelencephalic"); the indirect path is predominantly innervated by collaterals of projections that descend through the pyramidal tract, and whose corticocortical collaterals are not reciprocal (Lei *et al.* 2004; Morishima and Kawaguchi 2006). As noted earlier (§6.3), projections from interconnected cortical areas systematically converge on striatal FSIs at the single unit level (Ramanathan *et al.* 2002), and FSIs show a substantial preference for direct path SPNs (Gittis *et al.* 2010). Thus, the innervation of the direct path is distinguished by systematic patterns of reciprocal long range connectivity and corresponding striatal convergence, whereas indirect path corticostriatal inputs are predominantly collaterals of descending fibers such as corticopontine motor output, whose cells of origin do not reciprocate with each other, and as reviewed below, show markedly less striatal convergence.

### *9.3. BG output to thalamus arises from activity in relatively superficial cortical layers, and passes exclusively through striatal matrix, while striosomes receive input from relatively deeper layers, with areal distinctions.*

The direct path through the BG to thalamus implicates SPNs in the matrix compartment exclusively (Rajakumar *et al.* 1993), and the corticostriatal innervation of the matrix is differentiated from that of the striosomes in important ways. While the striosomes and matrix are both broadly targeted by most cortical areas, the striosomes preferentially receive projections from L6 and deep L5, while the matrix is preferentially targeted by superficial L5, and by L2 and L3 (Gerfen 1989; Kincaid and Wilson 1996).

Ascending projections from the densely direct-path-recipient PF thalamic nucleus pervasively and diffusely innervate the matrix compartment of associative striatum, while largely avoiding striosomes; CM projections to sensorimotor striatum are less pervasive but similarly prefer matrix (Sadikot *et al.* 1992b). The CL and PC nuclei also project densely to the caudate striatum (Kaufman and Rosenquist 1985a). The striatal projections of these intralaminar nuclei appose the dendrites of SPNs, with varying physiological and morphological properties (Lacey *et al.* 2007), and evidence also suggests that they innervate striatal FSIs (Sidibé and Smith 1999). As proposed earlier (§7.12), thalamostriatal projections may position the striatum to monitor (and therefore optimize and rapidly sequence) the synchronies that its output produces in thalamus, and thus presumptively in cortex, via BG output structures.

Intriguing areal distinctions in cortex have also been identified. In primate, dorsolateral PFC (DLPFC) targets matrix densely and broadly, largely avoiding striosomes, while orbitofrontal and anterior cingulate cortex preferentially target striosomes (Eblen and Graybiel 1995). Matrix appears specialized to project to the pallidal segments and the SNr, while striosomes appear specialized to project to midbrain dopamine centers such as the substantia nigra *compacta* part (SNc), to whose densocellular zone they are reciprocally linked (Jiménez-Castellanos and Graybiel 1989; Crittenden *et al.* 2016).

Activity in matrix reflects immediate prior reward, suggesting reward-guided generation of ongoing cognition and behavior, while that in striosomes reflects anticipated outcomes, and is less reflective of immediate prior reward, suggesting implication in the generation of signals that drive or mediate motivation and the expression of plasticity (Bloem *et al.* 2017).

Striosomes strongly influence the SNc and VTA through a pallidohabenular circuit (Rajakumar *et al.* 1993; Herkenham and Nauta 1979; Hikosaka 2010; Hong and Hikosaka 2008; Balcita-Pedicino *et al.* 2011), while dopaminergic projections from the midbrain preferentially target striatal matrix (Graybiel *et al.* 1987). The involvement of striosome circuitry in motivational processing, and of dopamine in modulating responses to afferent activity, is reviewed later (§10). In particular, their roles in modulating the dynamics of superficial cortical microcircuits in PFC (Yang and Seamans 1996; Towers and Hestrin 2008), introduced earlier (§7.9), are crucial.

### *9.4. Cholinergic, serotonergic, and dopaminergic localization to striatal matrix suggest specialization for dynamic, high fidelity processing of oscillatory signals.*

The classic technique for differentiating striosomes from matrix is to stain the striatum to visualize distribution of the enzyme acetylcholinesterase (AChE) (Graybiel and Ragsdale 1978), rendering the striosomes as pale poorly stained patches. Serotonergic projections to striatum also preferentially innervate the matrix compartment (Lavoie and Parent 1990). As reviewed in detail later (§10), dopamine, ACh, and serotonin are potent modulators of oscillatory neuronal responsiveness. Thus, differential prominence of these neurotransmitters in the matrix compartment suggests specialization for the relay of oscillatory activity.

Most striatal ACh arises from an intrinsic population of interneurons comprising 2-3% of striatal neurons (Contant *et al.* 1996), which is believed to be identical to the electrophysiologically identified tonically active neurons (TANs) of the striatum (Aosaki *et al.* 1995). These neurons discharge tonically at 2-10 Hz in the absence of sensorimotor activity, and are differentially localized to the matrix, particularly to the matrix border regions adjoining striosomes (Aosaki *et al.* 1995).

The PPN, itself profusely targeted by the GPi and SNr (Semba and Fibiger 1992; Grofova and Zhou 1998; Parent *et al.* 2001; Huerta-Ocampo *et al.* 2021), provides an additional, extrinsic, supply of ACh to the striatum, and this too preferentially targets the matrix compartment (Wall *et al.* 2013). Moreover, the striatally projecting neurons of the midline and intralaminar thalamus are targeted by the PPN (Erro *et al.* 1999), and as noted earlier (§5.7), preferentially target the TAN population, participating intimately in goal-directed learning (Bradfield *et al.* 2013). FSIs, noted above for their selective and robust innervation of direct path SPNs and their putative high fidelity relaying of oscillatory activity, are extensively modulated by cholinergic inputs (Koós and Tepper 2002). Thus, the matrix compartment of the striatum is distinguished by participation in multiple, coordinated cholinergic circuits.

### *9.5. A pattern of differential innervation in the direct path suggests specialization for integration and motivated action.*

According to the BGMS model, the direct path of the BG establishes task-appropriate long range effective connections, while the indirect path largely serves to damp or desynchronize competing activity, to further secure the selected connections. Wall *et al.* (2013) identified instructive differences between afferents to these two intermingled populations of SPNs in mouse: The direct path was found to receive significantly heavier projections from primary somatosensory, ventral orbitofrontal, cingulate, frontal association, prelimbic, perirhinal, and entorhinal cortex, and to receive essentially the entire striatal projections from the amygdalar nuclei, STN, and DRN. The indirect path was found to receive a significantly heavier projection from primary motor cortex. Preferential targeting of the direct path by primary somatosensory, and of indirect path by primary motor, comports with a model in which the direct path establishes connections and facilitates actions consistent with context and task requirements, while the indirect path inhibits completed, competing, ineffective, and irrelevant activity and functional connectivity.

Direct path SPNs show higher activation thresholds and more extensive dendritic processes (~25% more dendrites) than indirect path SPNs, suggesting greater integration through the direct path (Gertler *et al.* 2008). When synchronized cortical activity is confined to a single focus in primary motor cortex, the consequent striatal activation strongly prefers the indirect path (Berretta *et al.* 1997). This disparity is a natural consequence of the indirect path preference of the corticostriatal projection originating in primary motor cortex, but might also be explained in part by a preferential responsiveness in the direct path to conditions of multi-areal activity, suggested by the role proposed in the BGMS model implicating it in the induction of selective synchronies between distant areas that typically already harbor activity.

### *9.6. Cortical inputs to the direct path appear to be an exquisitely context sensitive sparse code, with relatively high divergence-convergence.*

The information borne by the intratelencephalic corticostriatal projection appears to be distinct from that borne by the corticostriatal collaterals of the corticopontine projection from the same area. Turner and DeLong (2000) showed that in primate primary motor activity, corticopontine neurons consistently show activity associated with movement execution and, particularly, the muscular contractile command stream, whereas activity in intratelencephalic neurons is often independent of muscle activity, is exquisitely context- and feature-dependent, and is usually confined to a particular aspect of current conditions (sensory context, movement preparation, or movement underway). They suggested that these patterns of direct path input to the striatum are a sparse code, of the sort demonstrated in temporal and visual cortex (Rolls and Tovee 1995; Vinje and Gallant 2000).

Wright *et al.* (1999, 2001) showed in rat that intratelencephalic corticostriatal afferents from primary sensory areas have diffuse, convergent, and bilateral terminal patterns, implicitly raising opportunities for information integration. In contrast, they showed that corticopontine collateral input is ipsilateral, and preserves topographic specificity and organization, terminating in discrete varicosities without convergence, with thicker and faster axons. Moreover, they showed that the intratelencephalic and corticopontine projections enter the striatum almost at right angles to each other, which appears to further cultivate information integration.

Earlier studies identified the differential pattern of corticostriatal arborizations, finding that those of the intratelencephalic collaterals in the striatum are ~1.5 mm in diameter, with sporadic branching and varicosities, while the corticopontine collateral arborizations are dense, focused within a volume with longest dimension ~500 µm, and do not cross boundaries of adjacent striosomes (Cowan and Wilson 1994; Kincaid and Wilson 1996). Another investigation found that the corticopontine projection originates chiefly in lower L5, while the intratelencephalic projection originates chiefly in upper L5 and in L3 (with L3 predominating slightly in sensory cortex), and that the striatal terminal boutons of the former are roughly twice the size of the terminals of the latter (Reiner *et al.* 2003). More recent studies have confirmed that intratelencephalic afferents exhibit a higher prevalence of numerous and widely distributed terminals than do corticopontine afferents (Hooks *et al.* 2018; Morita *et al.* 2019).

### *9.7. The direct and indirect paths, and striatal matrix and patch compartments, are neither crisply distinct nor mutually exclusive.*

Preferential projection by classes of corticostriatal neurons is a matter of tendencies, not rules. Intratelencephalic corticostriatal axons prefer direct path SPNs by a 4:1 ratio, while corticopontine collateral axons prefer indirect path SPNs by a 2.5:1 ratio (Lei *et al.* 2004). Recent findings using genetically manipulated mice have shown that the cytological and hodological compartmentation of the striatum into striosomes and matrix is not crisp, with both striosomal and matriceal SPNs receiving both limbic and sensorimotor inputs, and projections to SNc arising from both striosomal and matriceal SPNs (Smith *et al.* 2016). Earlier studies demonstrated similar minor projections of sensorimotor cortex to striosomes, and revealed sparse projections from striosomal neurons to the pallidal segments (Flaherty and Graybiel 1993). Motivational specificity and contextualization are apparent in striatal matrix activity (Donahue *et al.* 2018‡), and this intermodal convergence-divergence may be related.

The canonical marker for direct and indirect path SPNs is expression of dopamine receptors from the $D_1$ and $D_2$ receptor families, respectively (Gerfen and Surmeier 2011), but SPNs express DA receptors from the opposing family at low levels (Smith and Kieval 2000), and BG microcircuits intermingle the effects of DA receptors from both families (Gerfen and Surmeier 2011). Indeed, the axons of individual SPNs in primate frequently branch to both direct and indirect path targets (Parent *et al.* 1995; Levesque and Parent 2005). Moreover, in the ventral pallidum, neurons with projection patterns characteristic of the GPi/SNr and the GPe are closely intermingled, receiving projections from direct and indirect path SPNs (Groenewegen *et al.* 1993; Smith and Kieval 2000).

Voluntary behavior is preceded by simultaneous activation of both direct and indirect path SPNs (Cui *et al.* 2013; Donahue *et al.* 2018‡). This coactivation, while typically antagonistic, is not symmetric (Oldenburg and Sabatini 2015), and evidence suggests that these asymmetric dynamics are central to the sequential elaboration of precisely timed motor commands (Markowitz *et al.* 2018). Moreover, there is evidence that concurrent and asymmetric activity of SPNs in the direct, indirect, and striosomal paths collectively represents all aspects, phases, and structure of a task, with SPNs tiling the task space with activity to form a continuous representation of the task in all its particulars (Weglage *et al.* 2021; Arcizet and Krauzlis 2018). These dynamics are consistent with the proposition that "matrisomes" consisting of closely intermingled direct and indirect path SPNs, with presumptively overlapping dendritic processes, facilitate coordination of direct and indirect path output (Flaherty and Graybiel 1993).

Beyond these complexities in the physiology and dynamics of the indirect, direct, and striosomal pathways, there are many additional pathways that bypass and supplement them. The hyperdirect path from frontal cortex to the subthalamic nucleus (Nambu *et al.* 2002) is implicated in stopping actions (Schmidt *et al.* 2013), while the pallidostriatal projection is thought to be involved in the cancellation of stopped actions (Mallet *et al.* 2016). There is also evidence of pathways directly linking the cerebral cortex to the globus pallidus (Milardi *et al.* 2015; Smith and Wichmann 2015; Karube *et al.* 2019), and the globus pallidus to the cerebral cortex (Van Der Kooy and Kolb 1985; Zheng and Monti 2019‡).

The presumptive function of these various cross-channel, cross-receptor, and bypass paths is to enrich the range of dynamics and pool of information available to the implicated individual neurons, by which they might more rapidly and appropriately respond to ever-changing context.

# 10. Dopamine, Acetylcholine, Serotonin, and the Reticular Nucleus in Oscillatory Regulation

**In this section:**

### *10.1. The basal ganglia influence central neurotransmitter sources in the brainstem and basal forebrain, modulating thalamocortical activity.*

Beyond their GABAergic projections to thalamic matrix and intralaminar nuclei, the BG are positioned to influence cortical and thalamic activity by targeting modulatory centers in the brainstem and basal forebrain. Indeed the dopaminergic centers of the midbrain are considered integral components of the basal ganglia, and project to frontal cortex and associated nuclei of the thalamus. Additional prominent targets are the cholinergic basal forebrain (including the nucleus basalis of Meynert (NBM)), the cholinergic tegmentum (the pedunculopontine and laterodorsal tegmental nuclei (PPN and LDT)), the brainstem serotonergic centers (dorsal and median raphe nuclei (DRN and MRN)), and the thalamic reticular nucleus. Neuromodulator channels are thought to be central to the orchestration of network dynamics, the disambiguation of circuits with a surfeit of synaptic connectivity (Bargmann 2012), and the control of neuronal dynamics to impart contextually appropriate receptive fields (Guarino *et al.* 2025 ‡ ); roles centrally attributed to the basal ganglia in the BGMS model.

Bidirectional control of these modulatory influences has been demonstrated — specifically, these mechanisms are associated with the direct and indirect path structures and populations, though not in canonical patterns. Phasic elevation of dopamine, which promotes locomotion, is through projections from D2-expressing SPNs in the striatum's striosomes to a specialized population in the GPe, that in turn target midbrain DA neurons (Lazaridis *et al.* 2024). Phasic inhibition of dopamine, inhibiting locomotion, is through D1-expressing striosomal SPNs projecting directly to midbrain DA neurons (Lazaridis *et al.* 2024). Thus the behavioral associations of the D1 and D2 populations in the DA-targeting striosomes are opposite those of the D1 and D2 populations in the thalamus-targeting striatal matrix, but their anatomical targeting is similar, likely owing to the nearly opposite effects of phasic GABA and phasic DA.

BG modulation of acetylcholine release has been similarly shown to be bidirectional, through opposed effects of the D1 and D2 striatal populations targeting the basal forebrain (Chen *et al.* 2024 ‡ ) and cholinergic midbrain tegmentum (Fallah *et al.* 2024‡).

Similarly, the intralaminar and reticular nuclei of the thalamus are targeted by D1 and D2 SPNs, respectively, with the GPe interposed in the latter projection (Parent *et al.* 2001; Van der Werf *et al.* 2002; Nakajima *et al.* 2019).

These arrangements suggest a brain-wide system for articulate modulatory control by the basal ganglia, pivoting on cell populations in the striatum.

Dopamine (DA) is a key modulatory neurotransmitter intrinsic to the BG, where it raises the excitability of direct path SPNs by activating their $D_1$-class receptors, and reduces the excitability of indirect path SPNs by activating their $D_2$-class receptors (Gerfen and Surmeier 2011). For reviews, see for example Schultz (1998), Bromberg-Martin *et al.* (2010), and Yetnikoff *et al.* (2014). Roles for DA in the control of oscillatory activity, in and beyond the BG, have been described that bear directly on the BGMS model, and are reviewed below.

Despite comprising less than one percent of neurons, cholinergic cells perform crucial roles in, and indeed beyond, the nervous system (Woolf and Butcher 2011). They are proposed to play a key role in orienting attention (Sarter and Bruno 1999), in induction of vigilance and fast sleep rhythms (Steriade 2004), in induction of plasticity (Rasmusson 2000), and in the formation of memories (Hasselmo 2006).

Serotonin (5-hydroxytryptamine, 5-HT) is implicated in regulation of sleep and wakefulness (Pace-Schott and Hobson 2002; Monti 2011), cognitive and behavioral flexibility (Clarke *et al.* 2006), and signaling of reward magnitude (Daw *et al.* 2002; Nakamura *et al.* 2008).

The thalamic reticular nucleus has crucial roles in attention and oscillatory regulation (Pinault 2004), and is also crucially involved in sleep processes (Contreras *et al.* 1997).

These roles of the DA, ACh, and 5-HT systems, and the TRN, are evidently closely related to each other. Indeed, the supplies of DA, ACh, and 5-HT, are closely coupled, as detailed below.

### *10.2. Dopamine communicates cognitive and motivational significance.*

The effects of DA are complex. Through broad projections to BG and amygdalar nuclei, frontal cortex, and associated thalamic nuclei, the release of DA arising from BG-controlled neurons in the ventral midbrain (chiefly SNc, VTA, and the retrorubral field, RRF) and other areas has been proposed to have a crucial role in motivational control, by signaling reward, surprise, novelty, even aversiveness, and in general, salience (Schultz *et al.* 1997; Bromberg-Martin *et al.* 2010; Ioanas *et al.* 2022). Recent results show that DA particularly signals causal associations (Jeong *et al.* 2022; Hart *et al.* 2024).

Midbrain DA projections have systematic topography (Fallon 1988), and evidence suggests separable correlates in subpopulations of DA neurons for distinct functions, and for various separable aspects of reward prediction error, such as timing vs. magnitude (Lau *et al.* 2017). Activation of DA projections from the VTA to the basal amygdala has been shown in mice to be associated with the formation of fear memories (Tang *et al.* 2020), which are archetypically aversive. There is evidence that distinct clusters of DA neurons in the ventral midbrain are specialized for an assortment of reinforcement roles, particularly motivational reward and motor invigoration (Saunders *et al.* 2018), and that distinct BG circuits bear reinforcement signals for distinct functional domains, only some of which entail plainly motivational signaling (Pascucci *et al.* 2017).

DA projections to the striatum have been shown in mice to be functionally heterogeneous and selective, exhibiting topographic structure, with activation in wave-like spatiotemporal sweeps across regions of functionally related striatum, showing particular and stereotyped heterogeneity along the mediolateral axis (Hamid *et al.* 2021). These spatiotemporal waves showed a strong relationship between propagation direction and instrumental agency: as learning progressed, a task entailing strong instrumental contingency showed progressively more well-defined mediolateral propagation, while a simpler Pavlovian variant of the task showed lateromedial propagation, consistent with established roles for dorsal medial and dorsal lateral striatal functional specialization, and suggesting a crucial role in the dynamics of credit assignment (Hamid *et al.* 2021).

DA has been proposed to signal disparities between expected and actual outcomes, dipping phasically upon disappointment and rising phasically upon surprising reward, driving reinforcement learning mechanisms (Schultz 1998, 2013). In fact, evidence suggests that DA is crucial in signaling prediction errors *per se*, with or without reward associations (Sharpe *et al.* 2017).

DA release has been proposed to signal the expected value of work, in order to encourage continuation of efforts expected to culminate in a rewarding outcome, and discourage continuation of other efforts (Hamid *et al.* 2015). Indeed this neuroeconomic function has been ascribed to the BG as an ensemble (Goldberg and Bergman 2011). As noted earlier (§9.3), striosomes appear specialized to control ventral midbrain DA centers; medial PFC control of striosomes, and striosomal control of ventral midbrain DA, have been implicated in cost-benefit decision making (Friedman *et al.* 2015; Crittenden *et al.* 2016). Activity in striosomes, compared to that in striatal matrix, has been shown to preferentially encode reward-predicting cues in particular, and anticipated outcomes in general (Bloem *et al.* 2017).

Surprising sensory events can evoke prominent, short-latency DA bursts, regardless of reward association, in 60-90% of DA neurons throughout the full extent of the SNc and VTA, apparently constituting an alerting response serving to marshal attention; these bursts seem to correlate with the degree to which the stimulus captures attention by surprise, they diminish with predictability and familiarity, and they are fairly nonselective, triggered by sensory surprises that superficially resemble motivationally significant stimuli (Bromberg-Martin *et al.* 2010). This comports with the many studies that have found that the BG are integral to orientation of attention, and generation of responses, to motivationally relevant sensory stimuli (e.g. van Schouwenburg *et al.* 2010b; Cools *et al.* 2004; Leventhal *et al.* 2012).

#### *10.3. Dopamine in prefrontal cortex and associative thalamus augments responsiveness to afferent activity.*

*In vitro* studies on PFC pyramidal neurons have found that DA raises their excitability (Penit-Soria *et al.* 1987; Shi *et al.* 1997; Yang and Seamans 1996). Similarly, in the thalamic MD nucleus, DA has been shown *in vitro* to raise sensitivity to afferent activity (Lavin and Grace 1998). DA release in the MD largely derives from direct appositions arising from the VTA; indeed neurons in the VA and VL nuclei are also directly targeted by the midbrain DA centers (VTA, SNc, and RRF), as are the midline nuclei (Sánchez-González *et al.* 2005). $D_2$ receptors are found throughout the associative thalamus (Rieck *et al.* 2004), and while DA terminals only sparsely synapse on neurons in the intralaminar thalamus (Sánchez-González *et al.* 2005), $D_2$ receptors in the CM, PF, PC, and CL nuclei are particularly dense (Rieck *et al.* 2004), suggesting a large role there for volume-conducted DA action, with correspondingly less spatiotemporal specificity.

#### *10.4. Dopamine promotes oscillatory synchronization in and between the BG and motor cortex.*

Following observations of treated and untreated parkinsonian primates, human and non-human, it has been proposed that DA has a decisive role in the regulation of global beta synchrony in BG, with increases in DA providing for narrowly focused striatal responses to cortical beta activity and consequent facilitation of action, while decreases in DA promote broad propagation of cortical beta, concomitant global beta synchrony, and the retarding or arresting of action (Jenkinson and Brown 2011; Magill *et al.* 2001). As noted earlier (§6.15), the DA-depleted striatum is characterized by the spontaneous and pervasive formation of synchronized clusters of SPNs (Humphries *et al.* 2009).

A pattern of broad beta synchrony, focally disrupted in association with performance of rewarded tasks, has been found in healthy (non-parkinsonian) monkeys (Courtemanche *et al.* 2003). These patterns appear to be DA-dependent: In an experiment in which global DA levels were manipulated to ~500% and <0.2% of their natural baseline, the low-DA condition was accompanied by pervasive synchrony with locally prevailing LFP, while the high-DA condition showed widespread focal desynchronization from prevailing LFP in primary motor cortex and dorsolateral striatum (Costa *et al.* 2006). DA manipulation was not found to affect overall cortical firing rates, underscoring the primacy of synchrony (and not rate) in these dynamics. The pattern of the hyperdopaminergic condition resembles the “desynchronization” of focally synchronized gamma oscillations in activated thalamocortical ensembles (Steriade *et al.* 1996), which according to the BGMS model often involve synchronized oscillations propagating focally through the BG.

At the system level, beta oscillation frequency is strongly and directly coupled to dopamine level, while beta power, unit-LFP coherence, and phase-amplitude coupling, have a more complicated relationship with dopamine levels (Iskhakova *et al.* 2021).

Recent evidence suggests that DA acts to shift the size of responding SPN ensembles, rather than the rate of discharge of individual SPNs; acute DA blockade halves and triples the number of responding direct and indirect path SPNs respectively, generating a strong imbalance in favor of the indirect pathway, and significantly impairing spontaneous locomotion (Maltese *et al.* 2021). Earlier (§6.17), I suggested that SPNs cast votes for decisions, which are then tallied by downstream structures. According to this narrative, the effect of DA on the striatum can be viewed as expanding or contracting the pool of potential votes on either side of the decision.

#### *10.5. Dopamine promotes synchrony between the medial temporal lobe and prefrontal cortex, and stabilization of synchrony-mediated functional connectivity, promoting continuation and memorization of effective behaviors.*

Injection of DA into PFC has been seen to induce a spontaneous increase in synchrony between PFC and hippocampal LFPs, and to starkly alter the dynamics of PFC pyramidal neurons; activity shifts from in-phase with reciprocally associated interneurons (suggesting interneuronal inhibition) to opposite phase (suggesting interneuronal augmentation) (Benchenane *et al.* 2010). These effects of DA injection on PFC-hippocampal synchrony and PFC pyramidal neuron dynamics mimicked those seen without DA injection, in a well-trained behavioral task (Y maze navigation), at the choice point (the fork). DA released upon well-predicted reward, by inducing synchronization of PFC-hippocampal cell assemblies, might assure that effective behaviors are committed to long term memory, while ineffective ones are not (Benchenane *et al.* 2011). Naturally, counterproductive behaviors must also be remembered as such, implicating DA release associated with general salience (Bromberg-Martin *et al.* 2010).

As noted earlier (§7.9), DA in PFC has been found to attenuate receptivity to inputs on L1 apical dendrites (Yang and Seamans 1996), and to depress GABAergic lateral interactions among L2/L3 interneurons (Towers and Hestrin 2008), reducing the spatiotemporal coherence of oscillation there. As DA level rises, PFC neurons may thus become progressively less affected by superficial inputs from the BG-recipient thalamus and corticocortical feedback paths, so that effective behaviors are protected from disruption and distractions, and in particular, from induction of empirically extraneous functional connectivity. Indeed, DA release in PFC is suggested to stabilize working memory items there (Gruber *et al.* 2006). The effect of DA release on cortex may extend well beyond directly DA-recipient frontal cortex: an integrative theory has been proposed by van Schouwenburg *et al.* (2010a) and Bloemendaal *et al.* (2015) that DA release in PFC induces it to influence interconnected posterior cortex to stabilize goal-relevant representations and protect them from distractions, even while DA release in the BG promotes flexible adaptive responses to new information. Consistent with these accounts, evidence from resting state fMRI studies in DA-manipulated humans suggests that local activity and large scale functional networks are stabilized and reinforced by systemic DA elevation, while systemic DA depletion results in elevated variability of local activity, and the dissolution of large scale networks, particularly impacting between-module connectivity while largely sparing within-module connectivity (Shafiei *et al.* 2019).

### *10.6. Acetylcholine supply to cortex and thalamus is centralized and specific.*

The ACh supply for the cortex and thalamus arises from the basal forebrain, particularly the NBM, and from the PPN and LDT nuclei in the brainstem reticular activating system. Comprehensive direct cholinergic projections from the NBM to cerebral cortex (Mesulam *et al.* 1983; Mesulam 2004) are posited to modulate the predisposition of the targeted areas to robust afferent-driven oscillation, with fine spatiotemporal specificity (Muñoz and Rudy 2014). Each individual neuron in the NBM projects to a single small area of cortex confined to a diameter of 1-1.5 mm, prompting the proposal that the cholinergic population of the NBM is arranged to give arbitrary addressability of small areas of cortex, permitting activation of complex constellations subserving specific functions (Price and Stern 1983). fMRI evidence in humans suggests involvement of the NBM in the general orchestration of large scale cortical network dynamics, implicating both cholinergic and non-cholinergic projections (including coreleased glutamate and GABA) (Markello *et al.* 2018). The NBM's projections to TRN further position it to exert a wide-ranging influence over corticothalamic activity (Levey *et al.* 1987). BG control of the NBM is detailed below (§10.11), including profuse innervation of all sectors by the ventral striatum (Mesulam and Mufson 1984; Grove *et al.* 1986; Haber *et al.* 1990; Haber 1987).

The PPN and LDT have wide-ranging subcortical cholinergic projections, comprehensively innervating the thalamus, including its reticular nucleus (Hallanger *et al.* 1987; Satoh and Fibiger 1986; Steriade *et al.* 1988; Paré *et al.* 1988; Lavoie and Parent 1994). PPN targeting of the thalamus includes its primary sensory nuclei—the dorsolateral geniculate (DLG), medial geniculate (MG), and the ventrobasal complex (ventral posterolateral (VPL) and ventral posteromedial (VPM)) (Hallanger *et al.* 1987). It additionally projects densely to the NBM and nearly all BG structures (Lavoie and Parent 1994).

Underscoring their functional significance, these cholinergic supply centers have prominent roles in disease processes. PPN lesions result in akinesia, and PPN degeneration is associated with PD (Pahapill and Lozano 2000). Alzheimer's disease is associated with attrition of the magnocellular cholinergic population in the NBM, typically to less than 30% of normal (Arendt *et al.* 1983). In Sz, the concentration of choline acetyltransferase in PPN and LDT is markedly lower than normal, while the concentration of nicotinamide-adenine dinucleotide phosphate (NADPH) diaphorase appears to be roughly twice normal (Karson *et al.* 1996; German *et al.* 1999). Indeed, systemic cholinergic abnormality may be a frequent correlate of Sz, and atypical antipsychotics such as clozapine and olanzapine have a high affinity for muscarinic receptors (Raedler *et al.* 2006; Scarr and Dean 2008).

### *10.7. Acetylcholine promotes cortical responsiveness; cholinergic blockade in cortex drastically attenuates cortical activation, and when coupled with serotonergic blockade, resembles decortication.*

If the tonic supply of ACh to a cortical locus is interrupted, neurons there become dramatically less sensitive to their excitatory afferents, and correspondingly more prone to tonic synchrony with their neighbors; ACh modulates the propensity of these neurons to track high frequency afferent oscillation and generate corresponding efferent oscillation, particularly in the beta and gamma bands (Rodriguez *et al.* 2004). Phasic increase in ACh supply to an area, when coupled with afferent activity, induces profound plasticity within tens of minutes, persistently elevating the propensity of the targeted area to synchronize with afferent high frequency oscillation and consequently desynchronize with neighboring tonic oscillation (Rodriguez *et al.* 2004).

Most of the brainstem diffuse modulatory systems may act on cortex indirectly through the NBM ACh and raphe 5-HT systems; cortical electrocorticographic (ECoG) activation can be completely abolished by concurrent blockade of ACh and 5-HT (Dringenberg and Vanderwolf 1997, 1998). Rats subjected to this concurrent blockade, and exhibiting complete loss of ECoG activation, nonetheless engage in active locomotion, with normal posture and open eyes; however their behavior is disorganized and aimless like that of decorticated rats, including repeated, unhesitating walking plunges over precipices, and insensate behavior in swim-to-platform tests (Vanderwolf 1992).

### *10.8. Acetylcholine in cortex shows complex facilitatory effects, bearing some similarities to those of dopamine.*

In cortex, ACh is modulatory, neither excitatory nor nonselectively disinhibitory; its presynaptic release does not by itself induce postsynaptic activity (Sillito and Kemp 1983). When coupled with excitatory afferent activity, ACh has a dramatic facilitatory effect on most cortical neurons, while maintaining or narrowing their respective receptive fields; tonic activity (discharges attributable to background afferent activity) is also reduced, so the overall effect is a marked increase in signal/noise ratio (Sillito and Kemp 1983).

The effect of ACh on cortical interneurons is more diverse, with fast spiking inhibitory (FSI) interneurons in L5 hyperpolarized via muscarinic receptors, disinhibiting the L5 pyramidal neurons they target, while low threshold spiking (LTS) inhibitory interneurons are excited via nicotinic receptors, raising inhibitory output to their more superficial targets in L1-L3 (Xiang *et al.* 1998).

Cholinergic hyperpolarization of cortical FSIs may relax the coincidence detection window for perisomatic inputs to pyramidal neurons (Pouille and Scanziani 2001), effectively increasing their receptive field, even while the direct effect of ACh on them is a narrowing of their receptive fields as described above. Moreover, the coherent lateral spread of oscillatory activity in L2/L3 (Tamás *et al.* 2000) may be depressed by ACh hyperpolarization of FSIs (as by DA (Towers and Hestrin 2008)), spatially focusing activity in cortex. *In toto*, these effects appear to stabilize working memory attractor networks (Qi *et al.* 2021).

It has been shown in behaving rats that short latency ACh release, through effects mediated by a diversity of receptor types, is crucial to the generation and synchronization of performance-correlated oscillation in PFC (Howe *et al.* 2017). In task trials in which the animal detected a sensory cue, significantly elevated PFC ACh levels were detected within 1.5 s of cue presentation, and remained elevated until reward delivery. Gamma oscillation in the same area, measured by LFP, was found to be significantly elevated, at ~90 Hz from ~200-400 ms after cue presentation, then at ~50 Hz from ~400-1300 ms after the cue. Local infusion of an M1 muscarinic antagonist attenuated these gamma responses in trials in which the animal detected the cue, and was associated with a trend toward more missed cues. Infusion of a nicotinic antagonist attenuated the initial high gamma response to detected cues, and similarly had no effect on oscillatory power in trials in which the animal missed the cue. Detected cues, but not missed cues, were associated with significant cross-frequency coupling of the 50 Hz gamma response, to local theta oscillation detected by LFP. This coupling was abolished by infusion of the M1 antagonist, and was attenuated by the nicotinic antagonist.

### *10.9. Acetylcholine promotes thalamic responsiveness and high frequency thalamocortical synchrony.*

The effects of ACh on thalamic neurons have been found to be similar to those in cortex, facilitating responsiveness of excitatory neurons to afferent activity via M1 and M3 muscarinic receptors, as well as via nicotinic receptors, and having an opposite effect on inhibitory interneurons, where it induces hyperpolarization via M2 receptors, indirectly facilitating responsiveness (Parent and Descarries 2008; Steriade 2004). The cholinergic projection to PF (representing intralaminar nuclei) densely terminates in exclusively direct synapses (Parent and Descarries 2008), and PPN/LDT stimulation in the anesthetized cat, causing cholinergic activation of the thalamus, produces sustained, synchronized high frequency oscillation in intralaminar neurons and reciprocally connected cortical neurons, resembling patterns seen in the waking and REM sleep states (Steriade *et al.* 1996).

The terminal pattern of the cholinergic projection to the dorsal lateral geniculate nucleus (DLG, representing primary sensory nuclei) is almost entirely extrasynaptic (Parent and Descarries 2008), and this relatively diffuse pattern is likely to have markedly less spatiotemporal specificity than synaptic paths, so the diffuse ACh innervation of DLG comports with the expectation (according to the "binding by synchrony" hypothesis, briefly discussed later (§11.2)) that modulatory inputs to early sensory areas are arranged to not disrupt the fine time structure of activity therein. ACh inputs to the TRN are both synaptic (Parent and Descarries 2008) and extrasynaptic (Pita-Almenar *et al.* 2014), and are reported to hyperpolarize TRN neurons through M2 muscarinic receptors, disinhibiting their targets in the thalamus (Steriade 2004; Lam and Sherman 2010).

### *10.10. Acetylcholine has complex and often facilitatory effects in the BG.*

ACh has a variety of effects on striatum, through a variety of receptors: it can directly induce SPN depolarization and spontaneous firing, and in particular, facilitate the excitability of NMDA (glutamate) receptors on SPNs, while simultaneously reducing glutamate and GABA release; corticostriatal long term potentiation (LTP) in SPNs is also dependent on ACh activation of $M_1$ muscarinic receptors (Calabresi *et al.* 1998, 2000). As noted earlier (§5.7), when intrinsic cholinergic interneurons in the striatum are subjected to synchronous spike volleys, their cholinergic action on dopaminergic axons promotes intrinsic DA release in the striatum (Threlfell *et al.* 2012).

Early experiments entailing injection of cholinergic agents into striatum, pallidal segments, and STN, showed dysregulatory effects that generally appeared to be pathological activations (DeLong and Georgopoulos 2011).

### *10.11. The cholinergic centers are tightly integrated with BG circuitry.*

It has been proposed that the PPN, briefly discussed earlier (§9.4), is so intimate with the BG as to constitute an inextricable component thereof (Mena-Segovia *et al.* 2004). The GPi, VP, and SNr strongly and systematically project high velocity axon collaterals to it (Semba and Fibiger 1992; Grofova and Zhou 1998; Haber *et al.* 1985; Parent *et al.* 2001; Harnois and Filion 1982), and cholinergic and glutamatergic cells in the PPN in turn profusely target dopaminergic cells in the SNc, with at least some of the PPN cells that target SNc receiving projections from SNr (Grofova and Zhou 1998). While there is evidence that BG projections to PPN preferentially target non-cholinergic cells (Mena-Segovia and Bolam 2009), more recent evidence demonstrates that BG output, particularly from the SN, directly targets cholinergic cells in the PPN and LDT (Huerta-Ocampo *et al.* 2021), consistent with earlier reports that the GPi projects throughout PPN, most prominently to the central PPN (Shink *et al.* 1997), which in turn projects to the NBM (Lavoie and Parent 1994). Moreover, caudal PPN is targeted by the DRN, which itself is targeted by the BG, though the effect of 5-HT on the PPN is complex and unresolved (Vertes 1991; Steininger *et al.* 1997; Martinez-Gonzalez *et al.* 2011).

The ventral striatum projects profusely to all sectors of the NBM (Mesulam and Mufson 1984; Grove *et al.* 1986; Haber *et al.* 1990; Haber 1987), and the NBM receives substantial projections from the SNc and VTA, targeting cholinergic neurons (Záborszky and Cullinan 1996; Gaykema and Záborszky 1997). At least some VS afferents to NBM terminate directly on corticopetal cholinergic neurons; GABA input to these neurons is posited to dampen excitability, resulting in corresponding inattention in their cortical targets (Sarter and Bruno 1999). The GPe, like the NBM, but much less profusely, has direct cholinergic projections to cerebral cortex (Eid and Parent 2015), and both coexpress GABA in these projections (Saunders *et al.* 2015a, 2015b). And the GPe, like the NBM, projects directly to the TRN. The NBM may be an inextricable component of an extended BG system, as has been suggested of other areas of the substantia innominata (Heimer *et al.* 1997). Indeed a model has been proposed that integrates ACh projections from the NBM, the GPe, and the VP, with BG loop circuitry (Záborszky *et al.* 1991, Fig. 6).

### *10.12. Noradrenaline supply is centralized, and indiscriminately recruits attention and arousal.*

Noradrenaline (NA) originating in the locus coeruleus (LC) of the pontine tegmentum is implicated in the direct modulation of arousal throughout the forebrain; the LC responds to noxious, novel, and other highly salient stimuli, toward which attention is to be oriented, with low latency phasic responses time-locked to the stimulus (Berridge 2008; Sara and Bouret 2012). These phasic responses are posited to reset network connectivity to facilitate assembly of a new network oriented to the salient stimulus, and there is evidence that NA arising from LC has a more general role in set shifting, crucially implicating the reciprocal connectivity of LC with PFC (Sara and Bouret 2012).

However, the striatum is not an LC target (Aston-Jones and Cohen 2005), and descending inputs to the LC have been found to be highly restricted, excluding most BG and all thalamic structures; activation of LC by afferent activity has been found to be either generalized to its entirety, or generalized to an entire sensory domain; perhaps most tellingly, output from the LC has been found to be non-specific, with efferent populations in LC distributed throughout its extent, and only modest and partial segregation according to target structure (neocortex, thalamus, cerebellum, etc.) (Aston-Jones *et al.* 1986; Waterhouse *et al.* 1993; Loughlin *et al.* 1986).

Thus, while the LC is integral to the regulation of oscillatory activity and functional connectivity in the thalamocortical system, it seems clear that the LC is nonspecific in its mechanisms. It also seems clear that it is not substantially integrated into BG circuitry, notwithstanding evidence of a sparse projection from the ventral pallidum to rostral LC (Groenewegen *et al.* 1993). It seems likely that stimulus-related network formation facilitated by LC reset signals entails broad synchronies to which the striatum responds after the fact.

### *10.13. Serotonin supply to BG, cortex, and thalamus is centralized.*

5-HT supply to the telencephalon arises from the MRN and DRN, which project strongly to the midline, intralaminar, and mediodorsal thalamic nuclei, much of the BG, and to the entirety of cerebral cortex and the medial temporal lobe (Lavoie and Parent 1990; Vertes 1991; Vertes *et al.* 1999; Baumgarten and Grozdanovic 2000). Raphe projections exhibit complex specificity, with the DRN projecting to cortex with various topographies, while the MRN projects to cortex more diffusely (Wilson and Molliver 1991); the two nuclei generally target complementary forebrain regions (Vertes *et al.* 1999).

### *10.14. Serotonin has facilitatory effects beyond those of dopamine and acetylcholine.*

5-HT has an effect on its cortical targets much like that of ACh, facilitating responses to afferents, yielding ECoG desynchronization (Neuman and Zebrowska 1992), though the effect on individual neurons is complex, with most cells depolarized via 5-HT$_2$ receptors but some hyperpolarized via other receptors (Davies *et al.* 1987).

5-HT$_{2A}$ receptors are present on the apical dendrites of L5 pyramidal neurons, so 5-HT release facilitates responsiveness (Carter *et al.* 2005) precisely where it is inhibited by DA and ACh release. This effect apparently counteracts the posited focusing and stabilizing effects of DA and ACh described above; indeed almost all known hallucinogenic drugs act through this channel, and activation of 5-HT$_{2A}$ receptors is necessary and sufficient for their hallucinogenic effects (Glennon *et al.* 1984; González-Maeso *et al.* 2007; Fiorella *et al.* 1995; but see Maqueda *et al.* 2015).

The notion arising from the BGMS model is that 5-HT$_{2A}$ agonists (even including, rarely, SSRIs for treatment of never-before-hallucinating patients (Bourgeois *et al.* 1998; Waltereit *et al.* 2013)) open cortical columns more broadly to induction of effective connections via spike-timing-dependent gain control by corticocortical feedback and BG-thalamocortical output, and hallucinogens thereby induce spurious information flow and associations that would not normally reach the implicated pyramidal somata. Evidence suggests that psychedelic facilitation of spurious effective connections is not uniform, but rather entails abnormal enhancement of connectivity in sensory and somatomotor areas, simultaneous with abnormal attenuation of connectivity in associative areas, including the default mode network (Preller *et al.* 2018). This bears a striking resemblance to the large scale network dysconnectivity characteristic of Sz (Ji *et al.* 2019; Giraldo-Chica *et al.* 2018). Consistent with these accounts, evidence suggests that the abnormal functional topologies associated with lysergic acid diethylamide (LSD) and psilocybin entail a collapse of representational hierarchy, particularly a loss of the normal functional differentiation of unimodal sensory areas from association and executive areas (Girn *et al.* 2022).

Though the dysconnectivity of Sz may principally or frequently be rooted in GABAergic and dopaminergic dysfunction (discussed at greater length later (§13.5)), there is also a suggestion of 5-HT dysfunction (Geyer and Vollenweider 2008). Atypical antipsychotics such as clozapine, risperidone, and olanzapine show much higher affinity for 5-$HT_2$ receptors, which they usually occupy almost completely, than for the $D_2$ receptors targeted by earlier antipsychotics such as haloperidol (Kapur *et al.* 1999). Beyond this, common direct BG involvement is plausible. 5-$HT_{2C}$ receptors in the striatum, activated by hallucinogens (Fiorella *et al.* 1995), have been found to excite striatal FSIs (Blomeley and Bracci 2009), and direct striatal involvement in Sz has been posited (Graybiel 1997; Simpson *et al.* 2010; Wang *et al.* 2015). Indeed, recent evidence from an animal model directly implicates the visual striatum in the induction of visual hallucination (Schmack *et al.* 2021).

### *10.15. The dorsal and median raphe nuclei are multifariously coupled with the BG.*

All parts of the BG are innervated serotonergically by the raphe nuclei, with heterogeneous density within and between the organs of the BG, and highest density in the SN and GP (Lavoie and Parent 1990). The median and dorsal raphe nuclei (MRN and DRN) are targeted by the VP, SNr, and VTA (Peyron *et al.* 1997; Gervasoni *et al.* 2000; Levine and Jacobs 1992; Groenewegen *et al.* 1993). Coupling with BG DA centers and DA control structures is extensive. The VTA projects to the DRN and MRN; the DRN and MRN also project to DA cells in the SNc and VTA, and raphe projections to the SNr appear to be directed to the dendrites of DA neurons (Baumgarten and Grozdanovic 2000). The lateral habenula (LHb) projects strongly to all parts of the DRN (Peyron *et al.* 1997) and to the MRN (Herkenham and Nauta 1979), while the MRN projects massively throughout the extent of LHb (Vertes *et al.* 1999) and the DRN shows light but distinct targeting of LHb (Vertes 1991). The LHb is integral to BG DA circuitry—it is reciprocally linked with the VTA, directly and via the rostromedial tegmental nucleus (RMTg) (Herkenham and Nauta 1979; Hikosaka 2010; Balcita-Pedicino *et al.* 2011), and is profusely innervated by GPi and VP (Parent *et al.* 2001; Hong and Hikosaka 2008; Shabel *et al.* 2012; Groenewegen *et al.* 1993).

### *10.16. The cholinergic and serotonergic systems are tightly coupled.*

The MRN and DRN project densely to the PPN and LDT, and the DRN projects densely to the substantia innominata (including NBM, in primates) (Vertes 1991; Vertes *et al.* 1999; Steininger *et al.* 1997). The substantia innominata in turn projects to the DRN (Peyron *et al.* 1997), and PPN and LDT project to MRN and DRN (Semba and Fibiger 1992). The central 5-HT and ACh systems are thus directly and reciprocally coupled.

### *10.17. Projections from the nucleus basalis and dorsal raphe nucleus reflect corticocortical connectivity.*

As noted above (§10.6), the NBM projection to cortex is comprehensive and topographically organized. Single loci in NBM project jointly and specifically to interconnected areas of cortex, particularly frontal and posterior areas (Pearson *et al.* 1983; Ghashghaei and Barbas 2001; Záborszky *et al.* 2015). These loci of joint projection appear to entail distinct intermingled populations, with only a tiny minority (~3%) of cells collateralizing to both frontal and posterior areas (Záborszky *et al.* 2015), suggesting combinatorial flexibility. The raphe nuclei, particularly the DRN, are reported to exhibit similar organization, with small groups of dorsal raphe cells projecting to widely distributed, anatomically interconnected neocortical foci (Wilson and Molliver 1991; Molliver 1987). Evidently, these patterns of divergence are much like those of the thalamocortical projection, described earlier (§1.6).

### *10.18. Prefrontal control of cholinergic, serotonergic, and noradrenergic centers is extensive and orients attention.*

Direct and dense projections from PFC and other frontal cortical association areas to the NBM (Mesulam and Mufson 1984), PPN, and LDT (Semba and Fibiger 1992) thence to cortex and thalamus is a putative mechanism for sustained attention and inattention (Sarter *et al.* 2001; Záborszky *et al.* 1997). Indeed, PFC inactivation completely abolishes sensory-evoked ACh release in the sensory thalamus, and significantly reduces tonic ACh release in sensory cortex (Rasmusson *et al.* 2007). PFC projections to the DRN (Gonçalves *et al.* 2009) and LC (Jodoj *et al.* 1998; Aston-Jones and Cohen 2005) are thought to have similar and related functions. Because PFC is thoroughly and densely targeted by BG output via the thalamus and the midbrain DA centers, and projects directly and strongly to all BG input structures, PFC control of cholinergic and serotonergic centers implies BG influence on them, and suggests further coordination of output from these modulatory centers with BG output.

### *10.19. The thalamic reticular nucleus is implicated in oscillatory regulation, and is under BG and PFC control.*

The TRN, through GABAergic projections to other thalamic nuclei, is thought to act in a modulatory role, influencing activity and oscillations in the entire thalamus and cortex, particularly corticocortical functional connectivity (Pinault 2004). A crucial role in the generation of spindles during sleep is recognized (Contreras *et al.* 1997). Prefrontal projections to TRN are thought to play a prominent role in orientation of attention and suppression of distracters (Zikopoulos and Barbas 2006; Guillery *et al.* 1998), and dysfunction of the TRN, resulting in deficits in these and related functions, has been associated with Sz (Ferrarelli and Tononi 2011; Pinault 2011).

The GPe projects to the full rostrocaudal extent of the TRN (Hazrati and Parent 1991; Shammah-Lagnado *et al.* 1996), and this projection has been directly implicated in attentional control (Nakajima *et al.* 2019). Given evidence that some circuits through the TRN are open loops implicating more than one cortical area (Brown *et al.* 2020), this projection is positioned to directly modulate inter-areal cortical signaling. BG inputs to the TRN also target cells that project to the intralaminar thalamus (Kayahara and Nakano 1998), and experiments *in vitro* suggest that DA release in GPe inhibits its inputs to TRN (Gasca-Martinez *et al.* 2010), suggesting that these inputs conform to functions identified for the indirect path, dampening or disconnecting activity. Additionally, L5 of frontal cortex (the same layer most prominent in the corticostriatal projection) projects directly to TRN, with an important role in modulating general TRN-cortex synchrony (Hádinger *et al.* 2023), and glutamatergic nigroreticular projections have been demonstrated arising from striatum-recipient cells throughout the SN, both from the pars reticulata and the pars compacta, with roughly half of these fibers also found to release DA (Antal *et al.* 2014). And the close relationship of the BG with the TRN is bidirectional: as noted earlier (§5.7), evidence suggests that TRN-coordinated sleep spindles are crucial for corticostriatal plasticity (Lemke *et al.* 2021).

The interposition of the thalamic reticular nucleus in collaterals of L6 corticothalamic projections (Deschênes *et al.* 1994) is posited to produce nonlinearity, such that low frequency activity has a suppressive influence on thalamus via the TRN, while higher frequency activity is stimulative (Crandall *et al.* 2015). Modulation of the TRN (by the BG and PFC, in particular) might alter this dynamic, providing for adjustment of the threshold above which cortical activity stimulates activity in BG-recipient thalamus, and below which it is suppressive. This would gate the action of the BG on cortex, by controlling the supply of activity available for modulation at the implicated thalamocortical neurons. The BG and PFC are arranged to control this gate by adjusting the ACh supply to TRN, reducing or abolishing the suppressive influence of the TRN on corticothalamic targets (Lam and Sherman 2010).

# 11. Basal Ganglia Involvement in Sensory Processing

**In this section:**

### *11.1. The basal ganglia make selections in the sensory domains.*

The BG have been proposed to function in perceptual decision making in a fashion analogous to their function in behavioral decision making (Ding and Gold 2013). Indeed, the BG are pervasively involved in the phenomena and faculties underlying perception, attention, and awareness (Redinbaugh and Saalmann 2024). As reviewed earlier (§7.1), BG direct path output is arranged to influence activity not only in frontal cortex, but in posterior areas, including posterior sensory areas. Pathways described earlier (§10) by which the BG modulate central DA, ACh, and 5-HT supplies, and the TRN, imply a broad modulatory influence of the BG on sensory processing.

Motor control, long associated with the BG, has an inherent intimacy with attention, which entails selective perception; for example, common mechanisms and networks have been identified underlying attention and oculomotor control, both within and beyond the BG (Corbetta *et al.* 1998; Hikosaka *et al.* 2000). There is evidence that striatal activations continually track visuospatial attentional orientation, even in the absence of saccades and other overt actions (Arcizet and Krauzlis 2018). Attention is thought to reshape the representation of stimuli transmitted by sensory areas to align with the preferences of receivers, imparting to attended stimuli an advantage as they compete for influence over downstream structures (Ruff and Cohen 2019; Schünemann and Ernst 2023‡). The basal ganglia, through their powerful influences over the thalamus, frontal cortex, and neuromodulator channels, are positioned both to influence the shaping of representations in transmitters, and to adjust preferences in receivers.

When isochronous rhythmic visual events are presented to monkeys, in a task requiring saccades when oddballs are encountered in the sequence, firing of some striatal neurons is strongly entrained to the stimulus rhythm, and missing-stimulus oddballs evoke even stronger isochronous responses from those neurons, suggesting that perception of such phenomena involves cortico-basal ganglia ensembles (Kameda *et al.* 2019). Indeed, structural asymmetries of the globus pallidus correlate with alpha oscillatory power asymmetries in the visual cortex (Mazzetti *et al.* 2019), and activity in the BG-recipient central thalamus shows significant attention-related, performance-correlated upward shifts in power spectra (Schiff *et al.* 2013).

The basal ganglia are also extensively targeted by auditory cortex (Sitek *et al.* 2022 ‡ ), and are crucially involved in auditory discrimination (Znamenskiy and Zador 2013; Xiong *et al.* 2015). These pathways and mechanisms are fundamental to BG roles in speech production, discussed below. Auditory cortex shows a triangular pattern of structural connectivity, with reciprocally connected temporal and frontal loci projecting convergently to striatum (Van Hoesen *et al.* 1981). The same arrangement (discussed earlier (§4.3)) is seen with projections from parietal cortex to frontal cortex and striatum (Hwang *et al.* 2019). These are examples of a widespread motif (Yeterian and Van Hoesen 1978) that, in BGMS, is essential for phase-coherent integration of BG output into cortical activity.

### *11.2. Direct path influence on primary sensory thalamus is modulatory, not entraining or resetting.*

Primary sensory areas of the thalamus, and thalamic sensory areas in intimate topographic registration with the primary areas, are apparently avoided by direct path output (Percheron *et al.* 1996; Parent *et al.* 2001). This arrangement is an expected corollary of the “binding by synchrony” mechanism (von der Malsburg 1999; Singer and Gray 1995; Womelsdorf *et al.* 2007; Jia *et al.* 2013; Barth and MacDonald 1996; Siegel *et al.* 2008). Rigid BG-induced spike timing disruption of sensory processing pipelines at the thalamic level would derange the spatiotemporally precise registration by which ensembles of neurons representing a stimulus are proposed to be coherently bound together, and to be differentiated from neurons in the same area that are active but not associated with the stimulus. In fact, there is evidence for binding by synchrony in sensory nuclei of the thalamus: corticothalamic projections bearing synchronized oscillations associated with a visual stimulus entrain thalamocortical activity associated with that stimulus, increasing the effective neuronal gain for associated features (Sillito *et al.* 1994).

### *11.3. The basal ganglia project widely to sensory cortex, with notable exceptions.*

While the BG direct path output apparently avoids sensory thalamic nuclei, it does not avoid sensory areas at the cortical level. As noted earlier (§7.1), the BG-recipient intralaminar nuclei (CL, PC, CM, and PF) have been shown to project to visual, auditory, and somatosensory cortex (Van der Werf *et al.* 2002; Scannell 1999), and the MD, VA, and VL nuclei have been shown in primates to project to visual association cortex and the angular gyrus area of parietal cortex (Middleton and Strick 1996; Clower *et al.* 2005; Tigges *et al.* 1983). In the rostral intralaminar thalamus, the PC and CL nuclei show distinct intimacy with sensory areas, reaching all visual areas but the primary receptive fields; these projections show no apparent topographic pattern, but are accompanied by heavy projections to densely interconnected areas such as the frontal eye fields and posterior parietal association areas 5 and 7, with some axons found to collateralize multi-areally, e.g. to visual area 20a and areas 5 and 7 (Kaufman and Rosenquist 1985a; Van der Werf *et al.* 2002).

As noted earlier (§8.10), the caudal intralaminar nuclei in cat project to secondary and some associative auditory cortex, but avoid the primary, posterior, ventroposterior, and temporal auditory fields (Scannell 1999). This extensive lacuna suggests that the early stages of auditory processing are particularly sensitive to disruption of spike patterns. This might be attributable to the unique orientation of auditory perception to environmental phenomena that are typically oscillatory and momentary, so that the crucial phenomenological attributes of stimuli can only be faithfully represented in early processing stages with neural spikes that are precisely locked in time to occurrence of those attributes. While attended auditory stimuli must be faithfully represented, this imperative is relaxed with unattended stimuli. Indeed, evidence suggests that while attended auditory stimuli are processed in a continuous fashion, unattended ones are subject to rhythmic fluctuations in efficacy (Lui *et al.* 2025), and that early auditory structures are targeted by the basal ganglia to induce inattention (Nakajima *et al.* 2019). This suggests an arrangement in which activity associated with distracter stimuli is entrained to a control rhythm at a phase disfavored by downstream structures, explored below (§11.5).

Sensory representations may take the form of periodic codes, patterned to precisely reflect stimulus periodicity, which by coherent propagation and integration can enable precise sound localization (Brown and Curto 2022). Similar periodic coding principles might prevail in the grid cells of the hippocampal system (Brown and Curto 2022), which is similarly free of direct BG inputs, as outlined later (§14.2.4).

### *11.4. Basal ganglia output beyond the direct path projects to sensory areas at the cortical, thalamic, and brainstem levels.*

The projection systems associated with the GPe, SN *pars lateralis* (SNl), PPN, LDT, NBM, and DRN extensively target sensory areas, including those in thalamus. As noted earlier (§10.6), the PPN targets the primary sensory nuclei of the thalamus (Hallanger *et al.* 1987). The caudal GPe projects directly to the auditory and visual sensory sectors of the caudal TRN, to auditory cortex, the inferior colliculus, and through the SNl further influences visual and auditory processing via the latter's projections to the superior and inferior colliculi (Shammah-Lagnado *et al.* 1996; Moriizumi and Hattori 1991; Yasui *et al.* 1991). Projections of GABAergic cells in the NBM to TRN target the vision-specific portion of the latter, while cholinergic cells in NBM project to corresponding visual cortex (Bickford *et al.* 1994). While paths through the DRN from BG output structures to sensory cortex have yet to be directly demonstrated, the DRN comprehensively innervates cortex (Vertes 1991) and, as reviewed earlier (§10.15), is reciprocally coupled to the BG.

The superior colliculus is a key center for sensory (particularly visuospatial) processing: it is implicated not only in ocular saccades but in covert (i.e., non-motoric) orienting of attention (Robinson and Kertzman 1995), supplies powerful inputs to thalamic MD and pulvinar nuclei (Wurtz *et al.* 2005; Stepniewska *et al.* 2000) thence to visuocognitive cortex (Berman and Wurtz 2010; Lyon *et al.* 2010), and is under the direct influence of the SNr (Hikosaka and Wurtz 1983). The SC also projects to the thalamic intralaminar nuclei, and through them, the striatum, where it has strong attentional effects (Herman *et al.* 2020). The SC and BG thus influence each other strongly and reciprocally.

It is quite intriguing that BG output avoids the pulvinar, but extensively innervates the SC, given that the SC extensively innervates the pulvinar. Perhaps this relates to the proposed imperative to avoid deranging fine timing information in thalamocortical sensory modules. Nigrotectal terminals, while GABAergic, mostly appose medium or small dendrites (Behan *et al.* 1987) in a fashion similar to BG appositions in the intralaminar thalamus (Sidibé *et al.* 2002); enveloping perisomatic GABAergic appositions like those of the nigrothalamic projection to thalamic matrix (Bodor *et al.* 2008) are present in the same tectal population, but arise elsewhere (Behan *et al.* 1987).

### *11.5. The basal ganglia mediate attentional neglect of anticipated extraneous percepts.*

Nakajima *et al.* (2019) have demonstrated in mice that the BG are crucial mediators of corticothalamic regulation of inattention to distracting stimuli, building on earlier work substantiating roles for the TRN (Halassa *et al.* 2014; Wimmer *et al.* 2015) and PFC (Rikhye *et al.* 2018; Schmitt *et al.* 2017) in sensory attention mechanisms. In particular, they found that the prelimbic region of PFC induces inattention to distracting visual stimuli via a pathway through the dorsal caudal striatum, where projections from PFC and visual cortex converge, to the caudal GPe, thence to the visual sector of the TRN, thence to the primary visual thalamus (lateral geniculate body, LGB). They found that neglect of auditory distracters was mediated by a similar path ending in the medial geniculate body (MGB), and moreover, that in subtasks requiring auditory discrimination, auditory signal to noise ratio was executively enhanced via this PFC-BG-TRN-MGB path.

### *11.6. The basal ganglia may mediate attentional neglect of percepts arising predictably from intentional acts by the self.*

In healthy humans, speech production involves significant inhibition of responsiveness in auditory cortex to the sounds of self-produced speech; this inhibition is deficient in Parkinson's disease (Railo *et al.* 2020). This suggests that the BG are a key component of a mechanism whereby the organism avoids distraction by the perceptual correlates of successfully produced intentional behaviors. In short, the paths through the dorsal striatum to GPe, thence to TRN and thalamus, from the SNr to the SC, thence to thalamus, and from the ventral striatum to NBM, thence to cortex, may be crucial elements of a system that continually transforms behavioral output into selective, anticipatory inattention. The subthalamic nucleus (STN) is embedded within this circuitry. Activity there indeed encodes formants during speech production, with timing simultaneous with rather than preceding production, consistent with this posited role in which the BG mediate expectation-driven inattention (Lipski *et al.* 2024). The coherence of auditory-related activity in the STN is remarkable, reaching to frequencies well above those reached in cortex (Hnazaee *et al.* 2024 ‡ ). In the visual domain, the lateral geniculate body is involved in the inhibition of responses to self-generated behavior (Vega-Zuniga *et al.* 2025), which may involve pathways similar to those discussed above (Nakajima *et al.* 2019), crucially involving the basal ganglia. The STN has also been implicated in visual decision making, with experimental manipulation of STN activity causing changes in behavioral outcomes and response times (Rogers *et al.* 2024‡).

Corticostriatal input from primary motor cortex has been found to preferentially flow to the GPe (Wall *et al.* 2013), and corticostriatal input flowing to GPe is predominantly collaterals of axons destined for the pyramidal tract, bearing activity tightly correlated with executed motor commands (Lei *et al.* 2004; Morishima and Kawaguchi 2006). Collaterals of these axons also target the proximal dendrites of projection neurons in the intralaminar thalamic nuclei (Deschênes *et al.* 1998), from which these signals are relayed to all components of the BG. A key role posited for the signals carried by collaterals of motor output is as an “efference copy” or “corollary discharge”, primarily serving to contextualize sensory input, as suggested by projections from motor cortex to somatosensory cortex (DeFelipe *et al.* 1986). These signals may also serve to inform a system that differentiates between self-generated and other-generated percepts, attending the latter while disregarding the former (Crapse and Sommer 2008).

The BG might continually compute a dynamic template (in the sense described by the spatial computing proposal (Chen *et al.* 2025 ‡ )) imparted upon the thalamus via the TRN and SC, and upon the cortex via the NBM, so that sensory input that is the expected result of motor output, and is therefore cognitively extraneous and distracting, is functionally disconnected. Sparsity in direct path corticostriatal input, but not in indirect path input (Turner and DeLong 2000), comports with particular involvement of the indirect path in mediating this continual expectation-driven inattention. GPe facilitation of the TRN might act particularly by raising the threshold for corticothalamic inputs to transition from inhibitory to excitatory effects on their thalamic targets (Crandall *et al.* 2015).

The BG influence the SC directly (Hikosaka and Wurtz 1983), and if an effect of this influence is to induce neglect of expected percepts, then unexpected percepts will be salient. This comports with evidence that the SC, upon encountering unexpected sensory events, can reconfigure sensorimotor orientation in the thalamus via the zona incerta (Watson *et al.* 2015).

These arrangements suggest a combined dynamic in which the BG learn to minimize surprise, centrally implicating dopamine signaling (Schultz 1998, 2013), as noted earlier (§10.2). Indeed, some highly abstract models of cognition imply that the minimization of surprise is an organizing principle for the brain as a whole (Friston 2010), given evidence that feedback projections bear predictions, while feedforward projections bear “newsworthy” prediction errors (Friston 2018; Rao and Ballard 1999).

It has also been suggested that BG-mediated selection of an action jointly activates areas implicated in processing the expected perceptual correlates of that action (Colder 2015; but see Urbain and Deschênes 2007). The combined dynamic might consist of activation and effective connection of the executive and perceptual areas implicated in the action, culminating in execution of the action, whereupon an efference copy of the corticofugal motor output follows the paths through the BG to the TRN and NBM described here, in addition to corticocortical paths. By this narrative, if the action has the expected result, TRN and NBM outputs mask out the associated sensory inputs, presumably with a crucial role for collaterals of motor cortex output projecting to somatosensory areas of cortex and thalamus. If the results of the action deviate from expectations, then sensory inputs associated with the deviation are not masked out, but act as bottom-up drivers with particular salience due to the anticipatory recruitment of the associated cortical perceptual areas. The disparity is thus efficiently signaled, facilitating remediation.

That the BG systematically adjust sensory perceptions to track expectations and minimize surprise is suggested by the finding that the ventral striatum and ventral pallidum mediate prepulse inhibition of the acoustic startle response (mild auditory stimulus presented 30-500ms before startling stimulus) (Kodsi and Swerdlow 1995). This prepulse inhibition is deficient in many diseases associated with the BG, including OCD, Huntington's disease, and GTS (Swerdlow and Geyer 1998), is attenuated in Sz (Swerdlow and Geyer 1998; Quednow *et al.* 2008; Geyer and Vollenweider 2008), and is altered by hallucinogenic drugs (Vollenweider *et al.* 2007; Geyer and Vollenweider 2008). Anticipatory inhibition of the cortical response to self-generated speech is deficient not only in PD (Railo *et al.* 2020), but also in Sz and bipolar disorder (Ford *et al.* 2013).

There is evidence that corollary discharge underlying self-other differentiation is generally dysfunctional in Sz (Ford *et al.* 2001). Even basic coordination of motor output with sensory input is affected: smooth tracking of moving objects with the eyes is consistently impaired in Sz patients and their close relatives (Levy *et al.* 1994). This might be explained by dysfunction of the mechanisms of anticipatory inattention, if internally caused and therefore predictable sensory events are given spurious salience, prompting inappropriate actions (e.g., saccades). Mechanisms of attention implicating oscillatory activity exhibit graded dysfunction in Sz, and this dysfunction is associated with spurious salience for distracting stimuli, which would normally be subject to executive neglect by antisynchrony with the preferred phase of receivers (Lakatos *et al.* 2013). Deficient performance in Sz on stimulus-antisaccade tasks might be similarly rooted in dysfunction of the mechanisms of executive inhibition (Fukushima *et al.* 1988).

The accurate differentiation of self-generated from other-generated effects is crucial in the cognitive representation of agency (intentional action), and its dysfunction is likely intrinsic to Sz (van der Weiden *et al.* 2015; Ford *et al.* 2007). Evidence of projections in higher primates from the MD, VA, and VL nuclei to the angular gyrus (Tigges *et al.* 1983) is also suggestive, as this cortical area has been shown in humans to have a role in awareness of action consequences, and in particular, in the detection of disparities between intentions and results (Farrer *et al.* 2008), suggesting extensive BG involvement in the dynamics of agency. Agency in itself seems to influence the perception of results relative to the actions that produced them: Subjective intentionality significantly shortens the reported delay between action and results, compared to delays reported in involuntary action scenarios (such as an experimenter tugging an appendage with a fabric loop), and this shortening is lessened when the sense of agency in the action is disrupted by hypnosis (Lush *et al.* 2017) or coercion (Caspar *et al.* 2016). This too suggests BG involvement, as the BG have a prominent role in time-related perception (Buhusi and Meck 2005).

# 12. The Roles of the Basal Ganglia in General Cognitive Coordination

**In this section:**

### *12.1. The basal ganglia are implicated in the regulation of all large scale cortical networks.*

The BGMS model implicates the BG in the activation and coordination of large scale cortical networks spanning and pervading the sensory, motor, cognitive, and motivational domains. Evidently, general cognitive coordination is uniquely challenging in terms of combinatorial tractability. BG physiology reviewed earlier (§6) suggests that they are suited for such a role; below I explore this proposition more directly and in greater depth.

### *12.2. The combinatorially prodigious demands of cortical coordination are met by the combinatorial power of the basal ganglia.*

The connectedness of the cerebral cortex—the proportion of combinatorially possible direct long range links that are anatomically actualized—is quite high, at least 66% in macaques (Markov *et al.* 2014) and as much as 97% in mouse (Gămănuţ *et al.* 2018). 130-140 distinct cortical areas have been identified in the macaque (Markov *et al.* 2014), implying at least 5,500 (66% × (130 × 129) × ½) bidirectionally or unidirectionally interconnected pairs in each hemisphere and as many as 25,000 such pairs overall. More notionally, these figures imply at least $10^{78}$ ($2^{130 \times 2}$) distinct areal combinations, some substantial fraction of which might be both anatomically connected and usefully selected for momentary multi-areal effective connectivity. In humans, the number of distinct cortical areas is even larger, estimated at 180 (Glasser *et al.* 2016), implying more than 40,000 linkages and $10^{108}$ distinct areal combinations. The sheer scale of this network is also apparent in the estimated populations, with roughly $8 \times 10^9$ neurons and $6.6 \times 10^{13}$ synapses in the human cerebral cortex, connected by roughly 10 million kilometers (.07 of an astronomical unit) of axons (Murre and Sturdy 1995).

Implicit to the BGMS model is a proposal that the mammalian brain tames these combinatorial and population explosions with a mechanism combining spatiotemporally precise but mesoscopic connectivity selection with the stupendous topological flexibility of the BG. That the BG are in fact central to this facility is strongly implied by evidence that they are among the most connected regions of the brain (van den Heuvel and Sporns 2011), that they participate in a particularly wide variety of large scale synchronized networks (Keitel and Gross 2016), and that their dysfunction is directly associated with a graded contraction of the number of distinct areal combinations activated in the course of cognition (Sorrentino *et al.* 2021). Pathological atrophy of the BG is associated with reduction in “perturbational complexity index” in brain injury patients (Lutkenhoff *et al.* 2020), and traumatic disruption of projections from the BG-recipient mediodorsal thalamus to PFC is associated with a dynamical contraction of cortical EEG into fewer states, longer state dwelling, more predictable transitions, and repetitive cycles (Mofakham *et al.* 2022).

The corticostriatal projection, with its prodigious convergence and divergence (Flaherty and Graybiel 1994; Hintiryan *et al.* 2016), a total synapse population in humans of roughly $10^{12}$ (Kreczmanski *et al.* 2007; Kincaid *et al.* 1998; Zheng and Wilson 2002), and uncorrelated postsynaptic activity (Wilson 2013), is quantitatively suited to play a key role grappling with the immense dimensionality of the neocortex. Moreover, the unusual diversity of conduction delays through the BG, reviewed earlier (§5.4), may arrange for a population of “polychronous groups” of neurons that exceeds not only the number of neurons in the system, but perhaps even the number of *synapses*, providing for immense combinatorial coverage and dimensionality, and correspondingly stupendous representational capacities (Izhikevich 2006). In this view, each projection neuron in the striatum transiently participates in a vast array of contextually activated assemblies, each of which relates particular spatiotemporally dispersed inputs to corresponding learned spatiotemporally focused outputs, projecting to pallidal/nigral projection neurons, which in turn focally target the thalamus (and other areas). This arrangement is functionally homologous to an arrangement in the cerebellum, discussed at greater length later (§14.1), that is thought to provide for recognition of roughly $10^{82}$ distinct spatiotemporal patterns in the mouse (Sultan and Heck 2003). It is also similar in some important respects to arrangements of massive phased arrays of independent antennas, used and proposed for transmission and reception of signals in extremely flexible, high capacity, parallel, dynamically configurable communication links (Rusek *et al.* 2013) and, particularly, radar systems (Fuhrmann *et al.* 2010).

### *12.3. As controllers of corticocortical information routing, the basal ganglia are integral to higher mental function.*

Open circuits through the BG and thalamus, originating in one region and projecting to another one distant from the first, have long been appreciated (Joel and Weiner 1994). A comprehensive corticostriatal projection, and massive convergence through the BG, arrange for selections to reflect global, system-level information, including large scale dynamic network topology. In the BGMS model, this global information informs cortical routing decisions, and indeed allows for establishment of efficient directed routes even between topologically remote areas, via connectivity hubs.

Within the global workspace model of cognition (Dehaene and Naccache 2001; Baars 2005; Dehaene and Changeux 2011; Baars *et al.* 2013), the BG might arbitrate ephemeral access to and by specialized processors, and more generally, “dynamically mobilize” cortical areas for effective connection within the long range distributed network of conscious cognition. Equivalently, in the dynamic core model (Tononi and Edelman 1998), the BG might determine from moment to moment which corticothalamic modules are functionally well-connected.

Hybrid metaheuristics, a combinatorial optimization technique, involves such arrangements (Blum *et al.* 2011): densely and broadly connected areas may constitute a generic problem-solving (metaheuristic) mechanism, while more specialized and less widely connected areas are selectively integrated with the generic mechanism, when their respective domains of expertise are relevant to problems for which conscious intervention has occurred. Selective engagement also involves dynamic shifts of neural subspaces that orchestrate integrations and segregations, which may hinge on the spatial dynamics of oscillations (Miller *et al.* 2024). The BG figure prominently, because many of the cortical areas with the highest anatomical and functional connectedness—areas including the superior and lateral prefrontal, anterior cingulate, and medial orbitofrontal (Cole *et al.* 2010; van den Heuvel and Sporns 2011; Harriger *et al.* 2012; Elston 2000)—are particularly dense targets of BG output (Middleton and Strick 2002; Ullman 2006; Akkal *et al.* 2007). Indeed, the striatum itself has been found to contain the most connected brain regions by some measures (van den Heuvel and Sporns 2011), and to exhibit continual activation throughout the duration of a task, with shifting loci of activation that “tile” the span of the task, tracking context (Arcizet and Krauzlis 2018; Weglage *et al.* 2021).

Proponents of the global workspace theory of consciousness contemplate “auto-catalytic” organization of long range functional networks in cortex (Dehaene and Naccache 2001) to avoid a homuncular infinite regress (Dehaene and Changeux 2011). The BGMS model implies that these functional networks are self-organized by a coalition of cortical and subcortical mechanisms, with the BG in particular crucial to the selection and recruitment of cortical networks, and to the inhibition and isolation of areas not implicated in a selected network and the associated current context. The combined system of the PFC and BG has been proposed expressly to constitute a mechanistically complete explanation for coherent cognition, avoiding implications of a cognitive homunculus (Hazy *et al.* 2006). Curiously, the sensorimotor striatum contains many fragmentary sensorimotor homunculi in various configurations (Flaherty and Graybiel 1994), implying that the associative striatum contains many fragmentary cognitive maps, which through the looping architecture of the BG might be said to regress indefinitely, if not infinitely (see more below, regarding perturbative iteration).

### *12.4. The basal ganglia are arranged to control chaotic dynamics in cortex.*

The arrangement of the BG to influence cortical activity mesoscopically, without driving activity directly, has inspired the view that they dynamically modulate state attractors, shaping the evolution of cortical activity (Djurfeldt *et al.* 2001; Shine 2021). This view is particularly appealing in light of evidence and simulations that suggest that the cerebral cortex intrinsically maintains critical dynamics, supporting continually evolving dynamical activity (van Vreeswijk and Sompolinsky 1996; Haider *et al.* 2006, 2012; Okun and Lampl 2008; Ahmadian and Miller 2021; Ma *et al.* 2019; Rubin *et al.* 2015). These dynamics are shaped during development by input from the basal ganglia via the thalamus (Deemyad *et al.* 2024‡), and can be dynamically controlled by the BG-recipient thalamus (Müller *et al.* 2023). Criticality in networks has been shown to provide for optimal controllability of those networks: slight gain reductions in such networks shift them toward modular isolation and preservation of state, while slight gain increases shift them toward connectedness and integration (Li *et al.* 2019). This has direct implications for the relationship of the BG to cortex, because they are positioned to adjust the gain of thalamocortical and corticocortical circuits. Indeed, network dynamics can be manipulated largely independent of channel information contents and carrying capacity (Engelken *et al.*

2024 ‡ ), which can arrange to selectively gate in new information (Xu *et al.* 2024).

Critical dynamics characterize the cortex in the awake but resting state, while attention focused on a task induces broad subcriticality, reducing susceptibility to distracters (Fagerholm *et al.* 2015). Task-specific output from connectivity hubs in cortex may be crucial in maintaining task-appropriate effective connectivity despite the widespread subcriticality that accompanies task-focused cognition (Senden *et al.* 2018), and there is ample evidence that the BG are integral to hub circuitry and dynamics (e.g. Vatansever *et al.* 2016; Averbeck *et al.* 2014; Assem *et al.* 2020). Moreover, BG output to brainstem and basal forebrain neuromodulatory centers, detailed earlier (§10), position the BG to orchestrate shifts in balance between criticality (while in resting states) and subcriticality (while engrossed in tasks), supplying task-specific facilitation through the direct and neuromodulatory paths during task-focused, subcritical episodes. And as discussed later (§13.7), distraction and sleep deprivation are both associated with orchestrated subcriticality, entailing cognitive compromise.

Decisions may consist of dynamic and plastic reconfigurations of attractors in cortex that stabilize selected action representations against perturbations. Modeling of frontocortical neural activity underlying stimulus-response behavior in mice suggests that contextually appropriate response representations are actively stabilized against distracters during delay periods, with distance between attractor basins (associated with alternative actions) rising with the strength of the stimulus, and attractor basin depth increasing with learning (Finkelstein *et al.* 2021). Physiological evidence from human and non-human primates (fMRI and electrophysiologically measured spike rates, respectively) suggests that task performance entails global “quenching” of variability and long-range activity correlations, suppressing spontaneous activity, while associated computational modeling suggests that it entails stabilization of task-related attractors (Ito *et al.* 2020). These dynamics directly implicate the striatum. Modeling of the combined system of PFC and striatum in monkeys suggests they are central to stimulus-response learning, and that learning entails increasing the distance between, or the attractor basin depth of, the representations of candidate actions (Márton *et al.* 2020).

It has been proposed, under the rubric of “integrated information theory”, that consciousness in its essence is a series of selected states, each an ephemeral complex of informational relationships, within an internally well-connected system with massive dimensionality and the power to discriminate among the myriad possible states as wholes—in mammals, the system of the cerebral cortex and thalamus (Tononi 2004). The physical aspects of the brain directly implicated in consciousness are proposed to be those with maximal cause-effect power (Tononi *et al.* 2016). By these criteria, the global integration of cortical states implicit to the massive convergence of the BG, and predominant BG control of spike patterns in many of their thalamocortical targets, signify a central role in consciousness. Indeed, the integrated information measure Φ, evaluated for activity in a network spanning deep layers of lateral intraparietal cortex, caudate striatum, and intralaminar thalamus, is a particularly reliable indicator of state of consciousness (anesthesia, NREM sleep, resting wakefulness, or anesthesia interrupted by thalamic stimulation) (Afrasiabi *et al.* 2021). Later (§16); I explore this topic in greater depth.

### ***12.5. The basal ganglia are positioned to perform a central role in mental supervision and problem solving.***

Conscious cognition, and prefrontal cortex, are thought to be crucial for high level supervision, particularly the goal-motivated resolution of problems not resolved at lower (more local) levels (Dehaene and Naccache 2001; Miller and Cohen 2001). Evidence supports this proposition, and implicates the BG in these dynamics. It has been shown in rats that the learning of new goal-directed actions, adapted to changing scenarios, depends on the connection from PFC to striatum (Hart *et al.* 2018). In primates, the anterior cingulate cortex (ACC) and DLPFC are thought to subserve detection and mitigation of conflicts, through an iterative looping arrangement (Carter and van Veen 2007). As noted above, these areas densely reciprocate with the BG, with the ACC particularly targeting striosomes (Eblen and Graybiel 1995). As a dense target of mesencephalic DA, ACC has been proposed to form a loop with the BG subserving conflict management (Holroyd and Coles 2002). Evidence from electrophysiology in humans suggests that a key role of the ACC in conflict management is the selective amplification of useful task-relevant information in widely distributed functional constellations (Ebitz *et al.* 2020 ‡ ), which comports neatly with heavy projection of ACC to the striatum—to striosomes, through which cognitive salience might be selectively dopaminergically modulated, and to matrix, through which impulses and rhythms associated with the selection might be circulated broadly to cortex.

In terms of goal-motivated and iterative problem solving, a striking corollary of the BGMS model is that the BG both recognize and generate large scale patterns of synchrony in cortex, so that synchrony-oriented information processing in the BG is not just integrative, but recurrent. Iteration can provide for the formulation of solutions by a process of perturbative adjustments to representations (Lourenço *et al.* 2003; Czégel *et al.* 2020‡), constituting a metaheuristic algorithm. In this vein, closed circuits in cortex and BG have been proposed to underlie the dynamical emergence of valuations and decisions, with structural hierarchy and hidden layers enabling adaptation to varying timescales (Hunt and Hayden 2017).

Structural and functional recurrence in the PFC and BG in particular have been suggested to arrange for the progressive integration of evidence to drive decisions (Bogacz and Gurney 2007; Caballero *et al.* 2018; Yartsev *et al.* 2018), a facility which is deficient (relatively inefficient and inflexible) in Parkinson's Disease (O'Callaghan *et al.* 2017).

Another consequence of these looping arrangements is that they provide for neural network layering (depth) that is notionally infinite. In general, network recurrence allows for limited computational resources to be recycled and repurposed over successive iterations, allowing for situationally appropriate tradeoff of speed versus accuracy (Spoerer *et al.* 2020). Indeed it has been suggested that activation of thalamic matrix by BG inputs promotes integrative, serial processing (Shine 2021).

### *12.6. Noise in the basal ganglia and neocortex may crucially promote problem solving.*

Optimization without *a priori* expertise can benefit from random perturbations, whereby systems can escape impoundment in local optima to find broader or global optima (Lourenço *et al.* 2003; Findling and Wyart 2020‡; Palmer and O'Shea 2015; Palmer 2020). A related supposition is that the noise inherent to neural activity in cortex facilitates cognitive generalization, tending to make stimulus categorization maximally inclusive (Hu *et al.* 2019‡), and guarding against cognitive over-specialization (Hoel 2021). Remarkably, artificial random perturbation of cortical microcircuits in PFC has been shown to significantly and broadly enhance cognitive performance (Sheffield *et al.* 2020‡).

It is thus interesting that pallidal and nigral projection neurons appear to be arranged to continually inject noise into the thalamocortical system, through their tonic, rapid, independently rhythmic discharges (Brown *et al.* 2001; Stanford 2002; Wilson 2013). While this noise may improve signal fidelity in important respects, as suggested earlier (§3.2), it may also act to randomly perturb BG-thalamocortical state, via thalamocortical, thalamostriatal, thalamopallidal/thalamonigral, and thalamosubthalamic projections. At the striatal level, loops through the intralaminar nuclei, implicated in corticostriatal signaling (Ding *et al.* 2010) and execution of motor actions in uncertain environments (Mandelbaum *et al.* 2019), circulate this noise back to the striatum. In addition to generically introducing random perturbations, this might facilitate stochastic resonance, with adaptive aspects, as demonstrated in other systems with similar arrangements (Mitaim and Kosko 1998). And intrinsically, cortical microcircuits are perturbed by robustly irregular interneuron discharge patterns (Stiefel *et al.* 2013).

Learning to recognize coherent ensembles by their patterns of activity, which in BGMS is a fundamental role of the striatum, is notoriously difficult. The "quantum annealing" optimization technique, in which global minima are found by noisy quantum tunneling, appears to be particularly effective in solving this problem (Wierzbiński *et al.* 2023). Surprisingly, and speculatively, the BG might leverage quantum annealing to address the threefold challenge broached earlier (§2.2): (1) how are hidden useful trajectories first activated? (2) How is plasticity expressed to provide for their contextually appropriate subsequent reactivation? and (3) How is plasticity tapered off as the relation becomes adequately represented? The first of these is rooted partly in the noise bath of the basal ganglia reviewed above, the second is associated with striatal SPN activation concurrent with phasic dopamine and cholinergic pauses, with a pivotal role for the intralaminar thalamus (reviewed earlier (§5.7)), and (3) is rooted in progressively more synchronized BG output displacing progressively more noise, effectively lowering the annealing "temperature" (tunneling probability) after activation of consolidated stimulus-response relations, reflected in the transition from gamma to beta over the course of striatal learning (Howe *et al.* 2011; Leventhal *et al.* 2012). Phenomenological substantiation of functionally significant quantum phenomena in brains is a relatively recent development reviewed in detail later (§16.9).

Until a representation is robustly stabilized, random perturbations may contribute crucially to an organism's search for useful responses to environmental challenges and opportunities. Indeed, it has been suggested that brain dynamics are generally characterized by alternation between two modes, one chaotic, noisy, energetically diffuse, and prone to creativity, the other stable, deterministic, and energetically focused (Palmer and O'Shea 2015; Palmer 2020). If so, the BG are well-positioned to control these modes: as suggested above, BG influences over central dopamine, acetylcholine, and serotonin supplies, and modulation of thalamocortical activity targeting L1, position the BG to switch between broadly critical and broadly subcritical modes, which correspond to the chaotic and focused modes described by Palmer and O'Shea (2015).

### *12.7. The basal ganglia are intimately involved in the mechanisms of working memory.*

BG-controlled dopamine release in PFC is thought to stabilize working memory (WM) items, represented as attractor networks, against distracters and noise, simultaneous with its release in the BG enhancing targeted output to the implicated cortical loci (Gruber *et al.* 2006). This proposal comports neatly with findings, discussed earlier (§7.9), that dopamine increases overall PFC pyramidal neuron responsiveness to afferent activity, but reduces the responsiveness of their L1 apical dendrites (Yang and Seamans 1996), and depresses GABAergic lateral interactions in L2/L3 interneurons (Towers and Hestrin 2008). WM impairment in Sz has long been recognized as a cardinal symptom (Lee and Park 2005), and may be explained in large part by dysfunctions of DA regulation, excitatory-inhibitory balance, functional connectivity, and apical dendrite excitability (Uhlhaas 2013; van den Heuvel *et al.* 2013; Braver and Cohen 1999; Grace 2016; Goldman-Rakic 1999; Geyer and Vollenweider 2008; Dandash *et al.* 2017; Braun *et al.* 2021).

BG intimacy with the thalamic MD and VA nuclei underscores a multifarious role for the BG in managing working memory (Frank *et al.* 2001; McNab and Klingberg 2008; Chatham and Badre 2015; Kalivas *et al.* 2001; Haber and Calzavara 2009; Watanabe and Funahashi 2012; Mitchell and Chakraborty 2013; Xiao and Barbas 2004; Parnaudeau *et al.* 2013). In one influential model, the BG are thought to gate the establishment of items in WM and their inclusion in subsequent cognition, and to eject them from WM when they are no longer relevant to current context and goals, enabling reallocation of WM resources (Frank *et al.* 2001; O'Reilly and Frank 2006). Recent results show that the striatum, and the densely BG-recipient MD nucleus, are important for sustaining context-dependent WM-related activity in PFC during delay periods (Wilhelm *et al.* 2023; Bolkan *et al.* 2017). The MD in particular has been shown to regulate rule-contingent functional connectivity in PFC (Schmitt *et al.* 2017), and to be necessary for normal cortical dynamics associated with consciousness (Mofakham *et al.* 2022).

Constellations of PFC neurons activate persistently in sensorimotor stimulus-response scenarios (Haller *et al.* 2018). In self-paced tasks requiring complex and flexible responses to stimuli, such as antonym production, loci in human PFC were found to generate high gamma oscillations that immediately followed those in sensory cortex associated with the stimulus, and continued until, and during, high gamma generation in motor cortex associated with the response. Moreover, this PFC high gamma was essential for effective response production. These patterns of activity suggest that the PFC is crucial in the flexible organization of large scale functional networks, integrating relevant information and selecting appropriate responses (Haller *et al.* 2018).

Prefrontal gamma power rises as a function of WM load (Roux *et al.* 2012), and fine alignment of the phase angle of spikes in an oscillating cell subpopulation in PFC, relative to prevailing oscillation in the wider population, may serve to delimit and orthogonalize items represented by that subpopulation, readily explaining item capacity limitations (Siegel *et al.* 2009; Lisman and Idiart 1995). In an arrangement analogous to spike-timing-dependent selection among conflicting sensory inputs (Fries *et al.* 2002), BG-mediated selection among current WM items might entail synchronization of thalamocortical spiking with the corticocortical spiking associated with the selected item. BG-facilitated gamma bursts, securing effective connections, may thereby be produced, reintegrating the WM item into ongoing cognition. Indeed, gamma bursts in PFC accompany (putatively, induce) both the establishment of items in WM, and the subsequent activation of those items for inclusion in ongoing cognition (Lundqvist *et al.* 2016, 2022). The opposing prevalence of gamma and beta in WM neuron constellations has been implicated directly in top-down control of WM (Lundqvist *et al.* 2018a).

While WM entails persistent activity (Curtis and D'Esposito 2003; Wang 2001; Goldman-Rakic 1995), WM items may be further stabilized by, or even briefly exist only as, ephemeral synaptic potentiation and associated ephemeral attractor states in cortex (Lundqvist *et al.* 2018b, 2016; Miller *et al.* 2018; Rose *et al.* 2016). Some models of PFC-BG function in WM in fact necessitate such an intracellular state maintenance mechanism (Frank *et al.* 2001).

### *12.8. Functional parcellation of frontal cortex and basal ganglia is complex, and may emphasize intrinsically persistent and preparatory activity in frontal cortex, and impulsive and reactive activity in basal ganglia.*

It seems clear that the frontal cortex and the BG are functionally coextensive components of a single inextricable system (Frank *et al.* 2001; Miller and Cohen 2001; Calzavara *et al.* 2007; Miller and Buschman 2007). Their functional delineation is thus fraught with nuance and ambiguity. One interpretation is that the BG learn associations quickly, at a lower level of generality, and frontal cortex learns associations more slowly, at a higher level of generality and stability, trained by the BG, so that frontal cortex eventually takes the lead in responding to stimuli (Miller and Buschman 2007; Antzoulatos and Miller 2011). While it is intuitively obvious that representations reflecting a larger number of examples will tend to be more general and abstract, it is not clear that the BG and frontal cortex differ crucially on this count. Even after over-training of a task, at least in some scenarios activity in the BG still leads that in frontal cortex (Antzoulatos and Miller 2014; Banaie Boroujeni and Womelsdorf 2023), and there is strong evidence that at least in some forms of motor learning, cortex is crucial for initial acquisition but not necessary for performance subsequent to consolidation (Kawai *et al.* 2015).

According to Frank *et al.* (2001), the frontal cortex represents information with briefly persistent patterns of activation, while the BG fire selectively, usually impulsively, and only coincident with substantial afferent activity, to induce updates to those persistent patterns. This is also the view of Plenz and Aertsen (1994), and relates to models in which stability in cortex is punctuated by bursts of chaos associated with reconfiguration and the incorporation of new information (Xu *et al.* 2024; Engelken *et al.* 2024‡). This view is extended by the proposition noted above, whereby briefly persistent modulations of synaptic efficacy contribute to representation in frontal cortex (Lundqvist *et al.* 2018b; Rose *et al.* 2016).

Specialization of the basal ganglia for impulsive orchestration of updates is suggested by evidence that habit formation is accompanied by the emergence of sharp start and stop signals in the BG (Jog *et al.* 1999; Barnes *et al.* 2005; Jin and Costa 2010; Smith and Graybiel 2013). It also comports with the view adopted within the BGMS model, that thalamocortical feedback activity can facilitate transient cortical gamma bursting, establishing effective connections through synchronization (Larkum 2013; Womelsdorf *et al.* 2014). As noted above, brief gamma bursts in PFC have particularly been proposed to impulsively shift brain state to selectively integrate and manipulate working memory items (Lundqvist *et al.* 2018a).

The BGMS model, and the model of Frank *et al.* (2001), furthermore comport with the sparse pattern of task-specific activity exhibited by direct path corticostriatal neurons (Turner and DeLong 2000), noted earlier (§9.6), because the corticostriatal activation associated with a particular behavioral and environmental conjunction is thereby inherently spatiotemporally limited. Cortical inputs to the indirect path arise mainly from a different population (Wall *et al.* 2013) that is not at all sparse in its activation patterns, consistent with an inhibitory function.

There are hints to the functional delineation of frontal cortex and BG in the results of several fMRI studies. A study by Cools *et al.* (2004) arranged to separate the metabolic correlates of shifts in the task relevance of objects (effectively, polymodal sensory stimuli), from those of transiently operative abstract rules. The BG and PFC were both found to be integral to the former type of preparatory shift, allocating attention and responsiveness to the task-appropriate stimuli, but not to the latter, which was only associated with activity in PFC, subsequently biasing striatal responses appropriately. Nonetheless, evidence from another fMRI study suggests that the striatum may continually model the utility of abstract rules as such, activating in response to category prediction errors, as distinguished from reward prediction errors (Ballard *et al.* 2018).

Another study showed activation of the BG when unexpected sensory stimuli prompted reorienting of attention, but not in preparatory orienting and maintenance of attention (Shulman *et al.* 2009). However, a similar study identified a role for the ventral BG in mediating shifts of preparatory attention in response to unattended (but evidently detected) sensory cues, inducing appropriate changes in frontal-posterior functional connectivity (van Schouwenburg *et al.* 2010b).

### *12.9. The basal ganglia are implicated in cognitive flexibility and associated dysfunctions.*

Exploration of the physiological correlates of cognitive and attentional flexibility has substantiated a role for the BG (Leber *et al.* 2008; van Schouwenburg *et al.* 2010b, 2012; Weerasekera *et al.* 2023), leading van Schouwenburg *et al.* (2012, 2014) to propose that frontal cortex, particularly DLPFC, controls striatal function through dense topographically organized projections, and that these projections are crucial to cognitive flexibility. Similarly, in rodents, projections from orbitofrontal cortex to the dorsomedial striatum are indispensable for choices based on quantitative nuances of reward size (Gore *et al.* 2023). Magnetoencephalographic studies of the parkinsonian brain strongly support the proposition that the BG are directly implicated in cognitive flexibility: as the disease progresses, the variety of large scale functional constellations contracts correspondingly, with the striatum, pallidum, and thalamus, all significantly affected (Sorrentino *et al.* 2021). Similar effects accompany disruption of projections from the densely BG-recipient mediodorsal thalamus to PFC (Mofakham *et al.* 2022). And using fMRI, a correlation has been shown between cognitive flexibility and resting functional connectivity of hub areas with the BG (Vatansever *et al.* 2016). Convergence of broad PFC projections in the head of the primate caudate nucleus, and activation of this area in cognitive tasks irrespective of the implicated domains, suggest that the BG themselves contain hub areas that process information generically (Averbeck *et al.* 2014; Assem *et al.* 2020). Evidence also causally implicates this area in intuitive and insightful moves in competitive gameplay (Wan *et al.* 2011, 2012).

In adolescent development, the flexibility-related pathologies characteristic of OCD and attention deficit hyperactivity disorder (ADHD) are associated particularly with deficient myelination of projections from PFC to striatum (Ziegler *et al.* 2019). It may be particularly relevant that in humans, but not in other species, frontal and striatal regions continue to develop well into adulthood (Sowell *et al.* 1999).

The decision to pursue curiosity, despite risks inherent to that pursuit, is particularly associated with striatal activation (Lau *et al.* 2020), implicating the anterior cingulate cortex, central dorsal striatum, anterior globus pallidus, and adjacent ventral pallidum (White *et al.* 2019). Similarly, the ventral striatum and pallidum are particularly activated when an individual perseveres in self-motivated attempts at extremely difficult tasks (Sakaki *et al.* 2023). Multitasking, which entails both flexibility and perseverence, is most strongly associated with elevated information flow via the putamen to the inferior parietal sulcus and to a frontal area just anterior to the SMA; limits in the capacity of this path may underlie limits in the human capacity for multitasking (Garner *et al.* 2020; Garner and Dux 2015).

Even in the absence of PFC, the BG can maintain flexibility and curiosity. Lesion experiments in primates have demonstrated that flexible and contextually appropriate, "intellectual" behavior and curiosity are retained even in prefrontally decorticated animals, albeit with a surfeit of reactivity, provided the BG are preserved (Mettler 1945). This same series of experiments demonstrated that removal of striatal tissue produces hyperactivity and incuriosity, noting that "Animals lacking the striatum always display a certain fatuous, expressionless *facies* from which the eyes stare vacantly and with morbid intentness." Subsequent bilateral pallidotomy in these animals produced hypokinesia eventually "indistinguishable from periods of sleep"

# 13. Basal Ganglia Involvement in Cognitive Dysfunction and Collapse

**In this section:**

### *13.1. Deficits associated with basal ganglia damage are as diverse and profound as those associated with cortical damage.*

Evidence from the pathological brain supports the proposition that the BG are functionally coextensive with frontal cortex, implicating them extensively in action, awareness, and cognition. Throughout this paper I have noted implication of the BG system in schizophrenia, and related Parkinson's disease to various cognitive, perceptual, and behavioral syndromes. The BG have been implicated in GTS and OCD, as noted earlier (§8.10). Below, I briefly explore additional evidence of BG/frontal functional coextensiveness from clinical and lesion studies, and more thoroughly explore BG involvement in Sz.

### *13.2. Lesions and inactivations of the BG and associated structures are associated with severe impairments of consciousness and cognitive integrity, and therapies that target the BG can restore coherent consciousness.*

Lesions of the BG in humans, particularly of the caudate portion of the striatum, lead to cognitive and behavioral deficits commonly associated with frontocortical lesions—frequently, abulia (loss of mental and motor initiative), disinhibition, memory dysfunction, and speech disturbances including, rarely, aphasia (Bhatia and Marsden 1994). Similar deficits occur with BG-recipient thalamic infarcts involving the MD and VA nuclei (Stuss *et al.* 1988) and intralaminar nuclei (Castaigne *et al.* 1981; Van der Werf *et al.* 1999), with intralaminar infarcts also consistently associated with profound attention deficits. Accidental bilateral destruction of the GPi, incidental to treatment for Parkinson's disease, has resulted in akinetic mutism (Hassler 1982).

Severe disability following brain injuries is consistently associated with selective cell loss in central thalamic nuclei, and as noted earlier (§8.3), permanent vegetative state (PVS) is associated with loss spanning the rostrocaudal extent of the intralaminar nuclei and MD nucleus (Schiff 2010). PVS is invariably accompanied by diffuse subcortical white matter damage, and is usually accompanied by widespread or severe thalamic damage, but often presents with no apparent structural abnormalities in the cerebral cortex or brainstem (Adams *et al.* 2000).

PVS is also associated with significant impairment of backward connectivity from frontal to temporal cortex, relative to minimally conscious patients and normal controls (Boly *et al.* 2011), directly implicating the most densely BG-recipient areas and layers of cortex. Similarly, anesthesia-induced unconsciousness is associated with disruption of backward connectivity from frontal to parietal cortex (Ku *et al.* 2011) and, as noted earlier (§8.3), with inactivation of the intralaminar nuclei (Alkire *et al.* 2008). Moreover, artificial activation of the intralaminar thalamus can restore wakefulness, and restore fronto-parietal functional connectivity in forward and feedback directions (Redinbaugh *et al.* 2020; Müller *et al.* 2023). Evidence suggests that it is decoupling of modulatory feedback paths, disrupting the vertical integrator function of neocortical pyramidal neurons, that is the ultimate mechanism of a variety of anesthetics acting by a variety of proximal mechanisms (Suzuki and Larkum 2020). This comports with evidence that sensory pipelines are interrupted under propofol anesthesia at the primary cortical receptive field (Tauber *et al.* 2024). Indeed, restoration of long range alpha coherence in posterior cortex is an early indication of recovery from loss of consciousness (Zhou *et al.* 2024‡).

In some patients exhibiting akinetic mutism and other severe deficits associated with the minimally conscious state, administration of the GABAA agonist zolpidem has been found to reliably induce substantial but transient recovery, apparently by restoring normal function and oscillatory structure in frontal cortex, striatum, and thalamus (Brefel-Courbon *et al.* 2007; Schiff 2010). Similar transient recoveries in other patients exhibiting similar symptoms with BG involvement have been reported in response to administration of the DA agonist levodopa (McAuley *et al.* 1999; Berger and Vilensky 2014).

Perhaps the most remarkable discovery to emerge from various studies of the physiology of reduced or lost consciousness, is that the intralaminar thalamic nuclei, comprising a very small area indeed, are quite indispensable for consciousness (Bogen 1995; Baars 1995; Van der Werf *et al.* 2002). Recent evidence shows that these nuclei, and the neighboring mediodorsal nucleus, are selectively active at the earliest moment of conscious perception, and are likely the source of theta band oscillations associated with conscious perception that induce phase-coupled higher-band oscillations in cortex (Fang *et al.* 2024 ‡ ). These are the thalamic nuclei most intimate with the BG, bearing implications amply explored earlier (§8).

### *13.3. Schizophrenia involves significant disruption of frontostriatal connectivity and associated functionalities.*

Many diseases are associated with corticostriatal abnormalities (Shepherd 2013), and Sz in particular has been proposed to be fundamentally a dysfunction of cortico-striatal loops, particularly implicating DLPFC and its striatal targets (Robbins 1990; Simpson *et al.* 2010). It is associated with significant anatomical attenuation of the DLPFC-VS projection, observed in both patients and their asymptomatic siblings (de Leeuw *et al.* 2015; Weerasekera *et al.* 2024), simultaneous with abnormally elevated functional connectivity in the ventral frontostriatal system, and abnormally attenuated functional connectivity in dorsal frontostriatal systems, both of which are correlated with severity of symptoms, and are likewise apparent in both patients and their asymptomatic first-degree relatives (Fornito *et al.* 2013).

More generally, Sz is associated with deficient BG-mediated disengagement of the default mode network during directed task performance, simultaneous with striatal hyperactivity (Wang *et al.* 2015). Similarly, fMRI evidence demonstrates that Sz patients exhibit a characteristic pattern of significant differences in dynamical functional connectivity responses to sensory stimuli, with greater than normal connectivity established for some long range pairs, and less than normal for others (Sakoğlu *et al.* 2010). Consistent with those results, EEG evidence suggests perceptual deficiencies in Sz follow in part from dysfunction of top-down attention mechanisms, entailing abnormally low attention-related sensory gain, while bottom-up sensory processing is spared (Berkovitch *et al.* 2018).

### *13.4. Disruption of working memory in schizophrenia resembles that associated with frontocortical lesions and Parkinson's disease.*

Comparisons of spatial WM task performance by patients with frontocortical lesions, PD, and Sz, reveal related and often severe deficits (Pantelis *et al.* 1997). Sz patients show particularly severe deficits in set-shifting (Jazbec *et al.* 2007), and significantly attenuated WM capacity (Silver *et al.* 2003). If, as discussed earlier (§12.7), WM items are delimited by finely graded phase distinctions (Siegel *et al.* 2009), then the narrowness and accuracy of temporal discrimination imposes a limit on addressable item capacity. In Sz this selectivity is reduced by dysfunction of GABA-dependent cortical coincidence window mechanisms (Lewis *et al.* 2005; Gonzalez-Burgos *et al.* 2015), affecting the dynamics of corticocortical, BGMS, and non-BG-recipient transthalamic paths in similar measure. These deficient dynamics may lead to the localized structural deterioration characteristic of Sz: PFC and the thalamic mediodorsal nucleus are integral to WM (Bolkan *et al.* 2017; Schmitt *et al.* 2017), and atrophy of the circuitry linking these areas is significant in Sz (Giraldo-Chica *et al.* 2018).

### *13.5. Schizophrenia is characterized by multifarious abnormality of cortical physiology, particularly affecting connectivity hub areas, that may be the result of genetic factors implicating GABA signaling.*

Sz is associated with pervasive and progressive compromise of cerebral white matter integrity (Lim *et al.* 1999; Mori *et al.* 2007), decreased dendritic spine density (Glantz and Lewis 2000; Elston 2000), and pyramidal cell body atrophy (Rajkowska *et al.* 1998), particularly impacting long range links associated with connectivity hub areas of cortex (van den Heuvel *et al.* 2013; Collin *et al.* 2014; Crossley *et al.* 2014), which have been shown by mathematical argument to be particularly fragile (Gollo *et al.* 2018). Correspondingly, Sz is characterized by functional hypoconnectivity in the associative networks that implicate hub areas, simultaneous with functional hyperconnectivity in sensorimotor networks (Ji *et al.* 2019; Giraldo-Chica *et al.* 2018).

While abnormal hub area anatomical connectivity is most pronounced in individuals affected directly by the disease, the unaffected siblings of Sz patients also show significant attenuation of these links, relative to normal controls, even while connectivity in non-hub areas is unaffected in siblings, and is not significantly affected in Sz (Collin *et al.* 2014; de Leeuw *et al.* 2015). These patterns imply a large genetic component to the disease, and an etiology that implicates mechanisms of connectivity that are specific to hub areas, which as noted earlier (§12.3) include areas that are particularly dense targets of BG output.

Indeed, hereditary vulnerability to Sz may be a general hallmark of *Homo sapiens* distinguishing the species from other vertebrate taxa: many long range structural connections unique to the human cerebrum are also among those most affected by Sz, and may evidence a differentiating optimization for integrative long range connectivity via cortical hubs, resulting in a unique vulnerability to Sz, and also to other uniquely human pathologies such as autism, bipolar disorder, and Alzheimer's disease (van den Heuvel *et al.* 2019; Crossley *et al.* 2014; Vickery *et al.* 2024).

Cerebral disintegration in Sz may be rooted in GABAergic dysfunction, and consequent pervasive oscillatory deficits (Lewis *et al.* 2005; Ferrarelli and Tononi 2011; Gonzalez-Burgos *et al.* 2015; Uhlhaas and Singer 2010; Marissal *et al.* 2018). Indeed, gene expression and immunohistochemical patterns in the DLPFC of Sz patients show a reduction in abundance specific to GABAergic interneurons, particularly in L2 (Batiuk *et al.* 2022), and a recent genome-wide association study (GWAS) suggests that executive function, which as noted above is particularly compromised in Sz, is particularly associated with genes expressed in GABA pathways (Hatoum *et al.* 2023). Moreover, analysis of gene expression patterns in Sz and unaffected relatives also implicates abnormal glutamate signaling (Tiihonen *et al.* 2021). GABA and glutamate dysfunction can disrupt the synchronies to which the cortex and striatum respond, the mechanisms whereby the BG effect selections and modulate spike timing in their targets, and the time alignments between corticocortical and trans-thalamic spike volleys that are necessary for BGMS and other subcortically mediated synchronization mechanisms. Moreover, corticostriatal projections to striosomal SPNs, as to matriceal SPNs, synapse sparsely, with high thresholds for discharge, resulting in similar sensitivity to input synchronies (Kincaid *et al.* 1998; Carter *et al.* 2007; Zheng and Wilson 2002). As noted earlier (§12.5), striosomes and PFC are arranged in recurrent dopaminergic loops. Thus synchronal abnormalities in inputs to striatum likely produce dopaminergic dysregulation, and associated dynamical dysfunction and pathological expressions of plasticity, constituting a key mechanism for pathological progression.

Abnormal neural noise in Sz, which is a particularly strong marker for the condition, might also be rooted in abnormally elevated inhibitory conductances, themselves an adaptation to the inhibitory (GABAergic) interneuron degeneration characteristic of the condition (Peterson *et al.* 2023).

While therapies targeting GABA have thus far produced modest and mixed results, continued development may lead to effective prophylactic and genuinely curative drug treatments, with the potential to alleviate the negative and cognitive symptoms that have heretofore robustly resisted treatment (Carpenter *et al.* 1999; Gonzalez-Burgos *et al.* 2015; Keefe *et al.* 2007). Encouragingly, in the PCP animal model of Sz, artificial activation of fast spiking interneurons in mouse prelimibic cortex has been shown to ameliorate working memory function (Arime *et al.* 2024).

#### ***13.6. Schizophrenia may fundamentally be a dysfunction of subcortical synchronization mechanisms centered on the thalamus.***

The etiology of Sz, and even the epoch of its emergence as a disease in *Homo sapiens*, are notoriously obscure and controversial (Tandon *et al.* 2008). Sz patients exhibit a variety of seemingly contradictory symptoms, classified generally as positive, negative, and cognitive, with each subject exhibiting an idiosyncratic syndrome (Kay *et al.* 1987; Simpson *et al.* 2010). The explanation for this variety and obscurity is readily apparent, if the irreducible etiology of Sz is dysfunction of highly distributed and heterogeneous synchronization mechanisms centered on the thalamus—particularly BGMS, but also, largely analogous mechanisms implicating the cerebellum and hippocampus, described later (§14). This is implied by a recent GWAS study of Sz that particularly implicates hippocampal and neocortical pyramidal cells and interneurons, and striatal SPNs (Trubetskoy *et al.* 2022). A particular initiating syndrome within a particular mechanism or component thereof would likely result in Sz with a distinct symptomatology, but the cascading effects of the initiating syndrome disrupt the large scale dynamics of the brain in fundamentally similar ways, allowing etiologically distinct syndromes to be meaningfully grouped together under the rubric of “schizophrenias”.

Whatever its root causes, schizophrenia implicates most components of the BGMS system. Prominent are syndromes of the PFC, striatum, frontostriatal connectivity, and DA signaling, as noted above, and of the intralaminar nuclei and their connections to PFC, cortical FSI function, the TRN, GABA signaling generally, the PPN and LDT, the cholinergic system generally, and the 5-HT system, noted earlier. Also implicated are left-lateralized GP hyperactivity (Early *et al.* 1987), cytological and neurochemical anomalies in the BG-recipient and associative thalamus more broadly (Cronenwett and Csernansky 2010), aberrant functional connectivity of thalamus with cortex generally (Cheng *et al.* 2015), and abnormalities in the gross anatomy of the basal ganglia (Mamah *et al.* 2007). If any of these components is disrupted, the capacity for BGMS to appropriately establish and dissolve effective connections in cortex, and regulate the dynamics of existing connections, is disrupted in some fashion.

#### ***13.7. Madness and genius may both be manifestations of unusual BGMS.***

It is an old saw that madness and genius have much in common, even while evidence consistently demonstrates an inverse relationship between measures of intelligence and measures of psychopathology such as schizotypy (DeYoung *et al.* 2012). The “Openness/Intellect” trait in the “Big Five” personality model, and the concept of apophenia (the perception of patterns or connections where none exist), suggest a resolution of this paradox: schizotypy and genius both entail the perception of patterns and connections that are unusual and unfamiliar to others, the former as apophenia (confusion about reality), the latter as penetrating insight into reality (DeYoung *et al.* 2012; DeYoung 2015).

The Openness/Intellect trait is founded in cognitive exploration (DeYoung 2015), which is particularly associated with striatal activation (Lau *et al.* 2020) and BG networks (White *et al.* 2019). Generally, as noted earlier (§12.9), the striatum and its links with hub areas, particularly in frontal cortex, are integral to cognitive flexibility (Leber *et al.* 2008; van Schouwenburg *et al.* 2014; Vatansever *et al.* 2016; Mettler 1945).

Frontal cortex and striatum, according to genome-wide association study (GWAS) results, are particularly implicated in the expression of extremely high IQ (Coleman *et al.* 2019), and as reviewed above (§13.3), Sz is marked by extensive disruption of the functional relationship and structural connectivity between frontal cortex and striatum, with a strong hereditary component. The influence of genes on connectivity is most pronounced in hub areas, and GWASs have implicated many of those same genes in intelligence, schizophrenia, and energy metabolism (Arnatkeviciute *et al.* 2021).

While Parkinson's disease is marked by a graded contraction of cognitive flexibility (Sorrentino *et al.* 2021), it frequently involves complex and extended visual hallucinations marked by mind-wandering (a mental state in which thoughts are unguided and unconstrained), and this mind-wandering is associated with pathological coupling of hub areas with visual areas of cortex (Walpola *et al.* 2020). Psychosis in PD often involves pareidolia, a type of apophenia in which meaningless stimuli, such as clouds, are falsely recognized as meaningful patterns, such as cats; as PD progresses, these mild symptoms may grow in severity, culminating in psychotic delusions (ffytche *et al.* 2017). Similarly, as noted earlier; experimental elevation of local dopamine level in visual striatum causes hallucination-like perceptions (Schmack *et al.* 2021).

In the BGMS model, the striatum is the crux of flexible functional connectivity decisions. Given graded semantic spaces in neocortex (Rao *et al.* 1999; Huth *et al.* 2012; Simmons and Barsalou 2003; Rajalingham and DiCarlo 2019; Lettieri *et al.* 2019; Zhang *et al.* 2020), this implicates the striatum directly and uniquely in the formation of unusual connections characteristic of apophenia (as in Sz and PD) and penetrating insight (as in genius). Indeed evidence causally implicates the striatum in flashes of insight by experts in competitive gameplay (Wan *et al.* 2011, 2012), and also implicates it in schizophrenia and psychosis (Robbins 1990; Simpson *et al.* 2010; de Leeuw *et al.* 2015; Weerasekera *et al.* 2024; Fornito *et al.* 2013).

This also suggests a dichotomy that is perhaps surprising, between intelligence and genius, with conventional measures of the former less responsive to distinctions in BG function, and the latter more closely associated with BG function, and less captured by conventional measures of intelligence.

Brains may intrinsically be subject to a tradeoff between propensity for creativity and madness, on the one hand, and stolid stability, on the other. When cortical representations minimize distance in semantic space between alternatives, creativity might be heightened, because perturbative exploration then more readily covers the semantic space. Mental and motor dexterity are closely related (Silver *et al.* 2003), so evidence from studies on motor dexterity may inform this narrative. And indeed, performance in difficult dexterity-intensive motor tasks is significantly higher when the geometric distances between the cortical representations of adjacent fingers are smaller, even while perceptual confusion of those adjacent fingers is correspondingly elevated (Liu *et al.* 2021).

Perturbative exploration is largely grounded in the conjunction of homeostatic criticality (Haider *et al.* 2006; Ma *et al.* 2019; Ahmadian and Miller 2021), which inherently positions a universe of alternatives adjacently in configuration space, and neural noise, which in itself is thought to be an indispensable ingredient in the creative exploration of semantic space (Palmer and O'Shea 2015). Evidence suggests a direct correlation between the critical regime and fluid intelligence (Ezaki *et al.* 2020). However, with smaller distances between alternatives, and greater noise, there is a greater burden on regulatory and coordinative systems, both those intrinsic to cortex, and subcortical systems centered on the thalamus, particularly the BG and cerebellum. When these systems function deficiently, madness may result.

In Sz, spike coincidence windows are pathologically enlarged (Lewis *et al.* 2005; Gonzalez-Burgos *et al.* 2015), resulting in relatively indiscriminate signal gating, while noise is pathologically elevated (Winterer and Weinberger 2003; Peterson *et al.* 2023), presumptively causing and/or evidencing representational instability. These conditions plausibly push regulatory and coordinative systems past the breaking point, and because they are arranged in closed loops with cortex, the breakage feeds back on itself. Through plasticity mechanisms, the resulting confusion is likely to be progressively embodied in the physiology underpinning semantic maps and relations, as reviewed above (§13.5).

In principle, the brain could guard against this cascade of confusion by systematically increasing the semantic spatial distance between alternatives, for example by regulating cortical microcircuits to be subcritical, effectively reducing neural noise, at the cost of a proportional reduction in creativity and flexibility. Indeed, in the normal brain, sleep deprivation is accompanied by compensatory regulatory departure from criticality (Meisel *et al.* 2017), and graded cognitive and attentional deficits (Banks and Dinges 2007; Pesoli *et al.* 2022). Similarly, distractions entail cognitive deficits and burdens (Graydon and Eysenck 1989; Sörqvist *et al.* 2016), and as noted earlier (§12.4), focus on an ongoing task induces broad subcriticality, reducing susceptibility to distracters (Fagerholm *et al.* 2015).

Clearly, in Sz, regulatory compensation fails; analysis of network dynamics suggests that in Sz, top-down control of cognition entails higher than normal energy expenditure (is pathologically effortful), while small perturbations have an unusually large impact on network state (signifying pathological instability) (Braun *et al.* 2021). Relative to healthy controls, Sz patients tend to forgo rewards, to avoid cognitive effort associated with more optimal but more elaborate strategies requiring stable representation of more information (Gershman and Lai 2021). That Sz is still somewhat common in *Homo sapiens* (roughly 4.5 per 1000 (Tandon *et al.* 2008)), despite its devastating symptoms, attests to the irreducible evolutionary advantages of homeostatic criticality and endemic noise, and the creativity and flexibility they engender.

### *13.8. A classic diagnostic test for schizophrenia may demonstrate behavioral correlates of dysfunctional BGMS.*

Proverb comprehension was the basis of some early diagnostic tests for Sz, and while in clinical practice these tests have proved unreliable, a more recent study demonstrated a strong correlation between performance on a proverb comprehension task and performance on a theory of mind task, and much better performance on the proverb comprehension task among normal controls than among Sz patients (Brüne and Bodenstein 2005). Proverbs are metaphors, and the successful comprehension of a metaphor entails the recognition of certain abstract semantic relations (insight), simultaneous with the suppression of other, concrete, semantic relations (distractions). If these semantic relations are realized physiologically as long range effective connections, then impairment of the supervisory control of effective connectivity would manifest as impaired comprehension of metaphors.

### *13.9. Schizophrenia may be a condition of continual surprise.*

The impression that emerges from the various behavioral and physiological anomalies characteristic of Sz, is of a brain that is continually and indiscriminately surprised. Percepts that are normally anticipated or familiar, and ideas that are normally dismissed as absurd, are not appropriately neglected, but instead are given spurious salience, with associated hyperdopaminergia (Kapur 2003; Bromberg-Martin *et al.* 2010). This is a clear implication of a graded misalignment in schizophrenia between attention and stimulus-associated oscillatory neural activity, associated with indiscriminative perception (Lakatos *et al.* 2013).

There is evidence that activity in the default mode network and ventral striatum is selectively correlated with surprise (Brandman *et al.* 2021), suggesting a connection between pathologically prevalent surprise in Sz, and pathologically prevalent engagement of the default mode network during directed task performance, normally disengaged by the BG (Wang *et al.* 2015).

In Sz, subjective duration, causality, sequentiality, and simultaneity, are abnormal and distorted (Martin *et al.* 2013; Schmidt *et al.* 2011; Ciullo *et al.* 2016). These abnormal representations of reality intrinsically lead to surprise. To the degree that these representations depend on functional connections and associated coincidence windows and spike-timing-dependent gain mechanisms, they depend on normal GABA dynamics, which as noted above, are disrupted in Sz. Representational deficiencies inevitably lead to senseless surprise and ideation, and associated maladaptive attentional focus and expressions of plasticity. With expectations and impressions that are fundamentally untrustworthy (cognitive symptoms), paranoia and bizarre behavior (positive symptoms) and indiscriminate withdrawal (negative symptoms) naturally follow.

As discussed earlier (§7.3), cortical areas are arranged in hierarchies, with distinct laminar patterns for descending (top-down) and ascending (bottom-up) links. Information from the environment is thought to ascend such hierarchies, while information generated by internal predictive models descends. According to the predictive coding and predictive routing models, rational mentation entails rectification of these models as inconsistencies are encountered, so that descending activity better predicts ascending activity as learning progresses. Perhaps schizophrenia characteristically entails dysfunction at the crucial interface between the descending and ascending streams, with descending streams pathologically unresponsive to and unreflective of ascending streams. Indeed, many historical and current treatments for psychosis preferentially target mechanisms involved in the generation and effectuation of descending signals: prefrontal lobotomy and dopamine antagonist psychotropics, most obviously, but also experimental treatments noted earlier that target the intralaminar thalamic nuclei.

By these narratives, treatments that restore the trustworthiness of predictions, producing remission of cognitive symptoms, will naturally lead to remission of positive and negative symptoms. And indeed, just these sorts of results are apparent in experimental non-pharmacological therapies, consisting only of working memory exercises: cognitive and negative symptoms show significant and sustained improvement (Ramsay *et al.* 2017; Cella *et al.* 2017), and perhaps most remarkably, there is evidence that these exercises restore some of the functional connectivity between thalamus and PFC that is characteristically lost in Sz (Ramsay *et al.* 2017).

# 14. Comparisons to Parallel and Related Systems

**In this section:**

## *14.1. The Cerebellum*

**In this subsection:**

### *14.1.1. The cerebellum is ancient, structurally regular, and is central to predictive forward motor control.*

The cerebellum is an ancient structure, present at the base of vertebrate phylogeny (Bell 2002), and its pervasive involvement in precision motor learning and sensory-motor coordination was established generations ago (Ito 2002). The cerebellum learns associatively, and in particular, learns predictive relations underlying forward control in stimulus-response behaviors (Giovannucci *et al.* 2017) and dynamic proprioception (Weeks *et al.* 2017). The extreme regularity of its physiology, and an absence of intrinsic excitatory feedback paths, have long inspired mechanistic, computational models of its function (Heck 2016; Cheron *et al.* 2016).

### *14.1.2. The cerebellum is integrated with neocortical and basal ganglia circuitry, with a similarly broad functional domain.*

Across a wide variety of mammalian taxa, there is a consistent ratio of 3-4 cerebellar neurons for each neocortical neuron, suggesting a close relationship between these structures (Herculano-Houzel 2010). The cerebellum is reciprocally linked with cerebral cortex and the BG (Bostan and Strick 2010; Bostan *et al.* 2013; Bostan and Strick 2018; Milardi *et al.* 2016), and its closed loops with cortex (via the thalamus) resemble those of the BG (Schmahmann and Pandya 1997; Strick *et al.* 2009; Glickstein *et al.* 1985). Oscillatory synchronies between the cerebellum and the cerebrum are recognized and proposed to be functionally significant (Courtemanche *et al.* 2013; Courtemanche and Lamarre 2005; Cheron *et al.* 2016).

Like the BG, the cerebellum exhibits fractured and repeated somatotopy and modular divergence-convergence (Manni and Petrosini 2004; Apps and Garwicz 2000; Flaherty and Graybiel 1994), with inputs from widely distributed neocortical areas converging in various combinations (Brodal and Bjaalie 1997; Kincaid *et al.* 1998).

The cerebellum seems to be functionally nearly coterminous with the BG, including extensive and varied cognitive and other non-motor roles (Strick *et al.* 2009; Schmahmann and Pandya 1997) and generic ("multi-demand") roles (Assem *et al.* 2020). Roles have been identified for the cerebellum in fear memory formation and expression (Frontera *et al.* 2020), rhythmic perception (Kameda *et al.* 2019), and spatial attention (Craig *et al.* 2021) and inattention, including inattention to self-generated percepts (Kilteni and Ehrsson 2020), a role suggested earlier (§11.6) for the BG.

### *14.1.3. The cerebellum learns to predict the veridical consequences of motor output, and by those predictions, learns to adjust motor output to more precisely match context and intent.*

Learning to remove predictable features from sensory inflow is thought to be a fixture of cerebellar function throughout vertebrate phylogeny (Bell 2002; Ito 2001). While dopamine fluctuations signal prediction errors and induce plastic adaptations in the BG and wider forebrain (Schultz *et al.* 1997; Sharpe *et al.* 2017), in the cerebellum, powerful and highly specific inputs from the inferior olivary nuclei of the brainstem constitute prediction error signals; these signals induce plastic adaptations unless they are dynamically inhibited by accurate predictions relayed from the Purkinje projection cells to the olivary nuclei via the deep cerebellar nuclei (Schweighofer *et al.* 2013; Ito 2001), provided molecular layer interneuron inhibition of Purkinje cells is released (Zhang *et al.* 2023‡). These plastic adaptations, centered in the cerebellar cortex, result in the generation of refined motor output (Ito 2002; Giovannucci *et al.* 2017). Evidence suggests that the functional domain of these refinements is continuous transformations, as in movement and quantitative judgement, as distinguished from discrete representations, as in category classification and other symbolic processing (McDougle *et al.* 2022). This functional orientation is also suggested by evidence implicating the cerebellum in enhanced perceptual discrimination at a continuously variable target moment (Breska and Ivry 2021).

Sensorimotor learning consolidation demonstrates a dependence on the cerebellum that resembles the dependence of declarative memory formation on the hippocampal system (reviewed later (§14.2)) (Hadjiosif *et al.* 2024). Because the cerebellum is positioned to trim the phase relation between connected and synchronized areas (McAfee *et al.* 2019) (reviewed in detail below (§14.1.7)), and given results suggesting rapid expression of synaptic plasticity when directly connected projection cells are synchronized at high frequency with an appropriate phase relation (Cattani *et al.* 2024) (implicating fear memory formation, which also implicates the cerebellum (Frontera *et al.* 2020)), this suggests the cerebellum may have a general role facilitating the expression of plasticity in the forebrain, particularly in relation to learned tasks.

*14.1.4. Integration of the cerebellum into thalamocortical circuitry reflects its function, with a combination of topographic precision and convergence-divergence, and differs in key respects with basal ganglia integration.*

Each olivocerebellar axon branches to form about 7 “climbing fibers” in the cerebellum (Fujita and Sugihara 2013), each of which strongly and repeatedly apposes a single Purkinje cell with a 1:1 ratio (Reeber *et al.* 2013). The olivocerebellar system exhibits exquisitely precise topographic relations (Reeber *et al.* 2013), and its inputs span the central nervous system, from the lumbar spine, through various nuclei of the brainstem, the deep cerebellar nuclei, and superior colliculus, to layer 5 of the frontal and parietal neocortex, but notably exclude occipital and temporal cortex, the thalamus, and the BG in their entirety with the sole inclusion of the ventral tegmental area (Swenson and Castro 1983a, 1983b).

The pontocerebellar mossy fiber system, constituting the sole extrinsic input to the granule cells of the cerebellar cortex, relays neocortical input from frontal motor and eye field areas, parietal including posterior areas, the entire cingulate gyrus, and extra-striate occipital visual areas, with minor inputs from polysensory and auditory association areas in temporal cortex, the parahippocampal gyrus, and dorsolateral and some medial PFC; the corticopontine projection is topographically precise, while the pontocerebellar projection is convergent-divergent, providing opportunities for integration (Brodal and Bjaalie 1997; Glickstein *et al.* 1985; Schmahmann and Pandya 1997).

fMRI, electrophysiological, and tracer/degeneration studies demonstrate that cerebellar output targets the motor, association, sensory, and intralaminar nuclei of the thalamus, and the superior colliculus, and through them, widely distributed areas within motor, sensory, and association cortex, including skeletomotor, oculomotor, prefrontal, somatosensory, parietal, insular, temporal including primary auditory, and occipital including primary visual, with extremely high temporal fidelity (Schmahmann and Pandya 1997; Sultan *et al.* 2012; Strick *et al.* 2009; Kalil 1981; Katoh *et al.* 2000). This expansive, temporally precise, largely reciprocal influence of the cerebellum on motor, sensory, and association cortex is consistent with a central role in forward modeling of the anticipated sensory correlates of actions (Sultan *et al.* 2012; Blakemore *et al.* 1998; Lindner *et al.* 2006; Synofzik *et al.* 2008).

However, as noted earlier (§7.4), integration of the cerebellum into neocortical circuitry systematically differs from the arrangements that characterize the BG. Movement can be evoked by electrical stimulation of zones in the motor thalamus receiving input from the cerebellum (Vitek *et al.* 1996; Buford *et al.* 1996). Cerebellum-recipient neurons are consistently within the parvalbumin-positive “core” population, associated with specific and narrowly circumscribed topographic projections (Jones 2001; Kuramoto *et al.* 2009). And projections from these areas terminate in L2-L5, including L4 proper (Kuramoto *et al.* 2009; García-Cabezas and Barbas 2014; Clascá *et al.* 2012). Thus, as discussed earlier (§7.3), the laminar targets of paths through the cerebellum resemble corticocortical feedforward paths; in contrast, BG paths resemble corticocortical feedback paths.

*14.1.5. The structural attributes of the cerebellum suggest a restricted role in cognition.*

Underscoring its roles in perception and cognition, the cerebellum has been implicated in schizophrenia (Andreasen *et al.* 1998; Andreasen and Pierson 2008; Forlim *et al.* 2024).

However, the cerebellum is not necessary for coherent thought and behavior—these are preserved with manageable and finite deficits even in cases of complete cerebellar agenesis (Yu *et al.* 2015). In primates, the white to gray matter ratio is lower in cerebellum than in neocortex (in chimpanzee, 0.24 and 0.64 respectively), and across mammalian taxa the scaling exponent of that ratio is significantly lower in cerebellum than in neocortex (1.13 and 1.28 respectively) (Bush and Allman 2003). While roughly 80% of projection fibers in neocortex have terminals elsewhere in neocortex, the cerebellar cortex does not form direct links with itself (Heck and Sultan 2002). The main projections of the cerebellar cortex to the deep cerebellar nuclei are not directly reciprocated—indeed, cerebellar cortex forms no excitatory closed loops with the deep cerebellar nuclei or otherwise (Cheron *et al.* 2016).

These features suggest that the cerebellum intrinsically performs little or none of the intrinsically recurrent and self-associative processing characteristic of the corticothalamic system in particular and the forebrain in general. Unusually low variability of the dynamic functional connectivity of cerebellar loci, and uniquely high similarity of structural to functional connectivity in the posterior cerebellum (Fernandez-Iriondo *et al.* 2021), suggest a prevalence of relatively narrow, precise functions for cerebellar modules, with little dynamic flexibility. This is also the implication of evidence that functionally related regions of the cerebellum and cerebral cortex structurally covary, with a particular dissociation of less-abstract from more-abstract areas (Wang *et al.* 2024b). And it is consistent with a functional orientation toward fast, automatic, parallel, feed-forward, deterministic processing (Shine 2021).

*14.1.6. The cerebellum and basal ganglia share key physiological and hodological arrangements.*

Nonetheless, that the cerebellum and BG have similar topological relationships to cortex suggests that they may affect cortical activity by similar mechanisms. The physiological minutiae of the cerebellum bear a striking resemblance to key aspects of BG physiology and function:

The GABAergic Purkinje cells of the cerebellar cortex have high intrinsic firing rates of up to 90 Hz, and they modestly converge (~50:1) on projection neurons in the deep cerebellar nuclei, which are accelerated and entrained when that input is synchronous (Heck *et al.* 2013). The deep cerebellar nuclei receive collaterals of the excitatory mossy fibers that innervate the granule cells (Shinoda *et al.* 1992), and precisely follow stimuli to frequencies above 600 Hz, with little fatigue (Sultan *et al.* 2012), in an arrangement similar to that of the projection cells in intralaminar thalamic nuclei, reviewed earlier (§8.5).

Each Purkinje cell's flat fan-like dendritic tree receives excitatory inputs from ~175,000 weak appositions by thin, unmyelinated, slow-conducting (~0.5 m/s) "parallel fibers" traveling perpendicular to the Purkinje dendritic fans (Heck and Sultan 2002). In humans, the granule cells that originate these parallel fibers number about $10^{11}$ (Andersen *et al.* 1992), which comprises the vast majority of neurons in the brain as a whole (Azevedo *et al.* 2009). In mouse, each mossy fiber ends in roughly 150 small "rosettes", each of which apposes roughly 21 granule cells, so that each input fiber to the cerebellum is distributed to the cerebellar cortex through over 1000 parallel fibers (Sultan and Heck 2003), vastly expanding the dimensionality of the input pattern representation. The astronomically large population of granule cells, and the associated divergence-convergence of cerebellar circuitry, suggest enormous combinatorial power (Marr 1969; Sultan and Heck 2003).

Purkinje cells, like striatal SPNs, respond only to synchronized input (Sultan and Heck 2003), and have "Up" and "Down" states, with transitions triggered by impulsive input currents (Loewenstein *et al.* 2005). Similar to the slow and varied corticostriatal fiber population, and with a similar range of conduction delays, parallel fibers can function to align a constellation of non-coincident afferent spike volleys originating in diverse locations, so that their arrival at a Purkinje cell is precisely coincident, evoking a response; other Purkinje cells, at which these inputs are not coincident, are unresponsive (Braitenberg *et al.* 1997; Braitenberg 1961; Heck 1993, 1995; Heck *et al.* 2001; Heck and Sultan 2002; Sultan and Heck 2003). Braitenberg *et al.* (1997), noted earlier (§1.9), term this the "tidal wave" mechanism.

Similar to the BG, and with important implications discussed earlier (§8) regarding BG circuitry, the cerebellum targets both superficial and deep neocortical layers, via separate deep cerebellar nuclei; in particular, the fastigial nucleus targets superficial layers as part of a putative diffuse activating system (Steriade 1995).

*14.1.7. The cerebellum is positioned to tune and stabilize contextually appropriate large scale networks in the forebrain.*

These arrangements suggest that the cerebellum may affect cortical activity much as the BG do in BGMS. Indeed, this proposal has been previously suggested (Courtemanche *et al.* 2013), is implied by evidence of task-specific preparatory and reactive oscillatory synchronies between functionally related cerebellar and neocortical areas (Courtemanche and Lamarre 2005), and is strongly suggested by evidence that the cerebellum is necessary for normal activity-related gamma synchrony between sensory and motor cortex (Popa *et al.* 2013). It was recently reported that Purkinje cells explicitly represent the oscillatory phase difference between medial PFC and hippocampus, crucially implicating conduction delays in cerebellar parallel fibers (McAfee *et al.* 2019), and indeed evidence directly supports the proposition that the cerebellum modulates gamma coherence between these two cortical areas (Liu *et al.* 2022), and between sensory and motor cortex (Lindeman *et al.* 2021). By tuning the moment-to-moment phase relationship of oscillations in connected areas, the cerebellum is positioned to optimize and stabilize their directed functional connectivity (McAfee *et al.* 2022), purely by rate-coded modulation of oscillations in the targeted areas (McAfee *et al.* 2022; Herzfeld *et al.* 2023). Through integration with the basal ganglia (Bostan and Strick 2018), a combined dynamic emerges: the BG orchestrate topological shifts in functional connectivity, and the cerebellum optimizes and stabilizes the resulting connections, with both systems tracking detailed context supplied chiefly by cortical inputs.

As discussed earlier (§5.3) regarding the BG, and evidently with equal relevance to the cerebellum, recurrent paths with variable delays allow for selective reinforcement of particular oscillatory frequencies. Because each large scale functional constellation exhibits a characteristic profile of dominant frequencies (Keitel and Gross 2016; Becker and Hervais-Adelman 2020; Vezoli *et al.* 2021), this also suggests that the cerebellum, like the BG, can dynamically activate and stabilize specific large scale networks appropriate to context. Indeed, transcranial magnetic stimulation (TMS) of the cerebellum at theta and beta frequencies has been shown to facilitate performance of tasks whose functional networks are known to involve oscillations at frequencies that match the TMS, with double dissociation of the effects (Dave *et al.* 2020).

That oscillations in the cerebellum are functionally significant is clearly supported by evidence that cerebellar oscillations synchronized to meaningful rhythmic stimuli continue after cessation of the stimulus, coupled to frontal cortex, at the same rhythm, in what has been termed an “entrainment echo” (Zoefel *et al.* 2024‡).

Remarkably, individual Purkinje cells can learn to respond to temporally simple inputs with delayed, temporally complex, multiphasic outputs (Johansson *et al.* 2014; Majoral *et al.* 2020), which—like intrinsic oscillation in the BG—might allow the cerebellum to generate contextually appropriate oscillatory output even in the absence of oscillatory input.

### *14.2. The Hippocampal System*

**In this subsection:**

#### *14.2.1. The hippocampal system underlies the formation of episodic memories.*

The hippocampal system is thought to function as a persistent associative memory repository of first resort, capturing patterns of cortical network activation representing significant associations as they occur, in close coordination with prefrontal cortex (Eichenbaum and Cohen 2014; Battaglia *et al.* 2011; Rolls 2010; Squire *et al.* 2004; Damasio 1989; Meyer and Damasio 2009). Subsequent to initial encoded storage in the hippocampal formation, these associations are for a limited time available for retrieval (reactivation of the original cortical pattern), both to contribute to mental activity during wakefulness when relevant, and for migration to less labile (and more capacious) areas outside the hippocampus (Frankland and Bontempi 2005; Folkerts *et al.* 2018; Gilmore *et al.* 2021). This process of migration is believed to occur mostly or entirely during sleep (Wilson and McNaughton 1994; Battaglia *et al.* 2011; Rasch and Born 2013; Schapiro *et al.* 2019).

The special faculties of the hippocampal formation follow in part from its unique plasticity (Martin and Morris 2002; Deng *et al.* 2010; Snyder *et al.* 2005; Shors *et al.* 2001; Cameron and Mckay 2001; Hastings and Gould 1999) and its exceptional capacity for long-range functional connectedness (Lavenex and Amaral 2000; Mišić *et al.* 2014; Grandjean *et al.* 2017).

Note that, in this brief treatment, the term “hippocampal formation” refers to the collection of medial temporal lobe areas that are functionally and spatially contiguous with the hippocampus proper, namely the dentate gyrus, hippocampus, subiculum, presubiculum, parasubiculum, and entorhinal, perirhinal, and parahippocampal cortices (Lavenex and Amaral 2000). The “hippocampal system” comprises the hippocampal formation, the thalamic midline and anterior nuclear groups, the mammillary bodies, the septal nuclei and diagonal band of Broca, the circuitry interconnecting these loci, particularly the fornix, and the connections of these areas to the rest of the brain.

#### *14.2.2. Hippocampal lesions compromise the formation of episodic memories, while sparing other mental faculties.*

Bilateral lesions destroying or disabling the hippocampal formation are associated with severe anterograde amnesia and graded retrograde amnesia, but spare intellectual, attentional, and most working memory capacities, motor skill learning, and semantic and other non-episodic memory, and are not associated with any apparent progressive deterioration, neither of the initially unaffected mental faculties, nor of brain physiology outside that directly affected by the initial lesions (Schmolck *et al.* 2002; Corkin 2002; Annese *et al.* 2014; but see Schapiro *et al.* 2019). This pattern of deficits shows that the function of the hippocampal formation is highly specialized. Moreover, that function is not contingent on conscious engagement (Henke 2010).

Yet memory processing by the hippocampal formation entails sensitivity to and activation of widely distributed networks, close integration with PFC, profuse projections to the ventral BG (Brog *et al.* 1993), profuse innervation by midbrain DA centers (Gasbarri *et al.* 1996), and consolidation processes implicating widely synchronized thalamocortical signaling, all of which it shares with the highly generalized BGMS system described here. Moreover, long-term memory deficits in general, and hippocampal system dysfunction in particular, have been multifariously implicated in Sz as vulnerability indicators and primary symptoms (Holthausen *et al.* 2003; Harrison 2004; Seidman *et al.* 2003; Sigurdsson *et al.* 2010). Thus, though the functional role of the hippocampal system is circumscribed, many of its operating principles and physiological underpinnings are shared with the BG-thalamocortical system.

*14.2.3. The hippocampal system is organized around oscillations.*

Oscillatory activity, and sensitivity to phase, have been amply demonstrated in the hippocampal system and in long term memory processing (Fell and Axmacher 2011; Colgin 2011; Tort *et al.* 2008; Fernandez *et al.* 2013). Information flow in the hippocampal system is systematically organized around theta oscillation, with encoding of new incoming information at antiphase with retrieval of past information (Siegle and Wilson 2014; Hasselmo *et al.* 2002; Wilson *et al.* 2015). Rhythmic coordination of hippocampus and striatum has been demonstrated during learning (DeCoteau *et al.* 2007), and hippocampus and PFC exhibit increasingly synchronized oscillation as rules are acquired in a task framework, with peak coherence at the moment of decision; during subsequent sleep, hippocampal cell assemblies that participated in the coherent oscillation during performance are preferentially replayed (Benchenane *et al.* 2010). During tasks implicating navigation, large fractions of neurons, distributed widely in the forebrain, and well beyond directly connected areas, exhibit theta oscillations phase-locked to hippocampal theta (Schonhaut *et al.* 2024).

*14.2.4. The hippocampal system is largely arranged parallel to the BG system, with notable similarities and distinctions.*

It seems plausible, even likely, that reactivation of connectivity patterns by the hippocampal system entails a BGMS-related mechanism dependent on relays through the thalamus, particularly implicating the midline and anterior nuclear groups. Components of the hippocampal system project to all of the midline nuclei, which in turn project to superficial and deep layers of most cortical areas, and to the ventral striatum (Van der Werf *et al.* 2002). Notably, BG direct path and hippocampal system inputs are mutually exclusive in the midline and intralaminar nuclei, each nucleus innervated by one or the other, but not both (Van der Werf *et al.* 2002). The midline nuclei, like the intralaminars, are well-positioned to control cortical synchronies and associated effective connectivity (Saalmann 2014).

The anterior nuclear group is densely and reciprocally linked with the hippocampal formation, and projects extensively to neocortex, particularly to secondary motor, prefrontal, cingulate, retrosplenial, and some visual and temporal areas, but does not project to the BG (Jankowski *et al.* 2013), though many of these cortical targets are also targeted by BG-recipient thalamus, and indeed the GPi targets a restricted territory within the anteromedial nucleus (Xiao and Barbas 2002).

While inputs to the BG-recipient thalamus arise from the entire cortex, cortical inputs to the midline and anterior nuclei are highly restricted, confined almost entirely to the hippocampal system, despite projections from these nuclei encompassing nearly the entire cortex (Van der Werf *et al.* 2002; Jankowski *et al.* 2013). Thus, whereas BGMS is proposed to attend the control of arbitrary corticocortical connectivity, necessitating elaborate selection by the BG, involvement by the hippocampal system in the initial reactivation of a memory might entail only signals from the hippocampal formation to neocortex, corticocortically and via transthalamic paths through the midline and anterior nuclei. Notably, the hippocampal system possesses no mechanism for adjusting spike volleys for alignment in the manner of the cerebellum and BG; the central role of theta oscillation in hippocampal system dynamics might in part reflect a relative inflexibility of trans-hippocampal path delays, so that adequate time alignment can only be attained at lower working frequencies. This implies that the terminal patterns of hippocampal system projections to neocortex disfavor the interneurons that realize narrow coincidence windows there, though this is yet to be demonstrated empirically. Interestingly, hippocampally instigated cortical ripples thought to be involved in memory consolidation and recall show patterns of long range corticocortical phase locking in high gamma, without phase locking to hippocampal ripples (Dickey *et al.* 2022a, 2022b), again suggesting cruder temporal mechanisms characterize the hippocampal system proper.

Direct projections from the hippocampal formation to neocortex also parallel those of BG-recipient thalamus: evidence from genetically manipulated mice suggests that perirhinal projections to neocortical layer 1 gate the formation of memories, with a crucial role for selectively facilitated bursting of L5 pyramidal neurons, both in the formation and the retrieval of behavioral memories (Doron *et al.* 2020), much as projections from BG-recipient thalamus to superficial cortex influence the formation and persistence of functional connections, as discussed earlier (§7.6).

The obvious suggestion is that the midline nuclei and anterior nuclear group function within the hippocampal system the way the intralaminar nuclei and MD, VA, VL, and VM nuclei function within the system described by the BGMS model, with both systems operating chiefly by the spike-timing-dependent mechanisms endemic to the cerebral cortex, thalamus, and striatum. The PFC and ventral striatum, jointly targeted by the hippocampal formation and by thalamic and other nuclei in both systems, are then positioned to coordinate activity in these two vast and largely separate systems, particularly by incorporating motivation and behavioral relevance into the control of memory formation and activation. BG influence on hippocampal system activity is implied by projections from the BG-recipient PC and PF nuclei to perirhinal, entorhinal, prelimbic, and parahippocampal cortices (Van der Werf *et al.* 2002), and by the projection from the GPi to the anteromedial nucleus (Xiao and Barbas 2002). Additionally, the paraventricular and reuniens nuclei of the midline group receive dense projections from midbrain DA centers (Van der Werf *et al.* 2002), whereby the BG presumptively align memory dynamics with motivational and salience context, while the central medial nucleus of the intralaminar group receives hippocampal system inputs (uniquely among nuclei classified as "intralaminar"), and projects densely to the dorsal striatum (Van der Werf *et al.* 2002), suggesting episodic memory contextualization of dorsal BG inputs.

A notable architectural distinction between these two systems is that hippocampal formation input to thalamus is excitatory, like neocortical and cerebellar inputs to thalamus, whereas BG input is GABAergic. Thus the selection and timing of signals to be dispersed by the midline and anterior nuclei is presumptively determined before those signals arrive in thalamus, consistent with the much narrower collection of inputs compared to that of BG-recipient thalamus. However, TRN inputs to these nuclei (Jankowski *et al.* 2013; Kolmac and Mitrofanis 1997; Van der Werf *et al.* 2002), and pallidal collateral inputs to the anterior nuclear group (Parent *et al.* 2001; Xiao and Barbas 2002) provide paths whereby PFC and the BG might influence memory processes at the thalamic level (Pinault 2004; Zikopoulos and Barbas 2006; Guillery *et al.* 1998; Hazrati and Parent 1991; Shammah-Lagnado *et al.* 1996; Antal *et al.* 2014).

*14.2.5. The hippocampal system and PFC have crucial roles in the assignation of salience.*

Just as the PFC and BG may orchestrate BG-thalamocortical neglect of expected percepts (discussed earlier (§11.6)), PFC has been suggested to orchestrate neglect by the hippocampal formation of previously stored episodic information, inhibiting redundant memorization (Frankland and Bontempi 2005). Indeed, the suppression of contextually inappropriate memory activation has been proposed as a general role for the PFC in its relationship to the hippocampus (Eichenbaum 2017). PFC activity patterns have also been shown to represent associations between a stimulus and context, an action triggered by that stimulus and context, and an unexpectedly rewarded outcome associated with that action, allowing for accurate credit assignment and the formation of memories that subsequently guide behavior toward reward (Asaad *et al.* 2017). Novelty and surprise are associated with an increase in theta synchrony between PFC and the hippocampus, supporting learning of unexpected information (Gruber *et al.* 2018). Moreover, the hippocampal formation is itself sensitive to familiarity (Squire *et al.* 2004), and through its projections to PFC and the ventral striatum, may promote neglect of familiar perceptual minutiae that would otherwise be distracting. Sz is marked by deficiencies in these capabilities, and corresponding hippocampal abnormalities (Jessen *et al.* 2003; Weiss *et al.* 2004).

*14.2.6. The basal ganglia might control effective connection of cortical areas that are chiefly connected through the hippocampal system.*

The most studied and attested roles of the hippocampal system involve memory formation and recall, but the mechanism whereby it is understood to do this — rapid plasticity that establishes long range links between cortical areas — might be quite general. Evidence of hippocampal involvement in the performance and consolidation of skilled motor behavior, even without crucial involvement in initial acquisition of the skill (Schapiro *et al.* 2019; Burman 2019, 2018‡), suggests such a generality, and indeed suggests that the hippocampal system may orchestrate consolidation of long range links that do not themselves directly transit the hippocampal system. In terms of BGMS, it may not make any fundamental difference whether two areas are linked by direct, appropriately potentiated corticocortical projections, or by temporary routes through the hippocampal system. These two classes of long range linkage inevitably coexist according to the consolidation and reconsolidation theories of hippocampal function, and might indeed act synergistically. With hippocampal function centered on the rapid establishment of long range anatomical connections amenable to reactivation, BG function entailing the activation of long range anatomical connections, and PFC integral to the circuitry of both, it seems inevitable that these two systems are unified in their function. This proposition does, however, raise important questions about conduction delays associated with trans-hippocampal paths, compared to those of the corresponding corticocortical paths that are thought to be the ultimate destination of the relations migrated by consolidation.

### *14.3. The Zona Incerta*

The zona incerta (ZI) is an agglomeration of cytologically heterogeneous diencephalic nuclei below the thalamus, adjacent to the TRN and STN, connected with many of the areas and populations involved in BGMS (Mitrofanis 2005; Ricardo 1981; Shammah-Lagnado *et al.* 1985; McElvain *et al.* 2021). Multiple, overlapping somatotopic maps are found throughout the ZI, maintaining largely parallel segregated circuits between the neocortex, thalamus, superior colliculus, brainstem, and spinal cord (Nicolelis *et al.* 1992; Power *et al.* 1999), but with no apparent topographic structure in projections to intralaminar thalamus (Power *et al.* 1999). Like the striatum, the ZI is extensively innervated by cortical layer 5 (Mitrofanis and Mikuletic 1999). Like the GPi and SNr, many of its neurons contain parvalbumin (Trageser *et al.* 2006), and it has extensive GABAergic projections to thalamus, with giant terminals apposing the proximal dendrites of projection neurons in association nuclei (Barthó *et al.* 2002; Power *et al.* 1999).

The physiology of the ZI is unlike that of the BG and TRN in several important respects, such that its operating principles are clearly distinct. Unlike either the BG or the TRN, the ZI has prominent direct projections to cerebral cortex; these projections are GABAergic, predominantly appose outer L1, are topographically organized, are densest in somatosensory cortex, and also project sparsely to visual cortex (Lin *et al.* 1997; but see Saunders *et al.* 2015a). While BG projections to intralaminar nuclei preferentially appose smaller and more distal dendrites, ZI projections to the intralaminar thalamus appose larger and more proximal dendrites (Barthó *et al.* 2002). The tonic discharge rate of ZI neurons, averaging 2-4/s (Périer *et al.* 2000; Trageser *et al.* 2006), is a small fraction of that of GPi/SNr projection neurons.

Like the BG and thalamus, the ZI is targeted by cholinergic projections from the PPN and LDT (Trageser *et al.* 2006), and like the cortex, it is targeted by the basal forebrain (Kolmac and Mitrofanis 1999), but the effect of ACh on ZI is to silence it (Trageser *et al.* 2006). Thus the ZI response to ACh resembles that of the TRN (Steriade 2004; Lam and Sherman 2010). However, whereas the TRN receives collaterals of L6 axons and not of L5 axons, and has a reciprocal relationship with the rest of the thalamus, as noted above the ZI receives L5 collaterals, and it does not receive thalamic inputs (Barthó *et al.* 2002).

Through its widespread projections to thalamus, the ZI has been suggested to synchronize oscillations in large populations of projection cells, acting as a relay whereby signals from its afferents can selectively facilitate transmission of sensory signals by those projection cells (Barthó *et al.* 2002, 2007). Experimental and clinical results in PD show that hyperactivity and hypersynchrony in the ZI are associated with dyskinesia and bradyphrenia just as they are in the GPi and SNr (Merello *et al.* 2006; Périer *et al.* 2000, 2002), and indeed that deep brain stimulation (DBS) in ZI may be a more effective technique for alleviating medically refractory PD than DBS in STN (Plaha *et al.* 2006).

An intriguing proposal is that rhythmic GABAergic input to the sensorimotor and intralaminar thalamus from ZI relays activity from attentional orientation centers such as the superior colliculus, disrupting BG-related activity in the thalamus and replacing it with selective receptivity to unexpected sensory inputs deemed salient by attentional orientation centers (Watson *et al.* 2015). This fits well with the proposition that the general function of the ZI is to gate sensory receptivity (Trageser and Keller 2004; Trageser *et al.* 2006; Lavallée *et al.* 2005; Urbain and Deschênes 2007), and like BGMS, is a proposal that selections can be made in the thalamus by GABA-mediated spike-timing-dependent gain control. And with its GABAergic projections to upper L1, the ZI is positioned to adjust spike-timing-dependent gain in its targets with particular rapidity and thoroughness.

### *14.4. The Claustrum*

The claustrum may be functionally similar to the ZI and, by extension, to the BG, but with its own peculiarities. It has long been a subject of notoriously inconclusive study (Edelstein and Denaro 2004). Its function is murky, and like the ZI, it is something of a chimera, combining physiological and functional attributes of the cerebral cortex, the striatum, the thalamus, and the basolateral amygdala (Swanson and Petrovich 1998). Like the BG, the claustrum appears arranged to synchronize the cortical areas with which it is connected, but unlike the BG, it appears to do so only occasionally. In particular, unlike pallidal projection neurons, and like ZI projection neurons, claustral projection neurons have a low tonic firing rate, 0-10 spikes/s in awake animals (Edelstein and Denaro 2004 p.5).

The claustrum contributes to the recruitment of a generally well-adapted neural response to novel and unexpected situations (Badiani *et al.* 1998; Remedios *et al.* 2014), and to emotionally freighted stimuli (Redouté *et al.* 2000), by orienting attention toward immediate, external sensory specifics. It can inhibit PFC, implicitly releasing top-down regulation, by selectively targeting PFC interneurons (Jackson *et al.* 2018), and claustral activation during sensory processing is associated with elevated response variability in PFC responses, and elevated responsive network homogeneity, promoting network completion with homogeneous component-wise responses (Atilgan *et al.* 2024 ‡ ). Evidence also suggests that it is integral to a cognitive control network essential for attentional set-shifting (Fodoulian *et al.* 2020 ‡ ) and the maintenance of working memory contents (Bhattacharjee *et al.* 2024‡), and that its involvement is necessary for optimal performance in learning and behavioral scenarios that are cognitively demanding (White *et al.* 2020), with significant activation particularly at the onset of a demanding task phase (Krimmel *et al.* 2019).

Similar to the ZI, the claustrum reciprocates with topographic cortical maps for all of the exteroceptive senses; the implicated claustral neurons have quite large receptive fields, with some incidence of polymodality (Sherk 1986). The entire claustrum is modulated by afferents communicating situational salience (exceptionality) from VTA and SNc, the thalamic reuniens nucleus, the lateral hypothalamus, the locus coeruleus, the dorsal raphe nucleus (Słoniewski *et al.* 1986), and through some path yet to be fully anatomically elucidated, from a cholinergic source (Salerno *et al.* 1981; Nieoullon and Dusticier 1980).

Perhaps claustral neurons synchronize oscillations in the cortical areas connected to it, similar to the dynamic described earlier (§7.6) in the thalamocortical projection. For example, activity is relayed from anterior cingulate cortex, through the claustrum, to parietal and visual cortex, and this pathway may be a mechanism for top-down cognitive control (White and Mathur 2018). In the claustrum, however, the effect appears to be well-gated by the ascending modulatory afferents, resulting in the low tonic firing rate noted above. This is a limited role, similar to a general role proposed previously for the claustrum (Crick and Koch 2005), itself similar to the role ascribed to the BG in this paper.

The cortical areas with which the claustrum has been established to reciprocate, and which are therefore most likely to be subject to synchronization in the manner of thalamocortical circuits, are (1) various topographically mapped unimodal sensory areas for each of the senses (at least the exteroceptive ones) (Sherk 1986), (2) the frontal eye fields and supplementary motor area (SMA) (Sherk 1986), (3) several default mode network loci (orbitofrontal, cingulate) (Sherk 1986), and (4) the hippocampal system (Wilhite *et al.* 1986). The claustrum has unreciprocated projections to much of the rest of cortex, notably area 46 (DLPFC) (Sherk 1986), which might impart selective receptivity in the recipient areas to sensory and motivational activity that activates the claustrum, selectively boost cognitive activity that is already phase-locked with it, and relatively diminish other activity.

The claustrum has convergent afferents from thalamic midline and intralaminar nuclei (reuniens, CM, PF, PC, and CL) (Van der Werf *et al.* 2002). When the cholinergic, noradrenergic, dopaminergic, serotonergic, and other diffuse modulatory claustropetal afferents signal situational or anticipated salience (Salerno *et al.* 1981; Schultz 1998; Matsumoto and Hikosaka 2009; Nakamura *et al.* 2008), claustral neurons might attempt to synchronize activity in the sensory stream with itself, with oculomotor and skeletomotor activity, with the default mode network, and with the hippocampal system, so that the situation and its sensory correlates are well-attended, consciously integrated, and well-reflected by the memories that are activated and recorded.

# 15. Future Directions, Open Questions, and Closing Thoughts

**In this section:**

## *15.1. Theoretical Predictions and Proposed Experiments*

**In this subsection:**

### *15.1.1. Cortex and Striatum*

In the BGMS model, effective connections are established in cortex in response to striatal decisions. This phenomenon would likely be detectable in the relationships among cortical and striatal LFPs, and would inherently be detectable using large electrode arrays to simultaneously measure single unit activity in large populations of neurons in cortex and striatum. Specifically, in well-trained tasks, a highly significant relationship of consistent stimulus-response relations and delays should be found between spike volleys arising in a particular set of cortical loci, the generation and timing of subsequent spike volleys arising in one or more structurally connected striatal loci, and the establishment of functional connections between a consistent set of cortical loci (usually including the first set of cortical loci, through closed-loop circuits) as measured by LFP or individual spike activity in the latter.

In some experimentally accessible and reliably reproducible scenarios, the establishment of a long range corticocortical functional connection is predicted to be strongly contingent on the occurrence and precise timing of the striatal activation, as suggested by the results reported by van Schouwenburg *et al.* (2010b). The BGMS model predicts that if the striatal activation is absent or mistimed, the connection is unlikely to be established, else the connection is likely to be established. This class of experiment is readily accessible in animal models, and indeed is an implicit result of Monteiro *et al.* (2023), who showed an inverted-U relationship between striatal processing speed and behavioral proficiency, with natural conditions yielding best performance. Optogenetic manipulation is also an excellent technique for probing dynamics of this sort (Quintana *et al.* 2024 ‡ ). In humans, transcranial ultrasound stimulation (TUS) is non-invasive and sufficiently safe to use even in healthy populations, and has already shown promise as a spatially specific modulator of oscillatory activity in the striatum in a clinical population where efficacy could be confirmed directly through electrophysiological implants (Darmani *et al.* 2025). These results suggest that TUS can be used to locally perturb striatal activity, particularly in a closed-loop system in combination with EEG, to probe the causal role of the striatum in functional connectivity shifts, and in particular, whether coherence is crucial there.

### *15.1.2. Pallidal and Nigral Output to Thalamus*

The central prediction arising from the BGMS model is that relationships of entrainment characterize sparse ensembles of directly connected neurons spanning the entire BG direct path during activation. If the effect of pallidal and nigral output on the thalamus is probed in awake healthy (normal) animals, the prediction is that phasic activation in many cases entrains thalamic activity. Preliminary results reported by Schwab (2016) give evidence of ensemble phasic entrainment of motor thalamus by the GPi, while underscoring that spatiotemporal sparseness and stochasticity in this activation and entrainment greatly complicate characterization at the single unit level.

Notwithstanding methodological hurdles, oscillation in the output from an area of the intralaminar thalamus, measured in a fashion that carefully avoids conflation with activity associated with afferent corticothalamic activity, is predicted to be coherent with phasic oscillatory BG input to that area, characterized by LFP in structurally connected areas of BG output structures.

### *15.1.3. Stimulus-Locked Spiking From Cortex Through Basal Ganglia to Thalamus*

Another key prediction is that in an over-trained task, phasic pallidal spiking to a particular thalamic target associated with onset of a particular salient context within the task will exhibit, in aggregate, a very stable, narrowly distributed (±<2 ms) delay relative to the first cortical spike volley associated with onset, implicating a stable set of striatal and pallidal/nigral neurons, for environmental conditions and level of arousal similar to those that prevailed during training.

The causal relationship of trans-BG conduction delays to these timing relationships can of course be probed in simulation (Asadi *et al.* 2024), but might also be probed by selectively modulating activity in the implicated fibers, inducing oligodendrogenesis and associated myelination (Gibson *et al.* 2014).

#### *15.1.4. Correlation of Cortical Area to Focally Targeted Striatal Area as a Function of Oscillatory Band*

Because the average delay through dorsal trans-GPi paths is roughly one gamma period, while the average dorsal trans-SNr delay is roughly one beta period, BGMS predicts functional prominence of the gamma cycle in a cortical area and scenario when its inputs to BG flow primarily through the dorsal striatum to the GPi (particularly implicating the dorsal sensorimotor striatum), while the beta cycle is expected to dominate when activity flows primarily to the SNr (implicating the associative striatum). Even longer delays, commensurate with the theta cycle, may accompany paths through the ventral striatum and ventral pallidum, due to its intimacy with the medial temporal lobe (briefly reviewed below).

More generally, according to BGMS the dominant oscillatory band of activity in a given cortical locus and context should be the best predictor of the BG paths it activates, and the converse should hold similarly. This principle generalizes beyond thalamocortical-BG circuits, to encompass frequency bands characteristic of parallel and analogous systems, several of which were discussed earlier (§14). These domains and characteristic frequency dynamics can be associated with particular thalamic nuclear complexes; for example, the mediodorsal nucleus is associated with beta synchronies, the pulvinar with alpha synchronies, and the anterior thalamic group with theta synchronies (Ketz *et al.* 2015).

Cross-frequency coupling has been demonstrated in theta band interactions of the hippocampus with striatum (Tort *et al.* 2008), and is posited to be a general theme in BG-thalamocortical dynamics (Cannon *et al.* 2014; Brittain and Brown 2014). Putative cross-frequency BGMS operates by spike-time-dependent gain in cortex no less than in-band BGMS, suggesting the corollary prediction that cross-frequency coupled BG activity in over-trained tasks produces spike volleys in target areas that are spatiotemporally coincident, at a regular frequency ratio, with selected corticocortical spike volleys. Heavy projections from the hippocampal formation to the ventral striatum (Brog *et al.* 1993) suggest that the in-band relationship may hold in these scenarios, i.e. that the dominant band of the afferent determines the activated BG path at the striatal stage, with divergence to a parallel path in a subsequent stage.

### ***15.2. Some Notable Open Questions***

In terms of the statistics of conduction delays, what systematic patterns and topographies characterize the corticostriatal, striatonigral, and striatopallidal projection fiber populations? Modern techniques leveraging optogenetics and diffusion MRI allow the isolation and characterization of narrowly specified populations of axons (Skoven *et al.* 2023‡). These techniques could be used to explore BG fiber populations in unprecedented detail.

Lag-free long range synchronies in cortex (e.g. Vicente *et al.* 2008), with narrow pyramidal somatic coincidence windows (Pouille and Scanziani 2001; Volgushev *et al.* 1998), exist simultaneous with finite long range corticocortical delays (e.g. Gregoriou *et al.* 2009; Nowak and Bullier 1997). Exactly how does this work, at the level of cortical microcircuitry? How do the discharge and conduction delays of thalamocortical neurons and fibers compare as a function of nuclear origin? In particular, how do the delays of paths through the intralaminar and midline nuclei compare to those through other thalamic nuclei?

Gamma synchrony accompanies effective corticospinal activation (Schoffelen *et al.* 2005; Fries 2005), and the effects of single-pulse TMS stimulation in motor cortex on corticospinal activation depends on the phase of cortical beta at the moment of stimulation (Torrecillos *et al.* 2020). The pedunculopontine nucleus is integral to the control of voluntary movements (Tsang *et al.* 2010), and is profusely targeted by the GPi (Parent *et al.* 2001). Does this relationship entail BGMS? The same question applies to BG targeting of the superior colliculus, with regard to its attentional orientation and oculomotor functions.

Do the amygdala, hypothalamus, and other subcortical structures beyond those reviewed earlier (§14), use a BGMS-like mechanism to influence thalamocortical activity? The amygdala in particular has been construed as parallel to the ventral BG (Olmos and Heimer 1999), and indeed its central and medial nuclei are proposed to be continuous and homologous with the BG (Swanson and Petrovich 1998). Moreover, projections from the amygdala to PFC have been shown to convey signals that bias decision making (Burgos-Robles *et al.* 2017), similar to the role ascribed earlier (§7.3) to the BG.

The BGMS model implies an elaborate physiological arrangement of coordinated modularity, spanning all developmental levels, and many distinct neurotransmitter systems. How is this orchestrated? Rules governing the self-organization of projecting fiber populations and appositions must play a large part (e.g. Wedeen *et al.* 2012; Sanes and Yamagata 2009; de Wit and Ghosh 2015). But clearly, developmental exuberance, and activity-driven, correlation-sensitive plasticity must play a very large role. Exactly how are these development and plasticity mechanisms arranged to route and terminate long range fiber bundles appropriately, and optimize the timing of the stimulus-response functions of the BG as an ensemble?

How strong and broad is the BG influence on the cholinergic and serotonergic supplies to thalamus, cortex, and striatum? Are the BG arranged for bipolar control of these supplies, as they are for dopamine? And what are the topographies and microcircuitry of the BG inputs to TRN, NBM, PPN, LDT, DRN, MRN, and SC, by reference to the topographies and microcircuitry of their respective projections to and from thalamus and cortex?

The cytology and microcircuitry of the striatum are crucial to BGMS, and to basal ganglia dynamics in general. How do the cortico-FSI and cortico-SPN projections to striatum differ in cytological, laminar, and areal origin, in patterns of preference, apposition, and topography/ convergence/divergence, and by compartmental and hodological target (matrisome, striosome, border region, direct, indirect, etc.)? Similarly, how do thalamo-FSI and thalamo-SPN projections differ by these measures?

What are the functions of striatal neuron types beyond the SPNs, FSIs, and ACh interneurons, particularly as they relate to BGMS? In particular, what are the roles of somatostatin-positive LTS interneurons, and of calretinin-positive interneurons, which have yet to be classified physiologically (Kreitzer 2009)? Uniquely human adult neurogenesis of striatal calretinin interneurons (Ernst *et al.* 2014) is intriguing — what is the functional significance of this?

Is oscillatory phase preserved in the paths through the NBM, SNl, and PPN/LDT, and if so, do they entrain their targets? Do the BG, through some paths, entrain targets to antiphase, to quickly and decisively abolish connections? Is this one of the functions of FSIs that target indirect path SPNs? Indeed, is this one of the functions of GPe input to striatum (which preferentially innervates FSIs) and TRN, and of STN to GPi/SNr? Such arrangements seem plausible, but the evidence is as yet tenuous—albeit tantalizing (e.g. Schmidt *et al.* 2013).

The maximum conduction velocity in human corpus callosum is anomalously slow (Caminiti *et al.* 2009). Schizophrenia is also typified by abnormalities of the corpus callosum, and of interhemispheric coordination (Foong 2000; Whitford *et al.* 2010; Hoptman *et al.* 2012). Are interhemispheric dynamics in humans special, from a BGMS perspective, or otherwise?

### *15.3. Closing Thoughts*

This work was originally motivated by a deceptively simple notion, that a mechanistic explanation for cognitive problem solving capacities in mammals can be found in the conjunction of the cerebral cortex and basal ganglia, whose distinct information processing styles produce a synergy far greater than the sum of their parts. Exploring this notion led to the model presented here, placing the basal ganglia at the center of functional connectivity decisions and orchestration. At an intermediate level of detail, this notion pivots on several architectural features of the mammalian brain:

- Pattern recognition and well-timed activation by the striatum, as described by Plenz and Aertsen (1994);
- Control of functional connectivity by nonlinear spike-timing-dependent gain mechanisms, as discussed by von der Malsburg (1981), Singer (1993), Fries (2005, 2015), Larkum (2013), and Wang *et al.* (2021);
- Centrally (thalamically) mediated control of effective connectivity in cortex, as discussed by Jones (2001), Purpura and Schiff (1997), Saalmann (2014), Nakajima and Halassa (2017), and Salami *et al.* (2003);
- Graded semantic maps in cortex with highly regular long range links, as discussed by Huth *et al.* (2012), Simmons and Barsalou (2003), and Wedeen *et al.* (2012);
- A dense connectome with “rich club” organization, as discussed by van den Heuvel and Sporns (2011) and Markov *et al.* (2014);
- The primacy of information integration in cognition, as discussed by Tononi (2004);
- Convergence-divergence in the basal ganglia, as discussed by Flaherty and Graybiel (1994), Joel and Weiner (1994), Zheng and Wilson (2002), and Mailly *et al.* (2013);
- The basal ganglia construed as a central switching mechanism, as discussed by Redgrave *et al.* (1999);
- Spike volley and oscillatory coherence in the BG, as discussed by Berke *et al.* (2004), Leventhal *et al.* (2012), Schmidt *et al.* (2013), and Oberto *et al.* (2022);
- Slow and diverse CVs in paths through the striatum, as discussed by Tremblay and Filion (1989), Turner and DeLong (2000), and Jinnai and colleagues (Yoshida *et al.* (1993), Kitano *et al.* (1998)); and
- Basal ganglia output that entrains activity in its targets, as discussed by Goldberg *et al.* (2013, 2012), Antzoulatos and Miller (2014), and Kojima *et al.* (2013).

In the course of developing this conjunctive idea, which was of necessity quite vague at the outset, I encountered an array of significant implications, suggesting resolutions to long-standing mysteries and paradoxes in the physiology of the BG, and in the relationship of BG activity to thalamocortical activity.

My conclusion is that activity throughout the cerebral cortex is structured by large scale synchronies that are mesoscopically, globally, and continually influenced by the basal ganglia, which themselves respond selectively to large scale patterns of cortical synchrony, in an arrangement of continual iteration that is the mechanistic essence of flexible and directed cognition in mammals.

### *15.4. Acknowledgements*

Deepest appreciation to my wife Farah, whose support has made this work possible, and who has been richly rewarded with far more dinner table neuroscience than her appetite required. A special thank you to Dr. Daniel J. Gibson; some of the key ideas introduced here emerged from our stimulating discussions in the summer of 2015. His encouragement and editorial contributions to the initial version of this manuscript were absolutely invaluable. Thanks also to Prof. Kurtis Nishimura for his help and advice proofreading and formatting the second (2020) version of the manuscript. And heartfelt thank yous to Dr. Jason Sherfey and Prof. Nancy Kopell for their constructive feedback on the "nutshell" treatment (§2) added in the third (2025) edition. Thank you to Prof. Ann Graybiel for innumerable inspirations, insights, and discoveries before and while it has been my pleasure and privilege to contribute in my small way to the great success of her lab. And finally, thanks to all of the sage souls in our cherished Boston Cognitive Rhythms Collaborative, who have inspired many of the clarifications and improvements in the revisions. Naturally, any and all departures from concision, clarity, and solidity, are mine and mine alone.

# 16. Appendix: Consciousness, Integrated Information, and Selection Mechanisms

**In this section:**

### *16.1. Consciousness is tricky.*

Historically, the basal ganglia have not been implicated directly in the mechanistic underpinnings of consciousness (Koch *et al.* 2016), despite evidence outlined earlier implicating them in mental supervision (§12.5), working memory (§12.7), cognitive flexibility (§12.9), and disorders of consciousness (§13.2). Intuitively, there is a natural intimacy of the basal ganglia with consciousness, given winner-take-all dynamics in the basal ganglia (Redgrave *et al.* 1999; Mink 1996; Stocco 2018) and the singular unity of consciousness (Bayne 2010).

"Consciousness" is a notoriously slippery concept, even chimerical in many accounts. These narrative tribulations seem to be evidence of the fundamental qualities of conscious cognition. Where consciousness is broached above, most prominent are flexibility, integration and the breaching of modularity, intervention when modular strategies are flummoxed, and perhaps most unsettling, arrangements of notionally infinite recurrence.

Accordingly, consciousness is here construed to be an evolving pattern of relations—fundamentally, a kind of dynamic directed topology hosting ephemeral representations—featuring many degrees of freedom (high dimensionality), arbitrary information combination, self-acting and self-affecting selections (decisions, self-causation), and self-referential iteration (free-running state evolution, sometimes autonomous). More exhaustively, it is a mental mechanism that features the variously overlapping attributes and faculties of subjectivity, privacy, uniqueness, representation, genericness, ephemeral specificity, arbitrary associativity, intentionality, expectation, attention, perception, episodic continuity, transformation, and action. Evidently, conscious actions can be inwardly directed (chiefly, cognitive transformations, recollections, memorizations, and other decisions, all relative to current activity and potentiation) or outwardly directed (behaviors). Reportability and self-awareness, frequently attributed specially to human consciousness, are (by the present narrative) corollary. "Uniqueness" here signifies that there is only one consciousness in the normal waking brain (largely a corollary of its broadly integrative decision making, and the physiology underlying that faculty), and that the representations and core mechanisms of conscious cognition lack any architectural modularity dividing them into perceptive, cognitive, and active/agentive domains, but rather that these domains are all directly and irreducibly implicated by the same physiological substrate (as seen, for example, in the thalamic intralaminar nuclei, noted earlier (§8.9)). For further discussion of the capacities at issue, and of the inextricable entanglement within consciousness of the attributes and faculties attributed to it above, see Engle (2002). For a survey of theories of consciousness, in all their vast and wild diversity, see Kuhn (2024).

### *16.2. Associative areas of the thalamocortical system and basal ganglia plausibly underlie consciousness.*

Within the network of cortical connectivity hubs, particular areas and networks have been identified that are associated with faculties attributed above to consciousness. For example, Vincent *et al.* (2008) propose a "frontoparietal control network" comprising lateral PFC, anterior cingulate cortex, and the inferior parietal lobule, topographically and topologically separate from the hippocampal network and "dorsal attention" network. PFC and posterior parietal cortex (PPC) in particular have been implicated in theories of the physiological basis of fluid intelligence (Jung and Haier 2007). Hub networks are uniquely developed in humans, and genes identified as uniquely divergent in humans are highly expressed in these networks, are associated with individual variation in the function of these networks, and are associated with intelligence and schizophrenia (Wei *et al.* 2019).

The establishment of effective connections between dissociable networks, heralding a collapse of their mutual modularity, is associated with conscious awareness (Godwin *et al.* 2015), and particularly entails functional integration of task-specific networks with the resting state network (Fukushima *et al.* 2018), while pharmacologically induced loss of consciousness is associated with the pervasive breakdown of effective connectivity in cortex (Ferrarelli *et al.* 2010b). The modularity of an individual's resting state cortical networks, measured by fMRI, has been found to be predictive of task performance as a function of task complexity, with high modularity subjects exhibiting relatively high performance on simple tasks, and low modularity subjects exhibiting relatively high performance on complex ones (Yue *et al.* 2017).

The integrated system of the PFC and striatum, suggested to be central to cognitive flexibility (Leber *et al.* 2008; van Schouwenburg *et al.* 2010b, 2012, 2014; Hazy *et al.* 2006), includes many connectivity hubs and resting state network nodes (Cole *et al.* 2010; van den Heuvel and Sporns 2011; Harriger *et al.* 2012; van den Heuvel *et al.* 2009; Elston 2000). Indeed, correlated activity has been demonstrated between cortical resting state network nodes and loci distributed widely in the striatum (Di Martino *et al.* 2008; Vatansever *et al.* 2016; Wang *et al.* 2020). Consistent with these physiological arrangements, it has been proposed that BG direct path activation, through its influence on thalamic matrix and frontal cortex, dynamically biases cortex toward the formation of broader more integrative networks, with greater prevalence of feedback pathways, slower more deliberative cognition, and channeling of the contents of consciousness (Shine 2021). And the thalamus — both its BG-recipient "matrix" sectors and its "core" sectors — is thought to play a central role in conscious experience (Whyte *et al.* 2024; Fang *et al.* 2024‡).

Posterior cortex, and its relationship with striatum, are also clearly implicated in consciousness. As noted noted earlier (§12.4), LFP activity in a network spanning deep layers of lateral intraparietal cortex, caudate striatum, and intralaminar thalamus, evaluated with the integrated information measure Φ, reliably indexes state of consciousness (Afrasiabi *et al.* 2021). Functionally connected triads, in which posterior parietal cortex drives frontal cortex and the BG, are implicated in motor performance (Hwang *et al.* 2019), and the evidence from Afrasiabi *et al.* (2021), and related results from Redinbaugh *et al.* (2020, 2022), suggest that such triads may indeed be necessary for consciousness.

### *16.3. Highly abstract functional structure within the most associative areas of neocortex suggests the outlines of mental architecture.*

It has been proposed that PFC is organized into a spatially graded hierarchy, with the highest and most abstract representations located anteriorly, and the lowest and least abstract located posteriorly (Christoff and Gabrieli 2000; Badre and D'Esposito 2009). In the least abstract of these areas, numerous visuotopic maps (for example) have been identified (Silver and Kastner 2009), whereas the frontopolar cortex is concerned with highly abstract consideration and reconciliation of conflicting goals, and the management of competing and alternating cognitive sets (Mansouri *et al.* 2017). Functional specialization of the dorsomedial and dorsolateral

PFC has been proposed, with the former monitoring performance, and the latter guiding it; links between these areas exhibit mutual preferences according to position along the anterior-posterior axis (Taren *et al.* 2011). Within the DLPFC, subdivisions are apparent from their functional correlates and network connectivity—an anterior-ventral subregion is associated with attention and action inhibition processes, and is intimate with anterior cingulate cortex, while a posterior-dorsal one is associated with action execution and working memory, and is intimate with PPC (Cieslik *et al.* 2013).

In general, the primate cortex is characterized by gradients in average connectivity distance, with neurons in primary sensory and motor areas typified by the shortest average connection distances, while areas most remote from primary areas are typified by the most distant connections, particularly implicating lateral and medial frontoparietal cortex (Margulies *et al.* 2016; Oligschläger *et al.* 2019).

Goldman-Rakic (1988) described pervasive, systematic interdigitation of parallel circuits linking association cortex. Consistent with this account, several of the widely distributed networks identified in humans—resting state, frontoparietal control, and dorsal attention—have been shown to consist of at least two distinct parallel networks with similar gross structure, but spanning distinct interdigitated subregions, and with intriguing topological distinctions; e.g., one of the identified resting state subnetworks is intimate with the hippocampal system, while another is devoid of such intimacy (Braga and Buckner 2017; Braga *et al.* 2019).

Clearly this topographic and topological structure has consequences for mental architecture. Indeed, the microstructural characterization of projections between hub areas is among the most promising subjects for future investigation. Causally chained activity in the neuronal populations in hub areas, and the projections between them, are a plausible substrate for the most abstract conscious awareness and agency. Nonetheless, these projections might in fact have regular topographic structure, with recurrent finely parallel-segregated pathways supporting sustained activity, but no special convergence-divergence. Information integration would then inhere mostly in the moment-to-moment functional connectivities between each hub and the large arrays of specialized areas with which each is anatomically connected. Such an arrangement is central to the "global workspace" model (Dehaene and Naccache 2001; Dehaene and Changeux 2011).

### *16.4. Densely interconnected, highly associative frontal and posterior areas, with no intrinsic domain-specific functional topography, act as communication thoroughfares integrating activity with great flexibility.*

Evidence suggests that, indeed, anterior and medial PFC and cingulate cortex, the posterior medial and parietal cortex intimate with it, and the subcortical areas intimate with them, contain areas in which domain-specific functional topography is a transitory consequence of their effective connections from moment to moment, with highly abstract intrinsic topographies along lines of hierarchy and other generic aspects of cognition. These cortical regions are divisible into areas each consistently associated with a particular area of generic cognitive control operation (Wang *et al.* 2024a ‡ ), while activity in these same areas carries neural codes that are specific to the domain they ephemerally represent (Jackson *et al.* 2025; Xiang *et al.* 2024‡; Wentz *et al.* 2025‡). In a given task, activity in these areas shows lower representational dimensionality than that in the hierarchically lower sensory and motor cortices implicated in the task, suggesting compression to facilitate generalization; simultaneously, higher-dimensional representation, in these areas and in task-implicated sensory areas, is associated with greater cognitive flexibility, perhaps by allowing distinct treatment of a wider variety of task contexts, with reduced interference between them (Chakravarthula *et al.* 2025‡).

Using fMRI in humans, Assem *et al.* (2020) show that much of the frontoparietal control network, with adjacent areas within the default mode and dorsal attention networks, the head of the caudate nucleus, and substantial sectors of the cerebellum, are activated in cognitively demanding tasks regardless of the specific domains implicated by the task, suggesting that the role of these areas in cognition is generic, rather than domain-specific. Using tracers in monkeys, Averbeck *et al.* (2014) show that the head of the caudate nucleus is a zone of convergence for projections from all parts of the PFC, which they suggest is evidence that the striatum also contains topological hubs that perform computations across diverse domains.

Additional fMRI evidence supports the proposition that hub areas are engaged in cognition generically. Evidence suggests that functional representations in hub areas are dynamically oriented to task demands (Vromen *et al.* 2018 ‡ ), and that abstract relations in verbal comprehension are represented generically in hub areas (Zhang *et al.* 2020).

fMRI evidence also shows an inversely graded relationship between general intelligence and task-dependent large scale network reconfiguration in hub areas, suggesting a significant advantage in general intelligence for those individuals whose generic (e.g. default mode) cortical areas are intrinsically capable of ephemerally adjusting to task-specific requirements (Thiele *et al.* 2022), obviating reconfiguration engaging specialized areas with inherently less fluidity.

Versatility in hub areas seems to follow from the capacity of individual neurons, and indeed individual synapses, to participate in a vast array of distinct ephemeral assemblies of neurons with contextually appropriate conduction delays (Rigotti *et al.* 2013; Izhikevich 2006), even simultaneously (Naud and Sprekeler 2018; Caruso *et al.* 2018; Bernardi *et al.* 2020). Remarkably, the same neural subspaces are “recycled” within the same task for apparently completely different roles, first representing sensory information, and later representing behavioral response information (Wentz *et al.* 2025‡).

The proposition that the functional topography of hub areas is highly abstract and context-dependent also comports with the view of van den Heuvel *et al.* (2012) that a core network of connectivity hubs (a “rich club”) serves as a common, and therefore contentious, communication “backbone” subject to “greedy routing” strategies by more locally connected (and specialized) areas. Indeed, task-related activity in functionally connected PFC and PPC can be very similar, with almost identical tuning and time courses, throughout the performance of a task implicating working memory (Chafee and Goldman-Rakic 1998); beta band synchronies between these areas reflect only behaviorally relevant representations, with PFC neurons synchronized to PPC beta oscillation only if selective for contextually relevant categories (Antzoulatos and Miller 2016). Subcortical hub areas—particularly thalamus, striatum, claustrum, and hippocampus—exhibit similar “echo” relationships with cortex (Groot *et al.* 2023). The view that cortical areas with high abstraction and long range connectivity function as thoroughfares also follows from findings, noted earlier (§1.4), that frontal-posterior LFP synchrony accompanies attentional orientation, whether by top-down or bottom-up processes (Buschman and Miller 2007). Underscoring their central role in cognition generally, hub areas—in prefrontal, temporal, and parietal cortex, and also in thalamus, striatum, and hippocampus—are disproportionately implicated in systemic brain disease and mental illness (Crossley *et al.* 2014).

### *16.5. BG integration with cortical communication thoroughfares may underlie the versatility that is the hallmark of consciousness.*

Activity in particularly abstract areas of the PFC might arrange itself to impart nearly arbitrary patterns to the striatum, inducing highly flexible transformations by the BG of cortical activity and effective connectivity, and resolving backbone contention through selections, largely by the BGMS mechanism. This may largely underlie selective, task-related output from hub areas despite unselective inputs (Senden *et al.* 2018, 2017). By this narrative, the BG make available an enormous repertoire of neural gestures, that activity in hub areas can use to gate and operate on network activity, particularly activity within hub areas themselves. The corticostriatal projections from hub areas, and the input-output relations of the targeted areas of striatum, are then the essential substrate for cognitive flexibility, as suggested by van Schouwenburg *et al.* (2014). Indeed, as noted earlier (§12.9), cognitive flexibility is associated with functional connectivity of hub areas to BG (Vatansever *et al.* 2016). Moreover, significant dysfunction in this relationship, including both deficient cortical control of striatum, and deficient striatal control of cortex, has been shown in Sz (Wang *et al.* 2015).

One interpretation of these arrangements is that consciousness in its essence is an enormously flexible and agile mechanism for relating causes (stimuli) to effects (resulting thoughts and behaviors). By this narrative, conscious contemplation occurs when these ephemeral cause-effect relations are chained together, each stirring the next into existence, so that consciousness depends intrinsically on the physical causality of brain activity. This comports neatly with the view, noted earlier (§12.4), that the “physical substrate of consciousness” exhibits maximal cause-effect power (Tononi *et al.* 2016), with intriguing ontological implications (Findlay *et al.* 2024‡).

As suggested earlier (§12.7), if working memory items are patterns of activation in PFC, each characterized by a distinct phase angle (Siegel *et al.* 2009), then—through corticocortical feedback projections and BGMS—the PFC might establish and dissolve effective connections implicating a particular working memory item (effectively, a thought) with little or no interference from, or indeed to, latent items, except when a latent item is selected for integration with an active item. DA, ACh, and 5-HT under PFC and BG control are also crucial parts of this tool set, dynamically tuning the receptivity, contrast, focus, selectivity, and stability, of cortical signal paths.

Because inputs to the BG encompass the entire cortex, the BG can respond to activity in areas that are not functionally connected (not synchronized) with activity in conscious areas, and might act to synchronize the latter with the former, or the former with the latter. In this way, subconscious activity might be boosted into consciousness by BG selections, as proposed by Shine (2021). Indeed this likely describes any BG-mediated reorientation of attention in response to a sensory stimulus, particularly implicating the thalamic intralaminar nuclei.

### *16.6. Large scale plasticity in cortex implicates consciousness and BGMS.*

The propositions that the highest levels of PFC lack persistent domain-specific functional topographies, control the BG with great flexibility, and are directly implicated in conscious cognition, relate to the dynamics of skill learning and performance. Experience-driven skill acquisition entails the reorganization of cortical topography in sensory and motor areas (Buonomano and Merzenich 1998; Kleim *et al.* 2004). Topographic reorganization seems to necessitate hub areas without fixed functional topography, in order to maintain function while accommodating the shifting semantic correlates of the neurons comprising the implicated map. In principle, long range connections linking shifting maps to generic hubs might enable sensible integration into cognition at every stage of topographic reorganization.

The orchestration of topographic plasticity, and functional continuity during that process, likely implicate not only highly abstract areas of neocortex, but also the hippocampal formation, which is extremely labile and exceptionally well-connected (as briefly reviewed earlier (§14.2.1)), and has direct and transthalamic links to secondary motor cortex (Jankowski *et al.* 2013; Van der Werf *et al.* 2002). Perhaps the involvement of the hippocampal system in the domain of spatial navigation is just a special case of a general competence “navigating” the similarly 2 dimensional graded maps of neocortex—that is, the role of the hippocampal system in spatial navigation is actually to physically register associations between spatial locations and semantic objects, represented as specific large scale patterns of strengthened connectivity (mutual excitability) in neocortex, and this facility is readily suited to register associations among such semantic objects, with no particular association with physical space (Eichenbaum and Cohen 2014; Park *et al.* 2020; Kafkas *et al.* 2024). This facility seems perfectly suited to recruitment as scaffolding for the reorganization of graded maps in neocortex, including those underlying motor expertise (see e.g. Schapiro *et al.* 2019; Burman 2019, 2018‡).

Initial performance of a qualitatively new skill depends on the availability and engagement of working memory (Reber and Kotovsky 1997), and is aided by attention to the minutiae of performance (Beilock *et al.* 2002). Learning the skill does not entail topographic reorganization until late in the process, and initially pivots on activity and plasticity in the BG and cerebellum (Ungerleider *et al.* 2002), with a crucial role for corticostriatal SPN plasticity (Koralek *et al.* 2012). Once proficiency is attained, performance can in fact be significantly disrupted by attention (Beilock *et al.* 2002) and outsized incentives (Smoulder *et al.* 2024). If the BG learn precise sensorimotor sequences through practice (Graybiel 1998), and their performance involves finely tuned subcortical loops, then inapt engagement of high-order PFC, supplying disruptive signals to the striatum, is likely to disrupt overall performance. Similar disruption of input patterns to the cerebellum might have similar consequences, though the inherently lesser flexibility of the cerebellum (Fernandez-Iriondo *et al.* 2021) might effectively protect the cerebellum from this sort of dynamical disruption.

### *16.7. Mammalian consciousness is presumptively one family of instances among many, each family distinct but sharing a set of irreducible architectural features.*

It seems likely that the arrangement of high resolution spatially graded feature maps, with a "rich club" topology of dense high resolution interconnections, hub areas some of which are never plastically committed as feature maps, and dynamic timing-based mesoscopic control of effective connectivity and signaling characteristics by a recurrent, highly convergent-divergent multistage subsystem arranged for self-referential reinforcement learning, is not unique to mammals, or even to vertebrates, but rather is the essential architecture of many evolved conscious systems. Perhaps the resulting information processing *is* consciousness, in the sense meant by Tegmark (2015) in his proposition that "consciousness is the way information feels when being processed in certain complex ways".

And if the predicates of consciousness can be realized by a generic architecture, that also suggests that instances of this architecture are likely wherever there are organisms exhibiting complex and flexible behavior. Birds share major subcortical structures and connections with mammals, including the BG, with similarities and differences some of which were noted earlier (Luo and Perkel 1999; Kojima *et al.* 2013; Doupe *et al.* 2005). Capacities in corvids for flexible executive control, persistent strategic planning, and capacious working memory (Kabadayi and Osvath 2017; Balakhonov and Rose 2017) suggest that these structural commonalities are accompanied by functional ones. And there are intriguing homologies beyond the chordates, such as those between the arthropod central complex and the vertebrate basal ganglia (Strausfeld and Hirth 2013). Cephalopods are renowned for their adaptability and contextually appropriate problem-solving behavior (Mather 2008), yet their evolutionary history is quite separate from that of the vertebrates, but in many respects strikingly convergent (Packard 1972; Bullock 1984). Perhaps cognition in cephalopods is rooted in architectural features shared convergently with vertebrates? There are of course more general and fundamental questions about consciousness—for example, is there mentation and consciousness beyond (and before) brains (Rouleau and Levin 2025‡)?

### *16.8. The architecture of natural consciousness can inform the design of artificial problem-solving systems.*

Earlier (§12.3), I noted parallels between models of naturally evolved conscious cognition (particularly the "global workspace" model of Dehaene and Changeux (2011), the "dynamic core" model of Tononi and Edelman (1998), and the "integrated information" model of Tononi *et al.* (2016)), and successful machine learning architectures featuring recurrence, genericness, dynamic modularity, and stochasticity (hybrid metaheuristics (Blum *et al.* 2011) and iterated local search (Lourenço *et al.* 2003)). Related architectures developed for machine learning, inspired loosely by physiological features of the vertebrate brain, have been particularly successful. For example, artificial neural networks arranged for recurrence, convolutional transformation, adaptive competitive pooling, and hierarchical representation, have proved exceedingly effective in visual scene analysis (Pinheiro and Collobert 2014; Long *et al.* 2015; Spoerer *et al.* 2020) and the semantic analysis of verbal dialogues (Kalchbrenner and Blunsom 2013‡; Kalchbrenner *et al.* 2014‡). Palmer (2015) recommends a migration to computing machinery that integrates noise-prone, high-efficiency components, and this is particularly trenchant given evidence that noise is important in biological problem solving architectures, discussed earlier (§12.6) and at greater length below.

The limitations of existing mainstream AIs are striking. Large language models (LLMs) in particular struggle to keep their output consistent with rules and other constraints that have been articulated (Zhou *et al.* 2025), suggesting they are robustly unreliable to the degree that randomness is incorporated into their calculations (though this can be mitigated by incorporating more randomness in the training phase (Xuan *et al.* 2025‡). Indeed there are reasons to believe LLM architecture is inherently unreliable, due to an intractable tradeoff between capacity for learning and capacity for accuracy (Coveney and Succi 2025‡). And their reasoning is fundamentally an illusory simulation, that collapses when pushed beyond bounds established by training data, demonstrating a striking lack of generalization (Zhao *et al.* 2025‡). These deficiencies and awkward tradeoffs might be overcome by adopting architectural features described by physiologically grounded models like BGMS.

### *16.9. Exploitation of noise, including that with quantum origins, may contribute to consciousness.*

Earlier (§12.6), I discussed the notion that neural noise, cultivated generally in the central nervous system and particularly in the BG, may be crucial to neural problem solving and signal fidelity, with the BG positioned to shift brain dynamics between chaotic noise-perturbable and focused noise-immune modes (see also §3). Clearly, neural noise does have advantages in itself (McDonnell and Ward 2011). While some noise is physiologically unavoidable (Faisal *et al.* 2008; McDonnell and Ward 2011; Softky and Koch 1993; Stiefel *et al.* 2013), evolution might have arranged brains to be far less noisy than they are, and indeed some mechanisms in the nervous system are relatively noise-free (Kara *et al.* 2000; DeWeese *et al.* 2003).

Apparently gratuitous noisiness in the brain might be partly explained by evolutionary optimization for metabolic efficiency: the energetic cost for the generation, conduction, and response to action potentials is inversely proportional to the size of the implicated neurons and fibers, as is the propensity of those neurons and fibers to noisiness (Palmer and O'Shea 2015; Palmer 2020). By optimizing for energetic efficiency (unit of completed computation per unit of dissipated energy), evolution may have stumbled upon a physiological arrangement in which noise is pervasively and freely available as an input to computation, and is multifariously exploited (Palmer and O'Shea 2015; Palmer 2020).

In both the BG and the cerebellum, signals must transit fibers narrower than 200 nm (Difiglia *et al.* 1982; Sultan 2000), resulting in a high propensity for noise. The dimensions of these structures bring the remarkable implication that some of the noise in these paths is irreducibly quantum mechanical (Palmer and O'Shea 2015), so that the behavior of these neural systems is inherently non-computable in the Turing sense (Calude and Svozil 2008). In at least this narrow sense, biological brains are a kind of quantum computer. Indeed, in a broader sense, all iterative probabilistic phenomena are irreducibly dominated by quantum uncertainty (Albrecht and Phillips 2014; Bandak *et al.* 2024). The physical specifics of a system's causal structure are thought to have implications for that system's potential for consciousness (Findlay *et al.* 2024‡); these physical distinctions of real brains may thus be important. It is also intriguing that the semi-classical stochastic dynamics of neuron membrane potentials can be modeled using formalisms from quantum mechanics — formalisms such as the Schrödinger equation and the quantum of action — with Planck's constant $\hbar$ replaced by a neuronal quantum of action $\hat{h}$ (Ghose and Pinotsis 2025). Indeed these observations have implications beyond brains, given physiological features that are shared by neuronal and non-neuronal tissue (Levin 2014), and system architectures that subtly share important commonalities (Fields and Levin 2020). Below, I explore the proposition that brains exploit quantum mechanics (QM) more systematically, notwithstanding plausible articulations of skepticism (Koch and Hepp 2006; Baars and Edelman 2012; McKemmish *et al.* 2009; Litt *et al.* 2006).

Notoriously intractable computational problems are, in principle, made tractable by quantum computation (Kieu 2003, 2019). Many of these problems are clearly similar to those encountered by organisms in natural settings; thus there are clear evolutionary pressures selecting for arrangements that exploit quantum phenomena to optimize cognition. Indeed some of the exponential barriers to problem solving encountered with conventional computation (e.g. Coveney and Succi 2025‡) may reflect not just limitations of information architecture, but physical limitations that can only be overcome through fundamentally different physical arrangements.

Notably, simulations suggest that quantum phenomena are indispensable to the function of physiological ion channels—models of these mechanisms produce accurate predictions only if ions are represented by quantum mechanical wave functions (Summhammer *et al.* 2018), which by some measures may give the brain computational power orders of magnitude larger than that of modern supercomputers (Georgiev *et al.* 2020). The long-recognized capacity of retinal rod cells to detect single photons (Rieke and Baylor 1998) is canonically quantum mechanical, and the efficiency of photosynthesis in plants is now thought to depend in part on quantum coherent effects (Romero *et al.* 2014). More recently, evidence has been uncovered that the viscosity of aqueous poly-electrolyte solutions is strongly affected by atomic isotope substitutions (deuterium for hydrogen) (Dedic *et al.* 2019), suggesting that even nuclear quantum effects are plausibly subject to natural selection. Various experiments suggest QM underlying other biological phenomena (Jedlicka 2017). Perhaps the most significant evidence to date for quantum biology in nature relates to geomagnetic navigation by some birds, which has been shown to involve cultivation in the retina of prolonged coherent quantum entanglement (10s of microseconds) of pairs of electrons (Gauger *et al.* 2011). There is thus little remaining doubt that vertebrate nervous systems harbor some of the phenomena commonly referred to as "quantum strangeness", and in functionally causal roles. And the basal ganglia appear to be no exception. As noted above, narrow caliber fibers in striatal projections likely exhibit quantum effects. Below, I speculate that this and other aspects of basal ganglia physiology further implicate quantum phenomena.

Four relevant concepts in quantum theory are summarized below, with speculation on how they might help to more fully explain the mechanisms of mammalian cognition.

The first is **quantum superposition**, in which physical alternatives exist simultaneously as possibilities represented by a single complex-valued spatiotemporally extended quantum wave function, with the selection among those alternatives deferred unless and until a strong measurement disambiguates the history of the system, as in the classic dual-slit interference experiments (Silverman 2008).

The second is **quantum erasure**, in which the results of a strong measurement are irreversibly destroyed, thereby deferring disambiguation as though the strong measurement had not been performed (Walborn *et al.* 2002).

The third is **weak measurement**, in which only fragmentary and indecisive observations of a system are collected for some period, so that superposition in the system is only partly disrupted until a strong, disambiguating measurement result is subsequently unveiled (Aharonov *et al.* 1987, 2014; Dressel 2015; Tan *et al.* 2015).

The fourth is the quantum interpretation proposed by Zurek (1982) featuring environmentally induced selection, or ***einselection***, in which strong measurement is considered to be a selection among alternatives, on the basis of cumulatively overwhelming statistical support in favor of one of them at the expense of the others, as reflected by the proliferating impact of the winning alternative on the histories of nearby structures (Zurek 1982, 2003, 2022).

**Superposition** has a natural association with the conjunction in cortex of homeostatic criticality (Haider *et al.* 2006; Ma *et al.* 2019; Ahmadian and Miller 2021) and endemic neural noise (Faisal *et al.* 2008), some of which is intrinsically quantum (Palmer and O'Shea 2015). These conditions might arrange to put in superposition a family of local, microscopic perturbations of cortical state, positioning a universe of alternatives adjacently in configuration space, subject to neuromodulatory control (e.g. by dopamine, discussed earlier (§10.5)). In this way, cortex might exploit quantum parallelism to meaningfully increase the probability and rapidity with which better, or optimal, patterns can be detected, selected, and broadcasted by the BG. Notably, the simultaneous and momentary pattern of activation of an ensemble of frontal cortical maps can represent an extensive network of contingencies and associated selections, related indirectly but specifically to overt (observable) future actions and outcomes (Fine and Hayden 2022). Thus, modulation of frontal cortex by inputs from BG-recipient thalamus can select entire action plans extending arbitrarily far into the future. Because this modulation bears quantum noise, alternative action plans might thus be evaluated with quantum parallelism, with evident benefit to the organism's fitness.

These mechanisms for the parallel exploration of possibilities are related to the narrative advanced by Feynman and Hibbs (1965) and contextualized to the brain by Palmer (2020), in which ostensibly counterfactual (road-not-taken) trajectories in configuration space materially bear upon observable outcomes. Because superposition is purely a phenomenon of such wave functions (Silverman 2008), this would also imply that cortical state—or some important microscopic fragments thereof, perhaps shielded from decoherence—can only be adequately described by a quantum wave function. More precisely, the proposition here is that a constellation of shielded microscopic components, embedded sparsely within the neuropil, maintain coherence and the associated capacity for superposition and entanglement, for functionally relevant time spans (μs to ms), and ultimately impact the behavior of neurons as wholes. This notion, exemplified by but not limited to the proposal of Hameroff and Penrose (1996), reflects the irreducible non-locality of reality as described by QM. Large scale zero-lag synchronies in cortex, coordinated by the thalamus (noted earlier (§1.6)), might be crucial in coordinating these components on the timescale of their quantum mechanical coherence.

As discussed earlier (§12.6), the output nuclei of the BG inject noise into the thalamocortical system, with a plausible role in the cognitive search for solutions. Remarkably, it is these very noise-generating projection neurons that receive profuse, repetitive appositions along their dendrites from fine SPN axon collaterals with diameters of 100-200 nm (Difiglia *et al.* 1982). This subjects BG output projection cells to continual bombardment by noise, some of which is irreducibly quantum (Palmer and O'Shea 2015). This in turn can induce quantum jitter in their tonic discharge patterns, which is continually injected into the thalamocortical system, and indeed is subject to recirculation and synchrony detection.

Thalamostriatal loops through the intralaminar and other BG-recipient nuclei (Smith *et al.* 2004; McFarland and Haber 2000), previously implicated in corticostriatal signaling (Ding *et al.* 2010), execution of motor actions in uncertain environments (Mandelbaum *et al.* 2019), and expression of striatal plasticity (Bradfield *et al.* 2013), circulate this noise back to the striatum, where in addition to generically introducing random perturbations, it might facilitate stochastic resonance, with adaptive and quantum aspects, as demonstrated in other systems with similar arrangements (Mitaim and Kosko 1998; Goychuk and Hänggi 1999).

Synchronized striatal discharge—a decisive striatal selection—overwhelms these delicate effects, both by interrupting outflow from the GPi/SNr, and through generalized noise immunity associated with synchronized activity in nonlinear systems (Tabareau *et al.* 2010). But ultimately, it seems significant that the entire output of the striatum to the GP and SN, and indeed of the cerebellar granule cells to the Purkinje cells (Sultan 2000), is forced through a sieve-like array of narrow axon segments, intrinsically subject to quantum perturbation.

**Quantum erasure** and **weak measurement** have a natural association with the sparse receptive fields and tonically low firing rate (or complete quiescence) of corticostriatal neurons (Stern *et al.* 1997; Turner and DeLong 2000) and striatal SPNs (Sandstrom and Rebec 2003; Mahon *et al.* 2006). In principle, these arrangements might position the striatum to continually perform weak measurements of cortical activity. Until an input pattern triggers a striatal activation, these observations are largely discarded, which—again, in principle—would constitute quantum erasure, minimizing disruption of superpositions in cortex. The convergence and idiosyncrasy of corticostriatal targeting (Kincaid *et al.* 1998; Zheng and Wilson 2002), in concert with cortical criticality and noisiness placing many alternatives in superposition, might arrange for simultaneous and continuous exploration of the entire evolving neighborhood of alternatives, with the BG-thalamocortical loop only tightly closed for decisive selections.

**Einselection** has a natural association with selection by the BG. In BGMS, as detailed earlier (§6.2), selection entails the coherent and rapid (single cycle) routing of selected cortical activity to much of the brain. Indeed, synchronous oscillation protects neural circuits from random perturbations (Tabareau *et al.* 2010), so that basal ganglia mediated synchronization intrinsically and irreversibly displaces alternatives, while imparting maximum visibility and consequence to the selected activity. Perhaps the view of some early quantum theorists that conscious observation plays a special role in the "collapse" of the wave function (Henderson 2010) reflected an intuition that conscious attention to a phenomenon assures that it will have an especially proliferating and enduring impact, forcing disambiguation, according to Zurek (2003, 2022).

More generally, Zurek's "quantum Darwinism" is a proposal that physical reality at its most fundamental level is an evidence accumulation and decision engine. Perhaps this homomorphism with biological brains is real, and more than skin deep.

In any case, the proposition in broad outlines is that the wave function of the cortex speculatively explores competing decisions via ephemeral ambiguous superpositions—a form of quantum parallelism. One of the decisions is then stabilized at the expense of the others, according to the logic of einselection—particularly, through the formation of consensus by preponderant representational redundancy, again according to Zurek (2003, 2022).

BG circuits are arranged in loops, with the striatonigral and striatopallidal projections forced through fine caliber fibers subject to quantum perturbations. The output to the thalamus functions as a tonic noise generator, and the output of the BG-recipient thalamus, particularly the intralaminar thalamus, projects strongly back to the striatum. The cerebral cortex, moreover, is thought to be maintained near its dynamical critical point, with the BG-recipient thalamus modulating its dynamical regime (Müller *et al.* 2023). These arrangements suggest behavior like that of a Wilson cloud chamber or Geiger-Müller tube, in that a localized quantum perturbation can rapidly cascade to mesoscopic (and ultimately macroscopic) observable effects. Perhaps these arrangements can be viewed as realizing an iterated local search algorithm (Lourenço *et al.* 2003) with irreducibly quantum stochasticity, with as yet unexplored implications.

The proposition, in summary, is that BG selection involves einselection, that these irreducibly quantum aspects of neural noise are functionally significant, and that cortical configuration and history crucially involve a wave function in superposition. These are implications of the narrative advanced by Palmer (2020), who furthermore suggests that an inherently quantum awareness of the universe of alternatives is at the heart of the subjective sense of free will that canonically characterizes consciousness. These possibilities warrant focused, empirically grounded exploration, with an appreciation that the exploration may uncover evidence of surprising and even incongruous phenomena (Radin 1997; Radin *et al.* 2012; Silberstein and Bigelow 2024). Indeed, the savagely unruly mystical reality implied by these incongruous phenomena leads to the intuitive sense that they are genuinely dangerous, and likely explains their marginalization from modern science, despite empirical demonstrations that substantiate them (Radin 1997). Perhaps these phenomena suffuse reality with meaning, an ironic corollary to Steven Weinberg's notion that “The more the universe seems comprehensible, the more it also seems pointless.” (Weinberg 1977). In any case, notwithstanding attempts to salvage locality at any cost (Hossenfelder and Palmer 2020), the argument for fundamental non-locality is strong (Maudlin 2011; Rauch *et al.* 2018).

Some overarching notes of caution are warranted, beyond the recontextualization of QM to neurophysiology, itself a daunting challenge.

The “measurement problem” is an admission that current, experimentally vetted quantum theory — despite its unparalleled successes (e.g. Aoyama *et al.* 2019) — is not a complete description of reality, even in principle, and even in general terms. Spontaneous decay, as currently understood, is a Poissonian phenomenon occurring in continuous (infinitesimally subdivisible) time. In Everettian interpretations such as Zurek's, this temporal continuity bears the startling implication that the “branching” of physical reality is infinitely dense, i.e. that in every finite interval of time, an infinite number of alternative world-states is generated within a single configuration space. Maintaining the orthogonality of these parallel realities, as described by Everett's interpretation, would involve complete independence (and in Everett's account, *eternal* independence) of an infinitude of branches of a single wave function. This implies that the description of the quantum state of finite volumes of spacetime requires infinite information.

These sorts of descriptions have historically suggested theoretical shortcomings, here pointing toward a quantization of spacetime itself (and by implication, of gravity). Extrapolations from current theory are therefore likely to be misleading.

The existence of irreducibly quantum indeterminacy—of observables that are not derived from local information—gives further motivation for the intriguing view that reality in general, and causality in particular, are non-local with respect to metric spacetime (Maudlin 2011; Rauch *et al.* 2018), absent “superdeterminism” ((Hossenfelder and Palmer 2020)). Non-locality implies that there are no real isolated systems within the universe, even in principle, and the non-locality of quantum indeterminacy — i.e., that ostensibly bounded systems cannot be isolated from future influences that are not part of a complete description of the present bounded system — may be just another facet of a deeper non-locality that is also observed in the phenomena of wave function disambiguation (“collapse”, in Copenhagen parlance) and entanglement. Zurek's “einselection” implies that observation (measurement) consists of the entanglement of the observer with the observed, which ultimately implies that the observing mind is physically entangled with the subjects of its observations, with further implications for the ontology of neural plasticity. More generally, Big Bang cosmology and CMB isotropy imply that all matter and energy are entangled, even beyond the observational horizon, due to initial causal intimacy. The implications of these notions have been only occasionally and tentatively explored.

# 17. References

*Note that citations of preprints bear the annotation "‡" on the year of release, both above in the text and below in the bibliography.*